\documentclass[11pt]{article}
\usepackage{slashed}
\usepackage{amssymb}
\usepackage{axodraw}
\usepackage{amsmath}
\def\ifmath#1{\relax\ifmmode #1\else $#1$\fi}

\newcommand{\mchi}{m_{\chi}}

\def\Imag{\Im m\,}
\def\mg{m_{\tilde{G}}}
\def\msl{m_{\tilde{l}}}
\def\pstau{p_{\tilde{l}}}

\def\pZ{p_V}
\def\pg{p_{\tilde{G}}}
\def\pgpZ{\pg\cdot\pZ}
\def\pgpsl{\pg\cdot\pstau}
\def\pslpZ{\pstau\cdot\pZ}

\def\calO{{\cal{O}}}

\newcommand{\jxf}{J({\xf})}

\newcommand{\xf}{x_f}

\def\gm{\gamma}

\def\mz{m_Z}

\def\mh{m_h}

\def\gm{\gamma}
\def\Re{{\rm Re}}

\def\prop#1{s-m_{#1}^2+i \Gamma_{#1}m_{#1}}
\def\slp#1{\widetilde{\ell}_{#1}}

\def\neu#1{\chi_{#1}^0}

\def\mslp#1{m_{\widetilde{\ell}_{#1}}}
\def\msli{m_{\widetilde{\ell}_{a}}}
\def\mslj{m_{\widetilde{\ell}_{b}}}
\def\mslk{m_{\widetilde{\ell}_{c}}}

\def\ml{m_{\ell}}

\def\mnmn{m_{\chi^0_i}m_{\chi^0_j}}

\def\Dijqp#1{D^{abij}_{#1}}

\def\neu#1{\chi_{#1}^0}
\def\neu#1{\chi_{#1}^0}

\def\gm{\gamma}
\def\Re{{\rm Re}}

\def\prop#1{s-m_{#1}^2+i \Gamma_{#1}m_{#1}}
\def\slp#1{\widetilde{\ell}_{#1}}

\def\slsl#1{\mslp{a}^{#1}\mslp{b}^{#1}}
\def\slpsl#1{\mslp{a}^{#1}+\mslp{b}^{#1}}
\def\slmsl#1{\mslp{a}^{#1}-\mslp{b}^{#1}}
\def\T{\mathcal{T}}
\def\F{\mathcal{F}}

\def\Fu{{\mathcal F}^u}
\def\Ft{{\mathcal F}^t}
\def\Tu{{\mathcal T}^u}
\def\Tt{{\mathcal T}^t}
\def\Y{\mathcal{Y}}
\def\mnq{m_{\chi_i^0}}
\def\mnp{m_{\chi_j^0}}

\def\N1{\widetilde N_1}

\def\stilde{\widetilde}

\newcommand{\newc}{\newcommand}
\newc{\sigmabar}{\overline\sigma}
\newc{\MSbar}{\overline{{\rm MS}}}
\newc{\DRbar}{\overline{{\rm DR}}}

\def\lam{\lambda}

\def\eps{\epsilon}
\def\stilde{\widetilde}
\def\ra{\rightarrow}
\def\newcdot{\kern.06em{\cdot}\kern.06em}

\catcode`@11
\def\seceqaa{\@addtoreset{equation}{section}
\def\theequation{A\arabic{equation}}}
\def\seceqbb{\@addtoreset{equation}{section}
\def\theequation{B\arabic{equation}}}
\def\seceqcc{\@addtoreset{equation}{section}
\def\theequation{C\arabic{equation}}}
\def\seceqdd{\@addtoreset{equation}{section}
\def\theequation{D\arabic{equation}}}
\catcode`@11
\def\beq{\begin{equation}}
\def\eeq{\end{equation}}
\def\bea{\begin{eqnarray}}
\def\eea{\end{eqnarray}}

\def\eps{\epsilon}

\def\mgss{M_{\tilde{g}}}
\def\mchi{M_{\tilde{\chi}}}

\def \mij {m_{ij}}
\def \mjk {m_{jk}}
\def \mik {m_{ik}}
\def \mm {m_{3/2}}
\def \mi {m_{\nu_i}}
\def \mj {m_{e_j}}
\def \mk {m_{e_k}}

\def \mipp {m_{u_i}}
\def \mjpp {m_{d_j}}
\def \mkpp {m_{d_k}}
\def \aa {m_{\tilde \nu_{iL}}}
\def \bb {m_{\tilde e_{jL}}}
\def \cc {m_{\tilde e_{kR}}}
\def \aaps {m_{\tilde \nu_{iL}}}
\def \bbps {m_{\tilde d_{jL}}}
\def \ccps {m_{\tilde d_{kR}}}
\def \aapt {m_{\tilde e_{iL}}}
\def \bbpt {m_{\tilde u_{jL}}}
\def \ccpt {m_{\tilde d_{kR}}}
\def \aapp {m_{\tilde u_{iR}}}
\def \bbpp {m_{\tilde d_{jR}}}
\begin{document}
\begin{titlepage}
\begin{center}
{\Large\bf (N)LSP Decays and  Gravitino Dark Matter Relic Abundance in  Big Divisor (nearly) SLagy $D3/D7$ $\mu$-Split SUSY }
\vskip 0.1in Mansi Dhuria \footnote{email: mansidph@iitr.ernet.in}
and
 Aalok Misra\footnote{e-mail: aalokfph@iitr.ernet.in
}\\
Department of Physics, Indian Institute of Technology,
Roorkee - 247 667, Uttaranchal, India\\
 \vskip 0.5 true in
\date{\today}
\end{center}
\thispagestyle{empty}
\begin{abstract}

Using the (nearly) Ricci-flat Swiss-Cheese metric of
\cite{review-iii}, in the context of a mobile space-time filling
$D3$-brane restricted to a nearly special Lagrangian sub-manifold (in
the large volume limit, the pull-back of the K\"{a}hler form close to
zero and the real part of the pull-back of $e^{-i\theta},\theta=\frac{\pi}{2}$ times the nowhere-vanishing holomorphic three-form
providing the volume form on the three-cycle) of the ``big" divisor
with (fluxed stacks of) space-time filling $D7$-branes also wrapping
the ``big" divisor (corresponding to a local minimum), we provide an
explicit identification of the electron and the $u$-quark, as well as
their
$SU(2)_L$-singlet cousins, with fermionic super-partners of four Wilson line moduli; their superpartners turn out to be very heavy, the Higgsino-mass parameter turns out to be large, one obtains one light (with a mass of $125\ GeV$) and one heavy Higgs and the gluino is long lived (from a collider point of view) providing a possible realization of  ``$\mu$-Split Supersymmetry".
By explicitly calculating the lifetimes of decays of the co-NLSPs -
the first generation squark/slepton and a neutralino - to the LSP -
the gravitino - as well as gravitino decays, we verify that BBN
constraints relevant to the former as well as the requirement of the
latter to be (more than) the age of the universe, are satisfied. For
the purpose of calculation of the gravitino relic density in terms of
the neutralino/slepton relic density, we evaluate the latter  by evaluating the netralino/slepton
(co-)annihilation cross sections and hence show that the former
satisfies the requirement for a Dark Matter candidate.
\end{abstract}
\end{titlepage}

\section{Introduction}

It is challenging to provide a suitable choice of vacuum (local minimum) of string theory to infer various cosmological and phenomenological issues, and other fundamental physics. One of the most exciting unresolved issues in particle physics and cosmology is the nature of dark matter (DM) in the Universe. Over the years, astronomical and cosmological observations have put significant constraints on its expected properties. The recent Wilkinson Microwave Anisotropy Probe ( WMAP) observations   provide the relic abundance of a cold dark matter (CDM) \cite{C.L.Bennett} to be $\Omega_{CDM}h^2 = 0.1109\pm0.0056$. Theoretical status of DM generally emerges in the context of theories beyond Standard Model. It is well known that supersymmetric models with conserved R-parity contain one stable  neutralino which is a candidate for cold dark matter. However, in models coupled to gravity and even in other scenarios of supersymmetry breaking mediation including gauge mediation, the gravitino ( the supersymmetric partner of Graviton)  stands out as probably the most natural and attractive candidate for the LSP and DM although it generically suffers from a well-known ``cosmological gravitino problem" \cite{GMoreau} the resolution of which depends on whether one considers gravitino to be LSP or unstable particle.  The long-lived gravitino's are generated in the early Universe. In the standard big-bang cosmology, they were in thermal equilibrium and then because of their weak gravitational interactions, frozen out while they were relativistic. In this case their abundance might overclose the Universe. Even if during inflationary epoch of the universe, primordial abundance of gravitino is completely diluted, the problem can not be solved  because gravitinos are regenerated in the thermal bath after the reheating if the reheating temperature is high enough though overclosure of universe by dark matter constrain the exact value of reheating temperature \cite{MBolz, V.S.Rychkov, R.Rangarajan}. However, in addition to it, the production of gravitino also depends on non- thermal production mechanism, the abundance of which is independent of the reheating temperature.  i.e $\Omega^{total}_{3/2}= \Omega^{th}_{3/2}+ \Omega^{NLSP}_{3/2}$.  Generically in case of heavy gravitino, re-heating temperature is low enough to produce appropriate abundance of Gravitino.  Therefore sufficient number of gravitinos can be produced after NLSP decays to gravitino after its decoupling from the thermal plasma \cite{Fweng}. In other words, annihilation density does not depend on freeze out of gravitino (which causes a gravitino problem), instead depends on freeze out of NLSP.  Hence relic density of gravitino is given by relic density of NLSP \cite{Fweng,Jonathan.L.Feng} according to the relation $\Omega_{\tilde G}= \frac{m_{3/2}}{m_{\tilde G}}\Omega_{3/2}$.  Therefore it is interesting to study the thermal cross-section and hence relic density of the (co-)NLSP(s) to  get the the right order of relic density of gravitino.

In spirit of above discussion, the purpose of present work is to give the signatures of gravitino as a potential dark matter in the context of the large volume limit of type IIB (``big divisor") $D3/D7$ Swiss cheese phenomenology. The consistent compactification scheme and phenomenological features of the model were initiated in \cite{D3_D7_Misra_Shukla, ferm_masses_MS} and ``${\mu}$ split-SUSY scenario"-like realization of the same was shown to be possible to be realized in \cite{Dhuria+Misra_mu_Split SUSY} where in addition to getting one light Higgs (one of the primary motivation of $\mu$  split SUSY Scenario), we evaluated the life time of gluino which came out to be long and hence  satisfied one of the important  phenomenological features of ${\mu}$ split-SUSY scenario. In general, stability of LSP is governed by conserved R-parity in most of the supersymmetric models. However, in this paper,  we have considered stability of the gravitino as an LSP  in absence of R-parity conservation.  We argue that, even in the presence of non-zero R-parity violating couplings, large squark masses help to reduce the decay width and hence the lifetime of the gravitino decays become very long, typically  of or larger than the age of the Universe. The explicit calculation of  matrix amplitudes and hence life times of N(LSP) candidates  requires the complete identification of the SM particles. Therefore we build up a set up which is able to provide an explicit identification of the first generation $SU(2)_L$ leptonic and quark doublets as well as their $SU(2)_L$-singlets, with fermionic super-partners of four Wilson line moduli. Further,  soft SUSY parameters obtained in the context of gravity mediation provides sleptons and gaugino/lightest neutralino (being almost degenerate in mass)  as co-NLSPs while gravitino naturally appears as Lightest Supersymmetric particle (LSP) which traditionally can be considered as viable dark matter candidate. The more explicit  realization of the same requires life time of the LSP to be around and preferably more than the age of the universe.

The organization of the paper is as follows: In section {\bf 2}, we start off with details of  an improved version of the large volume scenario set up discussed in \cite{D3_D7_Misra_Shukla}. Here, we construct four harmonic distribution one-forms supported on a sub-locus of big divisor localized along the mobile space-time filling $D3$-brane, and by utilizing the geometric K\"{a}hler potential of \cite{review-iii} we show that the aforementioned sub-locus in the large volume limit is nearly a special Lagrangian sub-manifold of the ``big" divisor. By calculating the intersection matrix valued in the Wilson line moduli sub-space $a_{I=1,...,h^{(0,1)}_-(\Sigma_B)}$, appearing in K\"{a}hler coordinate $T_B$, we write (the K\"{a}hler sector of the) K\"{a}hler potential that includes  four Wilson line moduli $a_1, a_2, a_3, a_4$ and two position moduli of a mobile space-time filling $D3$-brane restricted to the abovementioned (nearly) SLag (corresponding to a local minimum); for the purposes of evaluation of ``physical"/normalized Yukawa couplings, soft supersymemtry breaking parameters, etc.,  this K\"{a}hler potential is diagonalized to produce the ${\cal A}_I, {\cal Z}_i$-basis. Further, the estimate of the Dirac mass terms appearing in the ${\cal N}=1$ supergravity action of \cite{Wess_Bagger} calculated from superpotential and K\"{a}hler potential suggest that the fermionic superpartners of ${\cal A}_1$ and ${\cal A}_3$ correspond respectively to  the first generation leptons: $e_L$ and $e_R$, and the fermionic superpartners of ${\cal A}_2$ and ${\cal A}_4$ correspond respectively to the first generation quarks: $u_L$ and $u_R$. We also provide explicit bi-fundamental representations for the four Wilson line moduli and two $D3$-brane position moduli (after having turned on appropriate two-form fluxes on the $D7$-brane world volume decomposing adjoint-valued matter fields into bi-fundamental matter fields) and speculate about the possible modification in the relevant ${\cal N}=1$ chiral coordinate in the presence of the same. We also show that the (effective) physical Yukawas change only by ${\cal O}(1)$ under an RG-flow from the string to the EW scale. Building up on this identification , we  have evaluated the contribution of  various three-point vertices in the context of gauged $N=1$ supergravity action, in section {\bf {3,4}} and {\bf 5}. In section {\bf 3}, in order to meet the  requirement of an appropriate DM candidate, we  calculate life time of three-body R-parity violating decays of gravitino LSP which comes out to be more than  the age of the Universe.  In {\bf 4.1} and {\bf 4.3}, we study decay width and life time of two-body and hadronic three-body decays of gauginos (NLSP) and co-NLSPs (sleptons/squarks), life times of which, except for gluino-to-Goldstino decay,  ensure that energy release from both gauge boson and hadronic decays do not spoil the the predictions arising from Big Bang Nucleosynthesis (BBN)  and cosmic microwave background, etc.; gluino is hence ruled out as an NLSP. In section {\bf 4.2}, we consider R-parity violating decays of (Wino/Bino-dominated)neutralino to ordinary particles (even R-parity) which do not affect the abundance of gravitino produced by neutralino decays into gravitino if the decay life time of the former is more than the life time of the latter. The more exact determination also requires that relic density of dark matter must be compatible with present observations and should not overclose the Universe. Therefore, in order to be able to perform a reliable comparison between theoretical predictions and improving measurements of the relic abundance from underground DM searches, in section {\bf 5}, we calculate the relic density of co-NLSP's and hence relic abundance of gravitino. Strictly speaking, we consider very conservative approach where one assumes that decays of NLSP's account for almost all of the dark matter of the Universe . Utilizing extensively the results of \cite{Takeshi_Leszek}, \cite{takeshi_coannihilation} for providing the exact analytical expressions of various annihilation channels of neutralino and slepton,  we obtain estimates of thermal cross-section and hence relic density of N(LSP) candidates. Section {\bf 6} has the conclusions. There are five appendices; appendix A has some details of the geometric K\"{a}hler potential calculation. Appendix B describes the evaluation of various intersection numbers $C_{I \bar J}$ corresponding to four Wilson line moduli. Appendix C has details of calculations of various soft supersymmetry breaking parameters and the fact that it is possible to obtain around $10^2\ GeV$ masses for $W/Z$ vector gauge bosons and $125\ GeV$ for the light Higgs. Appendix D includes expansions of moduli space metric and its derivatives w.r.t each moduli up to terms linear in the fluctuations about their extremum values of the Wilson line and $D3$-position moduli. Lastly, in appendix E, we have given the intermediate steps of calculation of matrix amplitudes corresponding to $\tilde{l}\rightarrow l^\prime \tilde{G}V$ decay.
\section{Setup}

From Sen's orientifold-limit-of-F-theory point of view  corresponding to type IIB compactified on a Calabi-Yau three fold $Z$-orientifold with $O3/O7$ planes, one requires an elliptically fibered Calabi-Yau four-fold $X_4$ (with projection $\pi$) over a 3-fold $B_3(\equiv CY_3-$orientifold)  where $B_3$ is taken be  an $n$-twisted ${\bf CP}^1$-fibration over ${\bf CP}^2$ such that pull-back of the divisors in $CY_3$ automatically satisfy Witten's unit-arithmetic genus condition.  For $n=6$ \cite{DDF}, the Calabi-Yau three-fold $Z$ then turns out to be a unique Swiss-Cheese Calabi Yau  in ${\bf WCP}^4_{[1,1,1,6,9]}[x_1:x_2:x_3:x_4:x_5]$ given by a smooth degree-18 hypersurface in ${\bf WCP}^4[1,1,1,6,9]$; the exceptional divisor corresponding to resolution of a ${\bf Z}_3$-singularity $x_1=x_2=x_3=0$, via the monomial-divisor map, is encoded as the $\phi x_1^6x_2^6x_3^6 (\phi$ being an involutively odd complex structure modulus) polynomial deformation in the defining hypersurface. Also, throughout we will be working in the non-singular coordinate patch $x_2=1$ in the large volume limit of the Swiss-Cheese Calabi-Yau.

In the presence of a space-time filling $D3$-brane and a space-time filling $D7$-brane wrapping $\Sigma_B$, the K\"{a}hler potential relevant to all the calculations in this paper (without being careful about ${\cal O}(1)$ constant factors):
\begin{eqnarray}
\label{eq:K_I}
 & & K\sim-2 ln\left[\left(\sigma_B+{\bar\sigma}_B - \gamma K_{\rm geom}\right)^{\frac{3}{2}} - \left(\sigma_S+{\bar\sigma}_S - \gamma K_{\rm geom}\right)^{\frac{3}{2}} \right] \nonumber\\
 & & + \frac{\chi}{2}\sum_{m,n\in{\bf Z}^2/(0,0)}
\frac{({\bar\tau}-\tau)^{\frac{3}{2}}}{(2i)^{\frac{3}{2}}|m+n\tau|^3} - 4\sum_{\beta\in H_2^-(CY_3,{\bf Z})} n^0_\beta\sum_{m,n\in{\bf Z}^2/(0,0)}
\frac{({\bar\tau}-\tau)^{\frac{3}{2}}}{(2i)^{\frac{3}{2}}|m+n\tau|^3}cos\left(mk.{\cal B} + nk.c\right)\Biggr],
\nonumber\\
& &
\end{eqnarray}
will be rewritten in terms of the ${\cal N}=1$ chiral coordinates: $T_{\alpha=B,S},{\cal G}^{a=1,2},\zeta^A,z^{i=1,2}$
(See \cite{Jockers_thesis,D3_D7_Misra_Shukla} for definitions of all) - in particular:
\begin{eqnarray}
\label{eq:sigmas_Ts}
& & \sigma_\alpha\sim T_\alpha -\left( i\kappa_{\alpha bc}c^b{\cal B}^c + \kappa_\alpha + \frac{i}{(\tau - {\bar\tau})}\kappa_{\alpha bc}{\cal G}^b({\cal G}^c
- {\bar {\cal G}}^c) \right.\nonumber\\
 & & \left.+ i\delta^B_\alpha\kappa_4^2\mu_7l^2C_\alpha^{I{\bar J}}a_I{\bar a_{\bar J}} + \frac{3i}{4}\delta^B_\alpha\tau Q_{\tilde{f}} + i\mu_3l^2(\omega_\alpha)_{i{\bar j}} z^i\bigl({\bar z}^{\bar j}-\frac{i}{2}{\bar z}^{\tilde{a}}({\bar{\cal P}}_{\tilde{a}})^{\bar j}_lz^l\bigr)\right),
\end{eqnarray}
where
\begin{itemize}
\item
$\kappa_4$ is related to four-dimensional Newton's constant, $\mu_3$ and $\mu_7$ are $D3$ and $D7$-brane tensions;

 \item
 $\kappa_{\alpha ab}$'s are triple intersection integers of the CY orientifold;

 \item
 $c^a$ and $b^a$ are coefficients of RR and NS-NS two forms expanded in odd basis of $H^{(1,1)}_{{\bar\partial},-}(CY)$;

 \item
 $C^{I{\bar J}}_\alpha=\int_{\Sigma^B}i^*\omega_\alpha\wedge A^I\wedge A^{\bar J}$, $\omega_\alpha\in H^{(1,1)}_{{\bar\partial},+}(CY_3)$ and $A^I$ forming a basis for $H^{(0,1)}_{{\bar\partial},-}(\Sigma^B)$ - immersion map is defined as: $i:\Sigma^B\hookrightarrow CY_3$, $a_I$ is defined via a Kaluza-Klein reduction of the $U(1)$ gauge field (one-form) $A(x,y)=A_\mu(x)dx^\mu P_-(y)+a_I(x)A^I(y)+{\bar a}_{\bar J}(x){\bar A}^{\bar J}(y)$, where $P_-(y)=1$ if $y\in\Sigma^B$ and -1 if $y\in\sigma(\Sigma^B)$;

 \item
 $z^{\tilde{a}}, \tilde{a}=1,...,h^{2,1}_-(CY_3),$ are $D=4$ complex structure deformations of the CY orientifold, $\left({\cal P}_{\tilde{a}}\right)^i_{\bar j}\equiv\frac{1}{||\Omega||^2}{\bar\Omega}^{ikl}\left(\chi_{\tilde{a}}\right)_{kl{\bar j}}$, i.e.,
${\cal P}:TCY_3^{(1,0)}\longrightarrow TCY_3^{(0,1)}$ via the transformation:
$z^i\stackrel{\rm c.s.\ deform}{\longrightarrow}z^i+\frac{i}{2}z^{\tilde{a}}\left({\cal P}_{\tilde{a}}\right)^i_{\bar j}{\bar z}^{\bar j}$, $z^i$ are scalar fields corresponding to geometric fluctuations of $D3$-brane inside the Calabi-Yau and defined via: $z(x)=z^i(x)\partial_i + {\bar z}^{\bar i}({\bar x}){\bar\partial}_{\bar i}$, and 

\item
 $Q_{\tilde{f}}\equiv l^2\int_{\Sigma^B}\tilde{f}\wedge\tilde{f}$, where $\tilde{f}\in\tilde{H}^2_-(\Sigma^B)\equiv{\rm coker}\left(H^2_-(CY_3)\stackrel{i^*}{\rightarrow}H^2_-(\Sigma^B)\right)$.

 \end{itemize}

 The closed string moduli-dependent K\"{a}hler potential, includes perturbative (using \cite{BBHL}) and non-perturbative (using \cite{Grimm}) $\alpha^\prime$-corrections. Written out in (discrete subgroup of) $SL(2,{\bf Z})$(expected to survive orientifolding)-covariant form, the perturbative corrections are proportional to $\chi(CY_3)$ and non-perturbative $\alpha^\prime$ corrections are weighted by $\{n^0_\beta\}$, the genus-zero Gopakumar-Vafa invariants that count the number of genus-zero rational curves $\beta\in H_2^-(CY_3,{\bf Z})$. In fact, the closed string moduli-dependent contributions are dominated by the genus-zero Gopakumar-Vafa invariants which using Castelnuovo's theory of moduli spaces can be shown to be extremely large for compact projective varieties \cite{Klemm_GV} such as the one used.

Based on the study initiated in \cite{ferm_masses_MS,Dhuria+Misra_mu_Split SUSY}, one us (AM), inspired by the Donaldson's algorithm - see \cite{Donaldson_i} -  had obtained in \cite{review-iii}, an estimate of a nearly Ricci-flat Swiss Cheese metric for points finitely separated from $\Sigma_B$ based on the
following ansatz:
\begin{eqnarray}
\label{eq:K}
& & \hskip-0.8in K=ln \Biggl[h^{z_4^2{\bar z}_4^2} {z_4}^2 {{\bar z}_4}^2+h^{z_4^2{\bar z}_4^2} \sqrt[3]{V} {z_4} {{\bar z}_4}+h^{z_4^2{\bar z}_4^2} V^{23/36}
   ({z_1}+{{\bar z}_1}+{z_2}+{{\bar z}_2})+h^{z_4^2{\bar z}_4^2} V^{11/18} ({z_1} {{\bar z}_1}+{z_2} {{\bar z}_1}+{z_1}
   {{\bar z}_2}+{z_2} {{\bar z}_2})\nonumber\\
   & &\hskip-0.8in +h^{z_4^2{\bar z}_4^2} V^{11/18} \left({z_1}^2+{z_2}
   {z_1}+{{\bar z}_1}^2+{z_2}^2+{{\bar z}_2}^2+{{\bar z}_1} {{\bar z}_2}\right) +h^{z_4^2{\bar z}_4^2} V^{7/12} \left({{\bar z}_2}
   {z_1}^2+{{\bar z}_1}^2 {z_1}+{{\bar z}_2}^2 {z_1}+{{\bar z}_1} {z_2}^2+{z_2} {{\bar z}_2}^2+{{\bar z}_1}^2
   {z_2}\right)\nonumber\\
   & &\hskip-0.8in +h^{z_4^2{\bar z}_4^2} V^{5/9} \left({z_1}^2 {{\bar z}_1}^2+ {z_2}^2 {{\bar z}_1}^2+ {z_1}^2
   {{\bar z}_2}^2+{z_2}^2 {{\bar z}_2}^2\right)+h^{z_4^2{\bar z}_4^2} \sqrt{V} ({z_4}+{{\bar z}_4})+ h^{z_4^2{\bar z}_4^2} V^{17/36} ({{\bar z}_1}
   {z_4}+{{\bar z}_2} {z_4}+{z_1} {{\bar z}_4}+{z_2} {{\bar z}_4})\nonumber\\
   & & \hskip-0.8in +h^{z_4^2{\bar z}_4^2} V^{4/9} \left({{\bar z}_4}
   {z_1}^2+{{\bar z}_1}^2 {z_4}+{{\bar z}_2}^2 {z_4}+{z_2}^2 {{\bar z}_4}\right)+h^{z_4^2{\bar z}_4^2} V^{5/12} \left({{\bar z}_1}
   {{\bar z}_4} {z_1}^2+{{\bar z}_2} {{\bar z}_4} {z_1}^2+{{\bar z}_1}^2 {z_4} {z_1}+{z_2} {{\bar z}_2}^2
   {z_4}+{{\bar z}_1} {z_2}^2 {{\bar z}_4}+{z_2}^2 {{\bar z}_2} {{\bar z}_4}\right)\nonumber\\
   & &\hskip-0.8in +h^{z_4^2{\bar z}_4^2} V^{5/18} ({z_1}
   {{\bar z}_1} {z_4} {{\bar z}_4}+ {{\bar z}_1} {z_2} {z_4} {{\bar z}_4}+ {z_1} {{\bar z}_2} {z_4}
   {{\bar z}_4}+{z_2} {{\bar z}_2} {z_4} {{\bar z}_4})+h^{z_4^2{\bar z}_4^2} V^{11/36} (({z_1}+{{\bar z}_1}) {z_4}
   {{\bar z}_4}+({z_2}+{{\bar z}_2}) {z_4} {{\bar z}_4})\nonumber\\
   & & \hskip-0.8in +h^{z_4^2{\bar z}_4^2} \sqrt[3]{V} \left({z_4}^2+{{\bar z}_4}^2\right)+h^{z_4^2{\bar z}_4^2}
   V^{11/36} \left({z_4}^2 {\bar z}_1+{z_1} {{\bar z}_4}^2+{z_2} {{\bar z}_4}^2+{\bar z}_2 {z_4}^2\right)+h^{z_4^2{\bar z}_4^2}
   V^{5/18} \left({{\bar z}_1}^2 {z_4}^2+{{\bar z}_2}^2 {z_4}^2+{z_1}^2 {{\bar z}_4}^2+{z_2}^2
   {{\bar z}_4}^2\right)\nonumber\\
   & & \hskip-0.8in +h^{z_4^2{\bar z}_4^2} \sqrt[6]{V} \left({{\bar z}_4} {z_4}^2+{{\bar z}_4}^2 {z_4}\right)+h^{z_4^2{\bar z}_4^2} V^{5/36}
   \left(({{\bar z}_1}+{{\bar z}_2}) {{\bar z}_4} {z_4}^2+({z_1}+{z_2}) {{\bar z}_4}^2 {z_4}\right)+h^{z_4^2{\bar z}_4^2} V^{4/9}
   ({z_1} {{\bar z}_2} {z_4}+{{\bar z}_1} {z_2} {{\bar z}_4}\nonumber\\
   & & \hskip-0.5in+|{z_1}|^2
   ({z_4}+{{\bar z}_4})+{z_1} {{\bar z}_2} ({z_4}+{{\bar z}_4})+|z_2|^2
   ({z_4}+{{\bar z}_4}))+\sqrt[3]{V}\Biggr]
\end{eqnarray}
From (\ref{eq:metsecinv}), one sees that $h^{11}\sim h^{z_4^2{\bar z}_4^2}{\cal V}^{\frac{2}{3}}$.
Using (\ref{eq:K}), one can show that:
\begin{equation}
\label{eq:R11bar_i}
R_{z_i{\bar z}_j}\sim\frac{\sum_{n=0}^8a_n\left(h^{z_4^2{\bar z}_4^2}\right)^n {\cal V}^{\frac{n}{3}}}{\left(1+{\cal O}(1)h^{z_4^2{\bar z}_4^2} {\cal V}^{\frac{1}{3}}\right)^2{\cal V}^{\frac{1}{18}}\left(\sum_{n=0}^3b_n\left(h^{z_4^2{\bar z}_4^2}\right)^n {\cal V}^{\frac{n}{3}}\right)^2}.
\end{equation}
Solving numerically: $\sum_{n=0}^8a_n\left(h^{z_4^2{\bar z}_4^2}\right)^n {\cal V}^{\frac{n}{3}}=0$, as was assumed, one (of the eight values of) $h^{z_4^2{\bar z}_4^2}$, up to a trivial K\"{a}hler transformation, turns out to be ${\cal V}^{-\frac{1}{3}}, {\cal V}\sim 10^6$. Curiously, from GLSM-based analysis, we had seen in \cite{D3_D7_Misra_Shukla} that on $\Sigma_B(z_4=0)$, the argument of the logarithm received the most dominant contribution from the FI-parameter $r_2\sim{\cal V}^{\frac{1}{3}}$ which using this value of $h^{z_4^2{\bar z}_4^2}$, is in fact, precisely what one obtains even now. Using this value of $h^{z_4^2{\bar z}_4^2}$, one obtains: $R_{z_i{\bar z}_4},R_{z_4{\bar z}_4}\sim10^{-1}$. Further, as has been assumed that the metric components $g_{z_{1,2}{\bar z}_4}$ are negligible as compared to $g_{z_i{\bar z}_j}$ - this was used in \cite{ferm_masses_MS} in showing the completeness of the basis spanning $H^{1,1}_-$ for a large volume holomorphic isometric involution restricted to (\ref{eq:near_slag_i}),
 $z_1\rightarrow-z_1,z_{2,3}\rightarrow z_{2,3}$, which is now born out explicitly, wherein the latter turn out to about 10$\%$ of the former. One does not need the aforementioned restriction on the geometric Swiss-Cheese metric for large volume involutions restricted to (\ref{eq:near_slag_i}), of the following type. In the $u_2\neq0$-coordinate patch, the defining degree-18 hypersurface in ${\bf WCP}^4_{1,1,1,6,9}[u_1:u_2:u_3:u_4:u_5]$ can be written as: $1 + z_1^{18} + z_2^{18} + z_3^3 - \psi z_1z_2z_3z_4 - \phi z_1^6z_2^6 = 0, z_1=\frac{u_1}{u_2}, z_2=\frac{u_3}{u_2}, z_3=\frac{u_4}{u_1^6}, z_4=\frac{u_5}{u_1^9}$. The same can be rewritten as: $\left(i z_4\right)^2 - \psi z_1z_2z_3z_4 =  z_3^3 + \phi z_1^6z_2^6 - z_1^{18} - z_2^{18}$. This is can therefore be thought of as the following Weierstrass variety:
 $\begin{array}{rrc}T^2(z_3,z_4(z_3)) & \rightarrow&{\bf WCP}^4_{1,1,1,6,9}(z_1,z_2,z_3)\\
 & & \downarrow\pi \\
 & & {\bf CP}^2(z_1,z_2)\\
  \end{array}$. Now, assuming that the complex structure modulus $\psi$ has been stabilized to an infinitesimal value so that one can disregard the polynomial deformation proportional to $\psi$, and defining $\chi_{1,2}\equiv z_{1,2}^6$, this elliptic fibration structure can thought also of as:
  $\begin{array}{rrc}T^2(z_3,z_4(z_3)) & \rightarrow&{\bf WCP}^4_{1,1,1,6,9}(\chi_1,\chi_2,z_3)\\
 & & \downarrow\pi \\
 \left({\rm Complex\ curve}\right){\cal C}(\chi_1)&\longrightarrow & {\bf CP}^2(\chi_1,\chi_2)\\
 & & \downarrow\pi^\prime\\
 & & {\bf CP}^1(\chi_2)
  \end{array}$, corresponding to $\left(i z_4\right)^2\approx z_3^3 + \phi \chi_1\chi_2 - \chi_1^3 - \chi_2^3$. Alternatively, one can also think of the Weierstrass variety as:
  $\begin{array}{rrc}T^2(-\chi_1,z_4(-\chi_1)) & \rightarrow&{\bf WCP}^4_{1,1,1,6,9}(\chi_1,\chi_2,z_3)\\
 & & \downarrow\pi \\
 \left({\rm Complex\ curve}\right){\cal C}(-z_3)&\longrightarrow & {\bf CP}^2(-z_3,\chi_2)\\
 & & \downarrow\pi^\prime\\
 & & {\bf CP}^1(\chi_2)
  \end{array}$, corresponding to $\left(i z_4\right)^2\approx\left(-\chi_1\right)^3 + \phi \chi_1\chi_2 + z_3^3 - \chi_2^3$. Therefore, near (\ref{eq:near_slag_i}), one can define a large volume holomorphic involution: $\chi_1\leftrightarrow-z_3, \phi\rightarrow-\phi$(the fact that complex structure modulus $\phi$ is involutively odd, is also used in (\ref{eq:sigmas_Ts})).

After $z_i\rightarrow z_i+{\cal V}^{\frac{1}{36}}$ and $a_I\rightarrow a_I+\langle a_I\rangle$, the EW symmetry is spontaneously broken to $U(1)_{\rm em}$. We therefore require to construct four harmonic distribution one-forms $A_I$ localized on a sub-locus of $\Sigma_B$:
\begin{equation}
\label{eq:near_slag_i}
C_3:|z_1|\sim {\cal V}^{\frac{1}{36}},\ |z_2|\sim{\cal V}^{\frac{1}{36}},\ |z_3|\sim{\cal V}^{\frac{1}{6}}.
\end{equation}
This is a toroidal three-cycle and the Calabi-Yau can be thought as a $T^3$(swept out by \\$(arg z_1,arg z_2,arg z_3)$)-fibration over a large base $(|z_1|,|z_2|,|z_3|)$; precisely apt for application of mirror symmetry as three T-dualities a la S(trominger) Y(au) Z(aslow) \cite{SYZ}. Interestingly, (\ref{eq:near_slag_i}) is almost a s(pecial) Lag(rangian) sub-manifold. The requirements for the same are that $f^*J=0,\ f^*\Omega=e^{i\theta}{\rm vol}(C_3),$ where $f:C_3\rightarrow CY_3$ \cite{BBS}. Let us see if these requirements hold up. Using the geometric K\"{a}hler potential of (\ref{eq:K}), the geometric metric near $C_3$ is estimated to be:
\begin{equation}
\label{eq:geom met}
g_{i{\bar j}}\sim\frac{{\cal O}(10^{-1})}{2}\left(
\begin{array}{lll}
 \frac{1}{ \sqrt[18]{V}} & \frac{1}{ \sqrt[18]{V}} & \frac{1}{ V^{7/36}} \\
 \frac{1}{ \sqrt[18]{V}} & \frac{1}{ \sqrt[18]{V}} & \frac{1}{ V^{7/36}} \\
 \frac{1}{ V^{7/36}} & \frac{1}{ V^{7/36}} & \frac{1}{ \sqrt[3]{V}}
\end{array}
\right),
\end{equation}
which justifies the assumption made in \cite{ferm_masses_MS} about the
block-diagonal form of $g_{i{\bar j}}$. The same was necessary in
explicitly demonstrating $h^{1,1}_-\neq0$ in \cite{ferm_masses_MS} for the large volume involution: $z_1\rightarrow -z_1, z_{2,3}\rightarrow z_{2,3}$. The most non-trivial example of involutions which are meaningful only at large volumes is mirror symmetry implemented as three T-dualities in \cite{SYZ} (Also see the first paragraph of section {\bf 4.3} of \cite{appl_confs_uplift_dasguptaetal} for implementation of SYZ proposal and importance of large base for doing so);  mirror symmetry could be thought of as an involution since the mirror of a mirror is the original manifold.

\begin{enumerate}
\item
Using (\ref{eq:geom met}), one sees that:
\begin{eqnarray}
\label{eq:J_pull back}
& & f^*(ds_6^2)\sim0.05\left[\sum_{i,j=1}^2{\cal V}^{-\frac{1}{18}}\left(dz_id{\bar z}_{\bar j}\right)+{\cal V}^{-\frac{1}{3}}|dz_3|^2\right]\Biggr|_{C_3}\nonumber\\
& & \sim 0.05\left[\sum_{i,j=1}^2d (arg z_i)d (arg{\bar z}_{\bar j}) + d (arg z_3) d (arg {\bar z}_3)\right].
\end{eqnarray}
This implies $f^*J\sim0.05$, which is small.

\item
In the $u_2\neq0$- coordinate patch with $z_1=\frac{u_1}{u_2}, z_2=\frac{u_3}{u_2}, z_3=\frac{u_4}{u_2^6}, z_4=\frac{u_5}{u_2^6}$, by the Griffiths residue formula, one obtains:
\begin{equation}
\label{eq:Griffiths}
\Omega=\frac{dz_1\wedge dz_2\wedge dz_4}{\frac{\partial P}{\partial z_3}}=\frac{dz_1\wedge dz_2\wedge dz_4}{3z_3^2-\psi z_1z_2z_4}.
\end{equation}
 The coordinate $z_4$ can be solved for using:
\begin{equation}
\label{eq:x4_i}
z_4=\frac{\psi z_1z_2z_3\pm\sqrt{\psi^2(z_1z_2z_3)^2-4(1+z_1^{18}+z_2^{18}+z_3^3-\phi (z_1z_2)^6)}}{2}\sim \psi {\cal V}^{\frac{2}{9}}\pm i{\cal V}^{\frac{1}{4}},
\end{equation}
which implies that the $T^3$ will never degenerate as the roots will never coincide:
\begin{eqnarray}
\label{eq:x4_ii}
& & \hskip -0.5in dz_4\sim dz_3\left(\frac{\psi z_1z_2}{2} \pm\frac{\left(2\psi^2z_1^2z_2^2z_3 - 12 z_3^2\right)}{4\sqrt{\psi^2(z_1z_2z_3)^2-4(1+z_1^{18}+z_2^{18}+z_3^3-\phi (z_1z_2)^6)}}\right) + (...)dz_1 + (...)dz_2\nonumber\\
& & \sim\left(\frac{\psi z_1z_2}{2} \pm
  1.5i\frac{z_3^2}{\sqrt{z_1^{18}+z_2^{18}+z_3^3}}\right)dz_3 +
(...)dz_1 + (...)dz_2,\ {\rm near}\ (\ref{eq:near_slag_i}),
|\psi|<<1,|\phi|<<1.\nonumber\\
& &
\end{eqnarray}
\end{enumerate}
Therefore, restricted to (\ref{eq:near_slag_i}) and substituting (\ref{eq:x4_i}) and (\ref{eq:x4_ii}) in (\ref{eq:Griffiths}), one obtains:
\begin{eqnarray}
\label{eq:Omega}
& & \Omega\sim\frac{\left(\psi {\cal V}^{\frac{1}{18}}\pm 1.5i{\cal V}^{\frac{1}{12}}\right)}{3{\cal V}^{\frac{1}{3}}-\psi {\cal V}^{\frac{1}{18}}\left[\psi {\cal V}^{\frac{2}{9}}\pm i{\cal V}^{\frac{1}{4}}\right]}dz_1\wedge dz_2\wedge dz_3\nonumber\\
& & \sim  \left(\frac{\psi{\cal V}^{-\frac{5}{18}}}{3}\pm 0.5i {\cal
    V}^{-\frac{1}{4}}\right)
dz_1\wedge dz_2\wedge dz_3.
\end{eqnarray}
Hence, if one could estimate  the pull-back of the nowhere vanishing
holomorphic three-form as:
\begin{eqnarray}
\label{eq:f*Omega}
& & f^*\Omega\sim\left(\frac{\psi}{3}{\cal
    V}^{-\frac{5}{18}+\frac{1}{18}+\frac{1}{6}}+
0.5i{\cal V}^{-\frac{1}{4}+\frac{1}{18}+\frac{1}{6}}\right)d(arg
z_1)\wedge d(arg z_2)\wedge d(arg z_3)\Biggr|_{{\cal V}\sim10^6}
\nonumber\\
& & \sim\left(0.2\psi\pm0.3i\right)d(arg z_1)\wedge d(arg z_2)\wedge d(arg z_3),
\end{eqnarray}
and:
\begin{eqnarray}
\label{eq:vol C_3}
& &  {\rm vol}(C_3)\sim\sqrt{f^*g}f^*(dz_1\wedge dz_2\wedge dz_3)\nonumber\\
& &  \sim\sqrt{(0.05)^3}{\cal V}^{\frac{1}{18}+\frac{1}{6}}d(arg
z_1)\wedge d(arg z_2)\wedge d(arg z_3)\Biggr|_{{\cal V}\sim10^6}\nonumber\\
& & \sim0.2d(arg z_1)\wedge d(arg z_2)\wedge d(arg z_3),
\end{eqnarray}
then relative to a phase $e^{-i\theta},\theta=\frac{\pi}{2}$,
$\Im m\left(f^*e^{-i\theta}\Omega\right)\sim 0.2\psi d(arg z_1)\wedge
d(arg z_2)\wedge d(arg z_3)$, which for $|\psi|<<1$, is close to
zero; (equivalently)$\Re\left(f^*e^{-i\theta}\Omega\right)\sim$
vol($C_3)$. This implies that $C_3$ is nearly/almost a special Lagrangian
sub-manifold
\cite{Kachru+McGreevy_slag}.

Following our previous constructions in \cite{D3_D7_Misra_Shukla,ferm_masses_MS}, the harmonic distribution one-forms can be constructed by integrating:
\begin{equation}
\label{eq:distr 1 forms}
dA_I=\left(P_{\Sigma_B}(z_{1,2,3})\right)^Idz_1\wedge dz_2,
\end{equation}
with $I=1,2$(done)$,3,4$ where
\begin{equation}
\label{eq:A I}
A_I\sim \delta\left(|z_3|-{\cal V}^{\frac{1}{6}}\right)\delta\left(|z_1|-{\cal V}^{\frac{1}{36}}\right)\delta\left(|z_2|-{\cal V}^{\frac{1}{36}}\right)\left[\omega_I(z_1,z_2)dz_1 + \tilde{\omega}_I(z_1,z_2)dz_2\right].
\end{equation}
From (\ref{eq:distr 1 forms}), one sees that $A_I$ is harmonic only on $\Sigma_B$ and not at any other generic locus in the Calabi-Yau manifold; (\ref{eq:A I}) shows that $A_I$ are distribution one-forms on $\Sigma_B$ localized along the $D3$-brane which is localized on the three-cycle $C_3$ of (\ref{eq:near_slag_i}).

Writing $A_I(z_1,z_2)=\omega_I(z_1,z_2)dz_1+\tilde{\omega}_I(z_1,z_2)dz_2$ where $\omega(-z_1,z_2)=\omega(z_1,z_2), \tilde{\omega}(-z_1,z_2)=-\tilde{\omega}(z_1,z_2)$ and $\partial_1\tilde{\omega}=-\partial_2\omega$, one obtains:
\begin{eqnarray}
\label{eq:A_123}
& & \hskip -0.5in A_1(z_1,z_2,z_3\sim{\cal V}^{\frac{1}{6}})\sim - z_1^{18}z_2^{19}dz_1 + z_1^{19}z_2^{18}dz_2,\nonumber\\
& & \hskip -0.5in A_2(z_1,z_2,z_3\sim{\cal V}^{\frac{1}{6}})\sim - \left(\frac{z_2^{19}}{19}+z_1^{18}z_2\right)dz_1 + \left(\frac{z_1^{19}}{19}+z_2^{18}z_1\right)dz_2,\nonumber\\
& &\hskip -0.85in A_3(z_1,z_2,z_3\sim{\cal V}^{\frac{1}{6}})\sim -\left(\frac{z_2^{55}}{55}+3\frac{z_2^{37}}{37}\left(\sqrt{\cal V}+z_1^{18}\right)+z_2\left(\sqrt{\cal V}+z_1^{18}\right)^3 + \frac{z_2^{19}}{19}\left(3{\cal V}+6\sqrt{\cal V}z_1^{18}+3z_1^{36}\right)\right)dz_1 +(1\leftrightarrow 2)\nonumber\\
& & \sim -z_1^{18}z_2^{37}dz_1 + (1\leftrightarrow 2),\nonumber\\
& & A_4(z_1,z_2,z_3\sim{\cal V}^{\frac{1}{6}})\sim -z_1^{36}z_2^{37}dz_1 + z_2^{36}z_1^{37}dz_2.
\end{eqnarray}
Utilizing above, one can calculate the intersection numbers $C_{I \bar J}$ corresponding to set of four Wilson line moduli. The resultant contribution of various $C_{I \bar J}$'s are given in appendix B.

Lets now look at the term quadratic in the $D3$-brane position moduli in $T_{\alpha=B,S}$ which is given by:
\begin{equation}
\label{eq:D3 term T}
\left(\omega_\alpha\right)_{i{\bar j}}z^i\left({\bar z}^{\bar j} - \frac{i}{2}\left({\cal P}_{\tilde{a}}\right)^{\bar j}_{\ l}{\bar z}^{\tilde{a}}z^l\right),
\end{equation}
where
\begin{equation}
\label{eq:P}
\left({\cal P}_{\tilde{a}}\right)^{\bar j}_{\ l}=\frac{\Omega^{{\bar j}{\bar k}{\bar m}}\left(\chi_{\tilde{a}}\right)_{{\bar k}{\bar m}l}}{||\Omega||^2}\sim\frac{g^{j_1{\bar j}}g^{k_1{\bar k}}g^{m_1{\bar m}}\Omega_{j_1k_1m_1}\left(\chi_{\tilde{a}}\right)_{{\bar k}{\bar m}l}}{g^{i_2{\bar i}_2}g^{j_3{\bar j}_3}g^{k_2{\bar k}_2}\Omega_{i_2j_2k_2}{\bar\Omega}_{{\bar i}_2{\bar j}_2{\bar k}_2}}.
\end{equation}
As $\omega_\alpha$ forms a basis of $H^{1,1}_+(CY_3)$ and $\chi_{\tilde{a}}$ forms a basis of $H^{2,1}_-(CY_3)$, this does not therefore depend on the choice of the divisor $\alpha$.
By Griffith's residue formula, $\Omega=\Omega_{124}dz_1\wedge dz_2\wedge dz_4$ where
\begin{equation}
\label{eq:Omega I}
\Omega_{124}=\frac{1}{\frac{\partial P(z_1,z_2,z_3,z_4)}{\partial z_3}}=\frac{1}{3z_3^2-\psi z_1z_2z_4}.
\end{equation}
From $P(z_1,z_2,z_3,z_4)=0$, one obtains:
\begin{eqnarray}
\label{eq:z_3}
& & \hskip -0.8in z_3\sim \frac{\psi z_1z_2z_4}{\left(-\left(z_1^{18}+z_2^{18}+z_4^2-\phi z_1^6z_2^6\right)+\sqrt{\left(\psi z_1z_2z_4\right)^3+\left(z_1^{18}+z_2^{18}+z_4^2-\phi z_1^6z_2^6\right)^2}\right)^{\frac{1}{3}}}\nonumber\\
& & \hskip -0.8in + \left(-\left(z_1^{18}+z_2^{18}+z_4^2-\phi z_1^6z_2^6\right)+\sqrt{\left(\psi z_1z_2z_4\right)^3+\left(z_1^{18}+z_2^{18}+z_4^2-\phi z_1^6z_2^6\right)^2}\right)^{\frac{1}{3}},
\end{eqnarray}
which for $z_{1,2}\sim {\cal V}^{\frac{1}{36}}, z_4\sim {\cal V}^{\frac{1}{6}}$ yields $z_3\sim {\cal V}^{\frac{1}{6}}$. Hence,
\begin{equation}
\label{eq:Omega II}
\Omega_{124}\sim{\cal V}^{-\frac{1}{3}}.
\end{equation}
The polynomial deformation coefficient $\psi$ is a complex structure deformation modulus and given that it must be of the type $z^{\tilde{a}}$, i.e., odd under the holomorphic isometric involution $\sigma:z_1\rightarrow -z_1,z_{2,4}\rightarrow z_{2,4}$, $\sigma:\psi\rightarrow-\psi$. So, given that $\sigma^*\Omega=-\Omega$, hence $\sigma:\Omega_{124}\rightarrow\Omega_{124}$.

Now,
\begin{equation}
\label{eq:P}
\frac{1}{2}\left({\cal P}_{\tilde{a}}\right)^{\bar j}_{\ l}\sim\frac{\left(\chi_{\tilde{a}}\right)_{{\bar 4}{\bar 2}1}}{{\bar\Omega}_{{\bar 1}{\bar 2}{\bar 4}}}\sim\frac{{\cal V}^{\frac{1}{3}}}{2}\Biggr|_{{\cal V}\stackrel{<}{\sim}10^6}\sim{\cal O}(10).
\end{equation}
Assuming the complex structure moduli $z^{\tilde{a}}$ are stabilized at values: ${\cal O}(10)z^{\tilde{a}}\sim{\cal O}(1)$ for ${\cal V}\stackrel{<}{\sim}10^6$, one sees that contribution quadratic in $z^iz^j$ goes like $(\omega_\alpha)_{i{\bar j}}(z^i{\bar z}^{\bar j})$. $\omega_{B,S}$ are Poincare-duals of $\Sigma_{B,S}$ respectively. Hence, $\omega_{B,S}=\delta (P_{\Sigma_{B,S}})dP_{\Sigma_{B,S}}\wedge \delta({\bar P}_{\Sigma_{B,S}})d{\bar P}_{\Sigma_{B,S}}$.

\begin{itemize}
\item
Noting that $i_B^*dz_3\sim\frac{\phi z_1^5z_2^5(z_2dz_1+z_1dz_2)-(z_1^{17}dz_1+z_2^{17}dz_2)}{(\phi z_1^6z_2^6-z_1^{18}-z_2^{18}-1)^{\frac{2}{3}}}(i_B:\Sigma_B(1+z_1^{18}+z_2^{18}+z_3^3=\phi z_1^6z_2^6)\hookrightarrow CY_3)$, which near $z_{1,2}\sim{\cal V}^{\frac{1}{36}}$ implies $dz_3\sim{\cal V}^{\frac{5}{36}}(dz_1+dz_2)$. Hence, near $z_{1,2}\sim\frac{{\cal V}^{\frac{1}{36}}}{\sqrt{2}}, \omega_B\sim\frac{{\cal V}^{\frac{17}{18}}}{2^{17}}(dz_1+dz_2)\wedge(d{\bar z}_1+d{\bar z}_2)\biggr|_{{\cal V}\sim10^6}\sim (dz_1+dz_2)\wedge(d{\bar z}_1+d{\bar z}_2)$.

\item
Noting that $i_S^*dz_4\sim\frac{\phi z_1^5z_2^5(z_2dz_1+z_1dz_2)-(z_1^{17}dz_1+z_2^{17}dz_2)}{\sqrt{(\phi z_1^6z_2^6-z_1^{18}-z_2^{18}-1)}}(i_S:\Sigma_S(1+z_1^{18}+z_2^{18}+z_4^2=\phi z_1^6z_2^6)\hookrightarrow CY_3)$, which near $z_{1,2}\sim{\cal V}^{\frac{1}{36}}$ implies $dz_4\sim{\cal V}^{\frac{2}{9}}(dz_1+dz_2)$. Again, hence $z_{1,2}\sim\frac{{\cal V}^{\frac{1}{36}}}{\sqrt{2}}, \omega_S\sim\frac{{\cal V}^{\frac{17}{18}}}{2^{17}}(dz_1+dz_2)\wedge(d{\bar z}_1+d{\bar z}_2)\biggr|_{{\cal V}\sim10^6}\sim (dz_1+dz_2)\wedge(d{\bar z}_1+d{\bar z}_2)$.
\end{itemize}
We hence see that $(\omega_B)_{i{\bar j}}\sim(\omega_S)_{i{\bar j}}\sim{\cal O}(1)$ near $z_{1,2}\sim\frac{{\cal V}^{\frac{1}{36}}}{\sqrt{2}}$.

Therefore, near $c^a,b^a\sim\frac{\pi}{n k^a}<1$(in units of $M_p=1$), $n\sim{\cal O}(1),k^a\sim{\cal O}(10)$,
\begin{eqnarray}
\label{eq:K}
& &\hskip-1.38in \frac{K}{M_p^2}\sim-2 ln\biggl(\Biggl[\frac{T_B+{\bar T}_B}{M_p} - \Biggl(\mu_3l^2\frac{\left\{|z_1|^2 + |z_2|^2 + z_1{\bar z}_2 + z_2{\bar z}_1\right\}}{M_p^2}+{\cal V}^{\frac{10}{9}}\frac{|a_1|^2}{M_p^2}+{\cal V}^{\frac{11}{18}}\frac{\left(a_1{\bar a}_2+h.c.\right)}{M_p^2}+{\cal V}^{\frac{1}{9}}\frac{|a_2|^2}{M_p^2} + {\cal V}^{\frac{29}{18}}\frac{\left(a_1{\bar a}_3+h.c.\right)}{M_p^2}\nonumber\\
 & & \hskip -1in + {\cal V}^{\frac{10}{9}}\frac{\left(a_2{\bar a}_3+h.c.\right)}{M_p^2} + {\cal V}^{\frac{19}{9}}\frac{|a_3|^2}{M_p^2} + {\cal V}^{\frac{19}{9}}\left(a_1{\bar a}_4 + a_4{\bar a}_1\right) + {\cal V}^{\frac{29}{18}}\left(a_2{\bar a}_4 + a_4{\bar a}_2\right)+ {\cal V}^{\frac{47}{18}}\left(a_3{\bar a}_4 + a_4{\bar a}_3\right) + {\cal V}^{\frac{28}{9}}|a_4|^2\Biggr)\Biggr]^{3/2}\nonumber\\
 & & \hskip-1in  -\left(\frac{T_S+{\bar T}_S}{M_p}-\mu_3l^2\frac{\left\{|z_1|^2 + |z_2|^2 + z_1{\bar z}_2 + z_2{\bar z}_1\right\}}{M_p^2}\right)^{3/2}+\sum n^0_\beta(...)\biggr),
\end{eqnarray}
One can argue that near\\
 $\left(|z_{1,2}|\sim{\cal V}^{\frac{1}{36}}M_p,|z_3|\sim{\cal V}^{\frac{1}{6}}M_p, |a_1|\sim{\cal V}^{-\frac{2}{9}}M_p,|a_2|\sim{\cal V}^{-\frac{1}{3}}M_p,|a_3|\sim{\cal V}^{-\frac{13}{18}}M_p,|a_4|\sim{\cal V}^{-\frac{11}{9}}M_p,\zeta^A=0;\right.$ \\
 $\left.{\cal G}^a\sim\frac{\pi}{{\cal O}(1)k^a(\sim{\cal O}(10))}M_p\right)$, one obtains a local meta-stable dS-like minimum corresponding to the positive minimum of the potential\\
$e^KG^{T_S{\bar T}_S}|D_{T_S}W|^2$ stabilizing
${\rm vol}\left(\Sigma_B\right)\sim \Re ({\sigma}_B)\sim{\cal V}^{\frac{2}{3}}, {\rm vol}\left(\Sigma_S\right)\sim\Re{\sigma}_S\sim{\cal V}^{\frac{1}{18}}$ such that $\Re(T)_S\sim{\cal V}^{\frac{1}{18}}$ and in the dilute flux approximation:
\begin{eqnarray}
\label{eq:1overgsquared}
 \frac{1}{g_{j{=SU(3)\ {\rm or}\ SU(2)}}^2}&  = & \Re(T_{B}) + ln\left(\left.P\left(\Sigma_S\right)\right|_{D3|_{\Sigma_B}}\right) + ln\left(\left.{\bar P}\left(\Sigma_S\right)\right|_{D3|_{\Sigma_B}}\right)\nonumber\\
 & & + {\cal O}\left({\rm U(1)-Flux}_j^2\right)\sim{\cal V}^{\frac{1}{18}}.
\end{eqnarray}
Indeed, near the aforementioned stabilized values of the open string moduli,
\begin{eqnarray}
\label{eq:Casquared}
& & \left|C_{1{\bar 1}}|a_1|^2\right| \sim {\cal V}^{\frac{2}{3}},\ \left|C_{1{\bar 2}}\left(a_1{\bar a}_2 + h.c.\right)\right| \sim {\cal V}^{\frac{1}{18}},\ \left|C_{1{\bar 3}}\left(a_1{\bar a}_3 + h.c.\right)\right| \sim {\cal V}^{\frac{2}{3}},\nonumber\\
& & \left|C_{1{\bar 4}}\left(a_1{\bar a}_4 + h.c.\right)\right| \sim {\cal V}^{\frac{2}{3}},\
\left|C_{2{\bar 2}}|a_2|^2\right| \sim {\cal V}^{-\frac{5}{9}},\
\left|C_{2{\bar 3}}\left(a_2{\bar a}_3 + h.c.\right)\right| \sim {\cal V}^{\frac{1}{18}},\nonumber\\
& & \left|C_{2{\bar 4}}\left(a_2{\bar a}_4 + h.c.\right)\right| \sim {\cal V}^{\frac{1}{18}},\
\left|C_{3{\bar 3}}|a_3|^2\right| \sim {\cal V}^{\frac{2}{3}},\
\left|C_{3{\bar 4}}\left(a_3{\bar a}_4 + h.c.\right)\right| \sim {\cal V}^{\frac{2}{3}},\nonumber\\
& & \left|C_{4{\bar 4}}|a_4|^2\right| \sim {\cal V}^{\frac{2}{3}},
\end{eqnarray}
we see that there is the possibility that:
\begin{equation}
\label{eq:order1YM}
Vol(\Sigma_B)+C_{I{\bar J}}a_I{\bar a}_{\bar J} + h.c.\sim {\cal V}^{\frac{1}{18}},
\end{equation}
and hence (\ref{eq:1overgsquared}) could be implemented.

To obtain  the ``physical"/normalized Yukawa couplings, soft SUSY breaking parameters, vertices later in the paper, etc., one needs to diagonalize the matrix generated from the mixed double derivative of the K\"{a}hler potential, the same given as under:
\begin{equation}
\label{eq:Kahler pot}
\left(
\begin{array}{llllll}
 a_{11} \frac{1}{{\cal V}^{2/3}}  & a_{12} \frac{1}{{\cal V}^{2/3}} & a_{13}\frac{1}{{\cal V}^{5/12}}  &
   a_{14}\frac{1}{{\cal V}^{11/12}} & a_{15}\sqrt[12]{{\cal V}}  & a_{16} {\cal V}^{7/12}  \\
 a_{12}\frac{1}{{\cal V}^{2/3}} & a_{22} \frac{1}{{\cal V}^{2/3}} & a_{23} \frac{1}{{\cal V}^{5/12}} &
   a_{24} \frac{1}{{\cal V}^{11/12}} & a_{25}\sqrt[12]{{\cal V}} & a_{26} {\cal V}^{7/12}\\
 a_{13} \frac{1}{{\cal V}^{5/12}} & a_{23} \frac{1}{{\cal V}^{5/12}} & a_{33} {\cal V}^{4/9}  & a_{34} \frac{1}{\sqrt[18]{{\cal V}}}& a_{35}
   {\cal V}^{17/18} & a_{36} {\cal V}^{13/9} \\
 a_{14} \frac{1}{{\cal V}^{11/12}} & a_{24} \frac{1}{{\cal V}^{11/12}} & a_{34} \frac{1}{\sqrt[18]{{\cal V}}} &
   {a_{44}} \frac{1}{{\cal V}^{5/9}} & a_{45} {\cal V}^{4/9} & a_{46}{\cal V}^{17/18} \\
 a_{15} \sqrt[12]{{\cal V}} & a_{25}\sqrt[12]{{\cal V}} & a_{35} {\cal V}^{17/18} & a_{45} {\cal V}^{4/9}& a_{55} {\cal V}^{13/9}  & a_{56}{\cal V}^{35/18} \\
 a_{16} {\cal V}^{7/12} & a_{26} {\cal V}^{7/12} & a_{36} {\cal V}^{13/9}& a_{46}{\cal V}^{17/18} & a_{56}{\cal V}^{35/18}  & a_{66} {\cal V}^{22/9}
\end{array}
\right)
\end{equation}
With ${\cal V}\sim10^5$ and some fine tuning in $a_{11,12,22,14,13,23,24,36,44,46,56,66}$, the numerical eigenvalues are estimated to be:
\begin{equation}
\label{eq:eigenvals}
10^{12}, 10^7, 10^4, 10^{-2}, 10^{-3}, 10^{-5},
\end{equation}
with the corresponding eigenvectors given by:
\begin{eqnarray}
\label{eq:eigenvectors}
& & {\cal A}_4\sim a_4 + {\cal V}^{-\frac{3}{5}}a_3 + {\cal V}^{-\frac{6}{5}} a_1+{\cal V}^{-\frac{9}{5}} a_2 + {\cal V}^{-2}\left(z_1+z_2\right) ;\nonumber\\
& & {\cal A}_3\sim -a_3 + {\cal V}^{-\frac{3}{5}}a_4 - {\cal V}^{-\frac{3}{5}}a_1 - {\cal V}^{-\frac{7}{5}}a_2 + {\cal V}^{-\frac{8}{5}}\left(z_1+z_2\right);\nonumber\\
& & {\cal A}_1\sim a_1 - {\cal V}^{-\frac{3}{5}}a_3 + {\cal V}^{-1}a_2 - {\cal V}^{-\frac{6}{5}}a_4+ {\cal V}^{-\frac{6}{5}}\left(z_1+z_2\right);\nonumber\\
& & {\cal A}_2\sim - a_2 - {\cal V}^{-1}a_1 + {\cal V}^{-\frac{7}{5}}a_3 - {\cal V}^{-\frac{3}{5}}\left(z_1+z_2\right);\nonumber\\
& & {\cal Z}_2\sim - \frac{\left(z_1+z_2\right)}{\sqrt 2}  - {\cal V}^{-\frac{6}{5}}a_1 + {\cal V}^{-\frac{3}{5}}a_2 + {\cal V}^{-\frac{8}{5}}a_3+ {\cal V}^{-2}a_4;\nonumber\\
& & {\cal Z}_1\sim \frac{\left(z_1-z_2\right)}{\sqrt 2}  - {\cal V}^{-\frac{6}{5}}a_1 + {\cal V}^{-\frac{3}{5}}a_2 + {\cal V}^{-\frac{8}{5}}a_3+ {\cal V}^{-2}a_4.
\end{eqnarray}
The system of equations (\ref{eq:eigenvectors}) can be solved to yield:
\begin{eqnarray}
\label{eq:a_I+z_i}
& & z_1\sim \frac{\left({\cal Z}_2-{\cal Z}_1\right)}{\sqrt 2} - {\cal V}^{-\frac{6}{5}}{\cal A}_1 - {\cal V}^{-\frac{3}{5}}{\cal A}_2 + {\cal V}^{-\frac{8}{5}}{\cal A}_3 + {\cal V}^{-2}{\cal A}_4;\nonumber\\
& & z_2\sim -\frac{\left({\cal Z}_2+ {\cal Z}_1\right)}{\sqrt 2} - {\cal V}^{-\frac{6}{5}}{\cal A}_1 - {\cal V}^{-\frac{3}{5}}{\cal A}_2 + {\cal V}^{-\frac{8}{5}}{\cal A}_3 + {\cal V}^{-2}{\cal A}_4;\nonumber\\
& & a_1\sim {\cal V}^{-\frac{6}{5}}{\cal Z}_1 + {\cal V}^{-\frac{7}{5}}{\cal Z}_2 - {\cal A}_1 - \frac{{\cal A}_2}{\cal V} + {\cal V}^{-\frac{3}{5}}{\cal A}_3 + {\cal V}^{-\frac{6}{5}}{\cal A}_4;\nonumber\\
& & a_2\sim {\cal V}^{-\frac{3}{5}}{\cal Z}_1 + \frac{{\cal Z}_2}{\cal V} - \frac{{\cal A}_1}{\cal V}
+ {\cal A}_2 - {\cal V}^{-\frac{7}{5}}{\cal A}_3 + {\cal V}^{-\frac{9}{5}}{\cal A}_4;\nonumber\\
& & a_3\sim {\cal V}^{-\frac{7}{5}}{\cal Z}_1 + {\cal V}^{-\frac{9}{5}}{\cal Z}_2 - {\cal V}^{-\frac{3}{5}}{\cal A}_1 - {\cal V}^{-\frac{7}{5}}{\cal A}_2 - {\cal A}_3 + {\cal V}^{-\frac{3}{5}}{\cal A}_4;\nonumber\\
& & a_4\sim \frac{{\cal Z}_1}{{\cal V}^2} + {\cal V}^{-\frac{11}{5}}{\cal Z}_2 + \frac{{\cal A}_1}{{\cal V}} - {\cal V}^{-\frac{9}{5}}{\cal A}_2 + {\cal V}^{-\frac{3}{5}}{\cal A}_3 + {\cal A}_4;\nonumber\\
\end{eqnarray}
The non-perturbative superpotential we will be using is given by:
\begin{equation}
\label{eq:W}
W\sim\left(1+z_1^{18}+z_2^{18} + \left(3\phi_0z_1^6z_2^6 - z_1^{18}-z_2^{18}\right)^{\frac{2}{3}}-3\phi_0z_1^6z_2^6\right)^{n^s}e^{-n^s vol(\Sigma_S) - \mu_3(\alpha_S z_1^2 + \beta_S z_2^2 + \gamma_S z_1z_2)}.
\end{equation}
For $n^s=2$, $vol(\Sigma_S)$ and $z_i$ were stabilized at around ${\cal V}^{\frac{1}{18}}$ and ${\cal V}^{\frac{1}{36}}$. Now, considering fluctuations in $z_i:z_i\rightarrow {\cal V}^{\frac{1}{36}}+\delta z_i$, with the fluctuations expressible in terms of $\delta {\cal Z}_{1,2},\delta{\cal A}_{1,2,3,4}$ using (\ref{eq:a_I+z_i}), and with the understanding that
${\cal V}^{\frac{1}{18}}\left(1+\alpha_S+\beta_S+\gamma_S\right)\sim ln{\cal V}$, one obtains the following non-zero physical effective Yukuwa couplings :$\hat{Y}^{\rm eff}_{C_i C_j C_k}\equiv\frac{e^{\frac{K}{2}}Y^{\rm eff}_{C_i C_j C_k}}{\sqrt{K_{C_i{\bar C}_i}K_{C_j{\bar C}_j}K_{C_k{\bar C}_k}}}, C_i$ being an open string modulus which for us is $\delta{\cal Z}_{1,2},\delta{\cal A}_{1,2,3,4}$, where, e.g., $Y^{\rm eff}_{{\cal Z}_i{\cal A}_{1/2}{\cal A}_{3/4}}$ is the ${\cal O}({\cal Z}_i)$-coefficient in the following mass term in the ${\cal N}=1$ SUGRA action of \cite{Wess_Bagger}: $e^{\frac{K}{2}}{\cal D}_{{\bar{\cal A}}_{1/2}}D_{{\bar{\cal A}}_{3/4}}{\bar W}{\bar\chi}^{{\cal A}_{1/3}}\chi^{{\cal A}_{2/4}}$. Using (\ref{eq:a_I+z_i}), and:
\begin{eqnarray}
\label{eq:expKover2_calD_DW}
& & e^{\frac{K}{2}} {\cal D}_{z_1} D_{z_1}W = {\cal V}^{-\frac{43}{72}},\  e^{\frac{K}{2}} {\cal D}_{z_1} D_{a_1}W = {\cal V}^{-\frac{89}{72}},\
 e^{\frac{K}{2}} {\cal D}_{z_1} D_{a_2}W = {\cal V}^{-\frac{125}{72}},\nonumber\\
& & e^{\frac{K}{2}} D_{z_1} D_{a_3}W = {\cal V}^{-\frac{25}{36}},\ e^{\frac{K}{2}} D_{z_1} D_{a_4}W = {\cal V}^{-\frac{17}{72}},\
e^{\frac{K}{2}} D_{a_1} D_{a_1}W = {\cal V}^{-\frac{71}{72}},\nonumber\\
& & e^{\frac{K}{2}} {\cal D}_{a_1} D_{a_2}W = {\cal V}^{-\frac{107}{72}},\  e^{\frac{K}{2}} {\cal D}_{a_1} D_{a_3}W = {\cal V}^{-\frac{35}{72}},\
e^{\frac{K}{2}} {\cal D}_{a_1} D_{a_4}W = {\cal V}^{\frac{1}{72}},\nonumber\\
& & e^{\frac{K}{2}} {\cal D}_{a_2} D_{a_2}W = {\cal V}^{-\frac{1}{72}},\ e^{\frac{K}{2}} {\cal D}_{a_2} D_{a_3}W = {\cal V}^{-\frac{3}{4}},\
e^{\frac{K}{2}} {\cal D}_{a_2} D_{a_4}W = {\cal V}^{-\frac{5}{9}},\nonumber\\
& & e^{\frac{K}{2}} {\cal D}_{a_3} D_{a_3}W = {\cal V}^{\frac{1}{72}},\ e^{\frac{K}{2}} {\cal D}_{a_3} D_{a_4}W = {\cal V}^{\frac{37}{72}},\  e^{\frac{K}{2}} {\cal D}_{a_4} D_{a_4}W = {\cal V}^{\frac{73}{72}},
\end{eqnarray}
 one can verify that
$e^{\frac{K}{2}}{\cal D}_{{\cal A}_I}D_{{\cal A}_J}\bar{W}\sim e^{\frac{K}{2}}{\cal D}_{a_I}D_{a_J}\bar{W}$. Now, $\left(e^{\frac{K}{2}}{\cal D}_{\bar{\lambda}}D_{\bar{\Sigma}}\bar{W}\right)\bar{\chi}_L^{{\lambda}}\chi_R^{\Sigma}=\left(e^{\frac{K}{2}}{\cal D}_{\bar{\cal I}}D_{\bar{\cal J}}\bar{W}\right)\bar{\chi}_L^{{\cal I}}\chi_R^{\cal J}$, where $\lambda/\Sigma$ index ${\cal A}_I, {\cal Z}_J$ and ${\cal I}/{\cal J}$ index $a_I,z_i$. What is interesting is that using (\ref{eq:eigenvectors}), (\ref{eq:a_I+z_i}), e.g.,
\begin{eqnarray}
\label{eq:eq_mass_terms_diag_non-diag_basis}
& &\hskip-0.8in \left(e^{\frac{K}{2}}{\cal D}_{\bar{\cal A}_{1,2}}D_{\bar{\cal A}_{3,4}}\bar{W}\right)\bar{\chi}_L^{{\cal A}_{1,2}}\chi_R^{{\cal A}_{3,4}}\sim\left(e^{\frac{K}{2}}{\cal D}_{\bar{a}_{1,2}}D_{\bar{a}_{3,4}}\bar{W}\right)\bar{\chi}_L^{{\cal A}_{1,2}}\chi_R^{{\cal A}_{3,4}}=\left(\frac{\partial\bar{\cal A}_{1,2}}{\partial\bar{\cal M}_{\cal I}}\right)\left(\frac{\partial\bar{\cal A}_{3,4}}{\partial\bar{\cal M}_{\bar J}}\right) \left(e^{\frac{K}{2}}{\cal D}_{\bar{a}_{1,2}}D_{\bar{a}_{3,4}}\bar{W}\right)\bar{\chi}_L^{{\cal M}_{\cal I}}\chi_R^{{\cal M}_{\cal J}}\nonumber\\
& & \sim \left(e^{\frac{K}{2}}{\cal D}_{\bar{a}_{1,2}}D_{\bar{a}_{3,4}}\bar{W}\right)\bar{\chi}_L^{{a}_{1,2}}\chi_R^{a_{3,4}}.
\end{eqnarray}
This implies that the mass terms, in the large volume limit, are invariant under diagonalization of the open string moduli.

\begin{itemize}
\item
\begin{equation}
\label{eq:Yhatz1a1a2} \frac{{\cal O}({\cal Z}_i)\ {\rm term\ in}\
e^{\frac{K}{2}}{\cal D}_{{\cal A}_1}D_{{\cal A}_3}W}{\sqrt{K_{{\cal Z}_i\bar{\cal Z}_i}K_{{\cal A}_1\bar{\cal A}_1}K_{{\cal A}_3\bar{\cal A}_3}}}\equiv
\hat{Y}^{\rm eff}_{{\cal Z}_i{\cal A}_1{\cal A}_3}\sim 10^{-3}\times {\cal V}^{-\frac{4}{9}},
\end{equation}
which implies that for ${\cal V}\sim 10^5, \langle {\cal Z}_i\rangle\sim 246 GeV$,
 $\langle {\cal Z}_i\rangle \hat{Y}_{{\cal Z}_1{\cal A}_1{\cal A}_3}\sim MeV$ - about the mass of the electron!

\item
\begin{equation}
\label{eq:Yhatz1a1a2}\frac{{\cal O}({\cal Z}_i)\ {\rm term\ in}\
e^{\frac{K}{2}}{\cal D}_{{\cal A}_2}D_{{\cal A}_4}W}{\sqrt{K_{{\cal Z}_i\bar{\cal Z}_i}K_{{\cal A}_2\bar{\cal A}_2}K_{{\cal A}_4\bar{\cal A}_4}}}\equiv
\hat{Y}^{\rm eff}_{{\cal Z}_i{\cal A}_2{\cal A}_4}\sim 10^{-\frac{5}{2}}\times {\cal V}^{-\frac{4}{9}},
\end{equation}
which implies that for ${\cal V}\sim 10^5,\langle {\cal Z}_i\rangle\sim 246 GeV$,
 $\langle z_i\rangle \hat{Y}_{{\cal Z}_i{\cal A}_1{\cal A}_2}\sim10MeV$ - close to the mass of the up quark!
\end{itemize}
There is an implicit assumption that the vev of the $D3$-brane position moduli, identified with the neutral components of two Higgs doublets, can RG-flow down to $246 GeV$ - that the same is possible was shown in \cite{ferm_masses_MS}. The RG-flow of the effective physical Yukawas are expected to change by ${\cal O}(1)$ under an RG flow from the string scale down to the EW scale. This can be motivated by looking at RG-flows of the physical Yukawas $\hat{Y}_{\Lambda\Sigma\Delta}$ in MSSM-like models:
\begin{equation}
\label{eq:Yhat_RG-1}
\frac{d\hat{Y}_{\Lambda\Sigma\Delta}}{dt} = \gamma_{\Lambda}^\kappa\hat{Y}_{\kappa\Sigma\Delta}
+ \gamma_{\Sigma}^{\kappa}\hat{Y}_{\Lambda\kappa\Delta} + \gamma_{\Delta}^\kappa\hat{Y}_{\Lambda\Sigma\kappa},
\end{equation}
where the anomalous dimension matrix $\gamma_\Lambda^\kappa$, at one loop, is defined as:
\begin{equation}
\label{eq:anom_dim_mat}
\gamma_\Lambda^\kappa = \frac{1}{32\pi^2}\left(\hat{Y}_{\Lambda\Psi\Upsilon}\hat{Y}^*_{\kappa\Psi\Upsilon} - 2 \sum_{(a)}g_{(a)}^2C_{(a)}(\Phi_\Lambda)\delta_{{\cal I}}^{\kappa}\right),
\end{equation}
where $(a)$ indexes the three gauge groups and $\Lambda$, etc. index
the diagonalized basis fields ${\cal A}_I,{\cal Z}_i$. Using
(\ref{eq:W}), one can show that at $M_s$(string scale):
\begin{eqnarray}
\label{eq:Yhats}
& & \hat{Y}_{{\cal Z}_i{\cal Z}_j{\cal Z}_k}\sim{\cal
  V}^{\frac{1}{8}},\ \hat{Y}_{{\cal Z}_i{\cal Z}_i{\cal
    A}_1}\sim{\cal
  V}^{-\frac{79}{40}},\
\hat{Y}_{{\cal Z}_i{\cal Z}_i{\cal A}_2}\sim{\cal
  V}^{-\frac{31}{40}},\ \hat{Y}_{{\cal Z}_i{\cal Z}_i{\cal A}_4}\sim{\cal
  V}^{-\frac{143}{40}},\nonumber\\
& &\hat{Y}_{{\cal Z}_i{\cal Z}_i{\cal A}_3}\sim{\cal
  V}^{-\frac{107}{40}},\ \hat{Y}_{{\cal Z}_i{\cal A}_1{\cal
    A}_1}\sim{\cal
  V}^{-\frac{163}{40}},\  \hat{Y}_{{\cal Z}_i{\cal A}_1{\cal
    A}_2}\sim{\cal
  V}^{-\frac{23}{8}},\nonumber\\
& &  \hat{Y}_{{\cal Z}_i{\cal A}_1{\cal
    A}_3}\sim{\cal
  V}^{-\frac{191}{40}},\  \hat{Y}_{{\cal Z}_i{\cal A}_1{\cal
    A}_4}\sim{\cal
  V}^{-\frac{227}{40}},\  \hat{Y}_{{\cal Z}_i{\cal A}_2{\cal A}_2}\sim{\cal
  V}^{-\frac{67}{40}},\nonumber\\
& &  \hat{Y}_{{\cal Z}_i{\cal A}_2{\cal A}_3}\sim{\cal
  V}^{-\frac{143}{40}},\  \hat{Y}_{{\cal Z}_i{\cal A}_2{\cal A}_4}\sim{\cal
  V}^{-\frac{179}{40}}, \  \hat{Y}_{{\cal Z}_i{\cal A}_3{\cal A}_3}\sim{\cal
  V}^{-\frac{219}{40}},\nonumber\\
& &  \hat{Y}_{{\cal Z}_i{\cal A}_3{\cal
    A}_4}\sim{\cal
  V}^{-\frac{51}{8}},\  \hat{Y}_{{\cal Z}_i{\cal A}_4{\cal
    A}_4}\sim{\cal
  V}^{-\frac{55}{8}},\ \hat{Y}_{{\cal A}_1{\cal A}_1{\cal
    A}_1}\sim{\cal V}^{-\frac{247}{40}},\nonumber\\
& & \hat{Y}_{{\cal A}_1{\cal A}_1{\cal A}_2}
\sim{\cal V}^{-\frac{199}{40}},\  \hat{Y}_{{\cal A}_1{\cal A}_1{\cal A}_3}
\sim{\cal V}^{-\frac{55}{8}},\  \hat{Y}_{{\cal A}_1{\cal A}_1{\cal A}_4}
\sim{\cal V}^{-\frac{311}{40}},\nonumber\\
& &  \hat{Y}_{{\cal A}_1{\cal A}_2{\cal A}_2}\sim
\sim{\cal V}^{-\frac{151}{40}},\  \hat{Y}_{{\cal A}_1{\cal A}_2{\cal A}_3}
\sim{\cal V}^{-\frac{227}{40}},\ \hat{Y}_{{\cal A}_1{\cal A}_2{\cal A}_4}
\sim{\cal V}^{-\frac{263}{40}},\nonumber\\
& &  \hat{Y}_{{\cal A}_2{\cal A}_4{\cal A}_5}
\sim{\cal V}^{-\frac{339}{40}},\   \hat{Y}_{{\cal A}_1{\cal A}_4{\cal A}_4}
\sim{\cal V}^{-\frac{75}{8}},\ \hat{Y}_{{\cal A}_2{\cal A}_2{\cal
    A}_2}\sim{\cal V}^{-\frac{103}{40}},\nonumber\\
& &  \hat{Y}_{{\cal A}_2{\cal A}_2{\cal A}_3}
\sim{\cal V}^{-\frac{179}{40}},\  \hat{Y}_{{\cal A}_2{\cal A}_2{\cal A}_4}
\sim{\cal V}^{-\frac{43}{8}},\  \hat{Y}_{{\cal A}_2{\cal A}_3{\cal A}_3}
\sim{\cal V}^{-\frac{299}{60}},\nonumber\\
& &  \hat{Y}_{{\cal A}_2{\cal A}_3{\cal A}_4}
\sim{\cal V}^{-\frac{263}{40}},\  \hat{Y}_{{\cal A}_4{\cal A}_3{\cal A}_3}
\sim{\cal V}^{-\frac{367}{40}},\  \hat{Y}_{{\cal A}_1{\cal A}_3{\cal A}_3}
\sim{\cal V}^{-\frac{283}{40}},\nonumber\\
& &  \hat{Y}_{{\cal A}_2{\cal A}_4{\cal A}_4}
\sim{\cal V}^{-\frac{327}{40}},\  \hat{Y}_{{\cal A}_3{\cal A}_4{\cal A}_4}
\sim{\cal V}^{-\frac{403}{40}},\  \hat{Y}_{{\cal A}_3{\cal A}_3{\cal A}_3}
\sim{\cal V}^{-\frac{331}{40}},\ \hat{Y}_{{\cal A}_4{\cal A}_4{\cal A}_4}
\sim{\cal V}^{-\frac{439}{40}}.
\end{eqnarray}
From (\ref{eq:Yhats}), one sees that $\hat{Y}_{{\cal Z}_i{\cal
    Z}_i{\cal Z}_i}(M_s)\sim{\cal V}^{\frac{1}{8}}$ is the most
dominant physical Yukawa at the string scale. Hence,
\begin{equation}
\label{eq:anom_dimen-2}
\gamma_\Lambda^\Gamma\hat{Y}_{\Gamma\Sigma\Delta}\sim\frac{1}{32\pi^2}\hat{Y}_{\Lambda{\cal
  Z}_i{\cal Z}_i}\hat{Y}^*_{{\cal Z}_i{\cal Z}_i{\cal
  Z}_i}\hat{Y}_{{\cal Z}_i\Sigma\Delta}
-\frac{2\sum_{(a)}g_{(a)}^2C_{(a)}(\Phi_\Gamma)\delta_{\Lambda}^{\Gamma}\hat{Y}_{\Gamma\Sigma\Delta}}{32\pi^2}.
\end{equation}
Now, the first term in (\ref{eq:anom_dimen-2}) is volume suppressed as
compared to the second term at $M_s$; let us assume that the same will
be true up to the EW scale. Using the one-loop solution for
$\frac{g_{(a)}^2(t)}{16\pi^2}
=\frac{ \frac{\beta_{(a)}}{b_{(a)}}}{1 + \beta_{(a)}t}$ in
(\ref{eq:anom_dimen-2}) and therefore (\ref{eq:Yhat_RG-1}), one
obtains:
\begin{equation}
\label{eq:Yhat_sol_1}
\frac{d ln\hat{Y}_{\Lambda\Sigma\Delta}}{dt}\sim -2\left(\sum_{(a)}
\frac{C_{(a)}(\Lambda)\frac{\beta_{(a)}}{b_{(a)}}}{1 + \beta_{(a)}t}
+ \sum_{(a)}
\frac{C_{(a)}(\Sigma)\frac{\beta_{(a)}}{b_{(a)}}}{1 + \beta_{(a)}t} +
\sum_{(a)}
\frac{C_{(a)}(\Delta)\frac{\beta_{(a)}}{b_{(a)}}}{1 + \beta_{(a)}t}\right),
\end{equation}
whose solution yields:
\begin{equation}
\label{eq:Yhat_sol-2}
\hat{Y}_{\Lambda\Sigma\Delta}(t)\sim\hat{Y}_{\Lambda\Sigma\Delta}(M_s)
\prod_{(a)=1}^3\left(1 + \beta_{(a)}t\right)^{\frac{-2\left(C_{(a)}(\Lambda)
    +C_{(a)}(\Sigma) + C_{(a)}(\Delta)\right) }{b_{(a)}}}.
\end{equation}
The solution (\ref{eq:Yhat_sol-2}) justifies the assumption that all
$\hat{Y}_{\Lambda\Sigma\Delta}$s change only by ${\cal O}(1)$ as one
RG-flows down from the string to the EW scale.

A similar argument for the RG-evolution of $\hat{Y}^{\rm eff}_{{\cal Z}_i{\cal A}_I{\cal A}_J}$s would proceed as follows. We will for definiteness and due to relevance to the preceding discussion on lepton and quark masses, consider $\frac{d\hat{Y}^{\rm eff}_{{\cal Z}_1{\cal A}_1{\cal A}_3}}{dt}$ and $\frac{d\hat{Y}^{\rm eff}_{{\cal Z}_1{\cal A}_2{\cal A}_4}}{dt}$. Now,
\begin{eqnarray}
\label{eq:Yeffzia1a3-I}
& & \frac{d\hat{Y}^{\rm eff}_{{\cal Z}_1{\cal A}_1{\cal A}_3}}{dt} = \gamma_{{\cal Z}_1}^\Lambda\hat{Y}^{\rm eff}_{\Lambda{\cal A}_1{\cal A}_3} + \gamma_{{\cal A}_1}^\Lambda\hat{Y}^{\rm eff}_{{\cal Z}_1\Lambda{\cal A}_3}
+ \gamma_{{\cal A}_3}^\Lambda\hat{Y}^{\rm eff}_{{\cal Z}_1{\cal A}_1\Lambda},
\end{eqnarray}
where:
\begin{eqnarray}
\label{eq:Yeffzia1a3-II}
\gamma_{{\cal Z}_1}^\Lambda\ni\hat{Y}_{{\cal Z}_1\Upsilon\Sigma}\hat{Y}^*_{\Lambda\Upsilon\Sigma}\sim
\hat{Y}_{{\cal Z}_1{\cal Z}_1{\cal Z}_1}\hat{Y}^*_{\Lambda{\cal Z}_1{\cal Z}_1},\ {\rm etc.},
\end{eqnarray}
implying:
\begin{eqnarray}
\label{eq:Yeffzia1a3-III}
& & \frac{d\hat{Y}^{\rm eff}_{{\cal Z}_1{\cal A}_1{\cal A}_3}}{dt}\ni \hat{Y}_{{\cal Z}_1{\cal Z}_1{\cal Z}_1}\hat{Y}^*_{{\cal Z}_1{\cal Z}_1{\cal Z}_1}\hat{Y}^{\rm eff}_{{\cal Z}_1{\cal A}_1{\cal A}_3}
+ \hat{Y}_{{\cal Z}_1{\cal Z}_1{\cal Z}_1}\hat{Y}^*_{{\cal A}_1{\cal Z}_1{\cal Z}_1}\hat{Y}^{\rm eff}_{{\cal A}_1{\cal A}_1{\cal A}_3}
+ \hat{Y}_{{\cal Z}_1{\cal Z}_1{\cal Z}_1}\hat{Y}^*_{{\cal A}_2{\cal Z}_1{\cal Z}_1}\hat{Y}^{\rm eff}_{{\cal A}_2{\cal A}_1{\cal A}_3}\nonumber\\
& & \hat{Y}_{{\cal Z}_1{\cal Z}_1{\cal Z}_1}\hat{Y}^*_{{\cal A}_3{\cal Z}_1{\cal Z}_1}\hat{Y}^{\rm eff}_{{\cal A}_3{\cal A}_1{\cal A}_3} +
\hat{Y}_{{\cal Z}_1{\cal Z}_1{\cal Z}_1}\hat{Y}^*_{{\cal A}_4{\cal Z}_1{\cal Z}_1}\hat{Y}^{\rm eff}_{{\cal A}_4{\cal A}_1{\cal A}_3},
\end{eqnarray}
and a similar equation for $\frac{d\hat{Y}^{\rm eff}_{{\cal Z}_1{\cal A}_2{\cal A}_4}}{dt}$.
Using (\ref{eq:Yhats}) and:
\begin{eqnarray}
\label{eq:yeff_I or i a1a3 and a2a4}
& & e^{\frac{K}{2}}{\cal D}_{a_{1/2}}D_{a_{3/4}}W\ni {\cal V}^{-\frac{4}{9}}\left(z_i-{\cal V}^{\frac{1}{36}}\right) + {\cal V}^{\frac{2}{3}}\left(a_1-{\cal V}^{-\frac{2}{9}}\right)
+ {\cal V}^{-\frac{5}{6}}\left(a_2-{\cal V}^{-\frac{1}{3}}\right)\nonumber\\
& &  +
{\cal V}^{\frac{1}{6}}\left(a_3-{\cal V}^{-\frac{13}{18}}\right)
+ {\cal V}^{\frac{7}{6}}\left(a_4-{\cal V}^{-\frac{11}{9}}\right),
\end{eqnarray}
and (\ref{eq:Yeffzia1a3-III}), one sees that at $M_s$:
\begin{eqnarray}
\label{eq:Yeffzia1a3+a2a4-IV}
& & \gamma_{{\cal Z}_1}^\Lambda\hat{Y}^{\rm eff}_{\Lambda{\cal A}_1{\cal A}_3} + \gamma_{{\cal A}_1}^\Lambda\hat{Y}^{\rm eff}_{{\cal Z}_1\Lambda{\cal A}_3}
+ \gamma_{{\cal A}_3}^\Lambda\hat{Y}^{\rm eff}_{{\cal Z}_1{\cal A}_1\Lambda}\ni \frac{{\cal V}^{\frac{1}{8}+\frac{1}{8} - \frac{4}{9}}}{\sqrt{K_{{\cal Z}_1\bar{\cal Z}_1}K_{{\cal A}_1\bar{\cal A}_1}K_{{\cal A}_3\bar{\cal A}_3}}}\Biggr|_{{\cal V}\sim10^5}\sim{\cal V}^{-\frac{143}{180}},\nonumber\\
& &  \gamma_{{\cal Z}_1}^\Lambda\hat{Y}^{\rm eff}_{\Lambda{\cal A}_2{\cal A}_4} + \gamma_{{\cal A}_2}^\Lambda\hat{Y}^{\rm eff}_{{\cal Z}_1\Lambda{\cal A}_4}
+ \gamma_{{\cal A}_4}^\Lambda\hat{Y}^{\rm eff}_{{\cal Z}_1{\cal A}_2\Lambda}\ni \frac{{\cal V}^{\frac{1}{8}+ \frac{1}{8} - \frac{4}{9}}}{\sqrt{K_{{\cal Z}_1\bar{\cal Z}_1}K_{{\cal A}_2\bar{\cal A}_2}K_{{\cal A}_4\bar{\cal A}_4}}}\Biggr|_{{\cal V}\sim10^5}\sim{\cal V}^{-\frac{25}{36}}.
\end{eqnarray}
Hence, at the string scale, the anomalous dimension matrix contribution is sub-dominant as compared to the gauge coupling-dependent contribution, and by a similar assumption (justified by the solution below)/reasoning:
\begin{equation}
\label{eq:Yeffhat_sol}
\hat{Y}^{\rm eff}_{\Lambda\Sigma\Delta}(t)\sim\hat{Y}^{\rm eff}_{\Lambda\Sigma\Delta}(M_s)
\prod_{(a)=1}^3\left(1 + \beta_{(a)}t\right)^{\frac{-2\left(C_{(a)}(\Lambda)
    +C_{(a)}(\Sigma) + C_{(a)}(\Delta)\right) }{b_{(a)}}}.
\end{equation}
This suggests that possibly, the fermionic superpartners of ${\cal A}_1$ and ${\cal A}_3$ correspond respectively to  $e_L$ and $e_R$ and the fermionic superpartners of ${\cal A}_2$ and ${\cal A}_4$ correspond respectively to the first generation $u_L$ and $u_R$.

We also calculate masses of all superpartners corresponding to each wilson line moduli, position moduli and gaugino/Higgsino's appearing in the set up. The details of calculations of the same are given in appendix C and results of masses of all SM as well as superpartners are summarized in table \ref{table:mass scales}.
 \begin{table}[htbp]
\centering
\begin{tabular}{|l|l|}
\hline
Quark mass & $ M_{q}\sim O(5) MeV $\\
Lepton mass & $ M_{l}\sim {\cal O}(1) MeV $\\  \hline
Gravitino mass &  $ m_{\frac{3}{2}}\sim{\cal V}^{-\frac{n^s}{2} - 1}m_{pl}; n_s=2$ \\
Gaugino mass & $ M_{\tilde g}\sim V^{\frac{2}{3}}m_{\frac{3}{2}}$\\
Neutralino mass & $ M_{\chi^0_3}\sim V^{\frac{2}{3}}m_{\frac{3}{2}}$\\  \hline
$D3$-brane position moduli  & $ m_{{\cal Z}_i}\sim {\cal V}^{\frac{59}{72}}m_{\frac{3}{2}}$ \\
(Higgs) mass & \\
Wilson line moduli mass & $ m_{\tilde{\cal A}_I}\sim {\cal V}^{\frac{1}{2}}m_{\frac{3}{2}}$\\
& ${I =1,2,3,4}$\\ \hline
A-terms & $A_{pqr}\sim n^s{\cal V}^{\frac{37}{36}}m_{\frac{3}{2}}$\\
& $\{p,q,r\} \in \{{{\tilde{\cal A}_I}},{{\cal Z}_i}\}$\\ \hline
Physical $\mu$-terms & $\hat{\mu}_{{\cal Z}_i{\cal Z}_j}$ \\
 (Higgsino mass) & $\sim{\cal V}^{\frac{37}{36}}m_{\frac{3}{2}}$ \\ \hline
Physical $\hat{\mu}B$-terms & $\left(\hat{\mu}B\right)_{{\cal Z}_1{\cal Z}_2}\sim{\cal V}^{\frac{37}{18}}m_{\frac{3}{2}}^2$\\ \hline
\end{tabular}
\caption{Mass scales of first generation of SM as well supersymmetric and soft SUSY parameters}
\label{table:mass scales}
\end{table}

We will consider four stacks of $D7$-branes: a stack of 3, a stack of 2 and two stacks of 1. The matter fields: L-quarks and their superpartners will be valued in the bifundamentals $(3,{\bar 2})$ under $SU(3)_c\times SU(2)_L$; the L-leptons and their superpartners will be valued in the bifundamentals
$(2,\bar{-1})$ of $SU(2)_L\times U(1)_Y$.  Before the  fluxes (\ref{eq:flux})  are turned on,
the Wilson line moduli are valued in the adjoint of $U(7)$. With the following choice of fluxes:
\begin{equation}
\label{eq:flux}
F = \left(\begin{array}{ccccccc}f_1 & 0 & 0 & 0 & 0 & 0&0\\
0 & f_1 & 0 & 0 & 0 & 0&0\\
0 & 0 & f_1 & 0 & 0 & 0&0\\
0 & 0 & 0 & f_2 & 0 & 0&0\\
0 & 0 & 0 & 0 & f_2 & 0&0\\
0 & 0 & 0 & 0 & 0 & f_3&0\\
0 & 0 & 0 & 0 & 0 & 0 & f_4
\end{array}\right),
\end{equation}
the $U(7)$ is broken down to $U(3)\times U(2)\times U(1)\times U(1)$. So, the bifundamental Wilson line super-moduli ${\cal A}_I$  will be represented as:
\begin{equation}
\label{eq:Q L_I}
{\cal A}_I=\sum a_I^{ab}e_{ab} + \sum\theta \tilde{a}_I^{ab}e_{ab},\ \left[(a,b)=1,...,7\right],
\end{equation}
where $\left(e_{ab}\right)_{ij}=\delta_{ai}\delta_{bj}$. Hence,
  \begin{equation}
 \label{eq:L_L}
{\cal A}_1 =
 \left(\begin{array}{ccccccc}
 0 & 0 & 0 & 0 & 0 & 0 & 0\\
 0 & 0 & 0 & 0 & 0 & 0 & 0\\
 0 & 0 & 0 & 0 & 0 & 0 & 0\\
 0 & 0 & 0 & 0 & 0 & \tilde{\nu}_e + \theta\nu_e   & 0\\
 0 & 0 & 0 & 0 & 0 & \tilde{e} + \theta e & 0\\
 0 & 0 & 0 & \bar{\tilde{\nu}}_e + {\bar\theta}{\bar\nu}_e  &  \bar{\tilde{e}} + {\bar\theta}{\bar e} & 0 & 0\\
 0 & 0 & 0 & 0 & 0 & 0 & 0
 \end{array} \right);
\end{equation}
  \begin{equation}
\label{eq:Q L}
{\cal A}_2 =
\left(\begin{array}{ccccccc}
0 & 0 & 0 & \tilde{u}  + \theta u & 0 & 0 & 0\\
0 & 0 & 0 & 0 & 0 & 0 & 0\\
0 & 0 & 0 & 0 & 0 & 0 & 0\\
\bar{\tilde{u}}  + {\bar\theta}{u}^\dagger &  0 & 0 & 0 & 0 & 0 & 0\\
0 &  0 & 0 & 0 & 0 & 0 & 0\\
0 & 0 & 0 & 0 & 0 & 0 & 0\\
0 & 0 & 0 & 0 & 0 & 0 & 0
\end{array}\right),
\end{equation}
and
\begin{equation}
\label{eq:L_R}
{\cal A}_3 =
\left(\begin{array}{ccccccc}
0 & 0 & 0 & 0 & 0 & 0 & 0\\
0 & 0 & 0 & 0 & 0 & 0 & 0\\
0 & 0 & 0 & 0 & 0 & 0 & 0\\
 0 &  0 & 0 & 0 & 0 & 0 & 0\\
0 &  0 & 0 & 0 & 0 & 0 & 0\\
0 & 0 & 0 & 0 & 0 & 0 & \tilde{e}_R + \theta e_R \\
0 & 0 & 0 & 0 & 0 & \bar{\tilde{e}}_R + {\bar\theta}{e}_R^\dagger & 0
\end{array}\right).
\end{equation}

 \begin{equation}
 \label{eq:Q_R}
 {\cal A}_4 =
\left(\begin{array}{ccccccc}
 0 & 0 & 0 & 0 & 0 & 0 & \tilde{u}_R + \theta u_R\\
 0 & 0 & 0 & 0 & 0 & 0 & 0\\
 0 & 0 & 0 & 0 & 0 & 0 & 0\\
 0 & 0 & 0 & 0 & 0 & 0 & 0 \\
 0 & 0 & 0 & 0 & 0 & 0 \\
 0 & 0 & 0 & 0 &  0 & 0 & 0 \\
 \bar{\tilde{u}}_R + {\bar\theta}u_R^\dagger & 0 & 0 & 0 & 0 & 0 & 0
 \end{array} \right).
\end{equation}
Unlike the ${\cal A}_I$  which correspond to matter fields
corresponding to open strings stretched between two stacks of $D7$
branes (with different two-form fluxes turned on their world
volumes), the Higgses would arise as a geometric moduli corresponding
to the fluctuations in the position of the mobile space-time filling
$D3$-brane.

Now, we will see if one can construct appropriate  $a_I$ (bi-fundamental)
and $z_i$(mimicking bi-fundamental fields) such that
(\ref{eq:eigenvectors}), (\ref{eq:L_L}) - (\ref{eq:Q_R}), as well as:
\begin{equation}
\label{eq:H_u}
{\cal Z}_1=\left(\begin{array}{ccccccc}
 0 & 0 & 0 & 0 & 0 & 0 & 0\\
 0 & 0 & 0 & 0 & 0 & 0 & 0\\
 0 & 0 & 0 & 0 & 0 & 0 & 0\\
 0 & 0 & 0 & 0 & 0 & H_u + \theta\tilde{H}_u   & 0\\
 0 & 0 & 0 & 0 & 0 & 0 & 0\\
 0 & 0 & 0 & \bar{H}_u + {\bar\theta}\tilde{H}_u^\dagger  &  0 & 0 & 0\\
 0 & 0 & 0 & 0 & 0 & 0 & 0
 \end{array} \right),
\end{equation}
and
\begin{equation}
\label{eq:H_d}
{\cal Z}_2=\left(\begin{array}{ccccccc}
 0 & 0 & 0 & 0 & 0 & 0 & 0\\
 0 & 0 & 0 & 0 & 0 & 0 & 0\\
 0 & 0 & 0 & 0 & 0 & 0 & 0\\
 0 & 0 & 0 & 0 & 0 & 0   & 0\\
 0 & 0 & 0 & 0 & 0 & H_d + \theta\tilde{H}_d & 0\\
 0 & 0 & 0 &0  &  \bar{H}_d + {\bar\theta}\tilde{H}_d^\dagger & 0 & 0\\
 0 & 0 & 0 & 0 & 0 & 0 & 0
 \end{array} \right),
\end{equation}
 are satisfied. Guided by (\ref{eq:Q L_I}), we will make the following
 ansatze for $a_I$ and $z_i$:
\begin{eqnarray}
\label{eq:a_I_z_i}
& & a_I=\left(\begin{array}{ccccccc}
0 & 0 & 0 & a_I^{14} & a_I^{15} & a_I^{16} & a_I^{17} \\
0 & 0 & 0 & a_I^{24} & a_I^{25} & a_I^{26} & a_I^{27} \\
0 & 0 & 0 & a_I^{34} & a_I^{35} & a_I^{36} & a_I^{37} \\
{\bar a}_I^{14} & {\bar a}_I^{24} & {\bar a}_I^{34} & 0 & 0 & a_I^{46} & a_I^{47} \\
{\bar a}_I^{15} & {\bar a}_I^{25} & {\bar a}_I^{35} & 0 & 0 & a_I^{56} & a_I^{57} \\
{\bar a}_I^{16} & {\bar a}_I^{26} & {\bar a}_I^{36} & {\bar a}_I^{46}
& {\bar a}_I^{56} & 0 & a_I^{67}\\
{\bar a}_I^{17} & {\bar a}_I^{27} & {\bar a}_I^{37} & {\bar a}_I^{47} & {\bar a}_I^{57} & {\bar a}_I^{67} & 0
\end{array}\right),\nonumber\\
& & z_i=\left(\begin{array}{ccccccc}
0 & 0 & 0 & z_i^{14} & z_i^{15} & z_i^{16} & z_i^{17} \\
0 & 0 & 0 & z_i^{24} & z_i^{25} & z_i^{26} & z_i^{27} \\
0 & 0 & 0 & z_i^{34} & z_i^{35} & z_i^{36} & z_i^{37} \\
{\bar z}_i^{14} & {\bar z}_i^{24} & {\bar z}_i^{34} & 0 & 0 & z_i^{46} & z_i^{47} \\
{\bar z}_i^{15} & {\bar z}_i^{25} & {\bar z}_i^{35} & 0 & 0 & z_i^{56} & z_i^{57} \\
{\bar z}_i^{16} & {\bar z}_i^{26} & {\bar z}_i^{36} & {\bar z}_i^{46}
& {\bar z}_i^{56} & 0 & z_i^{67}\\
{\bar z}_i^{17} & {\bar z}_i^{27} & {\bar z}_i^{37} & {\bar z}_i^{47} & {\bar z}_i^{57} & {\bar z}_i^{67} & 0
\end{array}\right).
\end{eqnarray}
Using (\ref{eq:a_I_z_i}) in (\ref{eq:eigenvectors}), one obtains
$17\times6=102$ equations in $102$ variables. The same can be solved
(using Mathematica) to yield the following solution:
\begin{eqnarray*}
\label{eq:sols_a_I}
& &\hskip-0.5in  a_1 = \left(\begin{array}{ccccccc}
0 & 0 & 0 & \xi_1^{14}{\cal V}^{-\frac{7}{5}} \tilde{u}_L & 0 & 0 &
\xi_1^{17}{\cal V}^{-\frac{11}{5}} \tilde{u}_R \\
0 & 0 & 0 & 0 & 0 & 0 & 0 \\
0 & 0 & 0 & 0 & 0 & 0 & 0 \\
\xi_1^{14}{\cal V}^{-\frac{7}{5}} {\bar{\tilde{u}}}_L & 0 & 0 & 0 & 0 &
\xi_1^{46}\left(\frac{\tilde{e}_L}{2} + {\cal V}^{-\frac{8}{5}}H_u\right) & 0 \\
0 & 0 & 0 & 0 & 0 & \xi_1^{56}{\cal V}^{-\frac{8}{5}} H_d & 0 \\
0 & 0 & 0 &{\bar\xi}_1^{46}\left( \frac{{\bar{\tilde{e}}}_L}{2} + {\cal V}^{-\frac{8}{5}}{\bar H}_u\right) &
{\bar\xi}_1^{56}{\cal V}^{-\frac{8}{5}} {\bar H}_d & 0 & -\xi_1^{67} {\cal V}^{-\frac{4}{5}} e_R
\\
{\bar\xi}_1^{17}{\cal V}^{-\frac{11}{5}} {\bar{\tilde{u}}}_R & 0 & 0 & 0 & 0 &
-{\bar\xi}_1^{67}  {\cal V}^{-\frac{4}{5}} {\bar{\tilde{e}}}_R
\end{array}\right),\nonumber\\
& &\hskip-0.5in a_2 = \left(\begin{array}{ccccccc}
0 & 0 & 0 & 0.3\xi_2^{14} \tilde{u}_L & 0 & 0 &
\xi_2^{17}{\cal V}^{-\frac{13}{5}} \tilde{u}_R \\
0 & 0 & 0 & 0 & 0 & 0 & 0 \\
0 & 0 & 0 & 0 & 0 & 0 & 0 \\
0.3\xi_2^{14} {\bar{\tilde{u}}}_L & 0 & 0 & 0 & 0 &
\xi_2^{46}\left(-{\cal V}^{-\frac{7}{5}}\tilde{e}_L + {\cal V}^{-\frac{4}{5}}H_u\right) & 0 \\
0 & 0 & 0 & 0 & 0 & 5\xi_2^{56}{\cal V}^{-\frac{6}{5}} H_d & 0 \\
0 & 0 & 0 &{\bar\xi}_2^{46}\left( -{\cal V}^{-\frac{7}{5}}{\bar{\tilde{e}}}_L +
  {\cal V}^{-\frac{4}{5}}{\bar H}_u\right) &
5{\bar\xi}_2^{56}{\cal V}^{-\frac{6}{5}} {\bar H}_d & 0 & -\xi_2^{67} {\cal V}^{-\frac{9}{5}} e_R
\\
{\bar\xi}_2^{17}{\cal V}^{-\frac{13}{5}} {\bar{\tilde{u}}}_R & 0 & 0 & 0 & 0 &
-{\bar\xi}_2^{67}  {\cal V}^{-\frac{9}{5}} {\bar{\tilde{e}}}_R
\end{array}\right),\nonumber\\
& &\hskip-0.5in a_3 = \left(\begin{array}{ccccccc}
0 & 0 & 0 & \xi_3^{14}\frac{\tilde{u}_L}{{\cal V}} & 0 & 0 &
-6\xi_3^{17} \frac{\tilde{u}_R}{{\cal V}} \\
0 & 0 & 0 & 0 & 0 & 0 & 0 \\
0 & 0 & 0 & 0 & 0 & 0 & 0 \\
{\bar\xi}_3^{14}\frac{{\bar{\tilde{u}}}_L}{{\cal V}} & 0 & 0 & 0 & 0 &
\xi_3^{46}\left(-{\cal V}^{-\frac{4}{5}}\tilde{e}_L + {\cal V}^{-\frac{8}{5}}H_u\right) & 0 \\
0 & 0 & 0 & 0 & 0 & \xi_3^{56}{\cal V}^{-\frac{11}{5}} H_d & 0 \\
0 & 0 & 0 &{\bar\xi}_3^{46}\left(-{\cal V}^{-\frac{4}{5}}\tilde{e}_L + {\cal V}^{-\frac{8}{5}}H_u\right) &
5{\bar\xi}_3^{56}{\cal V}^{-\frac{11}{5}} {\bar H}_d & 0 & 0.3\xi_3^{67} e_R
\\
{\bar\xi}_3^{17}{\cal V}^{-\frac{13}{5}} {\bar{\tilde{u}}}_R & 0 & 0 & 0 & 0 &
0.3{\bar\xi}_3^{67}  {\bar{\tilde{e}}}_R
\end{array}\right),
\end{eqnarray*}
\begin{eqnarray}
\label{eq:sols_a_I_ii}
& &\hskip-0.5in a_4 = \left(\begin{array}{ccccccc}
0 & 0 & 0 & 0.08\xi_4^{14}\tilde{u}_L & 0 & 0 &
0.2\xi_4^{17} \tilde{u}_R \\
0 & 0 & 0 & 0 & 0 & 0 & 0 \\
0 & 0 & 0 & 0 & 0 & 0 & 0 \\
0.08{\bar\xi}_4^{14}{\bar{\tilde{u}}}_L & 0 & 0 & 0 & 0 &
\xi_4^{46}{\cal V}^{-\frac{7}{5}}\tilde{e}_L & 0 \\
0 & 0 & 0 & 0 & 0 & \xi_4^{56}{\cal V}^{-\frac{14}{5}} H_d & 0 \\
0 & 0 & 0 &{\bar\xi}_4^{46}{\cal V}^{-\frac{7}{5}}{\bar{\tilde{e}}}_L &
{\bar\xi}_4^{56}{\cal V}^{-\frac{14}{5}} {\bar H}_d & 0 & -7\xi_4^{67} e_R
\\
0.2{\bar\xi}_4^{17} {\bar{\tilde{u}}}_R & 0 & 0 & 0 & 0 &
-7{\bar\xi}_4^{67}  {\bar{\tilde{e}}}_R
\end{array}\right),
\end{eqnarray}
and
\begin{eqnarray}
\label{eq:sols z_i}
& & z_1=\left(\begin{array}{ccccccc}
0 & 0 & 0 & \alpha_1^{14}\frac{\tilde{u}_L}{{\cal V}} & 0 & 0 &
5\alpha_1^{17}{\cal V}^{-\frac{14}{5}} \tilde{u}_R \\
0 & 0 & 0 & 0 & 0 & 0 & 0 \\
0 & 0 & 0 & 0 & 0 & 0 & 0 \\
{\bar\alpha}_1^{14}\frac{{\bar{\tilde{u}}}_L}{{\cal V}} & 0 & 0 & 0 & 0 &
\alpha_1^{46}\left({\cal V}^{-\frac{9}{5}}\tilde{e}_L - \frac{H_u}{\sqrt{2}}\right) & 0 \\
0 & 0 & 0 & 0 & 0 & \alpha_1^{56} \frac{H_d}{\sqrt{2}} & 0 \\
0 & 0 & 0 &{\bar\alpha}_1^{46}\left({\cal V}^{-\frac{9}{5}}{\bar\tilde{ e}}_L -
  \frac{{\bar H}_u}{\sqrt{2}}\right) &
{\bar\alpha}_1^{56} \frac{{\bar H}_d}{\sqrt{2}} & 0 & \alpha_1^{67} {\cal
  V}^{-\frac{11}{5}}e_R
\\
5{\bar\alpha}_1^{17}{\cal V}^{-\frac{14}{5}} {\bar{\tilde{u}}}_R & 0 & 0 & 0 & 0 &
{\bar\alpha}_1^{67} {\cal
  V}^{-\frac{11}{5}}  {\bar{\tilde{e}}}_R
\end{array}\right),\nonumber\\
& & z_2=\left(\begin{array}{ccccccc}
0 & 0 & 0 & \alpha_2^{14}{\cal V}^{-\frac{4}{5}}{\tilde{u}_L} & 0 & 0 &
5\alpha_2^{17}{\cal V}^{-\frac{13}{5}} \tilde{u}_R \\
0 & 0 & 0 & 0 & 0 & 0 & 0 \\
0 & 0 & 0 & 0 & 0 & 0 & 0 \\
{\bar\alpha}_2^{14}{\cal V}^{-\frac{4}{5}}{\bar{u}_L} & 0 & 0 & 0 & 0 &
\alpha_2^{46}\left(6{\cal V}^{-\frac{8}{5}}\tilde{e}_L - \frac{H_u}{\sqrt{2}}\right) & 0 \\
0 & 0 & 0 & 0 & 0 & -\alpha_2^{56} \frac{H_d}{\sqrt{2}} & 0 \\
0 & 0 & 0 &{\bar\alpha}_2^{46}\left({\cal V}^{-\frac{8}{5}}\bar\tilde{ e}_L -
  \frac{{\bar H}_u}{\sqrt{2}}\right) &
-{\bar\alpha}_2^{56} \frac{{\bar H}_d}{\sqrt{2}} & 0 &
\alpha_2^{67}\frac{\tilde{e}_R}{{\cal V}^{-2}}
\\
5{\bar\alpha}_2^{17}{\cal V}^{-\frac{13}{5}} {\bar{\tilde{u}}}_R & 0 & 0 & 0 & 0 &
{\bar\alpha}_2^{67}\frac{\bar\tilde{e}_R}{{\cal V}^{-2}}
\end{array}\right)\nonumber\\
& &
\end{eqnarray}
In (\ref{eq:sols_a_I_ii}) and (\ref{eq:sols z_i}), $\xi_{I,i}^{ab},
1\leq a\leq 6, 1\leq b\leq7$ are ${\cal O}(1)$ numbers. Now, from (\ref{eq:CIJbar_I}), the intersection matrix can be written as:
\begin{equation}
\label{eq:CIJbar}
C^{I{\bar J}}=\left(
\begin{array}{cccc}
 c_{11} {\cal V}^{10/9} & c_{12} {\cal V}^{11/18} & c_{13} {\cal V}^{29/18} &
   c_{14} {\cal V}^{19/9} \\
 c_{12} {\cal V}^{11/18} & c_{22} \sqrt[9]{V} & c_{23} {\cal V}^{10/9} &
   c_{24} {\cal V}^{29/18} \\
 c_{13} {\cal V}^{29/18} & c_{23} {\cal V}^{10/9} & c_{33} {\cal V}^{10/9} &
   c_{34} {\cal V}^{47/18} \\
 c_{14} {\cal V}^{19/9} & c_{24} {\cal V}^{29/18} & c_{34} {\cal V}^{47/18} &
   c_{44} {\cal V}^{28/9}
\end{array}
\right),
\end{equation}
where $c_{ab}, 1\leq a,b\leq4$ are ${\cal O}(1)$ numbers. Assuming
that the complex structure moduli
$z^{\tilde{a}=1,...,h^{2,1}_-(CY_3)}$ are stabilized at very small
values, which is in fact already assumed in writing
(\ref{eq:sigmas_Ts}) which has been written upon inclusion of terms up
to linear in the complex structure moduli, let us define a modified
intersection matrix in the $a_I-z_i$ moduli space:
\begin{eqnarray}
\label{eq:CIJbar mod}
& & {\cal C}^{{\cal I}{\cal J}}=C^{I{\bar J}},\ {\cal I}=I, {\cal
  J}={\bar J};\nonumber\\
& &  {\cal C}^{{\cal I}{\cal J}}
=\mu_3\left(2\pi\alpha^\prime\right)^2\left(\omega_\alpha
\right)^{i{\bar j}},\ {\cal I}=i, {\cal
  J}={\bar j};\nonumber\\
& &  {\cal C}^{{\cal I}{\cal J}}=0,\ {\cal I}=I, {\cal
  J}={\bar j},\ {\rm etc.}.
\end{eqnarray}
Now, from (\ref{eq:sols_a_I}) and (\ref{eq:sols z_i}), one can show
that for $|z_i|\sim0.8{\cal V}^{\frac{1}{36}},\ {\cal V}\sim10^5$:
\begin{eqnarray}
\label{eq:tr-1}
& & {\cal C}^{{\cal I}\bar{\cal J}}Tr({\cal M}_{\cal I} {\cal M}_{\cal
  J}^\dagger)\sim
C^{a_1\bar{a}_1}|\tilde{e}_L|^2 + C^{a_2\bar{a}_2}|\tilde{u}_L|^2 +
C^{a_3\bar{a}_3}|\tilde{e}_R|^2
+ C^{a_4\bar{a}_4}|\tilde{u}_R|^2 +\mu_3\left(2\pi\alpha^\prime\right)^2|H_u|^2,
\end{eqnarray}
where ${\cal M}_{\cal I}\equiv a_I, z_i$. Now, using (\ref{eq:eigenvectors}), one can show that in the large volume limit:
 ${\cal C}^{{\cal A}_I\bar{\cal A}_{\bar J}}\sim C^{a_I\bar{a}_{\bar J}}, {\cal C}^{{\cal Z}_i\bar{\cal Z}_{\bar j}}\sim {\cal C}^{z_i\bar{z}_j}$. Clubbing together the Wilson line moduli and the $D3$-brane position moduli into a single vector: ${\cal M}_{\Lambda}\equiv {\cal A}_I,{\cal Z}_i$, then one sees from (\ref{eq:L_L}) - (\ref{eq:H_d}), and (\ref{eq:tr-1}):
  \begin{equation}
  \label{eq:tr-2}
  {\cal C}^{\Lambda\bar{\Sigma}}Tr\left({\cal M}_\Lambda{\cal M}_{\Sigma}^\dagger\right)\sim {\cal C}^{{\cal I}\bar{\cal J}}Tr({\cal M}_{\cal I} {\cal M}_{\cal
  J}^\dagger).
  \end{equation}
  In the large volume and rigid limit of $\Sigma_B(\zeta^A=0$ which corresponds to a local minimum), perhaps $ {\cal C}^{\Lambda\bar{\Sigma}}Tr\left({\cal M}_\Lambda{\cal M}_{\Sigma}^\dagger\right)$ is invariant under moduli transformations in the $({\cal A}_I,{\cal Z}_i)/(a_I,z_i)$-subspace of the open-string moduli space, which would imply that $C^{I{\bar J}}a_I{\bar a}_{\bar J} + \mu_3\left(\alpha^\prime\right)^2\left(\omega_B\right)_{i{\bar j}}z^i{\bar z}^{\bar j}$ for multiple  $D7$-branes, in a basis that diagonalizes $g_{{\cal M}_{\cal I}\bar{\cal M}_{\bar J}}$ at stabilized values of the open string moduli,  is replaced by (\ref{eq:tr-1}).

For the purpose of evaluation of (N)LSP decays in the subsequent sections, we will be using the following terms (written out in four-component notation or their two-component analogs and utilizing/generalizing results of \cite{Jockers_thesis}) in the ${\cal N}=1$ gauged supergravity action of Wess and Bagger \cite{Wess_Bagger} with the understanding that $m_{{\rm moduli/modulini}}<<m_{\rm KK}\left(\sim\frac{M_s}{{\cal V}^{\frac{1}{6}}}\Biggr|_{{\cal V}\sim10^{5/6}}\sim10^{14}GeV\right),M_s=\frac{M_p}{\sqrt{{\cal V}}}\Biggr|_{{\cal V}\sim10^{5/6}}\sim10^{15}GeV$, and that for multiple $D7$-branes, the non-abelian gauged isometry group\footnote{As explained in  \cite{Jockers_thesis}, one of the two Pecci-Quinn/shift symmetries along the RR two-form axions $c^a$ and the zero-form axion $\rho_B$ gets gauged due to the dualization of the Green-Schwarz term $\int_{{\bf R}^{1,3}}dD^{(2)}_B\wedge A$ coming from the KK reduction of the Chern-Simons term on $\Sigma_B\cup\sigma(\Sigma_B)$ - $D^{(2)}_B$ being an RR two-form axion. In the presence of fluxes (\ref{eq:flux}) for multiple $D7$-brane fluxes, the aforementioned Green-Schwarz is expected to be modified to $Tr\left(Q_B\int_{{\bf R}^{1,3}}dD^{(2)}_B\wedge A\right)$, which after dualization in turn modifies the covariant derivative of $T_B$ and hence the killing isometry.}, corresponding to the killing vector $6i\kappa_4^2\mu_7\left(2\pi\alpha^\prime\right)Q_B\partial_{T_B}, Q_B=\left(2\pi\alpha^\prime\right)\int_{\Sigma_B}i^*\omega_B\wedge P_-\tilde{f}$ arising due to the elimination of of the two-form axions $D_B^{(2)}$ in favor of the zero-form axions
$\rho_B$ under the KK-reduction of the ten-dimensional four-form axion \cite{Jockers_thesis} (which results in a modification of the covariant derivative of $T_B$ by an additive shift given by $6i\kappa_4^2\mu_7\left(2\pi\alpha^\prime\right)Tr(Q_B A_\mu)$)  can be identified with the SM group (i.e. $A_\mu$ is the SM-like adjoint-valued gauge field \cite{Wess_Bagger}):
\begin{eqnarray}
\label{eq:WB_gSUGRA N=1}
& & {\hskip -0.3in}{\cal L} = g_{YM}g_{T_B {\bar{\cal J}}}Tr\left(X^{T_B}{\bar\chi}^{\bar{\cal J}}_L\lambda_{\tilde{g},\ R}\right) +ig_{{\cal I}\bar{\cal J}}Tr\left({\bar\chi}^{\bar{\cal I}}_L\left[\slashed{\partial}\chi^{\cal I}_L+\Gamma^i_{Mj}\slashed{\partial} a^M\chi^{\cal J}_L
+\frac{1}{4}\left(\partial_{a_M}K\slashed{\partial} a_M - {\rm c.c.}\right)\chi^{\cal I}_L\right] \right) \nonumber\\
& & {\hskip -0.3in}+\frac{e^{\frac{K}{2}}}{2}\left({\cal D}_{\bar{\cal I}}D_{\cal J}\bar{W}\right)Tr\left(\chi^{\cal I}_L\chi^{\cal J}_R\right) + g_{T_B\bar{T}_B}Tr\left[\left(\partial_\mu T_B - A_\mu X^{T_B}\right)\left(\partial^\mu T_B - A^\mu X^{T_B}\right)^\dagger\right] \nonumber\\
& & + g_{T_B{\cal J}}Tr\left(X^{T_B}A_\mu\bar{\chi}^{\cal J}_L\gamma^\nu\gamma^\mu\psi_{\nu,\ R}\right) + \bar{\psi}_{L,\ \mu}\sigma^{\rho\lambda}\gamma^\mu\lambda_{\tilde{g},\ L}F_{\rho\lambda} + \bar{\psi}_{L,\ \mu}\sigma^{\rho\lambda}\gamma^\mu\lambda_{\tilde{g},\ L}W^+_\rho W^-_\lambda\nonumber\\
 & & + Tr\left[\bar{\lambda}_{\tilde{g},\ L}\slashed{A}\left(6\kappa_4^2\mu_7(2\pi\alpha^\prime)Q_BK + \frac{12\kappa_4^2\mu_7(2\pi\alpha^\prime)Q_Bv^B}{\cal V}\right) \lambda_{\tilde{g},\ L}\right]\nonumber\\
 & & + \frac{e^K G^{T_B\bar{T}_B}}{\kappa_4^2}6i\kappa_4^2(2\pi\alpha^\prime)Tr\left[Q_BA^\mu\partial_\mu
 \left(\kappa_4^2\mu_7(2\pi\alpha^\prime)^2C^{I\bar{J}}a_I\bar{a}_{\bar J}\right)\right] + {\rm h.c.},
\end{eqnarray}

To evaluate the contribution of various 3-point interaction vertices in context of ${\cal N}=1$ supergravity defined above, one needs to evaluate the full moduli space metric in fluctuations linear in different kind of moduli corresponding to different SM particles . The expansion of moduli space metric, its inverse and derivatives w.r.t to each moduli in terms of fluctuations linear in $z_i\rightarrow z_1 + {\cal V}^\frac{1}{36}{M_p}, a_1\rightarrow a_1 + {\cal V}^{-\frac{2}{9}}{M_p},a_2\rightarrow a_2 + {\cal V}^-{\frac{1}{3}}{M_p}, a_3\rightarrow a_3 + {\cal V}^{-\frac{13}{18}}{M_p},a_3\rightarrow a_3 + {\cal V}^{-\frac{11}{9}}{M_p}$ are given in appendix D. One can show that near (\ref{eq:near_slag_i}) along which $|z_1|$ and $|z_2|$ are on the same footing.

\section{Gravitino Decays}

The generation of mass scales of superpartners and realization of $\mu$ split SUSY in the context of  Big Divisor $D3-D7$ type IIB string compactifications provides us with the gravitino as the L(ightest) S(upersymmetric) P(article) and sleptons/squarks as N(ext-to) L(ightest) S(upersymmetric) P(article)s with  (Bino/Wino-type)gaugino/(Bino/Wino-type)gaugino-dominant neutralino, in the dilute-flux approximation,  only differing in their masses  by an ${\cal O}(1)$ factor. This helps in shedding some light on identifying a  viable dark matter candidate in this framework. Decays of the gravitino, are, in general, driven by any of the trilinear R-violating couplings which  necessarily involve squark or slepton as propagators. In MSSM, stability of LSP is governed by conserved R-parity however same is not a restrictive condition in `` $\mu$ split SUSY''.  Some of the R-parity-violating interactions are not necessarily negligible  and the consideration of LSP to be a viable dark matter candidate needs the contribution of the same to be evaluated. Here, the large squark masses in $\mu$-split SUSY helps to suppress the decay width. In BSM/2HDM models, one considers R-parity violating coupling generated from R-parity violating superpotential. However, here we consider the possible R-parity violating couplings in the effective ${\cal N}=1$ gauged supergravity action and by evaluating the same, we explicitly calculate the decay width and hence life time of gravitino.

In this section, we discuss R-parity violating two-body and three-body gravitino decays and show that the corresponding lifetimes are either of the order or more than the age of the universe which supports the argument that gravitino can exist as a viable dark matter candidate.

\subsection{Two-Body Gravitino Decays}

We discuss the decays of the gravitino into neutrino and gauge bosons as well as the light Higgs and neutrinos.
\begin{figure}
   \begin{center}
    \begin{picture}(135,137) (15,-18)
   \Line(100,50)(115,24)
   \DashArrowLine(125,41)(115,24){4}
   \Vertex(115,24){2.5}
   \Text(136,19)[]{{{$q$}}}
   \LongArrow(126,24)(135,8)
   \ArrowLine(115,24)(130,-2)
   \Photon(100,50)(115,24){5}{2}
   \Text(102,84)[]{{{$k$}}}
   \LongArrow(103,73)(112,88)
   \Photon(100,50)(130,102){5}{4}
   \Text(30,50)[]{{{$\psi_{\mu}$}}}
   \Text(70,65)[]{{{$p$}}}
   \LongArrow(61,59)(79,59)
   \Line(40,52)(100,52)\Line(40,48)(100,48)
   \Vertex(100,50){2.5}
   \Text(130,50)[]{{{$\left\langle\tilde{\nu}\right\rangle$}}}
   \Text(135,-10)[]{{{$\nu$}}}
   \Text(135,111)[]{{{$\gamma$}}}
   \Text(97,30)[]{{{$\tilde{\lambda}^0$}}}
  \end{picture}
  \caption{Two-body gravitino decay: $\psi_\mu\rightarrow\nu+\gamma$}
  \end{center}
  \end{figure}
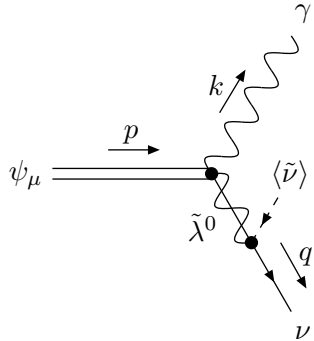
\newpage
In ${\cal N}=1$ SUGRA the gaugino-$\langle$sneutrino$\rangle$-neutrino vertex, Fig. 16,
\begin{figure}
\begin{center}
    \begin{picture}(135,137) (15,-18)
   \Line(100,50)(115,24)
   \Photon(100,50)(115,24){5}{2}
   \DashArrowLine(125,41)(115,24){4}
   \Vertex(115,24){2.5}
   \Text(130,50)[]{{{$\left\langle\tilde{\nu}\right\rangle$}}}
   \ArrowLine(115,24)(130,-2)
   \Text(135,-10)[]{{{$\nu$}}}
   \Text(97,30)[]{{{$\tilde{\lambda}^0$}}}
   \end{picture}
   \end{center}
   \caption{Gaugino-$\langle$sneutrino$\rangle$-neutrino vertex}
   \end{figure}
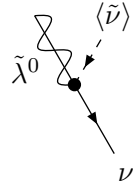
is given by:
\begin{equation}
\label{eq:gaugino-sneu-neu}
{\cal O}(a_1)-{\rm term\ in}\ \frac{g_{YM}g_{T^B{\bar a}_1}X^{T^B}}{\left(\sqrt{K_{{\cal A}_1{\bar{\cal A}}_1}}\right)^2}
\sim\frac{{\cal V}^{-\frac{11}{9}}}{10^4}.
\end{equation}
Similarly the gravitino-gauge-boson-$\langle\tilde{\nu}\rangle-\nu$-vertex in the ${\cal N}=1$ SUGRA Lagrangian is given by the ${\cal O}(a_1)$-term in $g_{YM}g_{a_1{\bar T}^{B}}X^{T^B}Z^0/A_\mu{\bar\nu}\gamma^\nu\gamma^\mu\psi_\nu$, which is: $\frac{g_{YM}g_{a_1{\bar T}^B}X^{T^B}}{\left(\sqrt{K_{{\cal A}_1{\bar{\cal A}}_1}}\right)^2}\sim\frac{{\cal V}^{-\frac{11}{9}}}{10^4}$.
Using \cite{Grefe_i}, one sees that the lifetime of the decay $\psi_\mu\rightarrow\gamma+\nu_e$ is given by:
\begin{eqnarray}
\label{eq:psi-nu+phot}
& & \Gamma\left(\psi_\mu\rightarrow\gamma+\nu_e\right)=\frac{1}{64\pi}\left(\frac{\langle {\cal A}_1\rangle}{\langle {\cal Z}_i\rangle}\right)^2\frac{m^3_{3/2}}{M_p^2}\left|U_{\tilde{\gamma}\tilde{Z}}\right|^2\left(\frac{V^{-\frac{11}{9}}}{10^4}\right)^2
\nonumber\\
& & \sim\frac{1}{64\pi}\left(\frac{\langle{\cal V}^{-\frac{2}{9}}\rangle}{{\cal V}^{\frac{1}{36}}}\right)^2\frac{m^2_{3/2}}{M_p^2}
m_Z^2\left(\frac{M_{\lambda_1}-M_{\lambda_2}}{M_{\lambda_1}M_{\lambda_2}}\right)^2sin^2\theta_W cos^2\theta_W
\left(\frac{V^{-\frac{11}{9}}}{10^4}\right)^2.
\end{eqnarray}
We have a gaugino mass degeneracy (up to ${\cal O}(1)$ factors) implying: $M_{\lambda_1}\sim M_{\lambda_2}$. One hence gets a very small decay width and an extremely enhanced lifetime.

   \begin{figure}
   \begin{center}
 \parbox{6.3cm}{
  \begin{picture}(135,137) (15,-18)
   \Line(100,50)(115,24)
   \DashArrowLine(125,41)(115,24){4}
   \Vertex(115,24){2.5}
   \Text(136,19)[]{{{$q$}}}
   \LongArrow(126,24)(135,8)
   \ArrowLine(115,24)(130,-2)
   \Photon(100,50)(115,24){5}{2}
   \Text(102,84)[]{{{$k$}}}
   \LongArrow(103,73)(112,88)
   \Photon(100,50)(130,102){5}{4}
   \Text(30,50)[]{{{$\psi_{\mu}$}}}
   \Text(70,65)[]{{{$p$}}}
   \LongArrow(61,59)(79,59)
   \Line(40,52)(100,52)\Line(40,48)(100,48)
   \Vertex(100,50){2.5}
   \Text(130,50)[]{{{$\left\langle\tilde{\nu}\right\rangle$}}}
   \Text(135,-10)[]{{{$\nu$}}}
   \Text(135,111)[]{{{$Z^0$}}}
   \Text(97,30)[]{{{$\tilde{\chi}^0$}}}
  \end{picture}
 }\;+\;\parbox{6.3cm}{
  \begin{picture}(155,137) (15,-18)
   \Text(30,50)[]{{{$\psi_{\mu}$}}}
   \Text(70,65)[]{{{$p$}}}
   \LongArrow(61,59)(79,59)
   \Line(40,52)(100,52)\Line(40,48)(100,48)
   \Vertex(100,50){2.5}
   \Text(135,-10)[]{{{$\nu_{\tau}$}}}
   \DashArrowLine(130,50)(100,50){4}
   \Text(142,50)[]{{{$\left\langle\tilde{\nu}_{\tau}\right\rangle$}}}
   \Text(128,32)[]{{{$q$}}}
   \LongArrow(118,36)(127,21)
   \ArrowLine(100,50)(130,-2)
   \Text(102,84)[]{{{$k$}}}
   \LongArrow(103,73)(112,88)
   \Photon(100,50)(130,102){5}{4}
   \Text(135,111)[]{{{$Z^0$}}}
  \end{picture}
 }.
 \caption{Two-body gravitino decay: $\psi_\mu\rightarrow Z^0+\nu$}
 \end{center}
 \end{figure}
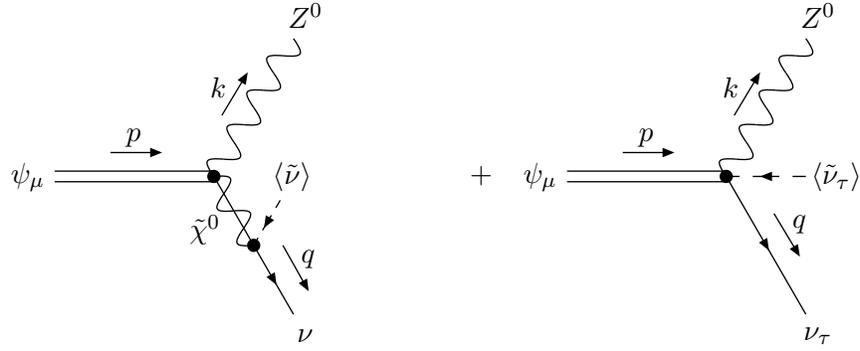
Again, using (\ref{eq:gaugino-sneu-neu}) and \cite{Grefe_i}, the decay width for $\psi_\mu\rightarrow Z+\nu$ is given by:
\begin{eqnarray}
\label{eq:gravitino_Z+nu}
& & \Gamma\left(\psi_\mu\rightarrow Z+\nu\right)\sim \frac{1}{64\pi}\left(\frac{\langle {\cal A}_1\rangle}{\langle {\cal Z}_i\rangle}\right)^2\frac{m^3_{3/2}}{M_p^2}\left(1-\frac{M^2_Z}{m^2_{3/2}}\right)^2\left(\frac{V^{-\frac{11}{9}}}{10^4}\right)^2
\nonumber\\
& &\hskip-1in \times\left[\left(\frac{M_Z}{M_{\lambda}}\right)^2\times{\cal O}(1) + {\cal O}(1)\left|1 + sin\beta cos\beta \left(\frac{M^2_Z}{M_\lambda\hat{\mu}_{{\cal Z}_1{\cal Z}_2}}\right)\right|^2 + \left(\frac{M_Z}{m_{3/2}}\right)\left(\frac{m_Z}{M_\lambda}\right)\left(1 + sin\beta cos\beta \left(\frac{M^2_Z}{M_\lambda\hat{\mu}_{{\cal Z}_1{\cal Z}_2}}\right)\right)\times{\cal O}(1)\right]\nonumber\\
& & \sim\frac{{\cal V}^{-\frac{1}{2}-6-\frac{22}{9}}}{64\pi\times 10^{8}}\times {\cal O}(1)M_p.
\end{eqnarray}
Using  $M_Z\sim 90 GeV$; $m_{ \lambda}\sim {\cal V}^{\frac{2}{3}} m_{\frac{3}{2}} $ and  $\hat{\mu}_{{\cal Z}_1{\cal Z}_2}\sim {\cal V}m_{\frac{3}{2}}$ from equations (\ref{eq:gaugino_mass}) and (\ref{eq:muhat_Z1Z2})
\begin{eqnarray}
\label{eq:gravitino_Z+nu1}
& & \Gamma\sim\frac{{\cal V}^{-\frac{1}{2}-6-\frac{22}{9}}}{64\pi\times 10^{8}}\times {\cal O}(1)M_p.
\end{eqnarray}
 This, for ${\cal V}\sim10^5$ yields $\tau\sim 10^{22}s$.
 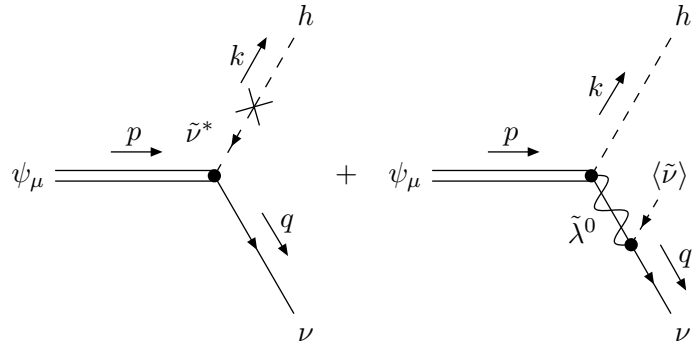
\begin{figure}
 \begin{center}
 \parbox{4.5cm}{
  \begin{picture}(135,137) (15,-18)
   \Text(30,50)[]{{{$\psi_{\mu}$}}}
   \SetWidth{0.5}
   \Text(70,65)[]{{{$p$}}}
   \LongArrow(61,59)(79,59)
   \Line(40,52)(100,52)\Line(40,48)(100,48)
   \Vertex(100,50){2.5}
   \Text(135,-10)[]{{{$\nu$}}}
   \Text(128,32)[]{{{$q$}}}
   \LongArrow(118,36)(127,21)
   \ArrowLine(100,50)(130,-2)
   \DashArrowLine(115,76)(100,50){4}
   \Text(109,96)[]{{{$k$}}}
   \LongArrow(110,85)(119,101)
   \DashLine(115,76)(130,102){4}
   \Line(108,74)(122,78)\Line(113,83)(117,69)
   \Text(135,111)[]{{{$h$}}}
   \Text(95,67)[]{{{$\tilde{\nu}^*$}}}
  \end{picture}
 }\;+\;\parbox{4.5cm}{
  \begin{picture}(135,137) (15,-18)
   \Line(100,50)(115,24)
   \DashArrowLine(125,41)(115,24){4}
   \Vertex(115,24){2.5}
   \Text(136,19)[]{{{$q$}}}
   \LongArrow(126,24)(135,8)
   \ArrowLine(115,24)(130,-2)
   \Photon(100,50)(115,24){5}{2}
   \Text(102,84)[]{{{$k$}}}
   \LongArrow(103,73)(112,88)
   \DashLine(100,50)(130,102){4}
   \Text(30,50)[]{{{$\psi_{\mu}$}}}
   \Text(70,65)[]{{{$p$}}}
   \LongArrow(61,59)(79,59)
   \Line(40,52)(100,52)\Line(40,48)(100,48)
   \Vertex(100,50){2.5}
   \Text(130,50)[]{{{$\left\langle\tilde{\nu}\right\rangle$}}}
   \Text(135,-10)[]{{{$\nu$}}}
   \Text(135,111)[]{{{$h$}}}
   \Text(97,30)[]{{{$\tilde{\lambda}^0$}}}
  \end{picture}
 }
\caption{Two-body gravitino decay: $\psi_\mu\rightarrow h + \nu$}
\end{center}
\end{figure}
The mass-like insertion is given by $m^2_{\tilde{\nu}h}$, which is the coefficient of $\tilde{\nu}h$ in the ${\cal N}=1$ SUGRA action. To estimate this, we note that near $|z_1|\sim|z_2|\sim0.7{\cal V}^{\frac{1}{36}},|a_1|\sim{\cal V}^{-\frac{2}{9}},|a_2|\sim{\cal V}^{-\frac{1}{9}},|a_3|\sim{\cal V}^{-\frac{13}{18}},|a_4|\sim{\cal V}^{-\frac{11}{9}}$, the potential can be approximated by
\begin{eqnarray}
\label{eq:Potential}
& & V\sim e^KG^{T_S{\bar T}_S}|D_{T_S}W|^2\nonumber\\
& & \hskip-1.6in\sim\frac{z^{72}e^{-2 n^s (vol(\Sigma_S) + \mu_3 z^2)}\sqrt{{\cal V}^{\frac{1}{18}}+\mu_3z^2}}{{\cal V}+\left({\cal V}^{\frac{2}{3}}+(\alpha_{12}{\cal V}^{\frac{5}{18}} + \alpha_{13}{\cal V}^{\frac{8}{9}} + \alpha_{14}{\cal V}^{\frac{8}{9}})a_1 + (\alpha_{11}{\cal V}^{\frac{10}{9}}a_1 + \alpha_{12}{\cal V}^{\frac{5}{18}}  + \alpha_{13}{\cal V}^{\frac{8}{9}} + \alpha_{14}{\cal V}^{\frac{8}{9}})|a_1| + \mu_3z^2\right)^{\frac{3}{2}} - \left({\cal V}^{\frac{1}{18}} + \mu_3 z^2\right)^{\frac{3}{2}}}.\nonumber\\
& &
\end{eqnarray}
The coefficient of the ${\cal O}\left((z - 0.6{\cal V}^{\frac{1}{36}})(a_1 - {\cal V}^{-\frac{2}{9}})\right)/\sqrt{K_{Z_1{\bar Z}_1}K_{{\cal A}_1{\bar{\cal A}}_1}}$ in
(\ref{eq:Potential}) turns out to be:
\begin{equation}
\label{eq:msnuhsquared}
m^2_{\tilde{\nu}h}\sim{\cal V}^{\frac{23}{18}}m^2_{3/2}\times10^{-16}.
\end{equation}

The gravitino-gaugino-Higgs vertex:
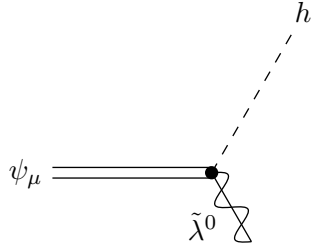
\begin{figure}
\begin{center}
\begin{picture}(135,137) (15,-18)
   \Line(100,50)(115,24)
   \Text(30,50)[]{{{$\psi_{\mu}$}}}
     \DashLine(100,50)(130,102){4}
  \Vertex(100,50){2.5}
   \Photon(100,50)(115,24){5}{2}
    \Line(40,52)(100,52)\Line(40,48)(100,48)
    \Text(135,111)[]{{{$h$}}}
    \Text(97,30)[]{{{$\tilde{\lambda}^0$}}}
   \end{picture}
   \end{center}
   \caption{Gravitino-gaugino-Higgs vertex}
   \end{figure}
will be given by the following term in the ${\cal N}=1$ SUGRA action:
\begin{equation}
\label{eq:grav-gaugino-h}
g_{YM}D^{T^B}{\bar\psi}_\mu\gamma^\mu\lambda=0,\ {\rm on\ shell\ gravitino}.
\end{equation}
Hence, the second diagram involving a gaugino NLSP, does not contribute. However, if the gaugino is replaced by a neutralino, then the Higgsino-component of this neutralino will contribute via the term:
\begin{equation}
\label{eq:grav-Higgsino in neutralino-Higgs-I}
g_{{\cal Z}_1{\bar Z}_1}\partial_\nu{\cal Z}_1{\bar\psi}_{\mu R}\gamma^\nu\gamma^\mu\tilde{H}^0_L + {\rm h.c.}
\end{equation}
in ${\cal N}=1$ SUGRA action. This yields a vertex:
\begin{equation}
\label{eq:grav-Higgsino in neutralino-Higgs-II}
\tilde{f}\frac{g_{{\cal Z}_1{\bar{\cal Z}}_1}\gamma_\mu\slashed{k}}{\left(\sqrt{K_{{\cal Z}_1{\bar{\cal Z}}_1}}\right)^2}\sim \tilde{f}\gamma_\mu\slashed{k}.
\end{equation}
Further, the Higgsino-$\langle\tilde{\nu}_e\rangle$-$\nu_e$ vertex will be given by ${\cal O}(a_1-z_i)$ term in
$e^{\frac{K}{2}}\left({\cal D}_{{\bar a}_1}D_{z_i}{\bar W}\right){\bar\nu}_{e\ L}\tilde{H}^0_R$, which after assigning $\langle z_i\rangle$ a value of $\sim{\cal V}^{\frac{1}{36}}$ yields:
\begin{equation}
\label{eq:vev z_i-vev nu_e-Higgsino-nu_e}
\tilde{f}{\cal V}^{-\frac{61}{36}}\langle\nu_e\rangle.
\end{equation}
Hence, using (\ref{eq:grav-Higgsino in neutralino-Higgs-II}), (\ref{eq:vev z_i-vev nu_e-Higgsino-nu_e}) and \cite{Grefe_i}:
\begin{eqnarray}
\label{eq:Gamma_gravtoh+nu_e}
& & \Gamma\left(\psi_\mu\rightarrow h+\nu_e\right)\sim\frac{1}{384\pi}\left(\frac{m^3_{3/2}}{M_p^2}\right)\left|\frac{m^2_{\tilde{\nu}_eh}}{m_h^2-m^2_{\tilde{\nu}_e}}
+sin\beta cos\beta \frac{m_Z^2}{M_{\tilde{g}}\hat{\mu}_{{\cal Z}_1{\bar{\cal Z}}_2}}\frac{\langle\nu_e\rangle}{\langle {\cal Z}_i\rangle}\tilde{f}^2{\cal V}^{-\frac{61}{36}}\right|^2\nonumber\\
& & \sim 10^{-3}{\cal V}^{-6+\frac{5}{9}}\times10^{-32}M_p\Biggr|_{{\cal V}\sim10^5}\sim10^{-60}M_p,
\end{eqnarray}
which yields a lifetime of around $10^{17}s$.

\subsection{Three-Body R-parity Violating Gravitino Decays}

In this subsection we will be considering gravitino decays involving three types of R-parity violating vertices that appear in R-parity violating superpotentials \cite{Gmoreau}: $$ W_{\slashed{R}_p} = {\lambda}_{ijk} L_iL_j E^c_k +
{\lambda}^{\prime}_{ijk} L_i Q_j D^c_k+   {\lambda}^{\prime \prime}_{ijk}U_i^cD_j^cD_k^c + \mu_i H L_i. $$

\underline{\bf {Decays involving $\lambda_{ijk}$ coupling}}

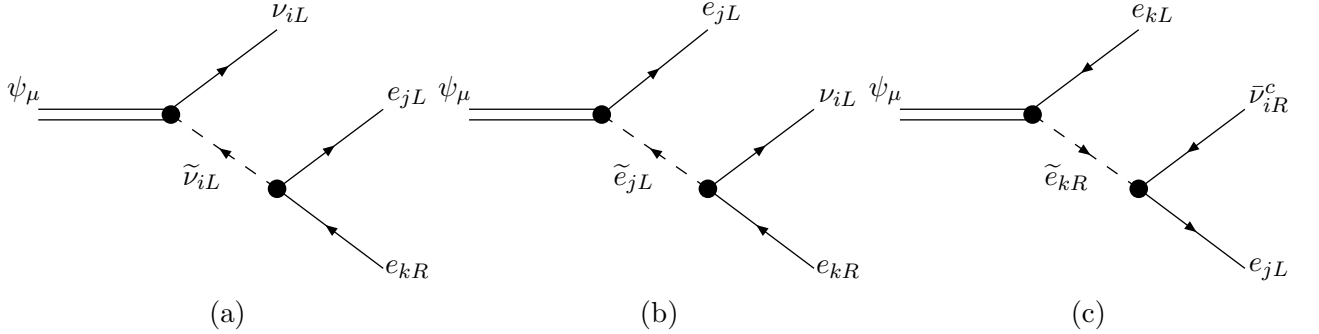
\begin{figure}[t!]
\label{decaydiag1}
\begin{center}
\begin{picture}(150,97)(100,100)
\Vertex(150,158){3.5}
\Line(100,160)(150,160)
\Line(100,156)(150,156)
\ArrowLine(150,160)(190,190)
\DashArrowLine (190,130)(150,158)5
\Vertex(190,130){3.5}
\ArrowLine(190,130)(230,160)
\ArrowLine(230,100)(190,130)
\Text(95,168)[]{$\psi_\mu$}
\Text(240,165)[]{$e_{jL}$}
\Text(240,100)[]{$e_{kR}$}
\Text(196,197)[]{${\nu_{iL}}$}
\put(155,132){${\stilde {\nu}}_{iL}$}
\put(165,80){(a)}
\end{picture}
\hspace{0.2cm}
\begin{picture}(150,97)(100,100)
\Line(100,160)(150,160)
\Line(100,156)(150,156)
\Vertex(150,158){3.5}
\ArrowLine(150,158)(190,190)
\DashArrowLine (190,130)(150,158)5
\Vertex(190,130){3.5}
\ArrowLine(190,130)(230,160)
\ArrowLine(230,100)(190,130)
\Text(95,168)[]{$\psi_\mu$}
\Text(240,165)[]{${\nu}_{iL}$}
\Text(240,100)[]{$e_{kR}$}
\Text(196,197)[]{${e_{jL}}$}
\put(155,132){${\stilde e}_{jL}$}
\put(165,80){(b)}
\end{picture}
\hspace{0.2cm}
\begin{picture}(150,97)(100,100)
\Line(100,160)(150,160)
\Line(100,156)(150,156)
\Vertex(150,158){3.5}
\ArrowLine(190,190)(150,160)
\DashArrowLine (150,158)(190,130)5
\Vertex(190,130){3.5}
\ArrowLine(230,160)(190,130)
\ArrowLine(190,130)(230,100)
\Text(95,168)[]{$\psi_\mu$}
\Text(240,165)[]{${\bar\nu}^c_{iR}$}
\Text(240,100)[]{$e_{jL}$}
\Text(196,197)[]{${e_{kL}}$}
\put(155,132){${\stilde e}_{kR}$}
\put(165,80){(c)}
\end{picture}
\phantom{xxx}
\end{center}
\caption{Three-body gravitino decays involving $\slashed{R}_p\ \lambda_{ijk}$ coupling}
\end{figure}

$\bullet$
Using the same approach as discussed to calculate R-parity violating interaction term in the previous (sub)section(s), the ${\tilde \nu_{iL}}-{e_{jL}}-{e_{kR}}$ vertex corresponding to Fig. 20(a) is given by considering the contribution of following term:
\begin{equation}
{\cal L }= \frac{e^{\frac{K}{2}}}{2}\left({\cal D}_{{\cal A}_1} D_{{\cal A}_3}W\right){\bar \chi}^{{\cal A}_1}\chi^{{\cal A}_3}, \nonumber\\
\end{equation}
where
$${\cal D}_{a_1}D_{a_3}W=\partial_{a_1}\partial_{a_3}W + \left(\partial_{a_1}\partial_{a_3}K\right)W+\partial_{a_1}KD_{a_3}W +
\partial_{a_3}KD_{a_1}W - \left(\partial_{a_1}K\partial_{a_3}K\right)W - \Gamma^k_{{a_1}{a_3}}D_kW,$$ where $a_3, z_i$ correspond to undiagonalized  moduli fields.
By expanding above in the fluctuations linear in $a_1\rightarrow a_1+{\cal V}^{-\frac{2}{9}}{M_p}$ , on simplifying
$$\frac{e^{\frac{K}{2}}}{2} {\cal D}_{a_1}D_{a_3}W \sim {\cal V}^{-\frac{19}{72}}\delta{a_1}$$
Utilizing above, the ${\tilde \nu_{iL}}-{e_{jL}}-{e_{kR}}$ vertex will be given by
\begin{equation}
e^{\frac{K}{2}}{\cal D}_{{\cal A}_1}D_{{\cal A}_3}W\chi^{{\cal A}_1}\chi^{{\cal A}_3} \sim e^{\frac{K}{2}}{\cal D}_{a_1}D_{a_3}W\chi^{{\cal A}_1}\chi^{{\cal A}_3} \sim  {\cal V}^{-\frac{19}{72}}\delta{\cal A}_1\chi^{{\cal A}_1}\chi^{{\cal A}_3}\nonumber\\
\end{equation}
and physical vertex will be given as
\begin{eqnarray}
\label{eq:2a2}
C^{{\tilde \nu_{iL}} {e_{jL}} {e_{kR}}} \sim \frac{{\cal V}^{-\frac{19}{72}}}{{\sqrt{\hat{K}_{{\cal A}_1{\bar{\cal A}}_1}{\hat{K}_{{\cal A}_1{\bar{\cal A}}_1}}{\hat{K}_{{\cal A}_3{\bar {\cal A}}_3}}}}}\sim \frac{{\cal V}^{-\frac{19}{72}}}{\sqrt {10^{15}}}\sim {\cal V}^{-\frac{7}{4}}, {\rm for}~{\cal V}\sim {10^5}.
\end{eqnarray}
$\bullet$

Similarly, ${\tilde e_{jL}}-{\nu_{iL}}-{e_{kR}}$ vertex of Fig. 20(b) is same as ${\tilde \nu_{iL}}-{e_{jL}}-{e_{kR}}$ vertex corresponding to Fig. 20(a) and given as:
\begin{eqnarray}
\label{eq:2b2}
C^{{\tilde e_{jL}} {\nu_{iL}}{e_{kR}}}\sim C^{{\tilde \nu_{iL}} {e_{jL}} {e_{kR}}} \sim {\cal V}^{-\frac{7}{4}}.
\end{eqnarray}
$\bullet$

The ${\tilde e_{kR}}-{\bar\nu^c_{iR}}-{e_{jL}}$ vertex corresponding to Fig. 20(c) comes from:
\begin{equation}
\frac{e^{\frac{K}{2}}}{2}\left({\cal D}_{{\cal {\bar A}}_1} D_{{\cal A}_1}W\right)\chi^{ {\cal A}^{c}_1}\chi^{{\cal A}_1}. \nonumber\\
\end{equation}
In terms of undiagonalized basis,
$${\cal D}_{\bar{a_1}}D_{a_1}W= \left(\partial_{\bar{a_1}}\partial_{a_1}K\right)W+\partial_{\bar{a_1}}KD_{a_1}W +
\partial_{a_1}KD_{\bar{a_1}}W - \left(\partial_{\bar{a_1}}K\partial_{a_1}K\right)W. $$
By expanding above in the fluctuations linear in $a_3\rightarrow a_3+{\cal V}^{-\frac{13}{18}}{M_p}$ by utilizing equations (\ref{eq:Kahler pot}) and (\ref{eq:W}), on simplifying
$$\frac{e^{\frac{K}{2}}}{2} {\cal D}_{a_1}D_{\bar {a_1}}W \sim {\cal V}^{-\frac{1}{3}}\delta{a_3},$$
implying that the lepton-squark-quark vertex is given by:
\begin{eqnarray}
& & e^{\frac{K}{2}}{\cal D}_{{\cal {\bar A}}_1}D_{{\cal A}_1}W\bar\chi^{{\cal A}^{c}_1}\bar\chi^{{\cal A}_1} \sim e^{\frac{K}{2}}{\cal D}_{\bar {a_1}}D_{a_1}W \chi^{{\cal A}^{c}_1}\chi^{{\cal A}_1}\sim  \left({\cal V}^{-\frac{1}{3}}\delta{\cal A}_3\right)\chi^{{\cal A}^{c}_1}\chi^{{\cal A}_1}, \nonumber\\
\end{eqnarray}
and physical vertex will be given as:
\begin{eqnarray}
\label{eq:2c2}
C^{{\tilde e_{kR}}{\bar\nu^c_{iL}}{e_{jL}}} \sim \frac{{\cal V}^{-\frac{1}{3}}}{{\sqrt{\hat{K}_{{\cal A}_1{\bar{\cal A}}_1}{\hat{K}_{{\cal A}_1{\bar{\cal A}}_1}}{\hat{K}_{{\cal A}_3{\bar {\cal A}}_3}}}}}\sim \frac{{\cal V}^{-\frac{1}{3}}}{\sqrt {10^{15}}}\sim {\cal V}^{-\frac{7}{4}}, {\rm for}~{\cal V}\sim {10^5}.
\end{eqnarray}
Now, the matrix amplitude for all three Feynman diagrams corresponding to Fig. 20 will be given as:
\begin{equation}
\label{ampfig1}
|M|^2= \vert {M_a +M_b+M_c} \vert^2
\end{equation}
The analytical results for the full matrix  amplitude summed over spins in terms of  pure and cross terms are given in \cite{Gmoreau}. Utilizing their results, we will estimate matrix  amplitude for all for both pure and cross terms to calculate decay width for the process $\psi_\mu{\to} \nu_i e_j \bar e_k$ in our set up. Strictly speaking, we will be neglecting fermion mass as compared to gravitino and squark masses.
 Introducing kinematic variables
 \begin{equation}
 2 p(\nu_i) \cdot p(e_j)
=
(1 - z_{e_k}) \mm^2 ,
\
 2 p(e_j) \cdot p(e_k)
=
(1 - z_{\nu_i}) \mm^2 ,
\ 2 p(\nu_i) \cdot p(e_k)  = (1 - z_{e_j}) \mm^2 .
\end{equation}
In view of above, we can express the following definitions in terms of kinematic variables as given below:
\begin{eqnarray}
\label{eq:mij}
m^2_{ij} &= & (p(\nu_i)+p(e_j))^2 \sim  2p(\nu_i).p(e_j)= (1-z_{e_k})\mm^2 \nonumber\\
m^2_{jk} & = & (p(e_j)+p(e_k))^2 \sim 2p(e_j).p(e_k)=(1-z_{\nu_i})\mm^2 \nonumber\\
m^2_{ik} & = & (p(\nu_i)+p(e_k))^2 \sim  2p(\nu_i).p(e_k)=(1-z_{e_j})\mm^2,
\end{eqnarray}
utilizing the form of expressions given in the appendix of \cite{Gmoreau}
\begin{eqnarray}
\vert M_a \vert^2 &=& {1 \over 3} {( C^{{\tilde \nu_{iL}} {e_{jL}} {e_{kR}}})^2\over M^2_{pl}
(m^2_{jk}-\aa^2)^2}
(\mm^2-\mjk^2+\mi^2) (\mjk^2-\mj^2-\mk^2) \cr &&
\bigg ( {(\mm^2+\mjk^2-\mi^2)^2 \over 4 \mm^2}-\mjk^2 \bigg ),
\label{amp1}
\end{eqnarray}
 After simplifying using (\ref{eq:mij}), we have
\begin{eqnarray}
\vert M_a \vert^2 &=& {1 \over 3} {( C^{{\tilde \nu_{iL}} {e_{jL}} {e_{kR}}})^2\over M^2_{pl}
((1-z_{e_i})\mm^2-\aa^2)^2}
[z_{\nu_i}(1-z_{\nu_i}) (\frac{z^2_{\nu_i}}{4})],\nonumber\\
& & \sim {1 \over 3} {(C^{{\tilde \nu_{iL}} {e_{jL}} {e_{kR}}})^2\over M^2_{pl}
\aa^4}
[z_{\nu_i}(1-z_{\nu_i}) (\frac{z^2_{\nu_i}}{4})]
\label{amp1a}
\end{eqnarray}
Here we have neglected gravitino mass as compared to sfermion mass.
With  same steps of similar procedure, we have
\begin{eqnarray}
\vert M_{b} \vert^2 &=& {1 \over 3} {(C^{{\tilde e_L} {\nu_L}{e_R}})^2 \over M^2_{pl}
(m^2_{ik}-\bb^2)^2}
(\mm^2-\mik^2+\mj^2) (\mik^2-\mi^2-\mk^2)  \cr &&
\bigg ( {(\mm^2+\mik^2-\mj^2)^2 \over 4 \mm^2}-\mik^2 \bigg )\nonumber\\
& & \sim  {1 \over 3} {(C^{{\tilde e_L} {\nu_L}{e_R}})^2 \over M^2_{pl}
(\bb^2)^2}[z_{e_j}(1-z_{e_j}) (\frac{z^2_{e_j}}{4} )]
\label{amp2}
\end{eqnarray}
\begin{eqnarray}
\vert M_{c} \vert^2 &=& {1 \over 3} {(C^{{\tilde e_R}{\bar\nu^c_R}{e_L}})^2 \over M^2_{pl}
(m^2_{ij}-\cc^2)^2}
(\mm^2-\mij^2+\mk^2) (\mij^2-\mi^2-\mj^2) \cr &&
\bigg ( {(\mm^2+\mij^2-\mk^2)^2 \over 4 \mm^2}-\mij^2 \bigg )\nonumber\\
& & \sim {1 \over 3}{(C^{{\tilde e_R}{\bar\nu^c_R}{e_L}})^2 \over M^2_{pl}
(\cc^2)^2}[z_{e_k}(1-z_{e_k}) (\frac{z^2_{e_k}}{4})]
\label{amp3}
\end{eqnarray}
\begin{eqnarray}
2 Re(M_a M^\dagger_{b})&=&{1 \over 3} {( C^{{\tilde \nu_{iL}} {e_{jL}} {e_{kR}}}.C^{{\tilde e_L} {\nu_L}{e_R}})
\over M^2_{pl} (m^2_{jk}-\aa^2) (m^2_{ik}-\bb^2)} \bigg [
(\mik^2 \mjk^2 - \mm^2 \mk^2 - \mi^2 \mj^2) \cr && \bigg (
(\mm^2+\mk^2-\mi^2-\mj^2)
-{1 \over 2 \mm^2}(\mm^2+\mjk^2-\mi^2) \cr && (\mm^2+\mik^2-\mj^2) \bigg )
+ {1 \over 2} (\mij^2 - \mi^2 - \mj^2) (\mjk^2 - \mj^2 - \mk^2) \cr &&
(\mik^2 - \mi^2 - \mk^2) - {\mi^2 \over 2} (\mjk^2 - \mj^2 - \mk^2)^2
- {\mj^2 \over 2} (\mik^2 - \mi^2 - \mk^2)^2 \cr &&
- {\mk^2 \over 2} (\mij^2 - \mi^2 - \mj^2)^2 + 2 \mi^2 \mj^2 \mk^2 \bigg ]\nonumber\\
& & {\hskip -1.0in}\sim  {2 \over 3} {( C^{{\tilde \nu_{iL}} {e_{jL}} {e_{kR}}}.C^{{\tilde e_L} {\nu_L}{e_R}})
\over M^2_{pl} \aa^2 \bb^2} (1-z_{\nu_i})(1-z_{e_j})(-1-z_{e_k}+2 z_{\nu_i}+2 z_{e_j}-z_{\nu_i}.z_{e_k})
\label{amp12}
\end{eqnarray}
\begin{eqnarray}
2 Re(M_{b} M^\dagger_{c}) &=& {1 \over 3} {(C^{{\tilde e_L} {\nu_L}{e_R}}.C^{{\tilde e_R}{\bar\nu^c_R}{e_L}})
\over M^2_{pl} (m^2_{ik}-\bb^2) (m^2_{ij}-\cc^2)} \bigg [
(\mij^2 \mik^2 - \mm^2 \mi^2 - \mj^2 \mk^2) \cr && \bigg (
(\mm^2+\mi^2-\mj^2-\mk^2)
-{1 \over 2 \mm^2}(\mm^2+\mik^2-\mj^2) \cr && (\mm^2+\mij^2-\mk^2) \bigg )
+ {1 \over 2} (\mij^2 - \mi^2 - \mj^2) (\mjk^2 - \mj^2 - \mk^2) \cr &&
(\mik^2 - \mi^2 - \mk^2) - {\mi^2 \over 2} (\mjk^2 - \mj^2 - \mk^2)^2
- {\mj^2 \over 2} (\mik^2 - \mi^2 - \mk^2)^2 \cr &&
- {\mk^2 \over 2} (\mij^2 - \mi^2 - \mj^2)^2 + 2 \mi^2 \mj^2 \mk^2 \bigg ]\nonumber\\
& & {\hskip -1.0in}\sim  {2 \over 3} {(C^{{\tilde e_L} {\nu_L}{e_R}}.C^{{\tilde e_R}{\bar\nu^c_R}{e_L}})
\over M^2_{pl} \aa^2 \bb^2} (1-z_{e_j})(1-z_{e_k})(-1-z_{\nu_i}+2 z_{e_j}+2 z_{e_k}-z_{e_j}.z_{e_k}),
\label{amp23}
\end{eqnarray}
\begin{eqnarray}
2 Re(M_a M^\dagger_{c}) &=& {1 \over 3} {( C^{{\tilde \nu_{iL}} {e_{jL}} {e_{kR}}}.C^{{\tilde e_R}{\bar\nu^c_R}{e_L}})
\over M^2_{pl} (m^2_{jk}-\aa^2) (m^2_{ij}-\cc^2)} \bigg [
(\mij^2 \mjk^2 - \mm^2 \mj^2 - \mi^2 \mk^2) \cr && \bigg (
(\mm^2+\mj^2-\mi^2-\mk^2)
-{1 \over 2 \mm^2}(\mm^2+\mjk^2-\mi^2) \cr && (\mm^2+\mij^2-\mk^2) \bigg )
+ {1 \over 2} (\mij^2 - \mi^2 - \mj^2) (\mjk^2 - \mj^2 - \mk^2) \cr &&
(\mik^2 - \mi^2 - \mk^2) - {\mi^2 \over 2} (\mjk^2 - \mj^2 - \mk^2)^2
- {\mj^2 \over 2} (\mik^2 - \mi^2 - \mk^2)^2 \cr &&
- {\mk^2 \over 2} (\mij^2 - \mi^2 - \mj^2)^2 + 2 \mi^2 \mj^2 \mk^2 \bigg ]\nonumber\\
& & {\hskip -1.0in}\sim  {2 \over 3} {( C^{{\tilde \nu_{iL}} {e_{jL}} {e_{kR}}}.C^{{\tilde e_R}{\bar\nu^c_R}{e_L}})
\over M^2_{pl} \aa^2 \bb^2} (1-z_{\nu_i})(1-z_{e_k})(-1-z_{e_j}+2 z_{\nu_i}+2 z_{e_k}-z_{\nu_i}.z_{e_k}),
\label{amp13}
\end{eqnarray}
Utilizing the results from  (\ref{amp1})- (\ref{amp13}), (\ref{ampfig1}) one gets the following form:
\begin{eqnarray}
\label{finampfig1}
& & |M|^2= {1 \over 3} {( C^{{\tilde \nu_{iL}} {e_{jL}} {e_{kR}}})^2\over M^2_{pl}
\aa^4}
[z_{\nu_i}(1-z_{\nu_i}) (\frac{z^2_{\nu_i}}{4})]+ {1 \over 3} {(C^{{\tilde e_L} {\nu_L}{e_R}})^2 \over M^2_{pl}
(\bb^2)^2}[z_{e_j}(1-z_{e_j}) (\frac{z^2_{e_j}}{4} )]\nonumber\\
& & + {1 \over 3}{(C^{{\tilde e_R}{\bar\nu^c_R}{e_L}})^2 \over M^2_{pl}
(\cc^2)^2}[z_{e_k}(1-z_{e_k}) (\frac{z^2_{e_k}}{4})] \nonumber\\
& & + {2 \over 3} {( C^{{\tilde \nu_{iL}} {e_{jL}} {e_{kR}}}\cdot C^{{\tilde e_L} {\nu_L}{e_R}})
\over M^2_{pl} \aa^2 \bb^2} (1-z_{\nu_i})(1-z_{e_j})(-1-z_{e_k}+2 z_{\nu_i}+2 z_{e_j}-z_{\nu_i}.z_{e_k})\nonumber\\
& & +  {2 \over 3} {(C^{{\tilde e_L} {\nu_L}{e_R}}\cdot C^{{\tilde e_R}{\bar\nu^c_R}{e_L}})
\over M^2_{pl} \aa^2 \bb^2} (1-z_{e_j})(1-z_{e_k})(-1-z_{\nu_i}+2 z_{e_j}+2 z_{e_k}-z_{e_j}.z_{e_k})+
\nonumber\\
& & {2 \over 3} {( C^{{\tilde \nu_{iL}} {e_{jL}} {e_{kR}}}\cdot C^{{\tilde e_R}{\bar\nu^c_R}{e_L}}).
\over M^2_{pl} \aa^2 \bb^2} (1-z_{\nu_i})(1-z_{e_k})(-1-z_{e_j}+2 z_{\nu_i}+2 z_{e_k}-z_{\nu_i}.z_{e_k}).
\end{eqnarray}
The differential decay rate follows:
\begin{eqnarray}
\frac{d^2 \Gamma}{dz_{e_j} d z_{e_k}}
 &=&
\frac{N_c \mm}{2^8 \pi^3}
\biggl (\frac{1}{2} \sum_{\rm spins} |{\cal M}|^2 \biggr )
\end{eqnarray}
Putting the result of $|M|^2$ from above,
\begin{eqnarray}
& & \frac{d^2 \Gamma}{dz_{e_j} d z_{e_k}}\sim
\frac{N_c \mm}{2^9 \pi^3}{1 \over 3} [{( C^{{\tilde \nu_{iL}} {e_{jL}} {e_{kR}}})^2\over M^2_{pl}
\aa^4}
[z_{\nu_i}(1-z_{\nu_i}) (\frac{z^2_{\nu_i}}{4})]+ {1 \over 3} {(C^{{\tilde e_L} {\nu_L}{e_R}})^2 \over M^2_{pl}
(\bb^2)^2}[z_{e_j}(1-z_{e_j}) (\frac{z^2_{e_j}}{4} )]\nonumber\\
& & + {1 \over 3}{(C^{{\tilde e_R}{\bar\nu^c_R}{e_L}})^2 \over M^2_{pl}
(\cc^2)^2}[z_{e_k}(1-z_{e_k}) (\frac{z^2_{e_k}}{4})] \nonumber\\
& & + {2 \over 3} {( C^{{\tilde \nu_{iL}} {e_{jL}} {e_{kR}}}\cdot C^{{\tilde e_L} {\nu_L}{e_R}})
\over M^2_{pl} \aa^2 \bb^2} (1-z_{\nu_i})(1-z_{e_j})(-1-z_{e_k}+2 z_{\nu_i}+2 z_{e_j}-z_{\nu_i}.z_{e_k})\nonumber\\
& & +  {2 \over 3} {(C^{{\tilde e_L} {\nu_L}{e_R}}\cdot C^{{\tilde e_R}{\bar\nu^c_R}{e_L}})
\over M^2_{pl} \aa^2 \bb^2} (1-z_{e_j})(1-z_{e_k})(-1-z_{\nu_i}+2 z_{e_j}+2 z_{e_k}-z_{e_j}.z_{e_k})+
\nonumber\\
& & {2 \over 3} {( C^{{\tilde \nu_{iL}} {e_{jL}} {e_{kR}}}\cdot C^{{\tilde e_R}{\bar\nu^c_R}{e_L}}).
\over M^2_{pl} \aa^2 \bb^2} (1-z_{\nu_i})(1-z_{e_k})(-1-z_{e_j}+2 z_{\nu_i}+2 z_{e_k}-z_{\nu_i}.z_{e_k})]
\end{eqnarray}
Using the same approach as given in (\ref{eq:limit}),
\begin{eqnarray}
0 < z_{e_j} < 1, 1 - z_j <  z_{e_k}  <1.
\end{eqnarray}
and using the numerical estimates of masses, ${\mm}\sim {\cal V}^{-2}M_p, \aa^2= \bb^2= \cc^2 \sim m_{{\cal A}_1}\sim{\cal V}^{\frac{1}{2}}\mm$  as given in equation (\ref{eq:m_A_1}), after  integrating, decay width reduces to
\begin{eqnarray}
& & \Gamma\sim
\frac{N_c \mm^7}{(2^9.3.120) \pi^3. M^2_{pl}.{\cal V}^2 \mm^4} \biggr[( C^{{\tilde \nu_{iL}} {e_{jL}} {e_{kR}}})^2
+  (C^{{\tilde e_L} {\nu_L}{e_R}})^2  + (C^{{\tilde e_R}{\bar\nu^c_R}{e_L}})^2 \nonumber\\
& &  + {3 \over 4}\bigg(( C^{{\tilde \nu_{iL}} {e_{jL}} {e_{kR}}}\cdot C^{{\tilde e_L} {\nu_L}{e_R}})
 + (C^{{\tilde e_L} {\nu_L}{e_R}}\cdot C^{{\tilde e_R}{\bar\nu^c_R}{e_L}})+ ( C^{{\tilde \nu_{iL}} {e_{jL}} {e_{kR}}}\cdot C^{{\tilde e_R}{\bar\nu^c_R}{e_L}})\bigg)].
\end{eqnarray}
Utilizing the set of results given in equation no (\ref{eq:2a2}) - (\ref{eq:2c2}), decay width simplifies to
\begin{eqnarray}
& & \Gamma\sim
\frac{N_c \mm^7}{(2^9.3.120) \pi^3. M^2_{pl}.{\cal V}^2 \mm^4} (
  {\cal V}^{-\frac{7}{2}}) \nonumber\\
& & \sim  \frac{1}{10^6} \frac{{\cal V}^{-\frac{11}{2}}.\mm^3}{M^2_{pl}}\sim 10^{-45.5} GeV ; {\rm for~{{\cal V}\sim10^5}}.
\end{eqnarray}
Life time  will be given as
\begin{eqnarray}
 \tau &=&\frac{\hbar}{\Gamma}\sim\frac{10^{-34} Jsec}{10^{-45.5} GeV}\sim O(10^{21}) sec.
 \end{eqnarray}

\underline{\bf {Decays involving $\lambda'_{ijk}$ coupling}}

\begin{figure}[t!]
\label{decaydiag1}
\begin{center}
\begin{picture}(150,97)(100,100)
\Line(100,160)(150,160)
\Line(100,156)(150,156)
\Vertex(150,158){3.5}
\ArrowLine(150,158)(190,190)
\DashArrowLine (190,130)(150,158)5
\Vertex(190,130){3.5}
\ArrowLine(190,130)(230,160)
\ArrowLine(230,100)(190,130)
\Text(95,168)[]{$\psi_\mu$}
\Text(240,165)[]{$d_{jL}$}
\Text(240,100)[]{$d_{kR}$}
\Text(196,197)[]{${\nu_{iL}}$}
\put(155,132){${\stilde \nu}_{iL}$}
\put(165,80){(a)}
\end{picture}
\hspace{0.2cm}
\begin{picture}(150,97)(100,100)
\Line(100,160)(150,160)
\Line(100,156)(150,156)
\Vertex(150,158){3.5}
\ArrowLine(150,158)(190,190)
\DashArrowLine (190,130)(150,158)5
\ArrowLine(190,130)(230,160)
\ArrowLine(230,100)(190,130)
\Vertex(190,130){3.5}
\Text(95,168)[]{$\psi_\mu$}
\Text(240,165)[]{${\nu}_{iL}$}
\Text(240,100)[]{$d_{kR}$}
\Text(196,197)[]{${d_{jL}}$}
\put(155,132){${\stilde d}_{jL}$}
\put(165,80){(b)}
\end{picture}
\hspace{0.2cm}
\begin{picture}(150,97)(100,100)
\Line(100,160)(150,160)
\Line(100,156)(150,156)
\Vertex(150,158){3.5}
\ArrowLine(190,190)(150,158)
\Vertex(190,130){3.5}
\DashArrowLine (150,158)(190,130)5
\ArrowLine(230,160)(190,130)
\ArrowLine(190,130)(230,100)
\Text(95,168)[]{$\psi_\mu$}
\Text(240,165)[]{${\bar\nu}^c_{iR}$}
\Text(240,100)[]{$d_{jL}$}
\Text(196,197)[]{${d_{kR}}$}
\put(155,132){${\stilde d}_{kR}$}
\put(165,80){(c)}
\end{picture}
\phantom{xxx}
\end{center}
\caption{Three-body gravitino decays involving $\slashed{R}_p\ \lambda^\prime_{ijk}$ coupling}
\end{figure}

$\bullet$
The  ${\tilde \nu_L}-{d_{jL}}-{d_{kR}}$ vertex corresponding to figure 3(a) and ${\tilde l_L}-{u_L}-{d^{c}_L}$ vertex calculated in section {\bf 4} get identified with same set of moduli space superfields and given by similar interaction vertex. Therefore contribution of ${\tilde \nu_L}-{d_{jL}}-{d_{kR}}$ vertex is same as ${\tilde l_L}-{u_L}-{d^{c}_L}$ given in (\ref{eq:quark-slepton-quark}).
\begin{eqnarray}
\label{eq:3a2}
C^{{\tilde \nu_L} {d_{jL}} {d_{kR}}}\sim {\cal V}^{-\frac{5}{3}}{{\tilde \nu_L} {d_{jL}} {d_{kR}}}, {\rm for}~{\cal V}\sim {10^5}
\end{eqnarray}

$\bullet$
The ${\tilde d_{jL}}-{\nu_{iL}}-{d_{kR}}$  vertex corresponding to figure 3(b) and ${\tilde u_L}-{l_L}-{d^{c}_L}$ vertex calculated in section {\bf 4} get identified with same set of moduli space superfields and given by similar interaction vertex. Therefore contribution of ${\tilde d_{jL}}-{\nu_{iL}}-{d_{kR}}$ vertex is same as ${\tilde u_L}-{l_L}-{d^{c}_L}$ given in (\ref{eq:lepton-squark-quark1})
\begin{eqnarray}
\label{eq:3b2}
C^{{\tilde d_{jL}} {\nu_{iL}} {d_{kR}}} \sim {\cal V}^{-\frac{5}{3}}, {\rm for}~{\cal V}\sim {10^5}
\end{eqnarray}

$\bullet$
This time,  ${\tilde d_{kR}}-{\bar\nu^c_{iR}}-{d_{jL}}$  vertex corresponding to Fig. 21(c) and ${\tilde d_R}-{l_L}-{u_L}$ vertex calculated in section {\bf 4} get identified with same set of moduli space superfields and given by similar interaction vertex . Hence,  contribution of ${\tilde d_{kR}}-{\bar\nu^c_{iR}}-{d_{jL}}$ vertex is a same as ${\tilde d_R}-{l_L}-{u_L}$ given in (\ref{eq:lepton-squark-quark2})
\begin{eqnarray}
\label{eq:3c2}
C^{{\tilde d_{kR}}{\bar\nu^c_{iR}}{d_{jL}}} \sim {\cal V}^{-\frac{5}{3}}, {\rm for}~{\cal V}\sim {10^5}.
\end{eqnarray}

The analytical result of matrix amplitude for all three Feynman diagrams corresponding to Fig. 21 is same as calculated in Fig. 20 by replacing  $\aa \to \aaps \ (\aapt)$, $\bb \to \bbps \ (\bbpt)$ and
$\cc \to \ccps \ (\ccpt)$ and $C^{{\tilde \nu_{iL}} {e_{jL}} {e_{kR}}} \to C^{{\tilde \nu_{iL}} {d_{jL}} {d_{kR}}}$, $C^{{\tilde e_{jL}} {\nu_{iL}}{e_{kR}}} \to C^{{\tilde d_{jL}} {\nu_{iL}} {d_{kR}}}$.
 $C^{{\tilde e_{kR}}{\bar\nu^c_{iR}}{e_{jL}}} \to C^{{\tilde d_{kR}}{\bar\nu^c_{iR}}{d_{jL}}}$. Doing so and  using same numerical estimates of masses, ${\mm}\sim {\cal V}^{-2}M_p, \aaps^2= \bbps^2= \ccps^2 \sim {\cal V}^{\frac{1}{2}}\mm$ in our set up, after  integrating, decay width reduces to
\begin{eqnarray}
& &{\hskip -0.4in}  \Gamma\sim
\frac{N_c \mm^7}{(2^9.3.120) \pi^3. M^2_{pl}.{\cal V}^2 \mm^4} \biggr[( C^{{\tilde \nu_{iL}} {d_{jL}} {d_{kR}}})^2
+  (C^{{\tilde d_{jL}} {\nu_{iL}} {d_{kR}}})^2  + (C^{{\tilde d_{kR}}{\bar\nu^c_{iL}}{d_{jL}}})^2 \nonumber\\
& &  + {3 \over 4}\bigg( (C^{{\tilde \nu_{iL}} {d_{jL}} {d_{kR}}} \cdot C^{{\tilde d_{jL}} {\nu_{iL}} {d_{kR}}})
 + (C^{{\tilde d_L} {\nu_L} {d_{kR}}}\cdot C^{{\tilde d_{kR}}{\bar\nu^c_{iL}}{d_{jL}}})+ ( C^{{\tilde \nu_{iL}} {d_{jL}} {d_{kR}}} \cdot C^{{\tilde d_{kR}}{\bar\nu^c_{iR}}{d_{jL}}})\bigg)\biggr].
\end{eqnarray}
Utilizing the set of results given in equation no (\ref{eq:2a2}) - (\ref{eq:2c2}), decay width simplifies to
\begin{eqnarray}
& & \Gamma\sim
\frac{N_c \mm^7}{(2^9.3.120) \pi^3. M^2_{pl}.{\cal V}^2 \mm^4} (
  {\cal V}^{-\frac{10}{3}}) \nonumber\\
& & \sim  \frac{1}{10^6} \frac{{\cal V}^{-\frac{16}{3}}.\mm^3}{M^2_{pl}}\sim 10^{-44} GeV ; {\rm for~{{\cal V}\sim10^5}}
\end{eqnarray}
Life time  will be given as
\begin{eqnarray}
 \tau &=&\frac{\hbar}{\Gamma}\sim\frac{10^{-34} Jsec}{10^{-44} GeV}\sim O(10^{20}) sec
 \end{eqnarray}

\underline{\bf {Decays mediated via $\lambda''_{ijk}$ coupling}}

\begin{figure}[t!]
\label{decaydiag1}
\begin{center}
\begin{picture}(150,97)(100,100)
\Line(100,160)(150,160)
\Line(100,156)(150,156)
\Vertex(150,158){3.5}
\ArrowLine(150,160)(190,190)
\DashArrowLine (190,130)(150,160)5
\ArrowLine(190,130)(230,160)
\ArrowLine(230,100)(190,130)
\Text(95,168)[]{$\psi_\mu$}
\Text(240,165)[]{$d_{jL}$}
\Text(240,100)[]{$d^{c}_{kL}$}
\Text(196,197)[]{${u_{iR}}$}
\put(155,132){${\stilde u}_{iR}$}
\put(165,80){(a)}
\end{picture}
\hspace{0.2cm}
\begin{picture}(150,97)(100,100)
\Vertex(150,158){3.5}
\Line(100,160)(150,160)
\Line(100,156)(150,156)
\ArrowLine(150,158)(190,190)
\DashArrowLine (190,130)(150,158)5
\Vertex(190,130){3.5}
\ArrowLine(190,130)(230,160)
\ArrowLine(230,100)(190,130)
\Text(95,168)[]{$\psi_\mu$}
\Text(240,165)[]{${u}_{iL}$}
\Text(240,100)[]{$d^{c}_{kL}$}
\Text(196,197)[]{${d_{jR}}$}
\put(155,132){${\stilde d}_{jR}$}
\put(165,80){(b)}
\end{picture}
\end{center}
\caption{Three-body gravitino decays involving $\slashed{R}_p\ \lambda^{\prime\prime}_{ijk}$ coupling}
\end{figure}

$\bullet$
The ${\tilde u_{iR}}-{d_{jR}}-{d^c_{kL}}$ vertex corresponding to Fig. 22(a) arises from:
\begin{equation}
\frac{e^{\frac{K}{2}}}{2}\left({\cal D}_{{\cal {\bar A}}_4} D_{{\cal A}_4}W\right)\chi^{ {\cal A}^{c}_4}\chi^{{\cal A}_4}. \nonumber\\
\end{equation}
In terms of undiagonalized basis,
$${\cal D}_{\bar{a_4}}D_{a_4}W= \left(\partial_{\bar{a_4}}\partial_{a_4}K\right)W+\partial_{\bar{a_4}}KD_{a_4}W +
\partial_{a_4}KD_{\bar{a_4}}W - \left(\partial_{\bar{a_4}}K\partial_{a_4}K\right)W. $$
By expanding above in the fluctuations linear in $a_4\rightarrow a_4+{\cal V}^{-\frac{11}{9}}{M_p}$ by utilizing equations (\ref{eq:Kahler pot}) and (\ref{eq:W}), on simplifying one obtains:
$$\frac{e^{\frac{K}{2}}}{2} {\cal D}_{a_4}D_{\bar {a_4}}W \sim {\cal V}^{\frac{13}{6}}\delta{a_4},$$
implying that the contribution of quark-squark-quark vertex is given as under:
\begin{eqnarray}
& & e^{\frac{K}{2}}{\cal D}_{{\cal {\bar A}}_4}D_{{\cal A}_4}W\bar\chi^{{\cal A}^{c}_4}\bar\chi^{{\cal A}_4} \sim e^{\frac{K}{2}}{\cal D}_{\bar {a_4}}D_{a_4}W \chi^{{\cal A}^{c}_4}\chi^{{\cal A}_4}\sim  \left({\cal V}^{\frac{13}{6}}{\cal A}_4\right)\chi^{{\cal A}^{c}_4}\chi^{{\cal A}_4}, \nonumber\\
\end{eqnarray}
and the physical vertex will be given as following:
\begin{eqnarray}
\label{eq:4a2}
C^{{\tilde u_{iR}} {d_{jR}} {d^c_{kL}}} \sim \frac{{\cal V}^{\frac{13}{6}}}{{\sqrt{\hat{K}_{{\cal A}_4{\bar{\cal A}}_4}{\hat{K}_{{\cal A}_4{\bar{\cal A}}_4}}{\hat{K}_{{\cal A}_4{\bar {\cal A}}_4}}}}}\sim \frac{{\cal V}^{\frac{13}{6}}}{\sqrt {10^{36}}}\sim {\cal V}^{-\frac{43}{30}}{{\tilde u_{iR}} {d_{jR}} {d^c_{kL}}}, {\rm for}~{\cal V}\sim {10^5}.
\end{eqnarray}

$\bullet$
Similarly, the ${\tilde d_{jR}}-{d^c_{kL}}-{u_{iR}}$ vertex of Fig. 21(b) is same as ${\tilde u_{iR}}-{d_{jR}}-{d^c_{kL}}$ vertex corresponding to Fig. 21(a) and is given as under:
\begin{eqnarray}
\label{eq:4b2}
C^{{\tilde d_{jR}} {d^c_{kL}} {u_{iR}}} \sim C^{{\tilde u_{iR}} {d_{jL}} {d^c_{kL}}} \sim {\cal V}^{-\frac{43}{30}}{{\tilde d_{jR}} {d^c_{kL}} {u_{iR}}}, {\rm for}~{\cal V}\sim {10^5}.
\end{eqnarray}
 Again, introducing  kinematic variable
 \begin{equation}
 2 p(u_i) \cdot p(d_j)
=
(1 - z_{d_k}) \mm^2 ,\
 2 p(d_j) \cdot p(d_k)
=
(1 - z_{u_i}) \mm^2 ,
\
 2 p(u_i) \cdot p(d_k)=(1 - z_{d_j}) \mm^2 .
\end{equation}
and
\begin{eqnarray}
\label{eq:mij}
m^2_{ij} &= & (p(u_i)+p(d_j))^2 \sim  2p(u_i).p(d_j)= (1-z_{d_k})\mm^2 \nonumber\\
m^2_{jk} & = & (p(d_j)+p(d_k))^2 \sim 2p(d_j).p(d_k)=(1-z_{u_i})\mm^2 \nonumber\\
m^2_{ik} & = & (p(u_i)+p(d_k))^2 \sim  2p(u_i).p(d_k)=(1-z_{d_j})\mm^2,
\end{eqnarray}
Here also, utilizing the form of expressions given in the appendix of \cite{Gmoreau} for ${\lambda^{\prime \prime}_{ijk}}$ coupling, the full analytical result of the squared amplitude summed over the
spins for the gravitino decay reaction
$\psi_\mu \stackrel{\lambda^{\prime \prime}_{ijk}}{\to} u_i d_j d_k$ is
the sum of the following squared amplitude and interference teramas:
\begin{eqnarray}
\vert M_a \vert^2 &=& { N_c ! \over 3}
{(C^{\tilde G {u_{iR}} {\tilde {u_{iR}}}} C^{{\tilde u_{iR}} {d_{jR}} {d^c_{kR}}})^2 \over M_\star^2 (m^2_{jk}-\aapp^2)^2}
(\mm^2-\mjk^2+\mipp^2) (\mjk^2-\mjpp^2-\mkpp^2) \cr &&
\bigg ( {(\mm^2+\mjk^2-\mipp^2)^2 \over 4 \mm^2}-\mjk^2 \bigg ) \sim {1 \over 3} {( C^{{\tilde u_{iR}} {d_{jR}} {d^c_{kR}}})^2\over M^2_{pl}\aaps^4}
[z_{u_i}(1-z_{u_i}) (\frac{z^2_{u_i}}{4})]
\label{amp1pp}
\end{eqnarray}
\begin{eqnarray}
\vert M_{b} \vert^2 &=& {4 N_c ! \over 3} {(C^{{\tilde d_{jR}} {u_{iR}}{d^c_{kR}}})^2 \over
M_\star^2
(m^2_{ik}-\bbpp^2)^2}
(\mm^2-\mik^2+\mjpp^2) (\mik^2-\mipp^2-\mkpp^2)  \nonumber\\
& & {1 \over 3} {(C^{{\tilde d_{jR}} {u_{iR}}{d^c_{kR}}})^2 \over M^2_{pl}
(\bb^2)^2}[z_{d_j}(1-z_{d_j}) (\frac{z^2_{d_j}}{4})],
\label{amp2pp}
\end{eqnarray}
\begin{eqnarray}
2 Re(M_a M^\dagger_{b})&=&{2 N_c ! \over 3} {( C^{{\tilde u_{iR}} {d_{jR}} {d^c_{kR}}}\cdot C^{{\tilde d_{jR}} {u_{iR}}{d^c_{kR}}})
\over M_\star^2 (m^2_{jk}-\aapp^2) (m^2_{ik}-\bbpp^2)} \bigg [
(\mik^2 \mjk^2 - \mm^2 \mkpp^2 - \mipp^2 \mjpp^2) \cr &&
\bigg ( (\mm^2+\mkpp^2-\mipp^2-\mjpp^2)
-{1 \over 2 \mm^2}(\mm^2+\mjk^2-\mipp^2) \cr && (\mm^2+\mik^2-\mjpp^2)
\bigg )
+ {1 \over 2} (\mij^2 - \mi^2 - \mj^2) (\mjk^2 - \mj^2 - \mk^2) \cr &&
(\mik^2 - \mi^2 - \mk^2) - {\mi^2 \over 2} (\mjk^2 - \mj^2 - \mk^2)^2
- {\mj^2 \over 2} (\mik^2 - \mi^2 - \mk^2)^2 \cr &&
- {\mk^2 \over 2} (\mij^2 - \mi^2 - \mj^2)^2 + 2 \mi^2 \mj^2 \mk^2 \bigg ]\nonumber\\
& & {\hskip -0.8in}\sim  {2 \over 3} {( C^{{\tilde u_{iR}} {d_{jR}} {d^c_{kR}}}\cdot C^{{\tilde d_{jR}} {u_{iR}}{d^c_{kR}}})
\over M^2_{pl} \aa^2 \bb^2} (1-z_{u_i})(1-z_{d_j})(-1-z_{d_k}+2 z_{u_i}+2 z_{d_j}-z_{u_i}.z_{d_k}).
\label{amp12pp}
\end{eqnarray}
Decay width will be given as:
\begin{eqnarray}
& &{\hskip -0.4in}  \Gamma\sim
\frac{N_c \mm^7}{(2^9.3.120) \pi^3. M^2_{pl}.{\cal V}^2 \mm^4} \biggr[( C^{{\tilde u_{iR}} {d_{jR}} {d^c_{kR}}})^2
+  (C^{{\tilde d_{jR}} {u_{iR}}{d^c_{kR}}})^2  + {3 \over 4}\bigg( C^{{\tilde u_{iR}} {d_{jR}} {d^c_{kR}}}\cdot C^{{\tilde d_{jR}} {u_{iR}}{d^c_{kR}}})
 \bigg)\biggr].
\end{eqnarray}
Utilizing the set of results given in equation no (\ref{eq:4a2}) - (\ref{eq:4b2}), decay width simplifies to
\begin{eqnarray}
& & \Gamma\sim
\frac{N_c \mm^7}{(2^9.3.120) \pi^3. M^2_{pl}.{\cal V}^2 \mm^4} (
{\cal V}^{-\frac{43}{15}} ) \nonumber\\
& & \sim  \frac{1}{10^6} \frac{{\cal V}^{-\frac{73}{15}}.\mm^3}{M^2_{pl}}\sim 10^{-42} GeV ; {\rm for~{{\cal V}\sim 10^5}},
\end{eqnarray}
and therefore the life time  will be given by:
\begin{eqnarray}
 \tau &=&\frac{\hbar}{\Gamma}\sim\frac{10^{-34} Jsec}{10^{-42} GeV}\sim O(10^{18}) sec.
 \end{eqnarray}

\section{NLSP Decays}

 With in the approach of considering non-thermal production of Gravitino(LSP), all of the dark matter density are produced by non thermal decays of NLSP decaying into LSP and therefore, it is important to determine whether life time of these decays do not affect standard abundance of D, 3He, 4He, and 7Li, and hence predictions of BBN. The Big bang nucleosynthesis  predicts the universal abundances of D, 3He, 4He, and 7Li, essentially fixed by average lifetime  $\tau \sim 180 sec$. In our study, we  are considering both radiative as well as hadronic decay modes of N(LSP)'s. Therefore, if life time of NLSP's are more than $ 10^2$ sec, the high energy photons emitted via radiative decay might destroy the abundance of light elements by inducing photo-dissociation of same, hence reducing the standard baryon-to-photon ratio and resulting in baryon-poor universe . Similarly, during the hadronic decay, the released hadronic energy can produce mesons/charged pions causing interconversions between background proton and neutron (p / n conversion) that alters the neutron-to proton ratio, resulting in the change of 4He abundance and hence destroying this success of standard BBN. The bounds on primordial abundance of late decaying particle, taking into account hadronic decay mode of the same, have been discussed in detail in \cite{kohri_BBN}.

In view of the above discussion, we explicitly calculate the
decay widths and hence life times for possible NLSP candidates. The table of mass spectrum given in table 1 suggests that sleptons/squarks and (Bino/Wino-type)gaugino/(Bino/Wino-type)gaugino-dominant neutralino exists as N(ext-to) L(ightest) S(upersymmetric) P(article)s in the dilute-flux approximation. The diagonalization of neutralino mass matrix  for the four-Wilson-line-moduli setup, similar to \cite{Dhuria+Misra_mu_Split SUSY} corresponds to smallest eigenvalue of neutralino of the order ${\cal V}^{\frac{2}{3}}m_{3/2}$ (similar as gaugino mass) given by the following eigenvector:
\begin{equation}
\label{eq:chi3}
\tilde{\chi}_3^0\sim-\lambda^0+\tilde{f}\left(\tilde{H}_1^0+\tilde{H}_2^0\right) + {\rm charge\ conjugate},
\end{equation}
where $\tilde{H}_{1,2}^0$ are the Higgsinos.
We will first evaluate the lifetimes corresponding to two- and three-body decays of Wino/Bino-type gauginos/Neutralino and gluino decay to gravitino(LSP) followed by the lifetime corresponding to two- and three-body decays of the slepton to gravitino(LSP). In addition to this, there is a possibility for neutralino to directly decay into ordinary quarks via R-parity violating three-body decays and hence affect the abundance of gravitino produced by neutralino. Due to the presence of two competing kind of decays, results of life time of both will decide, whether decay of neutralino into ordinary quarks effects the abundance of gravitino or not. To verify the same, we have also evaluated the three-body decay diagrams of neutralino in the context of ${\cal N}=1$ Supergravity action.

\subsection{Gaugino Decays}

In this sub-section, we discuss two- and three-body decay modes of the Wino/Bino-type gauginos as well as the gluino.

\subsubsection{Two- and Three-Body Gaugino(Wino/Bino) Decays}

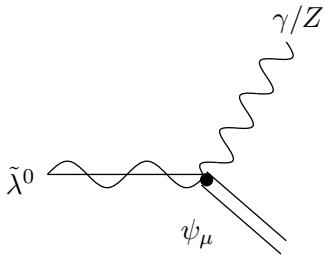
\begin{figure}
\begin{center}
    \begin{picture}(135,137) (15,-18)
   \Line(100,53)(130,27)
   \Line(98,48)(128,22)
   \Photon(40,52)(100,52){5}{2}
   \Photon(100,50)(130,102){5}{4}
   \Text(30,50)[]{{{$\tilde{\lambda}^0$}}}
   \Line(40,52)(100,52)
   \Vertex(100,50){2.5}
   \Text(135,111)[]{{{$\gamma/Z$}}}
   \Text(97,30)[]{{{$\psi_\mu$}}}
  \end{picture}
   \end{center}
   \vskip-0.3in
   \caption{Two-body gaugino-decay diagram}
   \end{figure}

The relevant gravitino-gaugino-photon/$Z$-boson vertex in the ${\cal N}=1$ SUGRA action of \cite{Wess_Bagger} is obtained from $\frac{1}{M_p}{\bar\psi}_\rho\sigma^{\mu\nu}\gamma^\rho\lambda_L^{(a)}F^{(a)}_{\mu\nu}, (a)$ corresponding to the three gauge groups. The decay width for $\tilde{B}\rightarrow\psi_\mu+\gamma$ is given by (See \cite{Hasenkamp}):
\begin{eqnarray}
\label{eq:Bino_grav+ph}
& & \Gamma\left(\tilde{B}\rightarrow\psi_\mu+\gamma\right)=\frac{cos^2\theta_W}{48\pi M_p^2}\frac{m^5_{\tilde{B}}}{m^2_{3/2}}
\left(1-\frac{m^2_{3/2}}{m_{\tilde{B}}^2}\right)^3\left(1+3\frac{m^2_{3/2}}{m_{\tilde{B}}^2}\right)\sim\frac{m^5_{\tilde{B}}}{m^2_{3/2}M_p^2}
\end{eqnarray}
For $m_{\tilde{B}}\sim m_{\tilde{g}}\sim{\cal V}^{\frac{2}{3}}m_{\frac{3}{2}}$ as given in equation (\ref{eq:gaugino_mass}) and $m_{3/2}\sim{\cal V}^{-2}M_p$ for $n^s=2$,  above decay process implies a lifetime of Bino ($\tilde{B}$) of around $10^{-30}s$.

Similarly, the decay width for $\tilde{B}\rightarrow\psi_\mu+Z$ is given by (See \cite{Hasenkamp}):
\begin{eqnarray}
\label{eq:Bino_grav+Z}
& & \Gamma\left(\tilde{B}\rightarrow\psi_\mu+Z\right)=\frac{cos^2\theta_W}{48\pi M_p^2}\frac{m^5_{\tilde{B}}}{m^2_{3/2}}
\sqrt{1-2\left(\frac{m_{\psi_\mu}^2}{m_{\tilde{B}}^2} + \frac{m^2_Z}{m^2_{\tilde{B}}}\right)+\left(\frac{m^2_{\psi_\mu}}{m_{\tilde{B}}}-\frac{m^2_Z}{m^2_{\tilde{B}}}\right)^2}\nonumber\\
& & \times\left[
\left(1-\frac{m^2_{3/2}}{m_{\tilde{B}}^2}\right)^2\left(1+3\frac{m^2_{3/2}}{m_{\tilde{B}}^2}\right)-\frac{m_Z^2}{m_{\tilde{B}}^2}
\left\{3+\frac{m^3_{3/2}}{m^3_{\tilde{B}}}\left(\frac{m_{3/2}}{m_{\tilde{B}}}-12\right)-\frac{x_Z^2}{x_{\tilde{B}}^2}\left(
3-\frac{m^2_{3/2}}{m^2_{\tilde{B}}}-\frac{m^2_Z}{m^2_{\tilde{B}}}\right)\right\}\right]\nonumber\\
& & \sim\frac{m^5_{\tilde{B}}}{m^2_{3/2}M_p^2}\sim{\cal V}^{-\frac{8}{3}}M_p.
\end{eqnarray}
So, (\ref{eq:Bino_grav+Z}) implies a lifetime of around $10^{-30}s$.

\begin{figure}
\begin{center}
\begin{picture}(150,97)(100,100)
\Line(100,160)(150,160)
\Photon(100,160)(150,160){5}{2}
\Vertex(150,160){4.6}
\Line(150,162)(190,190)
\Line(152,157)(192,185)
\Photon (150,160)(190,130){5}{4}
\Vertex(190,130){4}
\ArrowLine(230,160)(190,130)
\ArrowLine(190,130)(230,100)
\Text(95,168)[]{$\tilde{B}$}
\Text(240,165)[]{${\bar q}$}
\Text(240,100)[]{$q$}
\Text(196,197)[]{$\psi_\mu$}
\put(155,132){$\gamma/Z$}
\end{picture}
\end{center}
\caption{Three-body gaugino-decay diagram}
\end{figure}
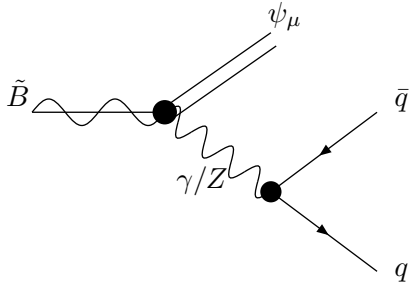

The gauge boson-quark-anti-quark vertex will be accompanied by $\frac{g_{{\cal A}_2{\bar {\cal A}}_2}g_{YM}\left(X^{T^B}K+iD^{T^B}\right)}{\left(\sqrt{K_{{\cal A}_2{\bar{\cal A}}_2}}\right)^2}$ and value of the same has been shown to be same as its SM value i.e ${\cal O}(1)$  in \cite{dhuria+misra_EDM}. Using further the results of \cite{Hasenkamp} and gaugino mass $\lambda^0\sim {\cal V}^{\frac{2}{3}}m_{\frac{3}{2}}$ from equation (\ref{eq:gaugino_mass}), one obtains the following result for the decay width $\lambda^0\rightarrow\psi_\mu+u+{\bar u}$:
\begin{eqnarray}
\label{eq:gaugino_grav+u+ubar}
& & \Gamma\left(\lambda^0\rightarrow\psi_\mu+u+{\bar u}\right)\sim\frac{g^2_{YM} m^5_{\lambda^0}}{32\left(2\pi\right)^3m^2_{3/2}M_p^2}\sim10^{-17}M_p \sim {\cal O}(10),
\end{eqnarray}
which yields a lifetime of around $10^{-25}s$.

We will now discuss three-body gaugino decays into gravitino mediated by squarks. To avoid channel overlap, for simplicity, we will assume that the gaugino decays mediated by vector bosons and those mediated by squarks, involve different gauginos.
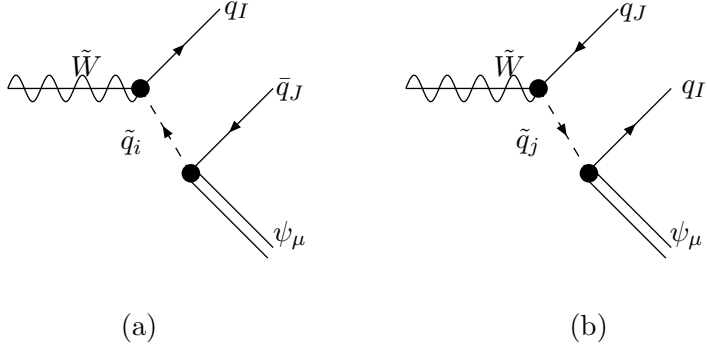
\begin{figure}
 \begin{center}
\begin{picture}(350,180)(50,0)
\Text(90,130)[]{$\tilde{W}$}
\Line(60,120)(110,120)
\Photon(60,120)(110,120){5}{4}
\Vertex(110,120){3.5}
\ArrowLine(110,120)(140,150)
\Text(147,150)[]{${q}_I$}
\DashArrowLine(129,88)(110,120){4}
\Text(107,100)[]{$\tilde{q}_i$}
\ArrowLine(160,120)(130,90)
\Text(167,120)[]{${\bar q_J}$}
\Vertex(129,88){3.5}
\Line(130,90)(160,60)
\Line(128,86)(158,56)
\Text(167,65)[]{$\psi_\mu$}
\Text(110,30)[]{(a)}
\Text(250,130)[]{$\tilde{W}$}
\Line(210,120)(260,120)
\Photon(210,120)(260,120){5}{4}
\Vertex(260,120){3.5}
\ArrowLine(290,150)(260,120)
\Text(297,150)[]{$\bar{q}_J$}
\DashArrowLine(260,120)(279,88){4}{}
\Text(257,100)[]{$\tilde{q}_j$}
\ArrowLine(280,90)(310,120)
\Text(320,120)[]{${q}_I$}
\Vertex(279,88){3.5}
\Line(280,90)(310,60)
\Line(278,86)(308,56)
\Text(317,65)[]{$\psi_\mu$}
\Text(280,30)[]{(b)}
\end{picture}
\end{center}
\vskip-0.5in
\caption{Three-body gaugino decays into the gravitino}
\end{figure}
Utilizing the conditions $\overline{\psi^{(+)}_\mu}\gamma^\mu=\gamma^\mu\psi_\mu^{(-)}=0$, the amplitudes for the above two diagrams can be written as:
\begin{eqnarray}
\label{eq:gaugino_to_grav_i}
& & M_{(a)}\sim2p_{\tilde{q}}^\mu\frac{1}{M_p}\left(\overline{\psi^{(+)}_\mu}(p_\psi)
v(p_{\bar{q}})\right)\left(\frac{i}{p_{\tilde{q}}^2-m_{\tilde{q}}^2 + i\epsilon}\right)
\left(\bar{u}(p_q)V_{\lambda-q-\tilde{q}}\lambda^{(+)}_{\tilde{g}}\right)\nonumber\\
& & M_{(b)}\sim2p_{\tilde{q}}^\mu\frac{1}{M_p}\left(\bar{u}(p_q){\psi^{(-)}_\mu}(p_\psi)
v(p_{\bar{q}})\right)\left(\frac{i}{p_{\tilde{q}}^2-m_{\tilde{q}}^2 + i\epsilon}\right)
\left(\overline{\lambda^{(-)}_{\tilde{g}}} V_{\lambda-q-\tilde{q}}v(p_{\bar{q}})\right),
\end{eqnarray}
$V_{\lambda-q-\tilde{q}}\sim\tilde{f}{\cal V}^{-\frac{4}{5}}$ being the gaugino-quark-squark vertex.
From (\ref{eq:gaugino_to_grav_i}), one obtains the following helicities averaged sum:
\begin{eqnarray}
\label{eq:gaugino_to_grav_ii}
& & \sum_{s_\psi=\pm\frac{3}{2},\pm\frac{1}{2},s_q,s_{\bar{q}},s_{\lambda{g}}=\pm\frac{1}{2}}
M_{a)} M_{(a)}^\dagger\sim|V_{\lambda-q-\tilde{q}}|^2\frac{p^\mu_{\tilde{q}}p^\nu_{\tilde{q}}}{M_p^2}
Tr\left[{\cal P}^{(+)}_{\mu\nu}\left(-\slashed{p}_{\bar{q}} + m_{\bar{q}}\right)\right]Tr\left[\left(\slashed{p}_{\lambda_{\tilde{g}}} + m_{\lambda_{\tilde{g}}}\right)\left(\slashed{p}_q + m_q\right)\right],\nonumber\\
& & \sum_{s_\psi=\pm\frac{3}{2},\pm\frac{1}{2},s_q,s_{\bar{q}},s_{\lambda{g}}=\pm\frac{1}{2}}
M_{b)} M_{(b)}^\dagger\sim|V_{\lambda-q-\tilde{q}}|^2\frac{p^\mu_{\tilde{q}}p^\nu_{\tilde{q}}}{M_p^2}
Tr\left[{\cal P}^{(-)}_{\mu\nu}\left(\slashed{p}_{{q}} + m_{{q}}\right)\right]Tr\left[\left(- \slashed{p}_{\lambda_{\tilde{g}}} + m_{\lambda_{\tilde{g}}}\right)\left(- \slashed{p}_{\bar{q}} + m_q\right)\right],\nonumber\\
\end{eqnarray}
and utilizing $\bar{u}(p_q){\psi^{(-)}_\mu}(p_\psi)=-\overline{\psi^{(+)}_\mu}(p_\psi)
v(p_{\bar{q}}), \overline{\lambda^{(-)}_{\tilde{g}}} v(p_{\bar{q}})=-\bar{u}(p_q)\lambda^{(+)}_{\tilde{g}} $ to rewrite $M_{(b)}$ and therefore obtain:
\begin{equation}
\label{eq:gaugino_to_grav_iii}
\sum_{s_\psi=\pm\frac{3}{2},\pm\frac{1}{2},s_q,s_{\bar{q}},s_{\lambda{g}}=\pm\frac{1}{2}}\Re M_{(a)}M_{(b)}^\dagger\sim |V_{\lambda-q-\tilde{q}}|^2\frac{p^\mu_{\tilde{q}}p^\nu_{\tilde{q}}}{M_p^2}
Tr\left[{\cal P}^{(+)}_{\mu\nu}\left(-\slashed{p}_{\bar{q}} + m_{\bar{q}}\right)\right]Tr\left[\left(\slashed{p}_{\lambda_{\tilde{g}}} + m_{\lambda_{\tilde{g}}}\right)\left(\slashed{p}_q + m_q\right)\right].
\end{equation}
In (\ref{eq:gaugino_to_grav_i}) - (\ref{eq:gaugino_to_grav_iii}), the positive and negative energy gravitino projectors are given by:
\begin{equation}
\label{eq:P_pm_grav}
{\cal P}^{(\pm)}\equiv -\left(\slashed{p}_{3/2} \pm m_{3/2}\right)\left[\left(\eta_{\mu\nu} - \frac{p_{3/2,\mu}p_{3/2,\nu}}{m_{3/2}^2}\right) - \frac{1}{3}\left(\eta_{\mu\sigma} - \frac{p_{3/2,\mu}p_{3/2,\sigma}}{m_{3/2}^2}\right)\left(\eta_{\nu\lambda} - \frac{p_{3/2,\lambda}p_{3/2,\lambda}}{m_{3/2}^2}\right)\gamma^\sigma\gamma^\lambda\right].
\end{equation}
A typical term that one would need to calculate is:
\begin{eqnarray}
\label{eq:gaugino_to_grav_iv}
& & p^\mu_{\tilde{q}} p^\nu_{\tilde{q}} Tr\left[{\cal P}^{(\pm)}_{\mu\nu}\left(\eta\slashed{p}_{\eta} + m_{\bar{q}}\right)\right],\ \eta=\pm\ {\rm corresponding\ to}\ p_{\eta}=p_q/p_{\bar{q}}\nonumber\\
& & \sim 4\left\{\left(m^2_{\tilde{q}} - \frac{\left(p_{\tilde{q}}\cdot p_{3/2}\right)^2}{m^2_{3/2}}\right)\left(\eta p_{3/2}\cdot p_\eta\pm m_{3/2}m_\eta\right)\right.\nonumber\\
 & & \left.- \left(p^{\tilde{q}}_\mu - \frac{p_{\tilde{q}}\cdot p_{3/2} p_\mu^{3/2}}{m^2_{3/2}}\right)\left(p_{\tilde{q}}^\mu - \frac{p_{\tilde{q}}\cdot p_{3/2} p^\mu_{3/2}}{m^2_{3/2}}\right)\left(\frac{\eta}{3}p_\eta\cdot p_{3/2} \pm m_{3/2}m_\eta\right)\right\}.
\end{eqnarray}
Using results of \cite{Manuel Toharia}, to get an estimate of the decay width, one sees that:
\begin{eqnarray}
\label{eq:gaugino_to_grav_v}
& &  \Gamma\left(\tilde{W}\rightarrow q + \bar{q} + \psi_\mu\right)\nonumber\\
& & \hskip-0.8in\sim
{\rm Max}\left[\frac{1}{m_{\lambda_{\tilde{g}}}^3}\int_{s_{23}=m^2_{3/2}}^{m^2_{\lambda_{\tilde{g}}}}ds_{23}
\int_{s_{13}=\frac{m_{3/2}^2m^2_{\lambda_{\tilde{g}}}}{s_{23}}}^{m_{3/2}^2 + m_{\lambda_{\tilde{g}}}^2 - s_{23}}  ds_{13}
\frac{(\ref{eq:gaugino_to_grav_iv})|V_{\lambda-\tilde{q}-q}|^2\times
\left(p_{\lambda_{\tilde{g}}}\cdot p_q + m_{\lambda_{\tilde{g}}}m_q\right)}{(s_{23}^2 - m_{\tilde{q}}^2)^2, (s_{13}^2 - m_{\tilde{q}}^2)^2,  (s_{23}^2 - m_{\tilde{q}}^2)(s_{13}^2 - m_{\tilde{q}}^2)}\Biggr|_{{\cal V}\sim10^5, m_{q,\bar{q}}=0}\right]\nonumber\\
& & \sim {\rm Max}\left(10^{-32},10^{-30}, 10^{-32}\right)M_p = 10^{-28}M_p,
\end{eqnarray}
implying that the corresponding lifetime would be $10^{-15}s$.

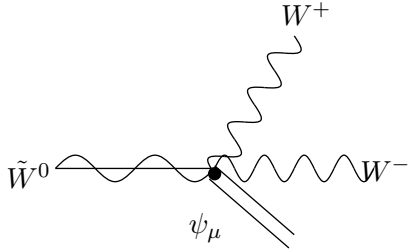
\begin{figure}
\begin{center}
    \begin{picture}(135,137) (15,-18)
   \Line(100,53)(130,27)
   \Line(98,48)(128,22)
   \Photon(40,52)(100,52){5}{2}
   \Photon(100,50)(130,102){5}{4}
   \Text(30,50)[]{{{$\tilde{W}^0$}}}
   \Line(40,52)(100,52)
   \Photon(100,52)(160,52){5}{4}
   \Text(165,52)[]{$W^-$}
   \Vertex(100,50){2.5}
   \Text(135,111)[]{{{$W^+$}}}
   \Text(97,30)[]{{{$\psi_\mu$}}}
  \end{picture}
   \end{center}
   \vskip-0.3in
   \caption{Contact-vertex three-body decay diagram}
   \end{figure}

In the approximation, $\frac{m_W}{m_{\tilde{W}}}\approx0,\frac{m_{3/2}}{m_{\tilde{W}}}\approx0$ (valid in the limit of massses $m_{\tilde{W}}\sim{\cal V}^{\frac{2}{3}}m_{\frac{3}{2}}$ and $m_{3/2}\sim{\cal V}^{-2}M_p$ as given in equation (\ref{eq:gaugino_mass})), the decay
width for $\tilde{W}^0\rightarrow\psi_\mu+W^++W^-$ is given by (See \cite{Hasenkamp}):
\begin{eqnarray}
\label{eq:Wino_grav+W+_W-}
& & \Gamma\left(\tilde{W}^0\rightarrow\psi_\mu+W^++W^-\right)\approx\frac{g^2_{YM}m_{\tilde{W}}^9}{34560\left(2\pi\right)^3m^4_WM_p^2}
\sim\frac{{\cal V}^{-8}M_p^5}{M_W^4},
\end{eqnarray}
which implies a lifetime of around $10^{-66}s$. Treating the Wino/Bino gauginos as the co-NLSPs, we hence conclude that the two- and three-body gaugino decays into the gravitino (LSP), respect the BBN bounds. If one were to calculate similar neutralino decays into the gravitino, then for the Higgsino contribution to the neutralino-gauge-gravitino vertex, one will have the following suppression factor:
$\tilde{f} \left(\tilde{f} {\cal V}^{-\frac{2}{3}}\right) \times {\cal O}(z_i)$ term in $g_{\rm YM}
\frac{g_{T_B \bar{z}_i}}{\sqrt{K_{{\cal Z}_i\bar{\cal Z}_i}}} \times \frac{\langle z_i\rangle}{M_p}$, which is $\tilde{f}^2 {\cal V}^{-\frac{2}{3} - \frac{5}{3}} \times 10^{\frac{5}{2}}.$ For ${\cal V}\sim10^5$, this is approximately, $10^{-17}$. The corresponding gaugino-gauge-gravitino vertex will, without worrying about the gamma matrices, be of ${\cal O}\left(\frac{p_Z}{M_p}\right)$. Now, even at low $Z$-momenta, this is around $10^{-16}$ for $M_Z\sim10^2 GeV$. Hence, the neutralino and gaugino decay (e.g., into gravitino and Z boson) widths, will approximately be the same.

\subsubsection{Gluino Decays Revisited}

 In this sub-section, we will readdress the gluino decays in the light of previous discussion done in \cite{Dhuria+Misra_mu_Split SUSY}. The earlier discussion was made with an ambiguity in identifying the quarks with right set of wilson line moduli. In the four Wilson line moduli framework where the first generation left handed quarks can be identified with Wilson line modulus $a_2$ (strictly speaking ${\cal A}_2$) and the first generation right handed quarks get identified with $a_4$ (strictly speaking ${\cal A}_4$) as mentioned above, by recalculating the moduli space metric, we briefly estimate gluino decay life time results for tree-level as well as one-loop decays into neutralino and Goldstino, the long life time of which not only provides another signature of $\mu$-split supersymmetry, the gluino decay into Goldstino (longitudinal component of gravitino) also demonstrates the invalidity of gluino as an appropriate NLSP from point of view of Dark Matter(LSP) production of Goldstino from neutralino .

$\underline{\bf{\tilde{g}\rightarrow q{\bar q}\chi_n}}$

 We first discuss gluino three-body decays that involve the process like $\tilde{g}\rightarrow q{\bar q}\chi_n$ - $\tilde{g}$ being a gaugino, $q/{\bar q}$ being quark/anti-quark and $\chi_n$ being a neutralino. More specifically, e.g., the gluino decays into an anti-quark and an off-shell squark and the off-shell squark decays into a quark and neutralino.
\begin{figure}
 \begin{center}
\begin{picture}(350,180)(50,0)
\Text(90,130)[]{$\tilde{g}$}
\Line(60,120)(110,120)
\Gluon(60,120)(110,120){5}{4}
\ArrowLine(110,120)(140,150)
\Text(147,150)[]{${q}_I$}
\DashArrowLine (130,90)(110,120){4}
\Text(107,100)[]{$\tilde{q}_i$}
\ArrowLine(160,120)(130,90)
\Text(167,120)[]{${\bar q_J}$}
\ArrowLine(130,90)(160,60)
\Text(167,65)[]{$\tilde{\chi}_3^0$}
\Text(110,30)[]{(a)}
\Text(250,130)[]{$\tilde{g}$}
\Line(210,120)(260,120)
\Gluon(210,120)(260,120){5}{4}
\ArrowLine(290,150)(260,120)
\Text(297,150)[]{$\bar{q}_J$}
\DashArrowLine(260,120)(280,90){4}{}
\Text(257,100)[]{$\tilde{q}_j$}
\ArrowLine(280,90)(310,120)
\Text(320,120)[]{${q}_I$}
\ArrowLine(310,60)(280,90)
\Text(317,65)[]{$\tilde{\chi}_3^0$}
\Text(280,30)[]{(b)}
\end{picture}
\end{center}
\vskip-0.5in
\caption{Three-body gluino decay diagrams}
\end{figure}
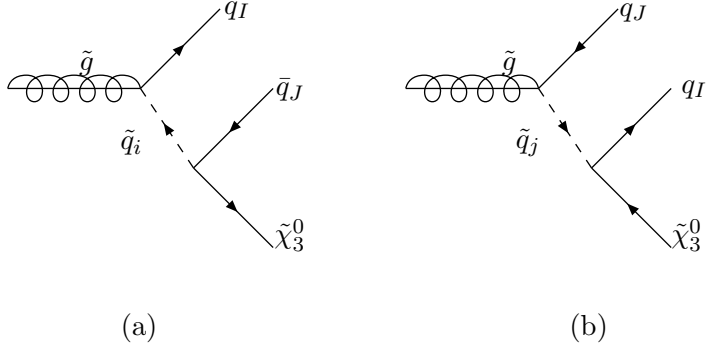
The Feynman diagrams involve gluino-quark/antivertex- squark vertex and neutralino-quark/antiquark-squark vertex. The gluino-quark/antiquark- squark in gauged supergravity action as also given in (\ref{eq:gaugino-quark-squark1}), is proportional to:
\begin{eqnarray}
\label{eq:G}
\ G^{{q/{\bar q}}_{{\cal A}_2}}_{\tilde{q}_{{\cal A}_2}}\sim\tilde{f}{\cal V}^{-\frac{4}{5}}{\tilde q}\chi^{q/ \bar q}\lambda_{\tilde{g}},\lambda_{\tilde{g}}~ {\rm corresponds~ to ~gluino}
\end{eqnarray}
and from (\ref{eq:neutralino-quark-squark1}), contribution of  physical neutralino-quark/antiquark-squark is proportional to
\begin{eqnarray}
\label{eq:X}
\ X^{{q/{\bar q}}}_{\tilde{q}}\sim \tilde{f}{\cal V}^{-\frac{4}{5}}{\tilde q}\chi^{q/ \bar q}{\tilde{\chi}_3^0}
\end{eqnarray}
Using RG
analysis of coefficients of the effective dimension-six gluino decay operators as given in
\cite{Guidice_et_al}, it was shown in \cite{Dhuria+Misra_mu_Split SUSY} that these coefficients at the EW scale are of the same order as that at the squark mass scale.

The analytical formulae to calculate decay width for three-body tree-level gluino decay channel as given in \cite{Manuel Toharia} is:-
\begin{eqnarray}
\label{eq:neutralinowidth}
\hspace{-.5cm}\Gamma(\tilde{g}\to\chi_{\rm n}^{o}q_{{}_I} \bar{q}_{{}_J} )
=  {g_s^2\over256 \pi^3 \mgss^3 } \sum_{i,j} \int ds_{13} ds_{23}\ {1\over2}
{\cal R}e\Big(A_{ij}(s_{23}) + B_{ij}(s_{13}) -  2 \eps_{n} \eps_{{g}}\
C_{ij}(s_{23},s_{13})\Big)
\end{eqnarray}
where the integrand is the square of the spin-averaged total amplitude
and $i,j=1,2,..,6$ are the indices of the squarks mediating the decay. The limits of integration in (\ref{eq:neutralinowidth}) are
given in  \cite{Dhuria+Misra_mu_Split SUSY}.
The $A_{ij}$ terms in (\ref{eq:neutralinowidth}) represent the contributions from the gluino decay channel involving gluino$\rightarrow$squark+quark and squark$\rightarrow$neutralino+anti-quark, whereas
 the $B_{ij}$ terms come from channel gluino $\rightarrow$squark+anti-quark and squark$\rightarrow$neutralino+quark. The same are defined in \cite{Dhuria+Misra_mu_Split SUSY}.
Utilizing the results as given in (\ref{eq:G}) and (\ref{eq:X}),
{\small $A_{ij}\Biggl(Tr\left[G^{{\bar q}}_{\tilde{q}}G^{{\bar q}}_{\tilde{q}}\ ^\dagger\right]
Tr\left[X^{q}_{\tilde{q}}X^{q}_{\tilde{q}}\ ^\dagger\right]\Biggr)\sim \tilde{f^4}{\cal V}^{-\frac{16}{5}}$}$\sim $
{\small $B_{ij}\Biggl(Tr\left[G^{q_{a_1}}_{\tilde{q}_{a_1}}G^{q_{a_1}}_{\tilde{q}_{a_1}}\ ^\dagger\right]
Tr\left[X^{{\bar q}_{a_1}}_{\tilde{q}_{a_1}}X^{{\bar q}_{a_1}}_{\tilde{q}_{a_1}}\ ^\dagger\right]\Biggr)\sim \tilde{f^4}{\cal V}^{-\frac{16}{5}}$},

{\small $C\Biggl(Tr\left[G^{{\bar q}}_{\tilde{q}_{a_1}}G^{q}_{\tilde{q}}\ ^\dagger X^{q}_{\tilde{q}}X^{{\bar q}}_{\tilde{q}}\ ^\dagger\right]\Biggr)\sim \tilde{f^4}{\cal V}^{-\frac{16}{5}}$}. Substituting these values for various vertex elements and solving equation (\ref{eq:neutralinowidth}) by using the same limits of integration as given in \cite{Dhuria+Misra_mu_Split SUSY}, dominating contribution of decay width  is:
\begin{eqnarray}
 \label{eq: Decay width 1}
 \hspace{-.5cm}\Gamma(\tilde{g}\to\chi_{\rm n}^{o}q_{{}_I} \bar{q}_{{}_J} )
&\sim&{g_s^2  \over256 \pi^3  {\cal V}^2 m_{\frac{3}{2}}^3 }\left[
{\tilde{f}^4  {\cal V}^{-\frac{16}{5}} m_{\frac{3}{2}}^4 }{\cal V}^{\frac{10}{3}}+ \tilde{f}^4{\cal V}^{-\frac{16}{5}}{\cal V}^{\frac{10}{3}} m_{\frac{3}{2}}^4 +\tilde{f}^4{\cal V}^{-\frac{16}{5}} {\cal V}^{\frac{10}{3}} m_{\frac{3}{2}}^4\right] \nonumber\\
& & \sim { g_s^2\over256 \pi^3  {\cal V}^2 m_{\frac{3}{2}}^3 }\left( \tilde{f}^4{\cal V}^{-\frac{16}{5}}{\cal V}^{\frac{10}{3}} m_{\frac{3}{2}}^4  \right)\sim O(10^{-4})\tilde{f}^4{\cal V}^{-\frac{19}{5}}{M_p} \nonumber\\
& & \sim O(10^{-5})\tilde{f}^4 GeV
 \end{eqnarray}
Considering   ${f^2}\sim{10}^{-10}$ as calculated in \cite{Dhuria+Misra_mu_Split SUSY} in dilute flux approximation , decay width of gluino i.e equation (\ref{eq: Decay width 1}) becomes:
$ \Gamma(\tilde{g}\to\chi_{\rm n}^{o}q_{{}_I} \bar{q}_{{}_J} )\sim O(10^{-5})\tilde{f}^4 < {\cal O}(10^{-25})GeV$.
 Further, Life time of gluino is given as
   \begin{eqnarray}
   \tau &=&\frac{\hbar}{\Gamma}\sim\frac{10^{-34} Jsec}{10^{-5}f^4 GeV}\sim\frac{10^{-19}}{f^4} > {10}^{1} sec
   \end{eqnarray}

 $\underline{\bf{\tilde{g}\rightarrow\tilde{\chi}_3^0+g}}$

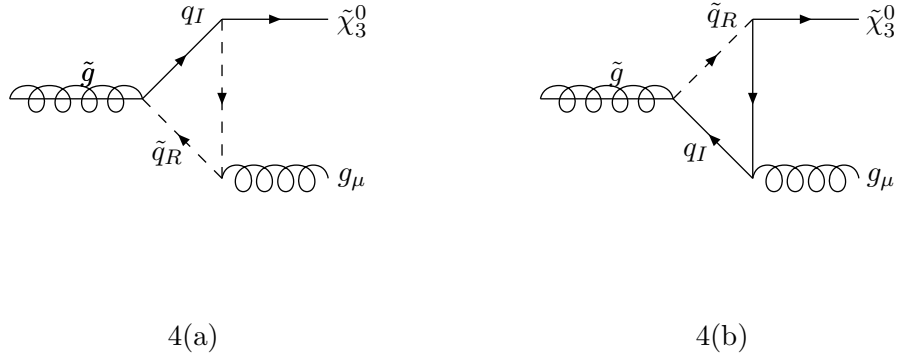
\begin{figure}
\begin{center}
\begin{picture}(1000,200)(50,0)
\Text(140,130)[]{$\tilde{g}$}
\Line(110,120)(160,120)
\Gluon(110,120)(160,120){5}{4}
\Text(140,130)[]{$\tilde{g}$}
\ArrowLine (160,120)(190,150)
\Text(180,152)[]{$q_I$}
\DashArrowLine (190,90)(160,120){4}
\Text(170,100)[]{$\tilde{q}_R$}
\DashArrowLine (190,150)(190,90){4}
\Gluon(190,90)(230,90){5}{4}
\Text(240,90)[]{$g_\mu$}
\ArrowLine(190,150)(230,150)
\Text(240,150)[]{$\tilde{\chi}_3^0$}
\Text(180,30)[]{4(a)}
\Line(310,120)(360,120)
\Gluon(310,120)(360,120){5}{4}
\Text(340,130)[]{$\tilde{g}$}
\DashArrowLine (360,120)(390,150){4}
\Text(380,152)[]{$\tilde{q}_R$}
\ArrowLine (390,90)(360,120)
\Text(370,100)[]{$q_I$}
\ArrowLine (390,150)(390,90)
\Gluon(390,90)(430,90){5}{4}
\Text(440,90)[]{$g_\mu$}
\ArrowLine(390,150)(430,150)
\Text(440,150)[]{$\tilde{\chi}_3^0$}
\Text(380,30)[]{4(b)}
\end{picture}
\end{center}
\caption {Diagrams contributing to one-loop two-body gluino decay}
\end{figure}

Relevant to Figs.4(a) and 4(b), the gluino-quark-squark vertex and the neutralino-quark-squark vertex will be given by (\ref{eq:G})- (\ref{eq:X}). As explained in \cite{Dhuria+Misra_mu_Split SUSY}, the quark-quark-gluon vertex relevant to figure 4, from \cite{Wess_Bagger} is given by:
  \begin{eqnarray}
  \label{eq:qqgl 1}
& &  g_{I{\bar J}}{\bar\chi}^{\bar J}{\bar\sigma}\cdot A\ {\rm Im}\left(X^BK + i D^B\right)\chi^I,\nonumber\\
& & \sim g_{YM}g_{{\cal A}_2}{\bar {{\cal A}_2}}{\bar\chi}^{\bar {{\cal A}_2}}{\bar\sigma}\cdot A\left\{6\kappa_4^2\mu_72\pi\alpha^\prime Q_BK+\frac{12\kappa_4^2\mu_72\pi\alpha^\prime Q_Bv^B}{\cal V}\right\}
  \end{eqnarray}
   $\chi^{{\cal A}_2}$ and ${\bar\chi}^{\bar {{\cal A}_2}}$ to quarks and antiquarks.
 $X=X^B\partial_B=-12i\pi\alpha^\prime\kappa_4^2\mu_7Q_B\partial_{T_B}$  corresponding to the killing isometry vector and  $D$ term generated
 is given by:
\begin{equation}
\label{eq:D_term}
D^B=\frac{4\pi\alpha^\prime\kappa_4^2\mu_7Q_Bv^B}{\cal V}
\end{equation}
Utilizing  value of $g_{{\cal A}_2{\bar {\cal A}}_{\bar 2}}\sim{\cal V}^{-\frac{5}{9}}, v^B\sim{\cal V}^{\frac{1}{3}}, Q_B\sim{\cal V}^{\frac{1}{3}}(2\pi\alpha^\prime)^2\tilde{f}$,  it has been shown in \cite{dhuria+misra_EDM} that quark-quark-gluon vertex will be of ${\cal O}(1)$.

  The gauge kinetic term for squark-squark-gluon vertex, relevant to Fig.4(b) will be given by $ {\frac{1}{\kappa_4^2{\cal V}^2}}G^{\sigma_B{\bar\sigma}_B}\tilde{\bigtriangledown}_\mu T_B\tilde{\bigtriangledown}^\mu {\bar T}_{\bar B}$.  Using the value of $G^{\sigma_B{\bar\sigma}_B}(EW)= {\cal V}^{\frac{7}{3}}$ (for detailed explanation, see \cite{dhuria+misra_EDM}), the given term generates the required squark-squark-gluon vertex as follows:
\begin{eqnarray}
\label{eq:sq sq gl}
& & \hskip-0.63in {\frac{6i\kappa_4^2\mu_72\pi\alpha^\prime Q_BG^{\sigma_B{\bar\sigma}_B}}{\kappa_4^2{\cal V}^2}}A^\mu\partial_\mu\left(\kappa_4^2\mu_7(2\pi\alpha^\prime)^2C_{2{\bar 2}}{\cal A}_2{\bar {\cal A}}_{\bar 2}\right)\xrightarrow[{\small \kappa_4^2\mu_7(2\pi\alpha^\prime)^2C_{2{\bar 2}}\sim{\cal V}^{\frac{1}{9}}}]{\small G^{\sigma_B{\bar\sigma}_B}\sim{\cal V}^{\frac{7}{3}},}\frac{{\cal V}^{\frac{7}{9}}\epsilon\cdot\left(2k-(p_{\tilde{\chi}_3^0}+p_{\tilde{g}})\right)}{\left(\sqrt{\hat{K}_{{\cal A}_2{\bar {\cal A}}_2}}\right)^2}\nonumber\\
& & {\hskip -0.5in} \sim O({10}^{2})\tilde{f}{\cal V}^{\frac{7}{9}}\left[ 2\epsilon\cdot k-\epsilon\cdot\left(p_{\tilde{\chi}_3^0}+p_{\tilde{g}}\right)\right]\sim \tilde{f}{\cal V}^{\frac{53}{45}}, {\rm for~{\cal V}\sim {10}^5}
\end{eqnarray}

\begin{figure}
\begin{center}
\begin{picture}(500,180)(-100,0)
\Text(280,120)[]{${\tilde f} {\cal V}^{\frac{53}{45}}$}
\DashArrowLine(0,120)(-50,120){5}
\Text(-80,120)[]{$\tilde{q}_R(k-p_{\tilde{g}})$}
\Text(65,153)[]{$\tilde{q}_R(k-p_{\tilde{\chi}_3^0})$}
\DashArrowLine(40,150)(0,120){5}
\Text(54,88)[]{$ g_\mu (\epsilon_\mu)$}
\Gluon(0,120)(40,90){5}{4}
\Text(0,81)[]{(\bf a)}
\end{picture}
\vskip-1.5in
\begin{picture}(500,200)(-100,0)
\Text(-57,120)[]{$\tilde{g}$}
\Text(280,120)[]{$\tilde{f}{\cal V}^{-\frac{4}{5}}$}
\Line(-50,120)(0,120)
\Gluon(-50,120)(0,120){5}{4}
\ArrowLine(0,120)(40,150)
\Text(47,150)[]{$\tilde{q}_R$}
\DashArrowLine(40,90)(0,120){4}
\Text(47,90)[]{$\bar{q}_I$}
\Text(0,81)[]{(\bf b)}
\end{picture}
\vskip-1.5in
\begin{picture}(500,220)(-100,0)
\Text(278,120)[]{${\tilde f}{\cal V}^{-\frac{4}{5}}$}
\ArrowLine(-50,120)(0,120)
\Text(-70,120)[]{$q_I(k)$}
\Text(57,150)[]{$\tilde{\chi}_3^0(p_{\tilde{\chi}_3^0})$}
\ArrowLine(0,120)(40,150)
\Text(63,83)[]{$ \tilde{q}_R(k-p_{\tilde{\chi}_3^0}) $}
\DashArrowLine(0,120)(40,90){5}
\Text(0,81)[]{(\bf c)}
\end{picture}
\vskip-1.1in
\begin{picture}(500,180)(-100,0)
\Text(280,120)[]{${\cal O}(1)$}
\ArrowLine(0,120)(-50,120)
\Text(-55,120)[]{$q_J$}
\Text(47,158)[]{$g_\mu(\epsilon_\mu)$}
\Gluon(40,150)(0,120){5}{4}
\Text(47,88)[]{$ q_I $}
\ArrowLine(40,90)(0,120)
\Text(0,81)[]{(\bf d)}
\end{picture}
\end{center}
\vskip-0.6in
\caption{Different vertices relevant to one-loop  gluino decay}
\end{figure}
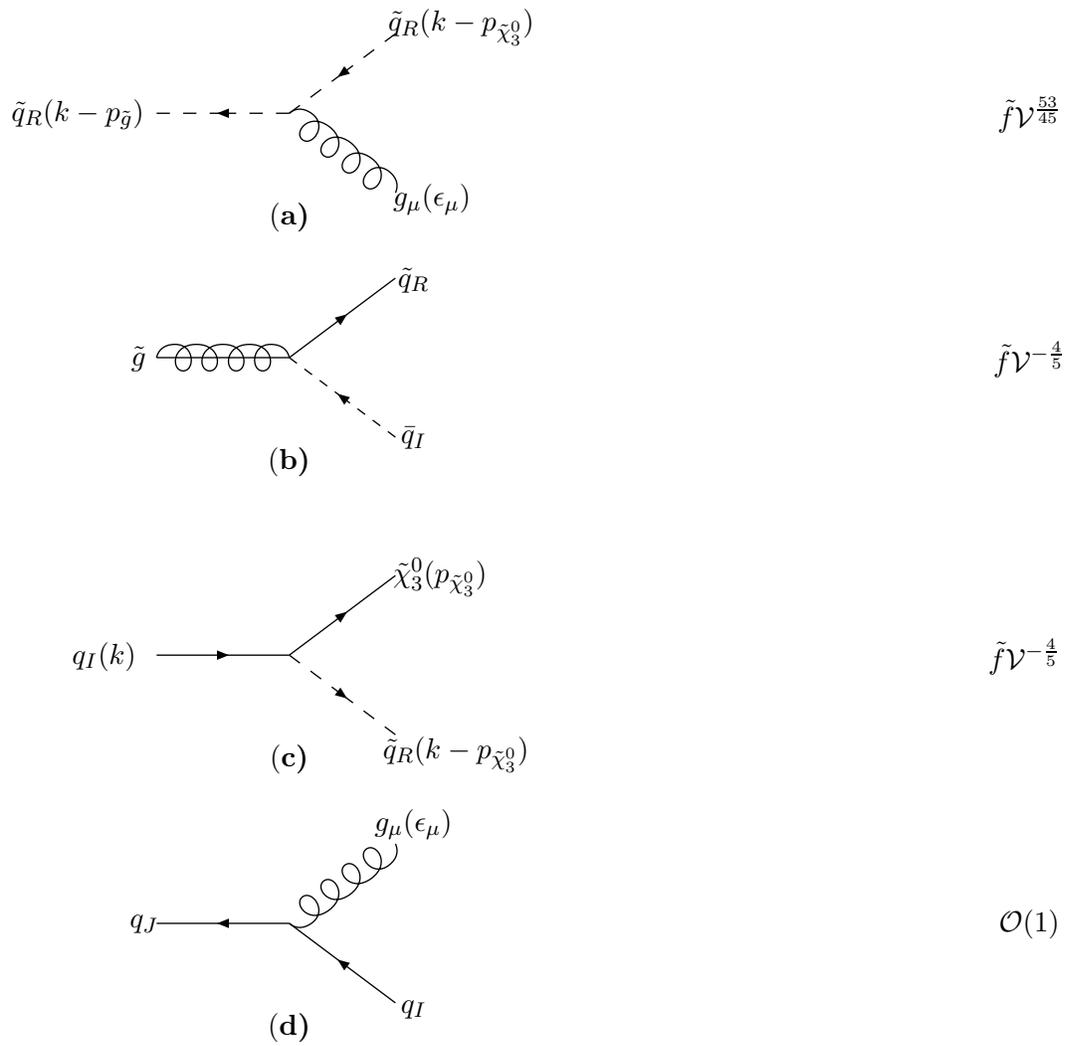
  In \cite{Dhuria+Misra_mu_Split SUSY}, it has been discusssed that behavior of Wilson coefficients corresponding to  two-body gluino decay does not change much upon RG evolution to EW scale.  Using the vertices calculated above relevant to Figs. 4(a) and 4(b), and the Feynman rules of \cite{2comp}, one obtains for the scattering amplitude:
\begin{eqnarray}
\label{eq:M_I}
& & {\cal M}\sim\tilde{f}^3M_p \int\frac{d^4k}{\left(2\pi\right)^4} \times{\cal V}^{-\frac{4}{5}}\left(\frac{i{\bar\sigma}\cdot k}{k^2-m_q^2+i\epsilon}\right)\left({\cal V}^{-\frac{4}{5}}
\right)\left(\frac{i}{\left[\left(k-p_{\tilde{G}}\right)^2-m^2_{\tilde{q}}+i\epsilon\right]}
\right)\nonumber\\
& & \times\left({\cal V}^{\frac{53}{45}}\right)\left(\frac{i}
{\left[\left(k-p_{\tilde{g}}\right)^2-m^2_{\tilde{q}}+i\epsilon\right]}\right)\nonumber\\
& & + \tilde{f}^2M_p \int\frac{d^4k}{\left(2\pi\right)^4} \times {\cal V}^{-\frac{4}{5}}\left(\frac{i}
{\left[\left(k+p_{\tilde{\chi}_3^0}\right)^2-m^2_{\tilde{q}}+i\epsilon\right]}\right)\left({\cal V}^{-\frac{4}{5}}
\right)\left(\frac{i{\bar\sigma}\cdot k}{k^2-m_q^2+i\epsilon}\right)\nonumber\\
& & \times\left(\frac{i{\bar\sigma}\cdot\left(k-p_{g_\mu}\right)}{\left[
\left(k-p_{g_\mu}\right)^2-m^2_q+i\epsilon\right]}\right)
\end{eqnarray}
Using the 1-loop integrals of \cite{Pass+Velt}:
\begin{eqnarray}
\label{eq:passvelt}
& & \frac{1}{i}\int\frac{d^4k}{\left(2\pi\right)^4}\frac{\left(k_\mu,\ k_\mu k_\nu\right)}{\left(k^2-m_1^2+i\epsilon\right)
\left[\left(k+p_1\right)^2-m_2^2+i\epsilon\right]\left[\left(k+p_1+p_2\right)^2-m_3^2+i\epsilon\right]}\nonumber\\
& & =4\pi^2\Biggl[p_{1\mu}C_{11}+p_{2\mu}C_{12}, p_{1\mu}p_{1\nu}C_{21}+p_{2\mu}p_{2\nu}C_{22}+\left(p_{1\mu}p_{2\nu}+p_{1\nu}p_{2\mu}\right)C_{23}+\eta_{\mu\nu}C_{24}\Biggr];
\nonumber\\
& & (a) m_1=m_q, m_2=m_3=m_{\tilde{q}};\ p_1=-p_{\tilde{\chi}_3^0}, p_2=-p_{g_\mu};\nonumber\\
& & (b) m_1=m_3=m_q, m_2=m_{\tilde{q}};\ p_1=p_{\tilde{\chi}_3^0}, p_2=-p_{\tilde{g}}.
\end{eqnarray}

  The one loop three point functions $C_{ij}$'s relevant for cases (a) and (b) have been calculated in \cite{Dhuria+Misra_mu_Split SUSY} and are given as under:
\begin{eqnarray}
\label{eq: three point functions}
& & \begin{array}{l}
C^{(a)}_{24}= O(10^6), C^{(b)}_{24}=  O(10^6) ;\nonumber\\
C^{(a)}_0= \frac{1}{4{\pi}^2}\times 10^{-21}\sim O(1)\times {10}^{-23}{GeV}^{-2};\nonumber\\
C^{(b)}_{11}=  O(10) \times {10}^{-16} {GeV}^{-2};\nonumber\\
C^{(b)}_{12}=  O(10)\times {10}^{-16} {GeV}^{-2};\nonumber\\
\end{array}\nonumber\\
& & \begin{array}{l}
C^{(b)}_{21}= C^{(b)}_{22}= C^{(b)}_{23}\sim O(1) \times {10}^{-10}{GeV}^{-2};\nonumber\\
C^{(b)}_0= \frac{1}{4{\pi}^2}\times 10^{-21}\sim O(1)\times {10}^{-23}{GeV}^{-2}.\nonumber\\
\end{array} \nonumber\\
\end{eqnarray}
Now, equation(\ref{eq:M_I}) can be evaluated to yield:
\begin{eqnarray}
\label{eq:M_II}
& & {\hskip -0.6in}{\bar u}(p_{\tilde{\chi}_3^0})\Biggl(\tilde{f}^3{\cal V}^{-0.4}\Biggl[\left\{{\bar\sigma}\cdot p_{\tilde{\chi}_3^0}C^{(a)}_{11}+{\bar\sigma}\cdot p_{g_\mu}C^{(a)}_{12}\right\}(2\epsilon\cdot p_{\tilde{\chi}_3^0})+\left\{{\bar\sigma}\cdot p_{\tilde{\chi}_3^0}\epsilon\cdot p_{\tilde{\chi}_3^0}C^{(a)}_{21}+{\bar\sigma}\cdot p_{g_\mu}\epsilon\cdot p_{\tilde{\chi}_3^0}C^{(a)}_{23}
+{\bar\sigma}\cdot\epsilon C^{(a)}_{24}\right\}\Biggr]\nonumber\\
& & {\hskip -0.3in}+ \tilde{f}^2{\cal V}^{-1.6}\Biggl[-\left\{{\bar\sigma}\cdot p_{\tilde{\chi}_3^0}C^{(b)}_{11}
+{\bar\sigma}\cdot p_{\tilde{g}}C^{(b)}_{12}\right\}{\bar\sigma}\cdot\epsilon{\bar\sigma}\cdot p_{g_\mu}+
{\bar\sigma}\cdot p_{\tilde{\chi}_3^0}{\bar\sigma}\cdot\epsilon {\bar\sigma}\cdot p_{\tilde{\chi}_3^0}C^{(b)}_{21} + {\bar\sigma}\cdot p_{\tilde{g}}{\bar\sigma}\cdot\epsilon {\bar\sigma}\cdot p_{\tilde{g}}C^{(b)}_{22}\nonumber\\
& & {\hskip -0.3in} -\biggl({\bar\sigma}\cdot p_{\tilde{\chi}_3^0}{\bar\sigma}\cdot\epsilon{\bar\sigma}\cdot p_{\tilde{g}} + {\bar\sigma}\cdot p_{\tilde{g}}{\bar\sigma}\cdot\epsilon {\bar\sigma}\cdot p_{\tilde{\chi}_3^0}\biggr)C^{(b)}_{23}+ {\bar\sigma}_\mu{\bar\sigma}\cdot\epsilon{\bar\sigma}^\mu C^{(b)}_{24}\Biggr]\Biggr)
u(p_{\tilde{g}}),
\end{eqnarray}
which equivalently could be rewritten as:
\begin{equation}
\label{eq:M_III}
 {\bar u}(p_{\tilde{\chi}_3^0})\Biggl[{\bar\sigma}\cdot{\cal A}+{\bar\sigma}\cdot p_{\tilde{\chi}_3^0}{\bar\sigma}\cdot\epsilon{\bar\sigma}\cdot{\cal B}_1 + {\bar\sigma}\cdot p_{g_\mu}
{\bar\sigma}\cdot\epsilon{\bar\sigma}\cdot{\cal B}_2 + D_3{\bar\sigma}_\mu{\bar\sigma}\cdot\epsilon{\bar\sigma}^\mu
C^{(b)}_{24} \Biggr]u(p_{\tilde{g}}),
\end{equation}
where
\begin{eqnarray}
\label{eq:AB1B2}
& & {\cal A}^\mu\equiv \tilde{f}^3{\cal V}^{-0.4}\left[p_{\tilde{\chi}_3^0}^\mu\epsilon\cdot p_{\tilde{\chi}_3^0}\left(2C^{(a)}_{11} + C^{(a)}_{21}\right) + p_{g_\mu}^\mu\epsilon\cdot p_{\tilde{\chi}_3^0}\left(C^{(a)}_{12} + C^{(a)}_{23}\right) + \epsilon^\mu C^{(a)}_{24}\right];\nonumber\\
& & {\cal B}_1^\mu\equiv  \tilde{f}^2{\cal V}^{-1.6}\left[-p_{g_\mu}^\mu\left(C^{(b)}_{11} + C^{(b)}_{12} + C^{(b)}_{23}-C^{(b)}_{22}\right) + p_{\tilde{\chi}_3^0}^\mu\left(C^{(b)}_{21} + C^{(b)}_{22} - 2C^{(b)}_{23}\right)\right];\nonumber\\
& & {\cal B}_2^\mu\equiv \tilde{f}^2{\cal V}^{-1.6}\left[p_{g_\mu}^\mu\left(C^{(b)}_{12}+C^{(b)}_{22}\right) + p_{\tilde{\chi}_3^0}^\mu\left( C^{(b)}_{22} - C^{(b)}_{23}\right)\right];\nonumber\\
& & {D_3}\equiv \tilde{f}^2{\cal V}^{-1.6}
\end{eqnarray}
Replacing ${\bar u}(p_{\tilde{\chi}_3^0}){\bar\sigma}\cdot p_{\tilde{\chi}_3^0}$ by $m_{\tilde{\chi}_3^0}{\bar u}(p_{\tilde{\chi}_3^0})$ and ${\bar\sigma}\cdot p_{\tilde{g}}u(p_{\tilde{g}})$ by $m_{\tilde {g}}$, and using $\epsilon\cdot p_{\tilde{\chi}_3^0}=0$,(\ref{eq:M_III}) be simplified to:
\begin{equation}
\label{eq:M_IV}
{\cal M}\sim {\bar u}(p_{\tilde{\chi}_3^0})\left[A{\bar\sigma}\cdot\epsilon + B_1{\bar\sigma}\cdot\epsilon
{\bar\sigma}\cdot p_{\tilde{\chi}_3^0} + B_2{\bar\sigma}\cdot p_{\tilde{g}}{\bar\sigma}\cdot\epsilon +
D_1{\bar\sigma}\cdot p_{\tilde{g}}{\bar\sigma}\cdot\epsilon{\bar\sigma}\cdot p_{\tilde{\chi}_3^0} + {D_3} {\bar\sigma}_\mu{\bar\sigma}\cdot\epsilon{\bar\sigma}^\mu C^{(b)}_{24}\right]u(p_{\tilde{g}}),
\end{equation}
where
\begin{eqnarray}
\label{eq:AB1B2D1 defs}
& & A\equiv {\cal V}^{-0.4} \tilde{f}^3 C^{(a)}_{24} - m_{\tilde{g}}m_{\tilde{\chi}_3^0} {\cal V}^{-1.6} \tilde{f}^2\Biggl\{
 C^{(b)}_{11} + 2 C^{(b)}_{12} + C^{(b)}_{23}-C^{(b)}_{22}\Biggr\},\nonumber\\
 & & B_1\equiv {\cal V}^{-1.6}\tilde{f}^2\left(C^{(b)}_{11} + 2 C^{(b)}_{12} +  C^{(b)}_{21}\right)m_{\tilde{\chi}_3^0},\nonumber\\
 & & B_2\equiv {\cal V}^{-1.6}\tilde{f}^2\left(C^{(b)}_{12}+C^{(b)}_{22}\right)m_{\tilde{g}},\ D_1\equiv {\cal V}^{-1.6}\tilde{f}^2\left(- C^{(b)}_{12}- C^{(b)}_{23}\right).
\end{eqnarray}
Utilizing values of C's calculated in (\ref{eq: three point functions}),
 \begin{eqnarray}
 \label{eq: A, B and D functions}
& &\hskip-0.6in  A \sim {\cal O}(10^{5})\tilde{f}^2, B_1\sim {\cal O}({10^{-7}})\tilde{f}^2{GeV}^{-2}, B_2\sim {\cal O}({10^{-7}})\tilde{f}^2{GeV}^{-2}, D_1\sim {\cal O}({10^{-18}})\tilde{f}^2{GeV}^{-2};{D_3}\sim {\cal V}^{-1.6}\tilde{f}^2; \nonumber\\
 \end{eqnarray}
\begin{eqnarray}
\label{eq:spinavGamma}
& &  {\hskip -1.0in}\sum_{\tilde{g}\ {\rm and}\ \tilde{\chi}_3^0\ {\rm spins}}|{\cal M}|^2
\sim Tr\Biggl(\sigma\cdot p_{\tilde{\chi}_3^0}\left[A{\bar\sigma}\cdot\epsilon + B_1{\bar\sigma}\cdot\epsilon
{\bar\sigma}\cdot p_{\tilde{\chi}_3^0} + B_2{\bar\sigma}\cdot p_{\tilde{g}}{\bar\sigma}\cdot\epsilon +
D_1{\bar\sigma}\cdot p_{\tilde{g}}{\bar\sigma}\cdot\epsilon{\bar\sigma}\cdot p_{\tilde{\chi}_3^0} + D_3 {\bar\sigma}_\mu{\bar\sigma}\cdot\epsilon{\bar\sigma}^\mu C^{(b)}_{24}\right]\nonumber\\
& & {\hskip -0.6in}\times\sigma\cdot p_{\tilde{g}}\left[A{\bar\sigma}\cdot\epsilon + B_1{\bar\sigma}\cdot\epsilon
{\bar\sigma}\cdot p_{\tilde{\chi}_3^0} + B_2{\bar\sigma}\cdot p_{\tilde{g}}{\bar\sigma}\cdot\epsilon +
D_1{\bar\sigma}\cdot p_{\tilde{g}}{\bar\sigma}\cdot\epsilon{\bar\sigma}\cdot p_{\tilde{\chi}_3^0} + D_3 {\bar\sigma}_\mu{\bar\sigma}\cdot\epsilon{\bar\sigma}^\mu C^{(b)}_{24}\right]^\dagger\Biggr),
\end{eqnarray}
which at:
\begin{equation}
\label{eq:kinematic_point}
p^0_{\tilde{\chi}_3^0}=\sqrt{m_{\tilde{\chi}_3^0}^2c^4+\rho^2},
p^1_{\tilde{\chi}_3^0}=p^2_{\tilde{\chi}_3^0}=p^3_{\tilde{\chi}_3^0}=\frac{\rho}{\sqrt{3}}
=\frac{1}{\sqrt{3}}\frac{c\left(m_{\tilde{g}}^2 - m_{\tilde{\chi}_3^0}^2\right)}{2m_{\tilde{g}}},
\end{equation}
yields:
\begin{eqnarray}
\label{eq:trace}
& &  \hskip -0.6in\frac{1}{256}{m_{\tilde{g}}}^2 \left[6 {m_{\tilde{g}}}^2 ({B_1}+{D_1} {m_{\tilde{g}}})^2+\left\{8
   {A_1}+16 D_3{C_{24}}+{m_{\tilde{g}}} \left(\left(5+\sqrt{3}\right) {B_1}+8 {B_2}+\left(5+\sqrt{3}\right) {D_1}
   {m_{\tilde{g}}}\right)\right\}^2\right],\nonumber\\
   & &
\end{eqnarray}
in the rest frame of the gluino.

Incorporating results of (\ref{eq:AB1B2D1 defs})in equation (\ref{eq:trace}), one gets
\begin{eqnarray}
\label{eq:spinavGamma result}
& &  \sum_{\tilde{g}\ {\rm and}\ \tilde{\chi}_3^0\ {\rm spins}}|{\cal M}|^2 \sim 0.2 {D_1}^2  m_{\tilde{g}}^6
\end{eqnarray}
Now, using standard two-body decay results (See \cite{Griffiths_particle}), the decay width $\Gamma$ is given by the following expression:
\begin{equation}
\label{eq:Gamma}
\Gamma=\frac{\sum_{\tilde{g}\ {\rm and}\ \tilde{\chi}_3^0\ {\rm spins}}|{\cal M}|^2\left(m_{\tilde{g}}^2-m_{\tilde{\chi}_3^0}^2\right)}{16\pi\hbar m_{\tilde{g}}^3}
\end{equation}

Using result of (\ref{eq: A, B and D functions}); $D_1\sim O({10^{-18}}){\tilde f}^2{GeV}^{-2}$  and $m_{\tilde{g}}\sim {\cal V}^{-\frac{4}{3}}{M_p}$ GeV, $m^{2}_{\tilde{\chi_3^0}}\sim m^{2}_{\tilde{g}}+ \frac{m_{\tilde g} {\tilde f}^2  v^2 {\cal V}^{\frac{2}{3}}}{m_{pl}}$ GeV, two body decay width is given as:
\begin{equation}
\label{eq:Gamma1}
\Gamma=  \frac{(0.2) {D_1^2} m_{\tilde {g}}^6} {16\pi m_{\tilde{g}}^3} \sim \frac{O({10}^{-2}) \tilde{f}^4.{10^{-36}}.{m_{\tilde {g}}}^6 (2\frac{m_{\tilde g} {\tilde f}^2  v^2 {\cal V}^{\frac{2}{3}}}{m_{pl}})} {m_{\tilde{g}}^3}\sim{10}^{-4} \tilde{f}^4 GeV
\end{equation}
Considering $\tilde{f} < {10}^{-5}$ as calculated above, $\Gamma < {10}^{-24}$ GeV.  Life time of gluino is given as:
   \begin{eqnarray}
   \tau &=&\frac{\hbar}{\Gamma}\sim\frac{10^{-34} Jsec}{10^{-4}f^4 GeV}\sim\frac{10^{-20}}{f^6} sec > {10}^{10}sec
   \end{eqnarray}

   {\bf Gluino($\tilde{g}$) decays into Goldstino($\tilde{G}$)}

We now consider the three-body decay of the gluino into Goldstino and a quark and anti-quark: $\tilde{g}\rightarrow\tilde{G}+q+{\bar q}$. The Feynman diagrams corresponding to this particular decay are shown in Fig. 8.
\begin{figure}
\begin{center}
\begin{picture}(350,180)(50,0)
\Text(90,130)[]{$\tilde{g}$}
\Line(60,120)(110,120)
\Gluon(60,120)(110,120){5}{4}
\Vertex(110,120){3.5}
\ArrowLine(110,120)(140,150)
\Text(149,150)[]{${q}_{a_2}$}
\DashArrowLine (130,90)(110,120){4}
\Text(107,100)[]{$\tilde{q}_{a_2}$}
\ArrowLine(160,120)(130,90)
\Text(167,120)[]{${\bar{q}_{a_2}}$}
\Line(130,90)(160,60)
\Line(128,86)(158,56)
\Vertex(129,88){3.5}
\Text(167,65)[]{$\psi_\mu$}
\Text(250,130)[]{$\tilde{g}$}
\Line(210,120)(260,120)
\Gluon(210,120)(260,120){5}{4}
\ArrowLine (290,150)(260,120)
\Vertex(260,120){3.5}
\Text(299,150)[]{$\bar{q}_{a_2}$}
\DashArrowLine(260,120)(280,90){4}{}
\Text(257,100)[]{$\tilde{q}_{a_4}$}
\ArrowLine(280,90)(310,120)
\Text(320,120)[]{$q_{a_2}$}
\Line(280,90)(310,60)
\Line(278,86)(308,56)
\Vertex(279,88){3.5}
\Text(317,65)[]{$\psi_\mu$}
\end{picture}
\vskip-0.4in
\caption{Three-body Gluino decay into Gravitino }
\end{center}
\end{figure}
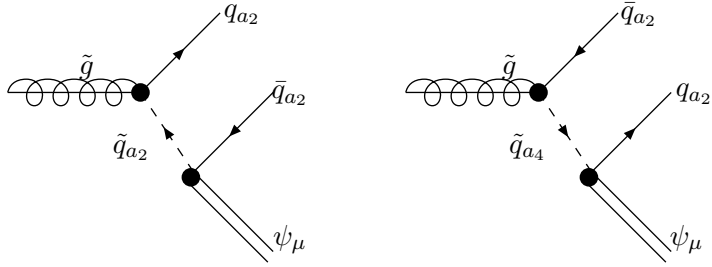

The gluino-(anti-)quark-squark vertex will again be given by (\ref{eq:gaugino-quark-squark1}). The gravitino-quark-squark vertex would come from a term of type  ${\bar\psi}_\mu
\tilde{q}_Lq_L$, which in ${\cal N}=1$ gauged supergravity lagrangian of \cite{Wess_Bagger} is given by:
\begin{equation}
\label{eq:gravitino-q-sq}
-g_{I{\bar J}}\left(\partial_\mu {\bar {\cal A}}^{\bar J}\right)\chi^I\sigma^\nu{\bar\sigma}_\mu\psi_\nu - \frac{i}{2}e^{\frac{K}{2}}\left(D_IW\right)\chi^I\sigma^\mu{\bar\psi}_\mu + {\rm h.c.}.
\end{equation}
From \cite{gravitinomodexp}, the gravitino field can be decomposed into the spin-$\frac{1}{2}$ Goldstino field $\tilde{G}$
via:
\begin{equation}
\label{eq:Goldstino_decomp}
\psi_\nu=\rho_\nu + \sigma_\nu\tilde{G},\ \tilde{G}=-\frac{1}{3}\sigma^\mu\psi_\mu,
\end{equation}
$\rho_\nu$ being a spin-$\frac{3}{2}$ field.
Hence, the Goldstino-content of (\ref{eq:gravitino-q-sq}), using $\sigma^\nu{\bar\sigma}^\mu\sigma_\nu=-2{\bar\sigma}^\nu$,
is given by:
\begin{equation}
\label{eq:Goldstino-s-sq_I}
2g_{I{\bar J}}\chi^I\left(\partial_\mu {\bar {\cal A}}^{\bar J}\right){\bar\sigma}^\mu\tilde{G} + \frac{3i}{2}e^{\frac{K}{2}}\left(D_IW\right)\chi^I\tilde{G} + {\rm h.c.}
\end{equation}
Now,
utilizing:
\begin{eqnarray}
\label{eq:Goldstino-s-sq_II}
& & g_{{\cal A}_2{\bar {\cal A}}_{2}}\sim{\cal V}^{-\frac{5}{9}}~ {\rm and}~ e^{\frac{K}{2}}D_{{\cal A}_2}W\Biggr|_{{\cal A}_2\rightarrow {\cal A}_{2} +{\cal V}^{-\frac{1}{3}}}\sim {\cal V}^{-\frac{37}{18}}{\cal A}_{2}
\end{eqnarray}
one obtains:
\begin{eqnarray}
\label{eq:Goldstino-s-sq_I}
& & {\hskip -0.5in}2g_{{\cal A}_{2}{\bar {\cal A}}_{2}}\chi^I\left({\bar\sigma}\cdot\frac{p_{q_{{\cal A}_2}}+ p_{\tilde{G}}}{M_p}\right){\bar\sigma}^\mu\tilde{G} + \frac{3i}{2}e^{\frac{K}{2}}\left(D_{{\cal A}_2}W\right)\chi^I\tilde{G} + {\rm h.c.}\nonumber\\
& & {\hskip -0.5in}\sim  {\cal V}^{-\frac{5}{9}}\chi^I\left(\frac{m_{G}}{M_p}\right){{\cal A}_2}\tilde{G} + {\cal V}^{-\frac{5}{9}}\chi^I\left(\frac{m_{q_{{\cal A}_2}}}{M_p}\right){{\cal A}_2}\tilde{G}+ {\cal V}^{-\frac{37}{18}}{\cal A}_2\chi^I\tilde{G} \sim {\cal V}^{-\frac{37}{18}}{\cal A}_2\chi^I\tilde{G}
\end{eqnarray}
for $m_{q_{{\cal A}_2}}\sim O(10)MeV, m_G \sim 0$ and ${\cal V}\sim 10^5$

The physical Goldstino-quark-squark vertex will be given as:
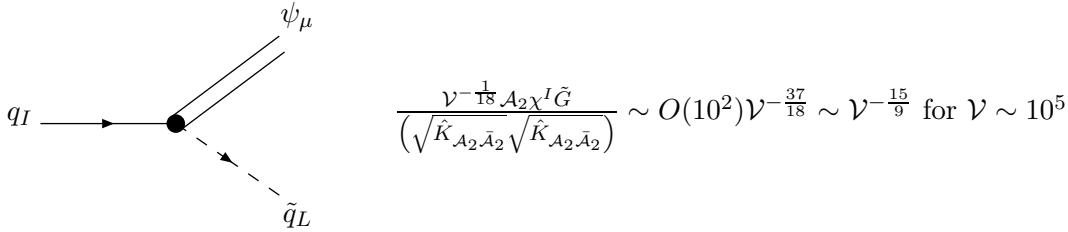
\begin{figure}
\begin{center}
\begin{picture}(500,140)(-50,80)
\Text(200,120)[]{$ {\hskip 0.3in}\frac{{\cal V}^{-\frac{1}{18}}{\cal A}_2\chi^I\tilde{G}}{\left(\sqrt{\hat{K}_{{\cal A}_2{\bar {\cal A}}_2}}\sqrt{\hat{K}_{{\cal A}_2{\bar {\cal A}}_2}}\right)}
\sim O({10}^{2}){\cal V}^{-\frac{37}{18}}\sim {\cal V}^{-\frac{15}{9}} ~{\rm for}~ {\cal V}\sim {10}^5 $}
\ArrowLine(-50,117)(1,117)
\Text(-58,120)[]{$q_I$}
\Text(47,158)[]{$\psi_\mu$}
\Vertex(1,117){3.5}
\Line(0,120)(40,150)
\Line(2,114)(42,144)
\Text(47,83)[]{$ \tilde{q}_L $}
\DashArrowLine(1,117)(40,90){5}
\end{picture}
\vskip-1in
\end{center}
\caption{The Goldstino-quark-squark vertex}
\end{figure}

For this particular case:-

{\small $A_{ij}\Biggl(Tr\left[G^{{\bar q}_{{\cal A}_2}}_{\tilde{q}_{{\cal A}_2}}G^{{\bar q}_{{\cal A}_2}}_{\tilde{q}_{{\cal A}_2}}\ ^\dagger\right]
Tr\left[\tilde{G}^{q_{{\cal A}_2}}_{\tilde{q}_{{\cal A}_2}}\tilde{G}^{q_{{\cal A}_2}}_{\tilde{q}_{{\cal A}_2}}\ ^\dagger\right]\Biggr)\sim \tilde{f}^2{\cal V}^{-\frac{8}{5}}.{\cal V}^{-\frac{10}{3}}\sim \tilde{f}^2{\cal V}^{-5}$},

 {\small $B_{ij}\Biggl(Tr\left[G^{q_{{\cal A}_2}}_{\tilde{q}_{{\cal A}_2}}G^{q_{{\cal A}_2}}_{\tilde{q}_{{\cal A}_2}}\ ^\dagger\right]
Tr\left[\tilde{G}^{{\bar q}_{{\cal A}_2}}_{\tilde{q}_{{\cal A}_2}}\tilde{G}^{{\bar q}_{{\cal A}_2}}_{\tilde{q}_{{\cal A}_2}}\ ^\dagger\right]\Biggr)\sim \tilde{f}^2{\cal V}^{-\frac{8}{5}}.{\cal V}^{-\frac{10}{3}}\sim \tilde{f}^2{\cal V}^{-5}$}

{\small $C\Biggl(Tr\left[G^{{\bar q}_{{\cal A}_2}}_{\tilde{q}_{{\cal A}_2}}G^{q_{{\cal A}_2}}_{\tilde{q}_{{\cal A}_2}}\ ^\dagger \tilde{G}^{q_{{\cal A}_2}}_{\tilde{q}_{{\cal A}_2}}\tilde{G}^{{\bar q}_{{\cal A}_2}}_{\tilde{q}_{{\cal A}_2}}\ ^\dagger\right]\Biggr)\sim \tilde{f}^2{\cal V}^{-\frac{8}{5}}.{\cal V}^{-\frac{10}{3}}\sim \tilde{f}^2{\cal V}^{-5}$}

Limits of integration as given in \cite{Dhuria+Misra_mu_Split SUSY} are:
$$s_{23 \max }= \left({\cal V}^{\frac{2}{3}}m_{\frac{3}{2}}-m_q\right){}^2,\ s_{23 \min }= m_q^2;$$
$$s_{13 \max }= {\cal V}^{\frac{4}{3}}m_{\frac{3}{2}}^2-s_{23}, s_{13 \min }= 0$$
where $m_{\tilde{g}}= {\cal V}^{\frac{2}{3}}m_{\frac{3}{2}} \sim {10}^{11} GeV, m_{\tilde{G}}= 0$.
Utilizing the values of vertex elements calculated above and from (\ref{eq:neutralinowidth}), we can calculate decay width for Gluino in this particular case.
\begin{eqnarray}
 \label{eq: Decay width goldstino}
 \hspace{-.5cm}\Gamma(\tilde{g}\to\chi_{\rm n}^{o}q_{{}_I} \bar{q}_{{}_J} )
&\sim&{g_s^2 \over256 \pi^3 {\cal V}^2 m_{\frac{3}{2}}^3 }\left[-
\tilde{f}^2{\cal V}^{-5} m_{\frac{3}{2}}^4 {\cal V}^{\frac{8}{3}}+  \tilde{f}^2{\cal V}^{-5}{\cal V}^{\frac{8}{3}} m_{\frac{3}{2}}^4 - \tilde{f}^2{\cal V}^{-5} {\cal V}^{\frac{8}{3}}m_{\frac{3}{2}}^4\right] \nonumber\\
& & \sim { g_s^2 \over256 \pi^3 {\cal V}^2 m_{\frac{3}{2}}^3 }( \tilde{f}^2 {\cal V}^{-5}{\cal V}^{\frac{8}{3}} m_{\frac{3}{2}}^4 ) \nonumber\\
& & \sim O({10}^{-4}){\cal V}^{-\frac{13}{3}}m_{\frac{3}{2}}\tilde{f}^2 GeV < 10^{-17}\tilde{f}^2 GeV < O(10^{-25})GeV
 \end{eqnarray}
The life time of gluino is given as:
   \begin{eqnarray}
   \tau &=&\frac{\hbar}{\Gamma}\sim\frac{10^{-34} Jsec}{10^{-17}f^2 GeV}\sim\frac{10^{-7}}{f^2}> 10^{3}sec
   \end{eqnarray}

We now consider the two-body decay of the gluino into a Goldstino and a gluon:
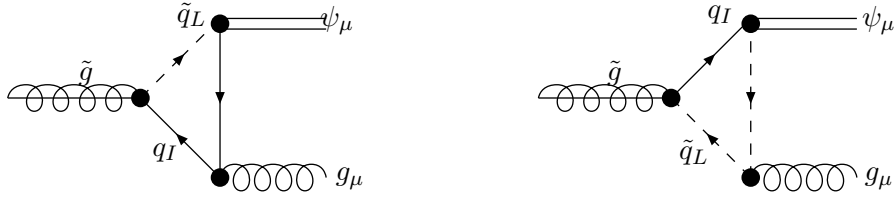
\begin{figure}
\begin{center}
\begin{picture}(1000,100)(50,80)
\Text(140,130)[]{$\tilde{g}$}
\Line(110,120)(160,120)
\Vertex(160,120){3.5}
\Gluon(110,120)(160,120){5}{4}
\DashArrowLine (160,120)(190,150){4}
\Text(180,152)[]{$\tilde{q}_L$}
\ArrowLine(190,90)(160,120)
\Text(170,100)[]{${q_I}$}
\ArrowLine(190,150)(190,90)
\Vertex(190,90){3.5}
\Gluon(190,90)(230,90){5}{4}
\Text(240,90)[]{$g_\mu$}
\Vertex(190,148){3.5}
\Line(190,150)(230,150)
\Line(190,146)(230,146)
\Text(235,150)[]{$\psi_\mu$}
\Line(310,120)(360,120)
\Vertex(360,120){3.5}
\Gluon(310,120)(360,120){5}{4}
\Text(340,130)[]{$\tilde{g}$}
\ArrowLine (360,120)(390,150)
\Text(380,152)[]{$q_I$}
\DashArrowLine (390,90)(360,120){4}
\Text(370,100)[]{$\tilde{q}_L$}
\DashArrowLine (390,150)(390,90){4}
\Vertex(390,90){3.5}
\Gluon(390,90)(430,90){5}{4}
\Text(440,90)[]{$g_\mu$}
\Vertex(390,148){3.5}
\Line(390,150)(430,150)
\Line(390,146)(430,146)
\Text(440,150)[]{$\psi_\mu$}
\end{picture}
 \end{center}
\caption{Diagrams contributing to one-loop Gluino decay into Goldstino and gluon}
\end{figure}

The matrix element for the above will be given by:
\begin{eqnarray}
\label{eq:MGold_I}
& & {\cal M}\sim {\tilde{f}}^2\int\frac{d^4k}{\left(2\pi\right)^4} \times{\cal V}^{-\frac{4}{5}}\left(\frac{i{\bar\sigma}\cdot k}{k^2-m_q^2+i\epsilon}\right)\left({\cal V}^{-\frac{15}{9}}\right)\left(\frac{i}{\left[\left(k-p_{\tilde{G}}\right)^2-m^2_{\tilde{q}}+i\epsilon\right]}
\right)\nonumber\\
& & \times\left({\cal V}^{\frac{53}{45}}\epsilon\cdot\left(2k-p_{\tilde{G}}-p_{\tilde{g}}\right)\right)\left(\frac{i}
{\left[\left(k-p_{\tilde{g}}\right)^2-m^2_{\tilde{q}}+i\epsilon\right]}\right) +\nonumber\\
& &{\tilde{f}}\int\frac{d^4k}{\left(2\pi\right)^4} \times{\cal V}^{-\frac{4}{5}}\left(\frac{i}
{\left[\left(k+p_{\tilde{G}}\right)^2-m^2_{\tilde{q}}+i\epsilon\right]}\right)\left({\cal V}^{-\frac{15}{9}}\right)\left( {\bar\sigma}\cdot\epsilon\right)\nonumber\\
& & \times\left(\frac{i{\bar\sigma}\cdot\left(k-p_{g_\mu}\right)}{\left[
\left(k-p_{g_\mu}\right)^2-m^2_q+i\epsilon\right]}\right)
\end{eqnarray}

As discussed in \cite{Dhuria+Misra_mu_Split SUSY}, the Wilson coefficients corresponding to Gluino-Goldstino- Gluon coupling do not change much upon RG evolution to EW scale. The matrix amplitude in equation (\ref{eq:MGold_I}) can be written as:
\begin{eqnarray}
\label{eq:MGold_II}
& & {\tilde{f}}^2 {\cal V}^{-1.2}\Biggl[\left\{{\bar\sigma}\cdot p_{\tilde{G}}C^{(a)}_{11}+{\bar\sigma}\cdot p_{g_\mu}C^{(a)}_{12}\right\}(2\epsilon\cdot p_{\tilde{G}})+\left\{{\bar\sigma}\cdot p_{\tilde{G}}\epsilon\cdot p_{\tilde{G}}C^{(a)}_{21}+{\bar\sigma}\cdot p_{g_\mu}\epsilon\cdot p_{\tilde{G}}C^{(a)}_{23}
+{\bar\sigma}\cdot\epsilon C^{(a)}_{24}\right\}\Biggr]\nonumber\\
& & + \tilde{f}{\cal V}^{-2.4}\Biggl[-\left\{{\bar\sigma}\cdot p_{\tilde{G}}C^{(b)}_{11}
+{\bar\sigma}\cdot p_{\tilde{g}}C^{(b)}_{12}\right\}{\bar\sigma}\cdot\epsilon{\bar\sigma}\cdot p_{g_\mu}+
{\bar\sigma}\cdot p_{\tilde{G}}{\bar\sigma}\cdot\epsilon {\bar\sigma}\cdot p_{\tilde{G}}C^{(b)}_{21} + {\bar\sigma}\cdot p_{\tilde{g}}{\bar\sigma}\cdot\epsilon {\bar\sigma}\cdot p_{\tilde{g}}C^{(b)}_{22}\nonumber\\
& & -\biggl({\bar\sigma}\cdot p_{\tilde{G}}{\bar\sigma}\cdot\epsilon{\bar\sigma}\cdot p_{\tilde{g}} + {\bar\sigma}\cdot p_{\tilde{g}}{\bar\sigma}\cdot\epsilon {\bar\sigma}\cdot p_{\tilde{G}}\biggr)C^{(b)}_{23}+ {\bar\sigma}_\mu{\bar\sigma}\cdot\epsilon{\bar\sigma}^\mu C^{(b)}_{24}\Biggr],
\end{eqnarray}
which equivalently could be rewritten as:
\begin{equation}
\label{eq:MGold_III}
{\bar u}(p_{\tilde{G}})\Biggl[{\bar\sigma}\cdot{\cal A}+{\bar\sigma}\cdot p_{\tilde{G}}{\bar\sigma}\cdot\epsilon{\bar\sigma}\cdot{\cal B}_1 + {\bar\sigma}\cdot p_{g_\mu}
{\bar\sigma}\cdot\epsilon{\bar\sigma}\cdot{\cal B}_2 + {D_4}{\bar\sigma}_\mu{\bar\sigma}\cdot\epsilon{\bar\sigma}^\mu
C^{(b)}_{24} \Biggr]u(p_{\tilde{g}}),
\end{equation}
where
\begin{eqnarray}
\label{eq:AB1B2}
& & {\cal A}^\mu\equiv {\cal V}^{-1.2}  {\tilde{f}}^2 \left[p_{\tilde{G}}^\mu\epsilon\cdot p_{\tilde{G}}\left(2C^{(a)}_{11} + C^{(a)}_{21}\right) + p_{g_\mu}^\mu\epsilon\cdot p_{\tilde{G}}\left(C^{(a)}_{12} + C^{(a)}_{23}-C^{(b)}_{22}\right) + \epsilon^\mu C^{(a)}_{24}\right];\nonumber\\
& & {\cal B}_1^\mu\equiv {\cal V}^{-2.4}  {\tilde{f}}\left[-p_{g_\mu}^\mu\left(C^{(b)}_{11} + C^{(b)}_{12} + C^{(b)}_{23}-C^{(b)}_{22}\right) + p_{\tilde{G}}^\mu\left(C^{(b)}_{21} + C^{(b)}_{22} - 2C^{(b)}_{23}\right)\right];\nonumber\\
& & {\cal B}_2^\mu\equiv {\cal V}^{-2.4}  {\tilde{f}}\left[p_{g_\mu}^\mu\left(C^{(b)}_{12}+C^{(b)}_{22}\right) + p_{\tilde{G}}^\mu\left( C^{(b)}_{22} - C^{(b)}_{23}\right)\right];\nonumber\\
& & D_4 \equiv {\cal V}^{-2.4} {\tilde{f}}.
\end{eqnarray}
This time around replacing ${\bar u}(p_{\tilde{G}}){\bar\sigma}\cdot p_{\tilde{G}}$ by  0 and ${\bar\sigma}\cdot p_{\tilde{g}}u(p_{\tilde{g}})$
by $m_{\tilde{g}}u(p_{\tilde{g}})$, (\ref{eq:MGold_III}) can be rewritten as:
\begin{equation}
\label{eq:MGold_IV}
 {\bar u}(p_{\tilde{G}})\left(A_2{\bar\sigma}\cdot\epsilon + B_3{\bar\sigma}\cdot p_{\tilde{g}}{\bar\sigma}\cdot\epsilon + D_2{\bar\sigma}\cdot p_{\tilde{g}}{\bar\sigma}\cdot\epsilon{\bar\sigma}\cdot p_{\tilde{G}} + D_4 {\bar\sigma}_\mu{\bar\sigma}\cdot\epsilon{\bar\sigma}^\mu C^{(b)}_{24}\right)u(p_{\tilde{g}}),
\end{equation}
where
\begin{eqnarray}
\label{eq:A2B3D2+defs}
& & A_2\equiv {\cal V}^{-1.2}{\tilde f}^2 C^{(a)}_{24}; B_3\equiv{\cal V}^{-2.4}{\tilde f}M_{\tilde{g}}(C^{(b)}_{12}+C^{(b)}_{22})\nonumber\\
& & D_2\equiv {\cal V}^{-2.4}{\tilde f} \left(-C^{(b)}_{12} + C^{(b)}_{23}\right); D_4 \equiv {\cal V}^{-2.4}{\tilde f}.
\end{eqnarray}

 Results of various C's functions required for this particular case similar to the ones given in \cite{Dhuria+Misra_mu_Split SUSY} are:
\begin{eqnarray}
\label{eq: three point goldstino functions}
& & \begin{array}{l}
C^{(a)}_{24}= C^{(b)}_{24}\sim  O(1);\nonumber\\
C^{(a)}_0 \sim O(10)\times {10}^{-22}{GeV}^{-2};\nonumber\\
C^{(b)}_{12}\sim  O(10)\times {10}^{-22} {GeV}^{-2};\nonumber\\
C^{(b)}_{11}\sim O(10)\times {10}^{-22} {GeV}^{-2};\nonumber\\
\end{array}\nonumber\\
& & \begin{array}{l}
 C^{(b)}_{22}= C^{(b)}_{23}\sim O(10)\times {10}^{-22} {GeV}^{-2};\nonumber\\
C^{(b)}_0= O(10)\times {10}^{-22} {GeV}^{-2}.\nonumber\\
\end{array} \nonumber\\
\end{eqnarray}

Utilizing (\ref{eq: three point goldstino functions}), one gets: $A_2\equiv  {10}^{-6} {\tilde f}^2 $, $B_3\equiv ({10}^{-23} {\tilde f}) {GeV}^{-1}$, $D_2\equiv  ({10}^{-34} {\tilde f}) {GeV}^{-2}$, $D_4 \equiv {\cal V}^{-2.4} {\tilde f}$
\begin{eqnarray}
\label{eq:spinavGammaGold}
& &  \sum_{\tilde{g}\ {\rm and}\ \tilde{G}\ {\rm spins}}|{\cal M}|^2
\sim   Tr\Biggl(\sigma\cdot p_{\tilde{G}}\left[A_2{\bar\sigma}\cdot\epsilon + B_3{\bar\sigma}\cdot p_{\tilde{g}}{\bar\sigma}\cdot\epsilon + D_2{\bar\sigma}\cdot p_{\tilde{g}}{\bar\sigma}\cdot\epsilon{\bar\sigma}\cdot p_{\tilde{G}} + D_4{\bar\sigma}_\mu{\bar\sigma}\cdot\epsilon{\bar\sigma}^\mu C^{(b)}_{24}\right]\nonumber\\
& & \times\sigma\cdot p_{\tilde{g}}\left[A_2{\bar\sigma}\cdot\epsilon + B_3{\bar\sigma}\cdot p_{\tilde{g}}{\bar\sigma}\cdot\epsilon + D_2{\bar\sigma}\cdot p_{\tilde{g}}{\bar\sigma}\cdot\epsilon{\bar\sigma}\cdot p_{\tilde{G}} + D_4{\bar\sigma}_\mu{\bar\sigma}\cdot\epsilon{\bar\sigma}^\mu C^{(b)}_{24}\right]^\dagger\Biggr),
\end{eqnarray}
which at:
\begin{equation}
\label{eq:kinematic_point}
p^0_{\tilde{G}}=m_{\tilde{g}}/2,p^1_{\tilde{G}}=p^2_{\tilde{G}}=p^3_{\tilde{G}}=\frac{m_{\tilde{g}}}{2\sqrt{3}},
\end{equation}
yields:
\begin{equation}
\label{eq:spinavgammaGold II}
 {m_{\tilde{g}}}^2 \left[{D_2}^2 {m_{\tilde{g}}}^4+\frac{1}{6} \left(6 {A_2}+12{D_4}
   {C_{24}^{(b)}}+{m_{\tilde{g}}} \left(6 {B_3}+\left(3+\sqrt{3}\right) {D_2} {m_{\tilde{g}}}\right)\right)^2\right]\sim  \tilde{f}^4 {A_2}^2 {m_{\tilde{g}}}^2.
\end{equation}
So, using results from \cite{Griffiths_particle}, the decay width comes out to be equal to:
\begin{equation}
\label{eq:GammaGold}
\Gamma=\frac{ \sum_{\tilde{g}\ {\rm and}\ \tilde{G}\ {\rm spins}}|{\cal M}|^2}{16\pi\hbar m_{\tilde{g}}}\sim O({10}^{-1}) {A^{2}_2} {m_{\tilde{g}}} \sim {10}^{-2}\tilde{f}^4  GeV.
\end{equation}
Considering value of $\tilde{f}^2 \sim{10}^{-10}$, $\Gamma < {10}^{-22}$ GeV.  Life time of gluino is given as:
   \begin{eqnarray}
   \tau &=&\frac{\hbar}{\Gamma}\sim\frac{10^{-34} Jsec}{10^{-2}f^4 GeV}\sim\frac{10^{-21}}{f^4} sec > {10^{-1}}sec
   \end{eqnarray}

\subsection{ R-Parity Violating Neutralino Decays}

 In the following we will be discussing the decay width of  tree level decay diagrams of neutralino mediated by sleptons/squarks, Higgs and gauge Bosons.

{\textbf{Decays mediated via squarks and sleptons}}

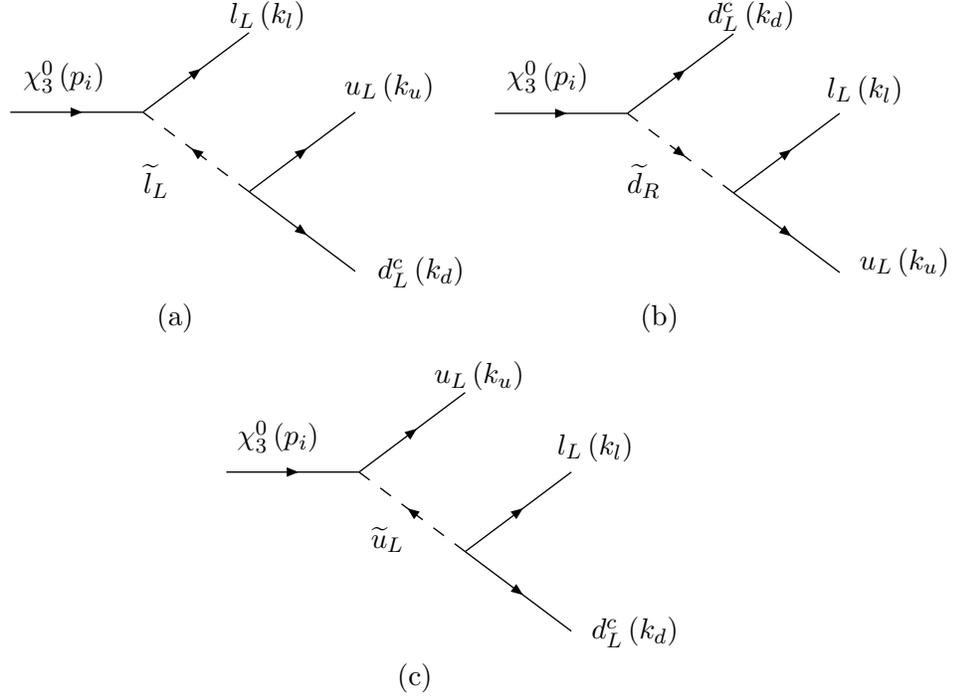
\begin{figure}[t!]
\label{decaydiag1}
\begin{center}
\begin{picture}(150,100)(-100,100)
\ArrowLine(-100,160)(-50,160)
\ArrowLine(-50,160)(-10,190)
\DashArrowLine(-10,130)(-50,160)5
\ArrowLine(-10,130)(30,160)
\ArrowLine(-10,130)(30,100)
\Text(-80,174)[]{${\chi^{0}_3}\,(p_i)$}
\Text(43,170)[]{${u_L} \,(k_{u})$}
\Text(55,100)[]{${d^{c}_L}\,(k_{d})$}
\Text(-3,197)[]{${l_L} \, (k_{l})$}
\put(-50,130){$\stilde {l}_L$}
\put(-45,80){(a)}
\end{picture}
\hspace{0.9cm}
\begin{picture}(150,100)(0,-25)
\ArrowLine(0,35)(50,35)
\ArrowLine(50,35)(90,65)
\DashArrowLine(50,35)(90,5)5
\ArrowLine(90,5)(130,35)
\ArrowLine(90,5)(130,-25)
\Text(20,49)[]{${\chi^{0}_3}\,(p_i)$}
\Text(140,44)[]{${l_L}\,(k_{l})$}
\Text(155,-21)[]{${u_L} \,(k_{u})$}
\Text(97,72)[]{${d}^c_L\,(k_{d})$}
\put(50,5){$\stilde {d}_R$}
\put(55,-45){(b)}
\end{picture}

\end{center}
\vspace{0.5cm}
\begin{center}
\begin{picture}(150,97)(100,100)
\ArrowLine(100,160)(150,160)
\ArrowLine(150,160)(190,190)
\DashArrowLine(190,130)(150,160)5
\ArrowLine(190,130)(230,160)
\ArrowLine(190,130)(230,100)
\Text(120,174)[]{${\chi^{0}_3} \,(p_i)$}
\Text(240,170)[]{$l_L\,(k_{l})$}
\Text(255,100)[]{$d^{c}_L\,(k_{d})$}
\Text(196,197)[]{$ u_L \,(k_{u})$}
\put(155,132){$\stilde {u}_L$}
\put(165,80){(c)}
\end{picture}
\phantom{xxx}
\end{center}
\caption{Feynman diagrams for the R-parity
violating decays of Neutralino ${\chi^{0}_3}$.}
\end{figure}

 Now, we  evaluate all possible neutralino decays which are mediated by squarks/sleptons and  also involve R-parity violating vertices as shown in Fig. 12. The life time calculation requires the evaluation of various matrix amplitudes corresponding to various decay channels. In particular, to evaluate the contribution of neutralino-squark-quark/neutralino-slepton-lepton, we will explicitly  work out squark-quark-gaugino vertex replacing the gaugino by $-\tilde{\chi}_3^0$ with mass half of that of the gluino and also the squark-quark-Higgsino vertex replacing the Higgsino by $\tilde{f}\tilde{\chi}_3^0$, and then add these contributions. To calculate the contribution of various interaction vertices shown in Fig.12 in the context of $N=1$ gauged supergravity, one needs to consider the following terms of gauged supergravity action  of Wess and Bagger \cite{Wess_Bagger}.
\begin{eqnarray}
\label{eq:fermion+fermion+sfermion}
& & {\hskip -0.3in}{\cal L} = g_{YM}g_{\alpha {\bar J}}X^\alpha{\bar\chi}^{\bar J}\lambda_{\tilde{g}} +ig_{i{\bar J}}{\bar\chi}^{\bar I}\left[{\bar\sigma}\cdot\partial\chi^i+\Gamma^i_{Lj}{\bar\sigma}\cdot\partial a^L\chi^j
+\frac{1}{4}\left(\partial_{a_L}K{\bar\sigma}\cdot a_L - {\rm c.c.}\right)\chi^i\right] \nonumber\\
& & {\hskip -0.3in}+\frac{e^{\frac{K}{2}}}{2}\left({\cal D}_iD_JW\right)\chi^i\chi^J+{\rm h.c.},
\end{eqnarray}
where, $X^\alpha$ corresponds to the components of a killing isometry vector. From here, one notes that $X^\alpha=-6i\kappa_4^2\mu_7Q_\alpha$, where $\alpha=B, Q_\alpha=2\pi\alpha^\prime\int_{T_B}i^*\omega_\alpha\wedge P_-\tilde{f}$ where $P_-$ is a harmonic zero-form on $\Sigma_B$ taking value +1 on $\Sigma_B$ and $-1$ on $\sigma(\Sigma_B)$ - $\sigma$ being a holomorphic isometric involution as part of the Calabi-Yau orientifold - and $\tilde{f}\in\tilde{H}^2_-(\Sigma^B)\equiv{\rm coker}\left(H^2_-(CY_3)\stackrel{i^*}{\rightarrow}H^2_-(\Sigma^B)\right)$. Also,
\begin{eqnarray}
\label{eq:DdW_def}
{\cal D}_iD_jW=\partial_i\partial_jW + \left(\partial_i\partial_jK\right)W+\partial_iKD_JW +
\partial_JKD_iW - \left(\partial_iK\partial_jK\right)W - \Gamma^k_{ij}D_kW.
\end{eqnarray}

\textbf{Higgsino-fermion/anitifermion-sfermion $+$ Higgsino-fermion/anitifermion-sfermion $+$ Fermion-sfermion-fermion/anitifermion vertices}

\begin{itemize}
\item
As discussed above, the neutralino-lepton-slepton  vertex corresponding to Fig. 11(a), is given by considering the contribution of gaugino-slepton-lepton vertex with a small admixture of higgsino-lepton-slepton vertex as given in equation (\ref{eq:chi3}). Since neutralino is of Majorana nature, in two-component notation, the contribution of  Higgsino-lepton-slepton vertex in gauged supergravity action of Wess and Bagger \cite{Wess_Bagger}, is given by  $$\frac{e^{\frac{K}{2}}}{2}\left({\cal D}_{{\cal Z}_1}D_{\bar {\cal A}_1}W\right)\chi^{{\cal Z}_i}{\chi^{c^{{{\cal A}}_1}}} +ig_{{\bar I}{\bar {A_1}}}{\bar\chi}^{Z_i}\left[{\bar\sigma}\cdot\partial{{\chi^{c^{{{\cal A}}_1}}}}+\Gamma^{{\cal A}_1}_{{\cal A}_1{\bar A_1}}{\bar\sigma}\cdot\partial {\cal A}_1{\chi^{c^{A_1}}}
+\frac{1}{4}\left(\partial_{{\cal A}_3}K{\bar\sigma}\cdot {\cal A}_1 - {\rm c.c.}\right){\chi^{c^{{\cal {A}}_1}}}\right]$$
 $\chi^{\cal {Z}}$ is $SU(2)_L$ higgsino, $\chi^{{\cal A}_1}$ corresponds to $SU(2)_L$ electron and ${\tilde {\cal A}_1}$ corresponds to left-handed squark and $g_{{\bar I}{\bar {\cal A}_1}}=0$.

 Strictly speaking, $SU(2)$ EW symmetry gets spontaneously broken for Higgsino-lepton-slepton vertex, however the effective Lagrangian respects $SU(2)$ symmetry. Therefore to calculate the contribution of same, basic idea is to generate a term of the type $l_L \tilde{l}_L{\tilde H^{c}}_L H_L$ wherein $\chi^{l_L}$ and $H_L$ are respectively the $SU(2)_L$ quark and Higgs doublets, $\tilde{l}_L$ is also an $SU(2)_L$ doublet and ${\tilde H^{c}}_L$ is $SU(2)_L$  Higgsino doublet . After spontaneous breaking of the EW symmetry when $H^0$ in $H_L$ acquires a non-zero vev $\langle H^0\rangle$, this term generates:$\langle H^0\rangle {\tilde H^{c}}_L{l_L}\tilde{l}_L$.
Now, in terms of undiagonalized basis, consider $${\cal D}_iD_{\bar a_1}W= \left(\partial_i\partial_{\bar a_1}K\right)W+\partial_iKD_{\bar a_1}W +
\partial_{\bar a_1}KD_iW - \left(\partial_iK\partial_{\bar a_1}K\right)W, $$ where $a_1, z_i$ correspond to undiagonalized moduli fields.
\begin{enumerate}
\item
Considering  $a_1\rightarrow a_1+{\cal V}^{-\frac{2}{9}}{M_p}$, using equations (\ref{eq:Kahler pot}),(\ref{eq:W}) and further picking up the component linear in $z_1$ as well as linear in fluctuation $ (a_1-{\cal V}^{-\frac{2}{9}}{M_p})$, we see that:
\begin{eqnarray}
\label{eq:DKDW1}
& &{\hskip -1in} e^{\frac{K}{2}}\left(\left(\partial_i\partial_{a_1}K\right)W+\partial_iKD_{a_1}W +
\partial_{a_1}KD_iW - \left(\partial_iK\partial_{a_1}K\right)W\right)\bar\chi^i\bar\chi^{a_1}\sim  {\cal V}^{-\frac{31}{18}} z_i (a_1-{\cal V}^{-\frac{2}{9}}{M_p}).
\end{eqnarray}

As shown in (\ref{eq:eq_mass_terms_diag_non-diag_basis})
$$ e^{\frac{K}{2}}{\cal D}_i D_{\bar {{\cal A}_1}}W \sim {\cal O}(1) e^{\frac{K}{2}}{\cal D}_i D_{{\bar a}_1}W,$$
Higgsino-lepton-slepton vertex will hence be given as:
$$e^{\frac{K}{2}}{\cal D}_i D_{{\cal A}_1}W \chi^{Z_i}{\chi^{c^{{\cal A}_1}}}\sim e^{\frac{K}{2}}{\cal D}_i D_{\bar {a_1}}W \chi^{{\cal Z}_i}\chi^{c^{{\cal A}_1}}\sim {\cal V}^{-\frac{31}{18}}{{\cal Z}_i}\delta{\cal A}_1 \chi^{{\cal Z}_i}\chi^{c^{{\cal A}_1}};$$
 ${\cal A}_1$ corresponds to left-handed slepton. The contribution of physical  Higgsino-lepton-slepton vertex after giving VEV to ${\cal Z}_I$  will be given as :
\begin{eqnarray}
\label{eq:Higgsino-lepton-slepton}
& &  C^{{\tilde H^{c}}_L {l_L} \tilde {l_L}}\sim\frac{{\cal V}^{-\frac{31}{18}}<{\cal Z}_i>}{{\sqrt{\hat{K}^{2}_{{\cal Z}_i{\bar{\cal Z}}_i}{\hat{K}_{{\cal A}_1{\bar {\cal A}}_1}}{\hat{K}_{{\cal A}_1{\bar {\cal A}}_1}}}}} \sim {\cal V}^{-\frac{3}{2}}.
\end{eqnarray}

\item
 The interaction vertex corresponding to gaugino-lepton-slepton vertex is given by $${\hskip -0.3in} g_{YM}g_{{ J}\bar{T}_B}X^{{*}B}{\bar\chi}^{\bar J}\lambda_{\tilde{g}}+ \partial_{a_1}T_B D^{B}{\bar\chi}^{\bar a_1}\lambda_{\tilde{g}}.$$ Utilizing (\ref{metric}) and $X^{B}=-6i\kappa_4^2\mu_7Q_{T_B}, \kappa_4^2\mu_7\sim \frac{1}{\cal V}, g_{YM}\sim{\cal V}^{-\frac{1}{36}}, Q_{T_B}\sim{\cal V}^{\frac{1}{3}}(2\pi\alpha^\prime)^2\tilde{f},
g_{YM}g_{{T_B} {\bar a}_1}\rightarrow-{\cal V}^{-\frac{2}{9}} a_1$
and $$g_{YM}g_{{T_B} {\bar {\cal A}}_1}\sim  {\cal V}^{-\frac{2}{9}} ({\cal A}_1-{\cal V}^{-\frac{2}{9}}M_p),$$ 
Using $T_{B}=Vol(\sigma_B)- C_{I{\bar J}}a_I{\bar a}_{\bar J} + h.c.$, (intersection matrices $C_{I{\bar J}}$ are given in {{\bf appendix B}}),  $\partial_{a_1} T_B\rightarrow {\cal V}^{\frac{10}{9}} (a_1-{\cal V}^{-\frac{2}{9}})$ and
the dominant contribution to the physical gaugino-lepton-slepton vertex is proportional to :
\begin{equation}
\label{eq:gaugino-lepton-slepton}
C^{\lambda_{\tilde{g}} {l_L}\tilde {l_L}}\sim \frac{g_{YM}g_{{T_B} {\bar {\cal A}_1}}X^{T_B}\sim {\cal V}^{-\frac{2}{9}} {\tilde f}}{{\left(\sqrt{\hat{K}_{{\cal A}_1{\bar {\cal A}}_1}}\right)}{\left(\sqrt{\hat{K}_{{\cal A}_1{\bar {\cal A}}_1}}\right)}}\sim  \tilde{f} {\cal V}^{-1}.
\end{equation}
\item
Keeping in mind the fact that physical neutralino eigenstate ${\chi}_3^0$ is largely a gaugino with a small admixture of Higgsino in our set up, by adding the contribution of (\ref{eq:Higgsino-lepton-slepton}) and (\ref{eq:gaugino-lepton-slepton}) as according to (\ref{eq:chi3}), the  physical \emph{neutralino-lepton-slepton} vertex is given as:
\begin{eqnarray}
\label{eq:neutralino-lepton-slepton}
& & {\hskip -0.5in} C^{\chi^0_3 {l_L}\tilde {l_L}}:   \tilde{f} {\cal V}^{-1}\delta{{\cal A}_1}\lambda_{\tilde{g}} {\bar \chi}^{\bar {\cal A}_1}+ \tilde{f} {\frac{v}{m_{pl}}}\left({\cal V}^{-\frac{2}{3}}\right)\delta{\cal A}_1 \chi^{I} {\bar\chi}^{\bar {{\cal A}_1}} \sim \tilde{f} \left({\cal V}^{-1}\right){\chi^0_3 {l_L}\tilde {l_L}} ~{\rm for}~ {\cal V}\sim {10}^5.
\end{eqnarray}
\end{enumerate}
\item
The \emph{quark-quark-slepton} vertex corresponding to Fig. 11(a) in gauged supergravity action is given by the following term: $\frac{e^{\frac{K}{2}}}{2}\left({\cal D}_{{\bar {\cal A}}_2}D_{{\cal A}_4}W\right)\chi^{{\cal A}^{c}_2}\chi^{{\cal A}_4}$. In terms of undiagonalized basis, consider
$${\cal D}_{\bar{a_2}}D_{a_4}W= \left(\partial_{\bar{a_2}}\partial_{a_4}K\right)W+\partial_{\bar{a_2}}KD_{a_4}W +
\partial_{a_4}KD_{\bar{a_2}}W - \left(\partial_{\bar{a_2}}K\partial_{a_4}K\right)W. $$
Using equations (\ref{eq:Kahler pot}) and (\ref{eq:W}):
\begin{eqnarray}
\label{eq:DKDW2}
& &{\hskip -1.0in}e^{\frac{K}{2}}\left(\left(\partial_{\bar{a_2}}\partial_{a_4}K\right)W+\partial_{\bar{a_2}}KD_{a_4}W + \partial_{a_4}KD_{\bar{a_2}}W - \left(\partial_{\bar{a_2}}K\partial_{a_4}K\right)W\right) \sim   {\cal V}^{-\frac{5}{9}}+ {\cal V}^{\frac{1}{3}}\delta a_1.
\end{eqnarray}
From (\ref{eq:DKDW2}), picking up the component of fluctuations linear in $a_1$ and from (\ref{eq:eq_mass_terms_diag_non-diag_basis}) $$e^{\frac{K}{2}}{\cal D}_{{\cal {\bar A}}_2}D_{{\cal A}_4}W\sim O(1) e^{\frac{K}{2}}{\cal D}_{\bar {a_2}}D_{a_4}W,$$
one gets the contribution of quark-slepton-quark vertex as :
\begin{eqnarray}
\label{eq:Da2a4W}
& & e^{\frac{K}{2}}{\cal D}_{{\cal {\bar A}}_2}D_{{\cal A}_4}W \chi^{{\cal A}^{c}_2} \chi^{{\cal A}_4} \sim e^{\frac{K}{2}}{\cal D}_{\bar {a_2}}D_{a_4}W \chi^{{\cal A}^{c}_2} \chi^{{\cal A}_4}\sim  \left({\cal V}^{-\frac{1}{3}}\delta{\cal A}_1\right)\chi^{{\cal A}^{c}_2}\chi^{{\cal A}_4}.
\end{eqnarray}
The physical quark-slepton-quark vertex is given by
\begin{eqnarray}
\label{eq:quark-slepton-quark}
C^{ d^{c}_L \tilde {l_L} u_L}\sim \frac{{\cal V}^{-\frac{1}{3}}}{{\sqrt{\hat{K}_{{\cal A}_2 {\cal A}^{c}_2}{\hat{K}_{{\cal A}_1{\bar {\cal A}}_1}}{\hat{K}_{{\cal A}_4{\bar {\cal A}}_4}}}}}\sim (10)^{-\frac{14}{2}}{\cal V}^{-\frac{1}{3}} \sim {\cal V}^{-\frac{5}{3}} {\rm for}~ {\cal V}\sim {10}^5.
\end{eqnarray}
\item The contribution of  Higgsino-quark-squark relevant to the \emph{Neutralino-quark-squark} vertex of Fig. 11(b) is given by: $$\frac{e^{\frac{K}{2}}}{2}\left({\cal D}_iD_{{\cal A}_4}W\right){\bar \chi}^ {i}\chi^{{\cal A}_4} +ig_{{\bar i }{{\cal A}_4}}{\bar\chi}^{I}\left[{\bar\sigma}\cdot\partial\chi^{ {{\cal A}_4}}+\Gamma^{{{\cal A}_4}}_{{\cal A}_4{{\cal A}_4}}{\bar\sigma}\cdot\partial {{\cal A}_4}{\chi}^{{{\cal A}_4}}
+\frac{1}{4}\left(\partial_{{\cal A}_4}K{\bar\sigma}\cdot {\cal A}_4 - {\rm c.c.}\right)\chi^{{{\cal A}_4}}\right].$$
Working in diagonalized basis, $g_{{\bar i }{{\cal A}_4}}$=0.

Now consider
$${\cal D}_iD_{a_4}W=\partial_i\partial_{a_4}W + \left(\partial_i\partial_{a_4}K\right)W+\partial_iKD_{a_4}W +
\partial_{a_4}KD_iW - \left(\partial_iK\partial_{a_4}K\right)W - \Gamma^k_{i{a_4}}D_kW,$$ where $a_4, z_i$ correspond to undiagonalized  moduli fields.
\begin{enumerate}
\item
Considering $a_4\rightarrow a_4+{\cal V}^{-\frac{11}{9}}{M_p}$ and utilizing (\ref{metricinv}),  (\ref{metricdervz1}) and (\ref{metricderva4}), one gets:
\begin{eqnarray}
\label{eq:affine3}
& & {\hskip -0.2in}  \Gamma^{z_i}_{z_ia_4}\sim {\cal V}^{\frac{11}{9}}-{\cal V}^{\frac{11}{6}}\delta a_4,\Gamma^{a_1}_{z_ia_4}\sim {\cal V}^{-\frac{13}{36}}+ {\cal V}^{\frac{19}{12}}\delta a_4, \Gamma^{a_2}_{z_ia_4}\sim {\cal V}^{\frac{31}{36}}+{\cal V}^{\frac{25}{12}}\delta a_4,  \Gamma^{a_3}_{z_ia_4}\sim {\cal V}^{-\frac{5}{36}}+ {\cal V}^{\frac{13}{12}}{\delta a_4}, \nonumber\\
& & \Gamma^{a_4}_{z_ia_4}\sim {\cal V}^{-\frac{23}{36}}+{\cal V}^{\frac{7}{12}}{\delta a_4}.
\end{eqnarray}
Next, using equations (\ref{eq:Kahler pot}), (\ref{eq:W}), (\ref{eq:affine3}), one obtains:
\begin{eqnarray}
\label{eq:DdW3}
& & {\hskip-1.3in}\frac{e^{\frac{K}{2}}}{2}(\Gamma^{z_i}_{z_ia_4}D_{z_i}W + \Gamma^{a_1}_{z_ia_4}D_{a_1}W + \Gamma^{a_2}_{z_ia_4}D_{a_2}W+\Gamma^{a_3}_{z_ia_4}D_{a_2}W  + \Gamma^{a_4}_{z_ia_4}D_{a_4}W)\sim \left({\cal V}^{-\frac{17}{72}}+ {\cal V}^{\frac{71}{72}}\delta a_4\right)
\end{eqnarray}
 and
\begin{eqnarray}
\label{eq:DKDW31}
& &{\hskip -1.3in} e^{\frac{K}{2}}\left(\left(\partial_i\partial_{a_4}K\right)W+\partial_iKD_{a_4}W +
\partial_{a_4}KD_iW - \left(\partial_iK\partial_{a_4}K\right)W\right)\chi^i\chi^{a_4}\sim {\cal V}^{-\frac{17}{72}}+ {\cal V}^{\frac{71}{72}}\delta a_4 \chi^i\chi^{a_4}.
\end{eqnarray}
Utilizing equation (\ref{eq:DdW3}) and (\ref{eq:DKDW31}), picking up the component linear in fluctuation $\delta a_4$, contribution of Higgsino-quark-squark vertex  will be
\begin{equation}
\label{eq:DIDa4W}
e^{\frac{K}{2}}{\cal D}_iD_{{\cal A}_4}W\chi^{{\cal Z}_i}\chi^{{\cal A}_4} \sim e^{\frac{K}{2}}{\cal D}_iD_{a_4}W\chi^i\chi^{{\cal A}_4} \sim  {\cal V}^{\frac{71}{72}}\delta {\cal A}_4\chi^I\chi^{{\cal A}_4}\nonumber\\
\end{equation}
and physical Higgsino-quark-squark takes the form as below :
\begin{eqnarray}
\label{eq:Higgsino-quark-squark}
& & {\hskip -0.6in}C^{H^{c}_L \bar d^{c}_L \tilde {d_R}} \sim\frac{{\cal V}^{\frac{71}{72}}}{{\sqrt{\hat{K}_{{\cal Z}_i{\bar{\cal Z}}_i}{\hat{K}_{{\cal A}_4{\bar {\cal A}}_4}}{\hat{K}_{{\cal A}_4{\bar {\cal A}}_4}}}}}  \sim \left[{(10)^{-\frac{19}{2}}}{\cal V}^{\frac{71}{72}}\right]\sim {\cal V}^{-\frac{11}{12}}.
\end{eqnarray}
 \item
 The interaction vertex corresponding to gaugino-quark-squark vertex is given by $${\hskip -0.3in}g_{YM}g_{{{{\bar {\cal A}}_4}}{\bar{T}_B}}X^{{*}B}{\chi^{c^{{\cal {A}}_4}}}\lambda_{\tilde{g}}.$$
For a K\"{a}hler moduli space, $g_{{{{\bar {\cal A}}_4}}{\bar{T}_B}}=0$
i.e the  gaugino-quark-squark vertex does not contribute to this particular vertex.
\item
 In view of above, the physical \emph{neutralino-quark-squark} vertex as according to equation (\ref{eq:chi3}) is given as:
\begin{eqnarray}
\label{eq:neutralino-quark-squark}
& & C^{{\chi^0}_3\bar d^{c}_L \tilde {d_R}}:   \tilde{f}{\frac{v}{m_{pl}}} [{\cal V}^{-\frac{1}{12}}\delta a_4 {\bar\chi}^{\bar {a_1}} \chi^{i}]\sim \tilde{f}{\frac{v}{m_{pl}}} {\cal V}^{-\frac{1}{12}}{{\chi^0}_3\bar d^{c}_L \tilde {d_R}}
\end{eqnarray}

\end{enumerate}
\item
The \emph{lepton-squark-quark}  vertex corresponding to Fig. 11(b) is given by considering the contribution $\frac{e^{\frac{K}{2}}}{2}\left({\cal D}_{\bar{\cal A}_1}D_{{\cal A}_2}W\right){\chi^{c^{{\cal {A}}_1}}}\chi^{{\cal A}_2}+ \frac{e^{\frac{K}{2}}}{2}\left({\cal D}_{{\cal A}_1}D_{\bar{{\cal A}_2}}W\right)\chi^{{\cal A}_1}{\chi^{c^{{\cal {A}}_2}}}$.
In terms of undiagonalized basis
$${\cal D}_{\bar {a_1}}D_{a_2}W= \left(\partial_{\bar {a_1}}\partial_{a_2}K\right)W+\partial_{\bar {a_1}}KD_{a_2}W +
\partial_{a_2}KD_{\bar {a_1}}W - \left(\partial_{\bar {a_1}}K\partial_{a_2}K\right)W $$
and
$${\cal D}_{a_1}D_{\bar{a_2}}W= \left(\partial_{a_1}\partial_{\bar {a_2}}K\right)W+\partial_{a_1}KD_{\bar{a_2}}W +
\partial_{\bar {a_2}}KD_{a_1}W - \left(\partial_{a_1}K\partial_{\bar {a_2}}K\right)W $$

Utilizing $a_4\rightarrow a_4+{\cal V}^{-\frac{11}{9}}$  and equations (\ref{eq:Kahler pot}), (\ref{eq:W}), on solving,
\begin{eqnarray}
\label{eq:DKDW4}
& &{\hskip -1.3in} e^{\frac{K}{2}}\left(\partial_{\bar {a_1}}\partial_{a_2}K\right)W+\partial_{\bar {a_1}}KD_{a_2}W +
\partial_{a_2}KD_{\bar {a_1}}W - \left(\partial_{\bar {a_1}}K\partial_{a_2}K\right)W\chi^{\bar {a_1}}\chi^{a_2}\sim \left({\cal V}^{-\frac{14}{9}}+ {\cal V}^{-\frac{1}{3}}\delta a_4\right)\chi^{a^{c}_1}\chi^{a_2}
\nonumber\\
& &{\hskip -1.3in} e^{\frac{K}{2}}\left(\partial_{a_1}\partial_{\bar {a_2}}K\right)W+\partial_{a_1}KD_{\bar{a_2}}W +
\partial_{\bar {a_2}}KD_{a_1}W - \left(\partial_{a_1}K\partial_{\bar {a_2}}K\right)W \chi^{a_1}\chi^{\bar {a_2}}\sim \left({\cal V}^{-\frac{14}{9}}+ {\cal V}^{-\frac{1}{3}}\delta a_4\right)\chi^{a_1}\chi^{a^{c}_2}.
\end{eqnarray}
From  (\ref{eq:DKDW4}), one gets the contribution of lepton-squark-quark vertex as :
\begin{eqnarray}
\label{eq:Da1a4W}
& & {\hskip -0.3in}e^{\frac{K}{2}}{\cal D}_{\bar {a_1}}D_{a_2}W\chi^{a^{c}_1}\chi^{a_2} + e^{\frac{K}{2}}{\cal D}_{a_1}D_{\bar {a_2}}W\chi^{a_1}\chi^{a^{c}_2}\sim  \left({\cal V}^{-\frac{1}{3}}\delta a_4\right)\chi^ {a^{c}_1}\chi^{a_2} +  \left({\cal V}^{-\frac{1}{3}}\delta a_4\right)\chi^{a_1}\chi^{a^{c}_2}
\end{eqnarray}
and the physical lepton-squark-quark vertex is given by
\begin{eqnarray}
\label{eq:lepton-squark-quark2}
& & {\hskip-0.7in} C^{l_L \tilde {d_R} u_L}\sim\frac{{\cal V}^{-\frac{1}{3}}}{{\sqrt{\hat{K}_{{\cal A}_1{\bar{\cal A}}_1}{\hat{K}_{{\cal A}_2{\bar {\cal A}}_2}}{\hat{K}_{{\cal A}_4{\bar {\cal A}}_4}}}}} + \frac{{\cal V}^{-\frac{1}{3}}}{{\sqrt{\hat{K}_{{\cal A}_1{\bar{\cal A}}_1}{\hat{K}_{{\cal A}_2{\bar {\cal A}}_2}}{\hat{K}_{{\cal A}_4{\bar {\cal A}}_4}}}}}\sim  {\cal V}^{-\frac{5}{3}}.
\end{eqnarray}

\item
The Neutralino-quark-squark vertex of Fig. 11(c)  is given as considering the contribution of
$$\frac{e^{\frac{K}{2}}}{2}\left({\cal D}_iD_{\bar A_2}W\right)\chi^{{\cal Z}_i}{\chi^{c^{{\cal {A}}_2}}} +ig_{{\bar {\cal Z}_i}{\bar {A_2}}}{\bar\chi}^{{\cal Z}_i}\left[{\bar\sigma}\cdot\partial+\Gamma^{{\cal A}_2}_{{\cal A}_2{\bar A_2}}{\bar\sigma}\cdot\partial {\cal A}_2{\chi^{c^{{\cal {A}}_2}}}
+\frac{1}{4}\left(\partial_{{\cal A}_2}K{\bar\sigma}\cdot {\cal A}_2 - {\rm c.c.}\right){\chi^{c^{{\cal {A}}_2}}} \right]$$
 $\chi^I$ is $SU(2)_L$ higgsino, $\chi^{A_2}$ corresponds to $SU(2)_L$ quark and ${{\cal A}_2}$ corresponds to left-handed squark.
and  $g_{{\bar I}{\bar {\cal A}_2}}=0$.

In terms of undiagonalized basis, consider $${\cal D}_iD_{\bar a_2}W= \left(\partial_i\partial_{\bar a_2}K\right)W+\partial_iKD_{\bar a_2}W +
\partial_{\bar a_2}KD_iW - \left(\partial_iK\partial_{\bar a_2}K\right)W ,$$ where $a_2, z_i$ correspond to undiagonalized moduli fields.
\begin{enumerate}
\item
Considering  $a_2\rightarrow a_2+{\cal V}^{-\frac{1}{3}}{M_p}$, using equations (\ref{eq:Kahler pot}),(\ref{eq:W}) and picking up the component linear in  $z_1$ as well as fluctuations of $a_2$ i.e ($ a_2- {\cal V}^{-\frac{1}{3}}{M_p}$) , one gets
\begin{eqnarray}
\label{eq:DKDW32}
& &{\hskip -0.7in} e^{\frac{K}{2}}\left(\left(\partial_i\partial_{\bar a_2}K\right)W+\partial_iKD_{\bar a_2}W +
\partial_{\bar a_2}KD_iW - \left(\partial_iK\partial_{\bar a_2}K\right)W\right)\sim ({\cal V}^{-\frac{20}{9}})z_i \delta a_2.
\end{eqnarray}
Using (\ref{eq:eq_mass_terms_diag_non-diag_basis}):
$$ e^{\frac{K}{2}}{\cal D}_i D_{\bar {{\cal A}_2}}W \sim O(1) e^{\frac{K}{2}}{\cal D}_i D_{{\bar a}_2}W,$$
the Higgsino-quark-squark vertex will be given as
$$e^{\frac{K}{2}}{\cal D}_i D_{{\cal A}_2}W \chi^{{\cal Z}_i}{\chi^{c^{{\cal {A}}_2}}} \sim e^{\frac{K}{2}}{\cal D}_i D_{\bar {a_2}}W \chi^{{\cal Z}_i}{\chi^{c^{{\cal {A}}_2}}}\sim {\cal V}^{-\frac{20}{9}}{\cal Z}_i\delta{\cal A}_2 \chi^{{\cal Z}_i}{\chi^{c^{{\cal {A}}_2}}};$$
 ${\cal A}_2$ correspond to left- handed squark and the contribution of physical  Higgsino-quark-squark vertex will be given as :
\begin{eqnarray}
\label{eq:Higgsino-quark-squark1}
& &  C^{{\tilde H^{c}}_L {u_L} \tilde {u_L}}\sim\frac{{\cal V}^{-\frac{20}{9}}<{\cal Z}_i>}{{\sqrt{\hat{K}^{2}_{{\cal Z}_i{\bar{\cal Z}}_i}{\hat{K}_{{\cal A}_2{\bar {\cal A}}_2}}{\hat{K}_{{\cal A}_2{\bar {\cal A}}_2}}}}} \sim {\cal V}^{-\frac{4}{5}}.
\end{eqnarray}
\item
 As discussed for gaugino lepton-slepton vertex in Figure (a), the gaugino- quark-squark vertex is given by $${\hskip -0.3in} g_{YM}g_{{ J}B^{*}}X^{{*}B}{\bar\chi}^{\bar J}\lambda_{\tilde{g}}+ + \partial_{a_2}T_B D^{B}{\bar\chi}^{\bar a_1}\lambda^{0}.$$. Utilizing \ref{metric} and $X^{B}=-6i\kappa_4^2\mu_7Q_{B}, \kappa_4^2\mu_7\sim \frac{1}{\cal V}, g_{YM}\sim{\cal V}^{-\frac{1}{36}}, Q_B\sim{\cal V}^{\frac{1}{3}}(2\pi\alpha^\prime)^2\tilde{f},$
 This time,
\begin{equation}
\label{eq:gYMgBIbar_zi_aI}
g_{YM}g_{B {\bar a}_2}\rightarrow-{\cal V}^{-\frac{5}{4}}(a_2-{\cal V}^{-\frac{1}{3}}), \partial_{a_2}T_B \rightarrow {\cal V}^{\frac{1}{9}} (a_2-{\cal V}^{-\frac{1}{3}})  \nonumber\\
\end{equation}
 The dominant contribution to the physical gaugino-quark-squark vertex is proportional to :
\begin{equation}
\label{eq:gaugino-quark-squark1}
C^{\lambda_{\tilde{g}} {u_L}\tilde {u_L}}\sim \frac{ {\cal V}^{-\frac{11}{9}} {\tilde f}}{{\left(\sqrt{\hat{K}_{{\cal A}_2{\bar {\cal A}}_2}}\right)}{\left(\sqrt{\hat{K}_{{\cal A}_2{\bar {\cal A}}_2}}\right)}}\sim \tilde{f}\left({\cal V}^{-\frac{4}{5}}\right).
\end{equation}
Following equation no (\ref{eq:chi3}) , the neutralino-quark squark vertex will be given by :
\begin{eqnarray}
\label{eq:neutralino-quark-squark1}
\hskip -1in C^{{\chi^{0}_3} {u_L}\tilde {u_L}}\sim C^{\lambda_{\tilde{g}} {u_L}\tilde {u_L}}+ {\tilde f}{\cal V}^{\frac{5}{6}}\frac{v}{m_{pl}} C^{H^{c}_L {u_L} \tilde {u_L}} \sim \tilde{f}{\cal V}^{-\frac{4}{5}}{{\chi^{0}_3} {u_L}\tilde {u_L}}.
\end{eqnarray}
\end{enumerate}
\item
 The {lepton-squark-quark} corresponding to  Fig. 11(c) is given by considering the contribution of  $\frac{e^{\frac{K}{2}}}{2}\left({\cal D}_{{\cal {\bar A}}_1} D_{{\cal A}_4}W\right){\chi^{c^{{\cal {A}}_1}}}\chi^{{\cal A}_4}$ in gauged supergravity action.
 In terms of undiagonalized basis,
$${\cal D}_{\bar{a_1}}D_{a_4}W= \left(\partial_{\bar{a_1}}\partial_{a_4}K\right)W+\partial_{\bar{a_1}}KD_{a_4}W +
\partial_{a_4}KD_{\bar{a_1}}W - \left(\partial_{\bar{a_1}}K\partial_{a_4}K\right)W. $$
Utilizing equations (\ref{eq:Kahler pot}) and (\ref{eq:W}), one obtains:
\begin{eqnarray}
\label{eq:DKDW3}
& &{\hskip -1.0in}e^{\frac{K}{2}}\left(\left(\partial_{\bar{a_1}}\partial_{a_4}K\right)W+\partial_{\bar{a_1}}KD_{a_4}W + \partial_{a_4}KD_{\bar{a_1}}W - \left(\partial_{\bar{a_1}}K\partial_{a_4}K\right)W\right) \sim   {\cal V}^{-\frac{1}{18}}+ {\cal V}^{-\frac{1}{3}}\delta a_2.
\end{eqnarray}
From (\ref{eq:DKDW3}), picking up the component of fluctuations linear in $a_2$ and from (\ref{eq:eq_mass_terms_diag_non-diag_basis}) using: $$e^{\frac{K}{2}}{\cal D}_{{\cal {\bar A}}_1}D_{{\cal A}_4}W\sim O(1) e^{\frac{K}{2}}{\cal D}_{\bar {a_1}}D_{a_4}W,$$
one gets the contribution of lepton-squark-quark vertex as :
\begin{eqnarray}
\label{eq:Da1a4W}
& & e^{\frac{K}{2}}{\cal D}_{{\cal {\bar A}}_1}D_{{\cal A}_4}W\bar{\chi^{c^{{\cal {A}}_1}}}\bar\chi^{{\cal A}_4} \sim e^{\frac{K}{2}}{\cal D}_{\bar {a_1}}D_{a_4}W {\chi^{c^{{\cal {A}}_1}}}\chi^{{\cal A}_4}\sim  \left({\cal V}^{-\frac{1}{3}}\delta{\cal A}_2\right){\chi^{c^{{\cal {A}}_1}}}\chi^{{\cal A}_4}.
\end{eqnarray}
The physical lepton-squark-quark vertex is given by
\begin{eqnarray}
\label{eq:lepton-squark-quark1}
C^{ d^{c}_L \tilde {l_L} l_L}\sim \frac{{\cal V}^{-\frac{1}{3}}}{{\sqrt{\hat{K}_{{\cal A}_2{\bar{\cal A}}_2}{\hat{K}_{{\cal A}_1{\bar {\cal A}}_1}}{\hat{K}_{{\cal A}_4{\bar {\cal A}}_4}}}}}\sim (10)^{-\frac{14}{2}}{\cal V}^{-\frac{1}{3}} \sim {\cal V}^{-\frac{5}{3}} {\rm for}~ {\cal V}\sim {10}^5.
\end{eqnarray}
\end{itemize}
\begin{figure}[htbp]
\begin{center}
\begin{picture}(500,180)(-100,0)
\Text(280,120)[]{$\tilde{f} {\cal V}^{-1}{\chi^0_3 {l_L}\tilde {l_L}}$}
\ArrowLine(-50,120)(0,120)
\Text(-70,120)[]{$\chi^0_3({p_i})$}
\Text(55,153)[]{${l}(p_{l})$}
\ArrowLine(0,120)(40,150)
\Text(54,88)[]{$ \tilde l_R$}
\DashArrowLine(0,120)(40,90){5}
\end{picture}
\vskip-1.3in
\begin{picture}(500,180)(-100,0)
\Text(280,120)[]{${\cal V}^{-\frac{5}{3}}{d^{c}_L \tilde {l_L} u_L} $}
\DashArrowLine(-50,120)(0,120){5}
\Text(-70,120)[]{$\tilde l_R$}
\Text(53,153)[]{${u}(p_{u})$}
\ArrowLine(0,120)(40,150)
\Text(54,88)[]{$ {\bar d}(p_{\bar d})$}
\ArrowLine(0,120)(40,90)
\end{picture}
\vskip-1.3in
\begin{picture}(500,180)(-100,0)
\Text(280,120)[]{$(\tilde{f}{\frac{v}{m_{pl}}} {\cal V}^{-\frac{1}{12}}){\chi^{0}_3\bar d^{c}_L \tilde {d_R}}$}
\ArrowLine(-50,120)(0,120)
\Text(-70,120)[]{$\chi^0_3({p_i})$}
\Text(53,153)[]{${\bar d}(p_d)$}
\ArrowLine(0,120)(40,150)
\Text(54,88)[]{$ \tilde d_R$}
\DashArrowLine(0,120)(40,90){5}
\end{picture}
\vskip-1.3in
\begin{picture}(500,180)(-100,0)
\Text(280,120)[]{${\cal V}^{-\frac{5}{3}}{l_L \tilde {d_R} u_L}$}
\DashArrowLine(-50,120)(0,120){5}
\Text(-70,120)[]{$\tilde d_R$}
\Text(53,153)[]{${u}(p_{u})$}
\ArrowLine(0,120)(40,150)
\Text(54,88)[]{$ {l}(p_{l})$}
\ArrowLine(0,120)(40,90)
\end{picture}
\vskip-1.3in
\begin{picture}(500,180)(-100,0)
\Text(280,120)[]{$\tilde{f} {\cal V}^{-\frac{4}{5}}{{\chi^{0}_3} {u_L}\tilde {u_L}}$}
\ArrowLine(-50,120)(0,120)
\Text(-70,120)[]{$\chi^0_3({p_i})$}
\Text(53,153)[]{$u(p_{u})$}
\ArrowLine(0,120)(40,150)
\Text(54,88)[]{$ \tilde u_R$}
\DashArrowLine(0,120)(40,90){5}
\end{picture}
\vskip-1in
\begin{picture}(500,200)(-100,0)
\Text(280,120)[]{${\cal V}^{-\frac{5}{3}}{\tilde{l_L}}\chi^{d^{c}_L}\chi^{l_L} $}
\DashArrowLine(-50,120)(0,120){5}
\Text(-70,120)[]{$\tilde u_R$}
\Text(53,153)[]{${u}(p_{u})$}
\ArrowLine(0,120)(40,150)
\Text(54,88)[]{$ {\bar d}(p_{\bar d})$}
\ArrowLine(0,120)(40,90)
\end{picture}
\caption{Different vertices in  tree level neutralino decay}
\end{center}
\end{figure}
Now, working in two component notation, in order to calculate the decay width, we will be using the decay width formula as given by H.Dreiner et al. in \cite{2comp}. Because of Majorana nature of neutralino, we are considering both right handed as well as left handed incoming neutralino though the dominant contribution occurs in case of vertices corresponding to right handed neutralino. Henceforth, we will be using incoming right handed neutralino in the matrix amplitude calculation.

Guided by their notations, right handed incoming and left handed outgoing are denoted by wave function $ y^\dagger_i \equiv
 y^\dagger (\boldsymbol{\vec p}_i ,\lam_i)$, $ x^\dagger_l \equiv  x^\dagger
(\boldsymbol{\vec p}_l ,\lam_l)$, $ x^\dagger_u \equiv  x^\dagger
(\boldsymbol{\vec p}_u ,\lam_u)$, and $ x^\dagger_d \equiv  x^\dagger
(\boldsymbol{\vec p}_d ,\lam_d)$, respectively . Using the numerical estimate of vertices as calculated in set of equations (\ref{eq:neutralino-lepton-slepton}),(\ref{eq:quark-slepton-quark}),(\ref{eq:neutralino-quark-squark}), (\ref{eq:lepton-squark-quark2}), (\ref{eq:neutralino-quark-squark1}),(\ref{eq:lepton-squark-quark1}) for all three feynman diagrams, the corresponding contributions to the decay amplitude is:
\begin{eqnarray}
& &{\hskip -0.5in}{\cal M}_1 =
{\tilde{f}{\cal V}^{-1}{\cal V}^{-\frac{5}{3}}}
\left (
\frac{i}{(p_i - k_l)^2 - m_{\tilde l_L}^2}
\right )   y^\dagger_i  x^\dagger_l  x^\dagger_u  x^\dagger_d\
\\
& & {\hskip -0.5in} i{\cal M}_2 =
{\tilde f}{\cal V}^{-\frac{1}{12}}\frac{v}{m_{pl}}{\cal V}^{-\frac{5}{3}}\left (
\frac{ i}{(p_i - k_d)^2 -  m_{\tilde d_R}^2}
\right )\nonumber
 y^\dagger_i  x^\dagger_d  x^\dagger_l  x^\dagger_u\
\\
 & & {\hskip -0.5in}i{\cal M}_3 =
{\tilde f}{\cal V}^{-\frac{4}{5}}{\cal V}^{-\frac{5}{3}} \left (
\frac{ i}{(p_i - k_u)^2 -  m_{\tilde u_L}^2}
\right ) y^\dagger_i  x^\dagger_u  x^\dagger_d  x^\dagger_l.
\end{eqnarray}
Neglecting all of the final state fermion masses, one can express kinematic variables in the form as given below:
\begin{eqnarray}
& & z_l\equiv  2 p_i \newcdot k_l/m_{\chi^0_3}^2
= 2 E_l/m_{\chi^0_3},  z_d \equiv  2 p_i \newcdot k_d/m_{\chi^0_3}^2
= 2 E_d/m_{\chi^0_3}, z_u \equiv  2 p_i \newcdot k_u/m_{\chi^0_3}^2
= 2 E_u/m_{\chi^0_3},
\end{eqnarray}
and the total matrix amplitude can be rewritten as:
\begin{eqnarray}
{\cal M} =
c_1  y^\dagger_i  x^\dagger_l  x^\dagger_u  x^\dagger_d
+
c_2  y^\dagger_i  x^\dagger_d  x^\dagger_l  x^\dagger_u
+
c_3  y^\dagger_i  x^\dagger_u  x^\dagger_d  x^\dagger_l,
\end{eqnarray}
where for $m_{\chi^0_3}\sim {\cal V}^{\frac{2}{3}}m_{\frac{3}{2}}$ and $m_{l_L}\sim m_{\tilde d_R} \sim m_{\tilde u_L}\sim m_{{\cal A}_1}\sim {\cal V}^{\frac{1}{2}}m_{\frac{3}{2}}$ as given in equations (\ref{eq:gaugino_mass}) and (\ref{eq:m_A_1})
\begin{eqnarray}
c_1 \equiv {\tilde{f}{\cal V}^{-\frac{3}{2}}{\cal V}^{-\frac{5}{3}}} /[
m_{\tilde l_L}^2 - m_{\chi^0_3}^2 (1 - z_l)] = -{\tilde{f}{\cal V}^{-1}{\cal V}^{-\frac{5}{3}}} /{m^2_{\frac{3}{2}}}[{\cal V}- {\cal V}^{\frac{4}{3}}
 (1 - z_l)] \nonumber
\label{eq:RPVNdecaycone}
\\
c_2 \equiv {\tilde f}{\cal V}^{-\frac{1}{12}}{\frac{v}{m_{pl}}}{\cal V}^{-\frac{5}{3}}/[
m_{\tilde d_R}^2 - m_{\chi^0_3}^2 (1 - z_d)]= -{\tilde f}{\cal V}^{-\frac{1}{12}}{\frac{v}{m_{pl}}}{\cal V}^{-\frac{5}{3}}/{m^2_{\frac{3}{2}}}[{\cal V}- {\cal V}^{\frac{4}{3}} (1 - z_d)] \nonumber
\\
c_3 \equiv {\tilde f}{\cal V}^{-\frac{4}{5}}{\cal V}^{-\frac{5}{3}}/[
m_{\tilde u_L}^2 - m_{\chi^0_3}^2 (1 - z_u)] =  -{\tilde f}{\cal V}^{-\frac{4}{5}}{\cal V}^{-\frac{5}{3}}/{m^2_{\frac{3}{2}}}[{\cal V}- {\cal V}^{\frac{4}{3}} (1 - z_u)]
\label{eq:RPVNdecaycthree}
\end{eqnarray}

As explained in \cite{2comp}, by applying Fierz identity, one can reduce the number of terms:
\begin{eqnarray}
{\cal M} =
(c_1 - c_3)  y^\dagger_i  x^\dagger_l  x^\dagger_u  x^\dagger_d
+
(c_2 - c_3)  y^\dagger_i  x^\dagger_d  x^\dagger_l  x^\dagger_u .
\end{eqnarray}
and
\begin{eqnarray}
|{\cal M}|^2 &=& |c_1 - c_3|^2  y^\dagger_i  x^\dagger_l x_\mu y_i
                x^\dagger_u  x^\dagger_d x_d x_u
+ |c_2 - c_3|^2  y^\dagger_i  x^\dagger_d x_d y_i
                 x^\dagger_l  x^\dagger_u x_u x_l
\nonumber \\ &&
- 2 {\rm Re}[(c_1 - c_3) (c_2^* - c_3^*)
 y^\dagger_i  x^\dagger_l x_l x_u  x^\dagger_u  x^\dagger_d x_d y_i]\,,
\end{eqnarray}
Summing over the fermion spins by using following identities,
\begin{eqnarray}
& & \sum_s x_\alpha({\boldsymbol{\vec p}},s)
x^\dagger_{\dot{\beta}}({\boldsymbol{\vec p}},s) =
+ p\newcdot\sigma_{\alpha\dot{\beta}}
\,,
\sum_s x^{\dagger\dot{\alpha}}({\boldsymbol{\vec p}},s)
 x^{\beta}({\boldsymbol{\vec p}},s) =
+ p\newcdot\sigmabar^{\dot{\alpha}\beta}
\,,
 \sum_s y^{\dagger\dot{\alpha}}({\boldsymbol{\vec p}},s)
y^{\beta}({\boldsymbol{\vec p}},s)
= + p \newcdot \sigmabar^{\dot{\alpha}\beta}
\nonumber\\
& & \sum_s y_\alpha({\boldsymbol{\vec p}},s)
y^\dagger_{\dot{\beta}}({\boldsymbol{\vec p}},s) =
+ p \newcdot \sigma_{\alpha\dot{\beta}}
\end{eqnarray}
 one obtain:
\begin{eqnarray}
\sum_{\rm spins} |{\cal M}|^2 &=&
|c_1 - c_3|^2 {\rm Tr}[k_l \newcdot \sigmabar p_i \newcdot \sigma]
              {\rm Tr}[k_d \newcdot \sigmabar k_u \newcdot \sigma]
+
|c_2 - c_3|^2 {\rm Tr}[k_d \newcdot \sigmabar p_i \newcdot \sigma]
              {\rm Tr}[k_u \newcdot \sigmabar k_l \newcdot \sigma]
\nonumber \\ &&
- 2 {\rm Re} \bigl [(c_1 - c_3) (c_2^* - c_3^*)
{\rm Tr}[ k_l \newcdot \sigmabar k_u \newcdot \sigma k_d \newcdot \sigmabar
          p_i \newcdot \sigma] \bigr ]\,.
\end{eqnarray}
Applying the trace formulae and using
\begin{eqnarray}
+ 2 k_l \newcdot k_d
=
(1 - z_u) m_{\chi^0_3}^2 ,
\quad\quad
+ 2 k_l \newcdot k_u
=
(1 - z_d) m_{\chi^0_3}^2 ,
\quad\quad
+ 2 k_d \newcdot k_u  &=& (1 - z_l) m_{\chi^0_3}^2 .
\phantom{xxxx}
\end{eqnarray}
\begin{eqnarray}
\label{eq:sumspins}
\sum_{\rm spins} |{\cal M}|^2
&=& 4 |c_1 - c_3|^2 p_i \newcdot k_l \, k_d \newcdot k_u
  + 4 |c_2 - c_3|^2 p_i \newcdot k_d \, k_l \newcdot k_u
\nonumber \\ &&
  -4  {\rm Re} \bigl [(c_1 - c_3) (c_2^* - c_3^*)]
  (k_l \newcdot k_u\, p_i \newcdot k_d
  + p_i \newcdot k_l \,k_d \newcdot k_u
  - k_l \newcdot k_d \, p_i \newcdot k_u)
\nonumber \\[3pt]
&=& m_{\chi^0_3}^4 \Bigl [
    |c_1|^2 z_l (1 - z_l)
    + |c_2|^2 z_d (1 - z_d)
    + |c_3|^2 z_u (1 - z_u)
\nonumber \\ &&
    - 2{\rm Re}[c_1 c_2^*] (1 - z_l)(1 - z_d)
    - 2{\rm Re}[c_1 c_3^*] (1 - z_l)(1 - z_u)
\nonumber \\ &&
    - 2{\rm Re}[c_2 c_3^*] (1 - z_d)(1 - z_u)
    \Bigr ]\,,
\end{eqnarray}

The differential decay rate follows:
\begin{eqnarray}
\frac{d^2 \Gamma}{dz_l d z_d}
=
\frac{N_c m_{\chi^0_3}}{2^8 \pi^3}
\biggl (\frac{1}{2} \sum_{\rm spins} |{\cal M}|^2 \biggr ) ,
\end{eqnarray}
where  $N_c = 3$  and the
kinematic limits are
\begin{eqnarray}
\label{eq:limit}
0 < z_l < 1, 1 - z_l <  z_d  <1.
\end{eqnarray}
Adopting the technique as used in \cite{2comp}, the total decay width for the above  Feynman diagrams is given as:
\begin{eqnarray}
\label{eq:gamma}
\Gamma = \frac{N_c m_{\chi^0_3}^5}{2^{8}\cdot3 \pi^3}
\left ( |c_1^{\prime *}|^2 + |c_2^{\prime *}|^2 + |c_3^{\prime *}|^2 - {\rm Re}[
c_1^{\prime *} c_2^{\prime *} + c_1^{\prime *} c_3^{\prime *}
+ c_2^{\prime *} c_3^{\prime *}] \right ),
\end{eqnarray}
where $c_i^{\prime *}, i=1,2,3$ are obtained from  $c_i$'s  by integrating $z_{l},z_{d}$ in the aforementioned limits. Therefore, on simplifying, the result comes out to be
\begin{eqnarray}
& & {\hskip-0.4in}\Gamma \sim \frac{3. {  {\cal V}^{\frac{10}{3}} m^5_{\frac{3}{2}}}{\tilde f}^2. {\cal V}^{-\frac{74}{15}}}{2^{8}\cdot3. \pi^3 {\cal V}^{\frac{8}{3}}m^4_{\frac{3}{2}}} \sim {10}^{-4}{\cal V}^{\frac{2}{3}-{\frac{74}{15}}}m_{\frac{3}{2}}{\tilde f}^2  \sim {10}^{-18}{\tilde f}^2
 \end{eqnarray}
 In dilute flux approximation, considering ${\tilde f}^2 < {10}^{-8}$, life time of neutralino:
\begin{eqnarray}
   \tau &=&\frac{\hbar}{\Gamma}\sim\frac{10^{-34} Jsec}{{10}^{-18}{\tilde f}^2}\sim\frac{10^{-34} Jsec}{10^{-26}GeV} > O(10^{1}) sec
   \end{eqnarray}


\subsection{Two-Body Slepton/Squark Decays}

\begin{figure}
\begin{center}
\begin{picture}(150,97)(100,100)
\DashArrowLine(100,157)(150,157)5
\Line(150,160)(190,190)
\Line(152,154)(192,184)
\Vertex(151,157){3.5}
\ArrowLine(150,157)(190,130)
\Text(95,168)[]{${\tilde{l}}$}
\Text(196,197)[]{${\psi_\mu}$}
\put(155,132){${l^\prime}$}
%
\end{picture}
\end{center}
\caption{Two-Body Slepton/Squark Decay}
\end{figure}
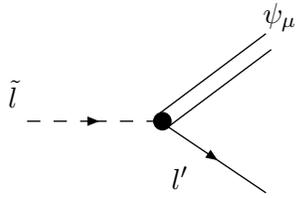
The decay width for $\tilde{l}/\tilde{q}\rightarrow l/q+\psi_\mu$ in our set up is given by:
\begin{equation}
\label{eq:sfermiontofermion+grav}
\Gamma\left(\tilde{l}/\tilde{q}\rightarrow l/q+\psi_\mu\right)\sim\frac{m_{\tilde{l}/\tilde{q}}^5}{m_{3/2}^2M_p^2}\sim{\cal V}^{-\frac{7}{2}}M_p,
\end{equation}
implying a lifetime of around $10^{-25.5}s$, satisfying the BBN constraints.

\subsection{Three-Body Slepton Decays}

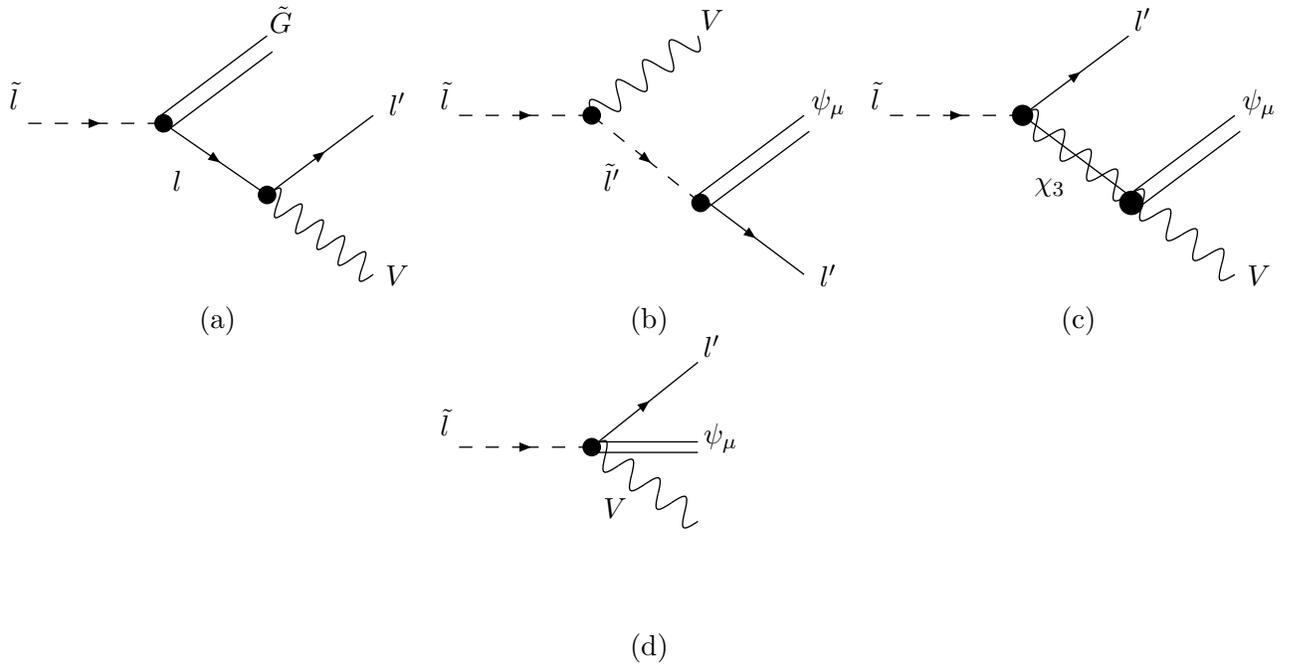
\begin{figure}[t!]
\label{decaydiag1}
\begin{center}
\begin{picture}(150,97)(100,100)
\DashArrowLine(100,157)(150,157)5
\Vertex(151,157){3.5}
\Line(150,160)(190,190)
\Line(152,154)(192,184)
\ArrowLine (151,157)(190,130)
\ArrowLine(190,130)(230,160)
\Vertex(190,130){3.5}
\Photon(190,130)(230,100){5}{5}
\Text(95,168)[]{${\tilde{l}}$}
\Text(240,165)[]{$l^\prime$}
\Text(240,100)[]{$V$}
\Text(196,197)[]{${\tilde{G}}$}
\put(155,132){${l}$}
\put(165,80){(a)}
\end{picture}
\hspace{0.2cm}
\begin{picture}(150,97)(100,100)
\DashArrowLine(100,160)(150,160)5
\Vertex(150,160){3.5}
\Photon(150,160)(190,190){5}{4}
\DashArrowLine (150,160)(191,127)5
\Line(190,130)(230,160)
\Line(192,124)(232,154)
\Vertex(191,127){3.5}
\ArrowLine(190,130)(230,100)
\Text(95,168)[]{$\tilde{l} $}
\Text(240,165)[]{$\psi_\mu$}
\Text(240,100)[]{$l^\prime$}
\Text(196,197)[]{$V$}
\put(155,132){$\tilde{l}^\prime$}
\put(165,80){(b)}
\end{picture}
\hspace{0.2cm}
\begin{picture}(150,97)(100,100)
\DashArrowLine(100,160)(150,160)5
\ArrowLine(150,160)(190,190)
\Line (150,160)(190,130)
\Vertex(150,160){4}
\Photon(150,160)(190,130){5}{4}
\Vertex(191,127){4.6}
\Line(190,130)(230,160)
\Line(192,124)(232,154)
\Photon(190,130)(230,100){5}{4}
\Text(95,168)[]{${\tilde{l}}$}
\Text(240,165)[]{$\psi_\mu$}
\Text(240,100)[]{$V$}
\Text(196,197)[]{$l^\prime$}
\put(155,132){$\chi_3$}
\put(165,80){(c)}
\end{picture}
\phantom{xxx}
\end{center}
\begin{center}
\begin{picture}(150,97)(100,100)
\DashArrowLine(100,158)(150,158)5
\ArrowLine(150,158)(190,190)
\Photon(150,158)(190,130){5}{4}
\Vertex(150,158){3.5}
\Line(150,160)(190,160)
\Line(150,156)(190,156)
\Text(95,168)[]{${\tilde{l}}$}
\Text(196,197)[]{$l^\prime$}
\put(193,160){$\psi_\mu$}
\put(155,132){$V$}
\put(165,80){(d)}
\end{picture}
\end{center}
\caption{Three-body slepton decays}
\end{figure}
As explained in \cite{Feng_et_al_sleptondecay}, the following set of effective operators are relevant to three-body slepton decays:
\begin{eqnarray}
\label{eq:Ois}
{\cal O}_1 &=& \overline{l'_h}\ p_{\tilde{l}}\cdot\tilde{G}^c\
p_{l'}\cdot\epsilon^*\nonumber\\
{\cal O}_2 &=& \overline{l'_h}\ p_l\cdot\tilde{G}^c \
p_{\tilde{l}}\cdot\epsilon^* \nonumber\\
{\cal O}_3 &=& \overline{l'_h}\ p_{\tilde{G}}\cdot\tilde{G}^c
\ p_{\tilde{l}}\cdot\epsilon^* \nonumber\\
{\cal O}_4 &=&i\ \overline{l'_h}\ \gamma\cdot\epsilon^*\
p_V\cdot \tilde{G}^c \nonumber\\
O_5 &=&\overline{l'_h}\ \gamma\cdot p_V\
\gamma\cdot\epsilon^*\ p_{\tilde{l}} \cdot \tilde{G}^c \nonumber\\
{\cal O}_6 &=&i\ \overline{l'_h}\ \gamma\cdot p_V
\ \gamma\cdot\epsilon^* \ p_V \cdot \tilde{G}^c \nonumber\\
{\cal O}_7 &=& i\ \overline{l'_h}\ \gamma\cdot p_{\tilde{G}}\
\gamma\cdot\epsilon^* \ p_V \cdot \tilde{G}^c \nonumber\\
{\cal O}_8 &=& i \ \overline{l'_h}\ \epsilon^*\cdot\tilde{G}^c \nonumber\\
{\cal O}_9 &=& i \ \overline{l'_h}\  \gamma\cdot p_V\
\epsilon^*\cdot \tilde{G}^c \nonumber\\
{\cal O}_{10} &=& i \ \overline{l'_h}\ \gamma\cdot p_{\tilde{G}}
\ \gamma\cdot p_V \epsilon^*\cdot\tilde{G}^c\ ,
\end{eqnarray}
where
$\epsilon_\mu$ is the polarization of gauge boson $V$.  Notice that
for an on-shell gravitino in the final state, ${\cal O}_3=0$ by using
the gravitino equation of motion.

Using the following volume-suppression factors in various vertices in Fig. 14(a)-(d) above:
\begin{eqnarray}
\label{eq:vertices}
& & \tilde{l}-\tilde{G}-l\ {\rm vertex}: \frac{g_{{\cal A}_1{\bar{\cal A}}_1}}{K_{{\cal A}_1{\bar{\cal A}}_1}}=1;\nonumber\\
& & l-l^\prime-V\ {\rm vertex}: \frac{g_{{\cal A}_1{\bar{\cal A}}_1}\tilde{f}{\cal V}^{-\frac{2}{3}} ln{\cal V}}{K_{{\cal A}_1{\bar{\cal A}}_1}};\nonumber\\
& & \tilde{l}-\tilde{l}-V\ {\rm vertex}: \frac{\left(\frac{\kappa_4^2\mu_7Q^BG^{T^B{\bar T}^B}}{{\cal V}}\right)\kappa_4^2\mu_7C_{a_1{\bar a}_1}}{\kappa_4^2K_{{\cal A}_1{\bar{\cal A}}_1}}\sim\frac{\tilde{f}{\cal V}^{-2}}{10^4};\nonumber\\
& & \tilde{l}-l^\prime-\tilde{G}\ {\rm vertex}: \frac{g_{{\cal A}_1{\bar{\cal A}}_1}}{K_{{\cal A}_1{\bar{\cal A}}_1}}=1,
\end{eqnarray}
and the operators of (\ref{eq:Ois}), the matrix elements for three-body
slepton decay [diagrams (a)- (d)] are
\begin{eqnarray}
\label{eq:Ms}
{\cal {M}}_V^a &\sim& \frac{\tilde{f}{\cal V}^{-\frac{11}{18}}}{M_p}
\frac{1}{s_{23}} \left(2{\cal O}_1-{\cal O}_5 \right)\nonumber\\
{\cal M}_V^b &\sim& \frac{\tilde{f}{\cal V}^{-2}}{10^4M_p}
\frac{1}{m^2_{\tilde{G}}+m_V^2-s_{13}-s_{23}}
\left({\cal O}_2+{\cal O}_3 \right)\nonumber \\
{\cal {M}}_V^c &\sim& \sum_i \frac{10\tilde{f}{\cal V}^{-\frac{3}{2}}}{M_p} \frac{1}{s_{13}-s_{\chi_3}}
\left[m_{\chi_3} \left({\cal O}_4-{\cal O}_9 \right)
- 4 \left({\cal O}_6+{\cal O}_7 - {\cal O}_{10} \right)
+ m_V^2{\cal O}_8 \right]
\nonumber\\
{\cal {M}}_V^d &\sim& i \frac{\tilde{f}{\cal V}^{-\frac{11}{9}}}{10^4M_p}{\cal O}_8 \ ,
\end{eqnarray}
where $\chi_3$ denotes the lightest neutralino in our set up, and
\begin{eqnarray}
s_{12}=(p_{\tilde{G}}+p_{l'})^2 \ , \quad
s_{13}=(p_{\tilde{G}}+p_V)^2 \ , \quad
s_{23}=(p_{l'}+p_V)^2 \ .
\end{eqnarray}
Notice that $s_{12}+s_{13}+s_{23}=\msl^2+\mg^2+m_Z^2$.

The differential decay width is
\begin{eqnarray}
 d\Gamma=\frac{1}{(2\pi)^3}\frac{1}{32m^3_{\tilde{l}}}
 |{\cal M}|^2 dm^2_{13}dm^2_{23} \ ,
\end{eqnarray}
where ${\cal M}={\cal M}_V^a+{\cal M}_V^b+{\cal M}_V^c+{\cal M}_V^d$.

The sum of the matrix elements can be written as
\begin{eqnarray}
{\cal{M}}(\tilde{l}_h\to l' V \tilde{G})
= {\cal M}_V^a+{\cal M}_V^b+{\cal M}_V^c + {\cal M}_V^d
= \sum_{i=1\ldots 10} {\cal{M}}_i \calO_i \ ,
\end{eqnarray}
where the ${\cal{M}}_i$ can be read off from
Eqs.~(\ref{eq:Ms}) above as:
\begin{eqnarray}
\label{eq:Mis}
& & {\cal M}_{1,5}\sim\frac{\tilde{f}{\cal V}^{-\frac{11}{18}}}{M_p s_{23}};\ {\cal M}_{2,3}\sim\frac{\left(\frac{\tilde{f}{\cal V}^{-2}}{10^4}\right)}{M_p(m_{3/2}^2 - s_{13} - s_{23})};\nonumber\\
& & {\cal M}_{4,9}\sim\frac{m_{\chi}\left(10\tilde{f}{\cal V}^{-\frac{3}{2}}\right)}{M_p(s_{13}-m_{\chi_3}^2)};\
{\cal M}_{6,7,10}\sim\frac{M_{4,9}}{m_{\chi_3}};\nonumber\\
& & {\cal M}_8\sim\frac{1}{M_p}\left(\frac{m_V^2\left(10\tilde{f}{\cal V}^{-\frac{3}{2}}\right)}{s_{13}-m_{\chi_3}^2} + \frac{\tilde{f}{\cal V}^{-\frac{11}{9}}}{10^4}\right).
\end{eqnarray}
The squared matrix
element is
\begin{eqnarray}
\left| {\cal{M}}(\tilde{l}_h \to l' V \tilde{G}) \right|^2
&=& \sum_{i=1\ldots 10} |{\cal{M}}_i|^2 \calO_{i,i} \nonumber \\
&& + \sum_{i,j=1\ldots 10}^{i<j} \Re({\cal{M}}_i {\cal{M}}_j^*)
\calO_{i,j}^{\text re}
+ \sum_{i,j=1\ldots 10}^{i<j} \Imag({\cal{M}}_i {\cal{M}}_j^*)
\calO_{i,j}^{\text im} \ .
\label{eq:square}
\end{eqnarray}
All the ${\cal{M}}_i$ are real except for ${\cal M}_8$, which has both
real and imaginary components.  The only non-zero contributions to the
last term in Eq.~(\ref{eq:square}) come from $\Imag({\cal{M}}_i
{\cal{M}}_8^*)$ ($i<8$) and $\Imag({\cal{M}}_8 {\cal{M}}_j^*)$
($j>8$). The resultant contribution of the same are evaluated in appendix E. Utilizing that,
\begin{equation}
\label{eq:Gamma_sltograv+lep+V}
\Gamma\left(\tilde{l}\rightarrow l^\prime+\tilde{G}+V\right)\sim
 \int_{m^2_{3/2}}^{2m^2_{\tilde{l}}}ds_{13}\int_{m_V^2}^{m^2_{\tilde{l}}}ds_{23}\frac{\Re\left({\cal M}_4{\cal M}_8^*\right){\cal O}^{\rm re}_{4,8}}{m^3_{\tilde{l}}}\sim10^{-15}M_p,
\end{equation}
implying that the associated life-time of this decay is about $10^{-28}$s, which respects the BBN constraints.
The summarized results of life time of various (N)LSP candidates are given in table 2.

\begin{table}[htbp]
\centering
\begin{tabular}{|l|l|}
\hline
{\bf Decay channels}   &  {\bf Average life time}  \\ \hline
  Gaugino decays:  $\tilde{B}\rightarrow\psi_\mu Z/{\gamma} $ & $ 10^{-30}s $\\
  $\tilde B\xrightarrow[]{Z}\psi_\mu u {\bar u} $  & $ 10^{-13}s $\\
  $\tilde {W}\xrightarrow[]{\tilde q}\psi_\mu u {\bar u} $  & $ 10^{-25}s $ \\

$\tilde{W}^0\rightarrow\psi_\mu W^+ W^- $ & $10^{-66}s $ \\ \hline
Gluino decays: $ \tilde{g}\rightarrow \chi_{\rm n}^{o}q_{{}_I} \bar{q}_{{}_J} $ & ${10}^{1} s$\\
$\tilde{g}\rightarrow\tilde{\chi}_3^0 g $ & ${10}^{10}s $ \\
 $\tilde{g}\rightarrow \psi_{\mu} q_{{}_I} \bar{q}_{{}_J}$ & $ {10}^{3}s $\\
$ \tilde{g}\rightarrow \psi_{\mu} g $ & $ {10}^{-1}s $\\ \hline
RPV Neutralino decay: $\chi^0_3\rightarrow u {\bar d} e^{-}$ & $ 10^{1}s $\\ \hline
Slepton decays: $\tilde{l}\rightarrow l^\prime \tilde{G} V $ & $ 10^{-28}s $\\
$ \tilde{l}/\tilde{q}\rightarrow l/q \psi_\mu $  & $ 10^{-25.5}s $\\ \hline
Gravitino decays: $\psi_\mu\rightarrow h \nu_e $ & $ 10^{17}s $ \\
$ \psi_\mu\rightarrow \nu\gamma,\nu Z  $ & $10^{22}s$ \\
$\psi_\mu \rightarrow L_iL_j E^c_k$ & $ 10^{21}s $ \\
$\psi_\mu \rightarrow L_i Q_j D^c_k$ & $ 10^{20}s $ \\
$\psi_\mu \rightarrow U_i^c D_j^c D_k^c$ & $ 10^{18}s $ \\  \hline
\end{tabular}
\label{table:decay_lifetime}
\caption{Life time estimates of various N(LSP) decay channels}
\end{table}

\section{Relic abundance of Gravitino dark matter candidate}

As discussed in section {\bf 4}, explicit calculations yielding the gravitino-decay life time of the same  order or greater than  age of the universe, justifies taking gravitino to be a viable dark matter candidate in our setup. Keeping in mind the fact that relic density of dark matter particle should be within the limits provided by recent WMAP observations and other direct and indirect experiments, this section is devoted to calculating the relic-abundance of the gravitino as a dark matter particle. The constraints on the cosmological gravitino abundance have been discussed in \cite{KHLOPOV_LINDE}. Assuming that re-heating temperature will be  low enough to produce an appropriate amount of dark matter in case of heavy gravitino, we focus on the case of gravitinos produced only in the decays of co-NLSPs and show that the same are produced in sufficient numbers to constitute all of non-baryonic dark matter.  The mass scales and  life time estimates of sleptons and neutralino discussed in section {\bf 2}  and {\bf 3} manifestly indicate the same to be valid co-NLSP's which freeze out with appropriate thermal relic density before decaying and then eventually decay into the gravitino. Therefore, gravitino then inherits much of the relic density of the neutralino/slepton.

\subsection{Neutralino density calculation}

The number density of cold relic density (CDM) from the early universe depends sensitively on the annihilation cross section of such particles. Amongst the various approaches to calculate the thermal cross section given in the literature, we rely on the partial-wave expansion approach used in \cite{Takeshi_Leszek} to calculate the annihilation cross section for each possible process. Since the $\mu$-split SUSY set up discussed in section {\bf 2}  includes only neutral light Higgs boson; heavy Higgs boson and similarly superpartners of neutral particles (i.e only Neutralino), we  consider possible annihilation channels which proceed via these particles only. The important channels that we will be discussing are: ${{\chi}^0_3 {\chi}^0_3\rightarrow hh}$, ${{\chi}^0_3 {\chi}^0_3\rightarrow ZZ}$, ${{\chi}^0_3 {\chi}^0_3\rightarrow ff}$.

The general procedure which is being followed in partial wave expansion approach, is given as follows.
As given in \cite{Keith.a.olive}, in QFT, the general cross-section is given as
\begin{equation}
\sigma v_{\rm M\o l}= \frac{1}{4E_1E_2}\int dLIPS {|\cal M|}^2
\end{equation}
where
\begin{equation}
dLIPS = (2\pi)^4 \delta^4(p_1 +p_2-\sum {p_j}) \prod \int {\frac{d^3{p_1}}{(2\pi)^3 2{\pi}_0}}.
\end{equation}
The thermally-averaged product of the neutralino pair-annihilation
cross section and their relative velocity $\langle \sigma v_{\rm M\o
l} \rangle$ is given as \cite{Takeshi_Leszek,Keith.a.olive}
\begin{equation}
\label{eq:thermalcrosc}
\langle \sigma v_{\rm M\o l} \rangle(T)=
\frac{\int d^3p_1 d^3p_2\, \sigma v_{\rm
M\o l}\, e^{-E_1/T} e^{-E_2/T}} {\int d^3p_1 d^3p_2\,  e^{-E_1/T} e^{-E_2/T} },
\end{equation}
where $p_1= (E_1, {\bf p}_1)$ and  $p_2= (E_2, {\bf p}_2)$
are the 4-momenta of the two colliding particles, and $T$ is the temperature of the bath.
Given the complexity of the general cross section, it is difficult to solve analytically.
Alternatively, one finds a way to get the solution by expressing $\langle \sigma v_{M\o l}\rangle$ in terms of $x=\frac{T}{M}$.
For this, one defines:
\begin{equation}
w(s)=\frac{1}{4}\int dLIPS {|\cal M|}^2=E_1E_2 \sigma v_{\rm M\o l}.
\end{equation}
Incorporating the value of $\sigma v_{\rm M\o l}$ in equation (\ref{eq:thermalcrosc}),
\begin{equation}
\label{eq:thermalcross}
\langle \sigma v_{\rm M\o l} \rangle(T)=
\frac{\int d^3p_1 d^3p_2\, w(s) e^{-E_1/T} e^{-E_2/T}} {\int d^3p_1 d^3p_2\, E_1E_2 e^{-E_1/T} e^{-E_2/T} }.
\end{equation}
By defining change of variables discussed in \cite{Keith.a.olive} i.e expressing momentum and energy in terms of x, carrying over the integration in terms of x,  $\langle\sigma v_{\rm M\o l}\rangle$ takes the form given below:
\begin{equation}
\label{eq:thermalcross}
\langle\sigma v_{\rm M\o l}\rangle=\frac{1}{m_\chi^2}
\left[w-\frac{3}{2}\left(2w-w'\right) x +
{\mathcal O} (x^2)\right]_{s=4m_\chi^2} \equiv a + bx + {\mathcal
O}(x^2).
\end{equation}
The coefficients $a$ and $b$ summed over all possible final states $f_1 f_2$ are defined in \cite{Takeshi_Leszek}, and given as:
\begin{eqnarray}
\label{eq:abf1f2}
& & \hskip -0.6in a=
\sum_{f_1 f_2}c \, \theta\left(4\mchi^2-(m_{f_1}+m_{f_2})^2 \right)\,
v_{f_1 f_2}\,\widetilde{a}_{f_1 f_2}, \nonumber\\
& & \hskip -0.6in b=
\sum_{f_1 f_2}c \, \theta\left(4\mchi^2-(m_{f_1}+m_{f_2})^2 \right)\,
v_{f_1 f_2} \Bigg\{ \widetilde{b}_{f_1 f_2}+  \widetilde{a}_{f_1 f_2}
\left[ -3+\frac{3}{4}v_{f_1 f_2}^{-2} \left(
      \frac{m_{f_1}^{2}+m_{f_2}^{2}}{2 \mchi^{2}}
   +\frac{(m_{f_1}^{2}-m_{f_2}^{2})^{2}}{8 \mchi^{4}}\right)\right]
                     \Bigg\}. \nonumber
\end{eqnarray}

The analytic expressions of a and b are given for the s, t and u channels of various possible annihilation processes are given in \cite{Takeshi_Leszek}. For the paper to be self-contained, we directly quote the forms of analytical expressions and  utilizing the same, we calculate the numerical estimates of a and b for all kinematically possible annihilation processes in our set up.

The idea is to first calculate the required vertices corresponding to different annihilation processes in the context of ${\cal N}=1$ gauged supergravity action and then use their estimates to calculate a and b coefficients . Following the same formalism as used in section {\bf 3} and {\bf 4}, utilizing the gauged supergravity action of Wess and Bagger, we obtain the numerical estimates of required vertices.

The physical eigenstates of neutralino mass matrix in the context of gauged supergravity action  are given as:
 \begin{eqnarray}
\label{eq:neutralinos_I}
& & \tilde{\chi}_1^0\sim\frac{-\tilde{H}_1^0+\tilde{H}_2^0}{\sqrt{2}};\ {\rm mass}\sim{\cal V}^{-\frac{35}{36}}M_p>m_{\frac{3}{2}},\nonumber\\
& & \tilde{\chi}_2^0\sim (\tilde{f}{\cal V}^{\frac{5}{6}}\frac{v}{m_{pl}})\lambda^0+\frac{\tilde{H}_1^0+\tilde{H}_2^0}{\sqrt{2}};\ {\rm mass}\sim{\cal V}^{-\frac{35}{36}}M_p>m_{\frac{3}{2}};\ CP:-,\nonumber\\.
& & \tilde{\chi}_3^0\sim-\lambda^0+ (\tilde{f}{\cal V}^{\frac{5}{6}}\frac{v}{m_{pl}})\left(\tilde{H}_1^0+\tilde{H}_2^0\right);\ {\rm mass}\sim
{\cal V}^{-\frac{4}{3}}M_p > m_{\frac{3}{2}}.
\end{eqnarray}
{\vskip 0.5in}
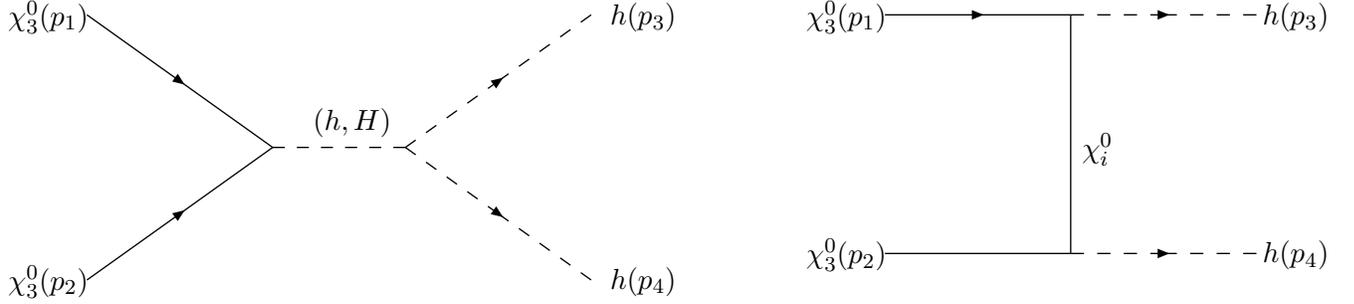
\begin{figure}
\begin{center}
\end{center}
{\vskip 1.0in}
\begin{center}
\begin{picture}(150,100)(-100,100)
\ArrowLine(-100,250)(-30,200)
\Text(-115,250)[]{$\chi^0_3(p_1)$}
\ArrowLine(-100,150)(-30,200)
\Text(-115,150)[]{${\chi}^{0}_3(p_2)$}
\DashLine(-30,200)(20,200)5
\Text(0,210)[]{$(h,H)$}
\DashArrowLine(20,200)(90,250)5
\Text(110,250)[]{$h(p_3)$}
\DashArrowLine(20,200)(90,150)5
\Text(110,150)[]{$h(p_4)$}
\end{picture}
{\hskip 2.0in}
\begin{picture}(150,100)(-100,100)
\ArrowLine(-100,250)(-30,250)
\Text(-115,250)[]{$\chi^0_3(p_1)$}
\DashArrowLine(-30,250)(40,250)5
\Text(55,250)[]{$h(p_3)$}
\Line(-30,250)(-30,160)
\Text(-20,200)[]{${\chi}^{0}_i$}
\Line(-100,160)(-30,160)
\Text(-115,160)[]{${\chi}^{0}_3(p_2)$}
\DashArrowLine(-30,160)(40,160)5
\Text(55,160)[]{$h(p_4)$}
\end{picture}
\caption{Feynman diagrams for
${\chi}^{0}_3 {\chi}^{0}_3\ra hh$ via $s$--channel Higgs exchange and t-channel ${\chi}^{0}_i$ exchange.}
\end{center}
\end{figure}

Utilizing (\ref{eq:neutralinos_I}),
\begin{eqnarray}
\label{eq:chichih}
& & C^{\chi^0_3 \chi^0_1 h}=  C^{\lambda^0 \tilde{H}^0 h}+  (\tilde{f}{\cal V}^{\frac{5}{6}}\frac{v}{m_{pl}})C^{\tilde{H}^0 \tilde{H}^0 h},\nonumber\\
& & C^{\chi^0_3 \chi^0_2 h}=  (\tilde{f}{\cal V}^{\frac{5}{6}}\frac{v}{m_{pl}})C^{\lambda^0 \lambda^0 h}+ (\tilde{f}^2{\cal V}^{\frac{5}{3}}\frac{v^2}{m^{2}_{pl}}) C^{\lambda^0 \tilde{H}^0 h}+  (\tilde{f}{\cal V}^{\frac{5}{6}}\frac{v}{m_{pl}})C^{\tilde{H}^0 \tilde{H}^0 h},\nonumber\\
& &  C^{\chi^0_3 \chi^0_3 h}=  C^{\lambda^0 \lambda^0 h}+  (\tilde{f}{\cal V}^{\frac{5}{6}}\frac{v}{m_{pl}})  C^{\lambda^0 \tilde{H}^0 h}+ (\tilde{f}^2{\cal V}^{\frac{5}{3}}\frac{v^2}{m^{2}_{pl}})  C^{\tilde{H}^0 \tilde{H}^0 h}
\end{eqnarray}
where $\tilde{H}^0 \sim \frac{\tilde{H}_1^0+\tilde{H}_2^0}{\sqrt{2}}$ is physical Higgsino and the physical light Higgs is defined as $h=  \frac{{H}_1^0-{H}_2^0}{\sqrt{2}}$\footnote{Working in the sublocus, where position moduli ${z_1}$ and ${z_2}$ are considered to be equivalent, for notational simplification, we will write Higgsino superfield  $\chi^{\frac{1}{\sqrt{2}}({{\cal Z}_1+ {\cal Z}_2})}\sim\chi^{{\cal Z}_i}$. }.

\underline{\bf {Higgsino-higgsino- Higgs vertex}}
\begin{equation}
{\cal L}=\frac{e^{\frac{K}{2}}}{2}\left({\cal D}_{{\cal Z}_1} D_{{\cal Z}_1}W\right) \chi^{{\cal Z}_1}_L \bar\chi^{{\cal Z}_i}_R +ig_{{\cal Z}_1{\bar {{\cal Z}_i}}}{\bar\chi}^{\bar {{\cal Z}_1}}_L \left[{\gamma}\cdot\partial\chi^{{\cal Z}_i}_L+\Gamma^{{\cal Z}_i}_{{\cal Z}_i{{\cal Z}_i}}{\gamma}\cdot\partial {{\cal Z}_i}\chi^{{\cal Z}_i}_L
+\frac{1}{4}\left(\partial_{{\cal Z}_i}K{\gamma}\cdot {\cal Z}_i - {\rm c.c.}\right)\chi^{{\cal Z}_i}_L\right];\nonumber\\
\end{equation}
$\chi^{{\cal Z}_i}_L/\bar\chi^{{\cal Z}_i}_R $ corresponds to left-/right-handed components of the Higgsino.
In terms of undiagonalized basis,
$${\cal D}_{z_1}D_{z_i}W= \left(\partial_{\bar{z_i}}\partial_{z_i}W\right) + \left(\partial_{\bar{z_i}}\partial_{z_i}K\right)W+\partial_{\bar{z_i}}KD_{z_i}W +
\partial_{z_i}KD_{\bar{z_i}}W - \left(\partial_{\bar{z_i}}K\partial_{z_i}K\right)W +  \Gamma^{K}_{z_i{z_i}}D_{K}W. $$
Since $SU(2)_L$ symmetry gets spontaneously broken for Higgsino-Higgsino-Higgs vertex, therefore the idea is to  expand above term  quadratic in $z_i$ such that one of the two $z_i$s acquires a VEV.
Utilizing $z_i\rightarrow z_i +{\cal V}^{\frac{1}{36}}{M_p}$ and thereafter solving with the help of equations (\ref{eq:Kahler pot}) and (\ref{eq:W}),  one has
$$\frac{e^{\frac{K}{2}}}{2} {\cal D}_{z_i}D_{\bar {z_i}}W \sim {\cal V}^{-\frac{16}{9}}\langle{z_i}\rangle \delta z_i $$
Following(\ref{eq:eq_mass_terms_diag_non-diag_basis}),
$e^{\frac{K}{2}}{\cal D}_{{\cal {\bar Z}}_i}D_{{\cal Z}_i}W\sim O(1) e^{\frac{K}{2}}{\cal D}_{\bar {z_i}}D_{z_i}W$, implying:
 \begin{equation}
 \label{eq:massterm}
 e^{\frac{K}{2}}{\cal D}_{{\cal {\bar Z}}_i}D_{{\cal Z}_i}W \chi^{{{\cal Z}_i}}_L \bar\chi^{{\cal Z}_i}_R \sim  \left({\cal V}^{-\frac{16}{9}}\langle{\cal Z}_i\rangle\right)\delta{\cal Z}_i \chi^{{{\cal Z}_i}}_L\bar\chi^{{\cal Z}_i}_R\sim \left({\cal V}^{-\frac{7}{4}}\langle{\cal Z}_i\rangle\right)\delta{\cal Z}_i \chi^{{{\cal Z}_i}}_L\bar\chi^{{\cal Z}_i}_R.
 \end{equation}
Using $\chi^{{\cal Z}_i} \sim {\cal V}m_{3/2}$ for the Higgsino mass  (see appendix B) and $m_{3/2}={\cal V}^{-2}{M_p}$,  one obtains:
\begin{eqnarray}
\label{eq:kinetic}
& & {\hskip -0.2in} g_{{\cal Z}_i{\bar {{\cal Z}_i}}}{\bar\chi}^{\bar {{\cal Z}_i}}_L {\gamma}\cdot\partial\chi^{{\cal Z}_i}_L \sim {\cal O}(1) g_{z_1{\bar{z_i}}}{\bar\chi}^{\bar {{\cal Z}_i}}_L {\gamma}\cdot\partial\chi^{{\cal Z}_i}_L \rightarrow \frac{{\cal V}^{-\frac{37}{36}}\langle{\cal Z}_i\rangle }{M_p}\delta{\cal Z}_i{\bar\chi}^{\bar {{\cal Z}_i}}_L{\gamma}\cdot p_{\chi^{{\cal Z}_i}}\chi^{{\cal Z}_i}_L\sim {\cal V}^{-2}{\bar\chi}^{\bar {{\cal Z}_i}}_L\delta{{\cal Z}_i}\chi^{{\cal Z}_i}_L; \nonumber\\
& & {\hskip -0.2in} g_{{\cal Z}_i{\bar {{\cal Z}_i}}}{\bar\chi}^{\bar {{\cal Z}_i}}_L\Gamma^{{\cal Z}_i}_{{\cal Z}_i{{\cal Z}_i}}{\gamma}\cdot\partial {{\cal Z}_i}\chi^{{\cal Z}_i}_L \sim {\cal O}(1) g_{z_i{\bar {z_i}}}\Gamma^{z_i}_{z_i{z_i}}{\bar\chi}^{\bar {{\cal Z}_i}}_L{\gamma}\cdot\partial {{\cal Z}_i}\chi^{{\cal Z}_i}_L \rightarrow \frac{{\cal V}^{-\frac{25}{36}}\langle{\cal Z}_i\rangle }{M_p}\delta{\cal Z}_i{\bar\chi}^{\bar {{\cal Z}_i}}_L{\gamma}\cdot (p_{\chi^{{\cal Z}_i}}+p_{\bar \chi^{{\cal Z}_i}})\chi^{{\cal Z}_i}_L\nonumber\\
& & \sim {\cal V}^{-\frac{5}{3}}{\bar\chi}^{\bar {{\cal Z}_i}}_L\delta{{\cal Z}_i}\chi^{{\cal Z}_i}_L ; \nonumber\\
& & {\hskip -0.8in} g_{{\cal Z}_i{\bar {{\cal Z}_i}}}{\bar\chi}^{\bar {{\cal Z}_i}}_L\frac{1}{4}\left(\partial_{{\cal Z}_i}K{\gamma}\cdot {\cal Z}_i - {\rm c.c.}\right)\chi^{{\cal Z}_i}_L  \sim  {\cal O}(1)g_{z_1{\bar {z_i}}}\partial_{z_i}K {\bar\chi}^{\bar {{\cal Z}_i}}_L{\gamma}\cdot\partial {{\cal Z}_i}\chi^{{\cal Z}_i}_L \rightarrow \frac{{\cal V}^{-\frac{4}{3}}\langle{\cal Z}_i\rangle }{M_p}\delta{\cal Z}_i{\bar\chi}^{\bar {{\cal Z}_i}}_L{\gamma}\cdot (p_{\chi^{{\cal Z}_i}}+p_{\bar \chi^{{\cal Z}_i}})\chi^{{\cal Z}_i}_L\nonumber\\
& & \sim {\cal V}^{-\frac{83}{36}}{\bar\chi}^{\bar {{\cal Z}_i}}_L\delta{{\cal Z}_i}\chi^{{\cal Z}_i}_L.
\end{eqnarray}
Incorporating results of  (\ref{eq:massterm}) and (\ref{eq:kinetic}), the physical Higgsino-Higgsino- Higgs vertex will be given as
\begin{eqnarray}
\label{eq:CHHh}
& & C^{\tilde {H}^0 \tilde {H}^0 h}= \frac{1}{{\sqrt{(\hat{K}_{{\cal Z}_i{\bar{\cal Z}}_i})^4}}}\left[({\cal V}^{-\frac{7}{4}})+ ({\cal V}^{-2}+{\cal V}^{-\frac{5}{3}} +{\cal V}^{-\frac{83}{36}})\right]\sim {\cal V}^{\frac{1}{4}}.
\end{eqnarray}

\underline{\bf{Gaugino-Higgsino-Higgs vertex}}
\begin{equation}
{\cal L}=g_{YM}g_{B {\bar {\cal Z}_i}}X^B \bar \chi^{{\cal Z}_i}_L \lambda_L+ \partial_{{\cal Z}_i}T_B D^{B}\tilde{H}^i\lambda^{i}.
\end{equation}
 Expanding the same in the fluctuations linear in ${\delta Z}_i$, we have
\begin{equation}
\label{eq:gYM_ReT_II}
g_{YM}g_{T^B{\cal Z}^i}X^{B}= {\tilde f} {\cal V}^{-2}\delta{{\cal Z}_i} , (\partial_{{\cal Z}_i}T_B) D^{B}\sim   {\tilde f} {\cal V}^{-\frac{4}{3}}\delta{{\cal Z}_i}.
\end{equation}
The physical Gaugino-Higgsino-Higgs vertex works out to yield:
\begin{eqnarray}
\label{eq:CgHh}
& & C^{h{\tilde H}^0\lambda_L}=  \frac{{\cal V}^{-\frac{4}{3}}\tilde{f}}{{\left(\sqrt{\hat{K}_{{\cal Z}_i{\bar {\cal Z}}_i}}\right)^2}}\sim \tilde{f}\left({10}^{5}{\cal V}^{-\frac{4}{3}}\right) \sim \tilde{f}{\cal V}^{-\frac{1}{3}}.
\end{eqnarray}

\underline{\bf{Gaugino-Gaugino-Higgs vertex}}
\begin{eqnarray}
& & {\cal L}= i{\bar {\lambda_L}}{\gamma}^m {\frac{1}{4}(K_{{\cal Z}_i}{\partial}_m {{\cal Z}_i}- c.c.)}{\lambda}_{L},
\end{eqnarray} where  ${\lambda}_{L}$ corresponds to gaugino. Here also, the aforementioned vertex  does not preserve $SU(2)_L$ symmetry - one has to obtain the term linear in $\langle z_i\rangle$. In terms of undiagonalized basis,
$\partial_{z_i}K\sim{\cal V}^{-\frac{2}{3}}\langle z_i\rangle$, and using
$\partial_{{\cal Z}_i}K\sim  {\cal O}(1) \partial_{z_i}K$, we have:
$\partial_{{\cal Z}_i}K\sim {\cal V}^{-\frac{2}{3}}\langle{\cal Z}_i\rangle$, incorporating the same
\begin{eqnarray}
\label{eq:Cggh}
 & & C^{h{\bar {\lambda_L}}{\lambda}_{L}}:  \frac{{\cal V}^{-\frac{2}{3}}\langle{\cal Z}_i\rangle{\bar {\lambda_L}}\frac{\slashed{\partial}{{\cal Z}_i}}{M_p}{\lambda_L}}{{\sqrt{(\hat{K}_{{\cal Z}_1{\bar {\cal Z}}_1}})^2}}\sim {10}^{5} {\cal V}^{-\frac{23}{36}}h {\bar {\lambda_L}}\frac{\slashed{p}_h}{M_p}{\lambda_L}\sim {\cal V}^{-\frac{23}{36}}{h}{\bar {\lambda_L}}\frac{{\gamma}\cdot({p_{{\bar {\lambda_L}}} + p_{{\lambda}_{L}}})}{M_p}{\lambda}_{L} \nonumber\\
 & & \sim{10}^{5} {\cal V}^{-\frac{23}{36}}h{\bar {\lambda_L}}\frac{m_{\tilde g}}{M_p}{\lambda}_{L}\sim {\cal V}^{-\frac{35}{36}}h{\bar {\lambda_L}}{\lambda}_{L}
 \end{eqnarray}

\underline{\bf{Higgs-Higgs-Higgs vertex:}} The value of effective Higgs triple interaction vertex in SM  is given by $ C^{h h h} \sim \frac{M^{2}_{H}}{v}$, v is electroweak VEV.


Now, using the set of results given in equation no (\ref{eq:CHHh}), (\ref{eq:CgHh}) and (\ref{eq:Cggh}), the contribution of vertices appearing in equation (\ref{eq:chichih}) are as follows:
\begin{eqnarray}
\label{eq:Nchichih}
& & C^{\chi^0_3 \chi^0_1 h}= {\tilde f}{\cal V}^{-\frac{1}{3}}, C^{\chi^0_3 \chi^0_2 h}\sim  \tilde{f} {\cal V}^{\frac{13}{12}}\frac{v}{m_{pl}}, C^{\chi^0_3 \chi^0_3 h}\sim {\cal V}^{-\frac{35}{36}}.
\end{eqnarray}

Since now we have got the estimates of coupling required to calculate partial wave coefficients for  $\chi^0_3\chi^0_3\rightarrow hh$ annihilation process,  we are in position to calculate the contribution of partial wave $a$ and $b$ coefficients in our set up just by using (and quoting verbatim below) the form of analytical results provided in \cite{Takeshi_Leszek}.

\underline{\bf{ s-channel Higgs-boson ($h,H$) exchange:}}

The analytic expressions of $\widetilde{a}_{hh}$ and $\widetilde{b}_{hh}$ corresponding to Fig. 23 are given as:
   \begin{eqnarray}
   \label{eq:ahh}
   \widetilde{a}_{hh}^{(h,H)}& = & 0, 
\\
   \widetilde{b}_{hh}^{(h,H)}& = & \frac{3}{64\,\pi} \left |
     \sum_{r=h,H} \frac{C^{hhr}\: C^{\chi^0_3\chi^0_3 r}}
       {4\, \mchi^{2}-m_{r}^{2}+i\, \Gamma_{r}\, m_{r}}
                 \right |^{2} .
    \end{eqnarray}
    \vspace{0.25cm}
Expanding the summation
\begin{eqnarray}
\widetilde{b}_{hh}^{(h,H)}& = & \frac{3}{64\,\pi} \left |
      \frac{C^{hhh}\: C^{\chi^0_3\chi^0_3 h}}
       {4\, m_{\chi^0_3}^{2}-m_{h}^{2}+i\, \Gamma_{h}\, m_{h}}+ \frac{C^{hhH}\: C^{\chi^0_3\chi^0_3 H}}
       {4\, m_{\chi^0_3}^{2}-m_{H}^{2}+i\, \Gamma_{H}\, m_{H}}
                 \right |^{2}  ;
\end{eqnarray}
Utilizing the value of mass $m_{h} =125 GeV$ and $ m_{H}\sim {\cal V}^{\frac{59}{72}} m_{\frac{3}{2}}$ as calculated in  appendix B, $m_{\chi^0_3}\sim {\cal V}^{\frac{2}{3}}m_{\frac{2}{3}}\sim {\cal V}^{-\frac{4}{3}}M_p$
and  $C^{hhh}\sim 10^2 GeV, C^{\chi^0_3 \chi^0_3 h}\sim  C^{\chi^0_3 \chi^0_3 H}\sim  {\cal V}^{-\frac{35}{36}} $ from above, after simplifying, we have
\begin{eqnarray}
\label{bhhh}
& & \widetilde{b}_{hh}^{(h,H)}\sim \frac{3}{64\,\pi} \left |\frac{10^{2}.{\cal V}^{-\frac{35}{36}}}{4.{\cal V}^{-\frac{8}{3}}m^{2}_{pl}}+ \frac{10^{2}.{\cal V}^{-\frac{35}{36}}}{{\cal V}^{-\frac{85}{36}}m^{2}_{pl} }\right |^{2} \sim \frac{3}{64\,\pi}\left (\frac{{\cal V}^{\frac{5}{3}}.10^2}{m^2_{pl}}\right )^{2} \nonumber\\
& & \sim \frac{3}{64\,\pi}\times O(10)^{-51}{GeV}^{-2}~{\rm for}~ {\cal V}\sim {10}^5.
\end{eqnarray}
Here we assume that $\Gamma_{h,H} < m_{h,H}$  in our set up.

$\bullet$
\underline{\bf{Neutralino\ ($\chi_{i}^{0}$) exchange:}}
    \begin{eqnarray}
    \widetilde{a}_{hh}^{(\chi^0)}& = & 0, 
\\
    \widetilde{b}_{hh}^{(\chi^0)}& = &  \frac{1}{16\,\pi}
      \sum_{i,j=1}^{3}  (C^{\chi_{i}^0 \chi^0_3, h })^2
        (C^{\chi_{j}^0 \chi^0_3, h * })^2
         \frac{1}{\Delta_{hi}^{2}\,\Delta_{hj}^{2} } \nonumber
\\
       & & \times \Big[4 \,m_{\chi^0_3}^{2}\,(m_{\chi^0_3}^{2}-{m_h}^{2})^{2}
           +4\,m_{\chi^0_3}\,(m_{\chi^0_3}^{2}-{m_h}^{2})\,
           (m_{\chi^0_3}+m_{\chi^0_i})\,\Delta_{hi} \nonumber \\
       & &\hspace{0.3in} +3\,(m_{\chi^0_3}+m_{\chi^0_i})\,(m_{\chi^0_3}+m_{\chi^0_j})\,
             \Delta_{hi}\,\Delta_{hj}\Big],
     \end{eqnarray}
where $\Delta_{hi}\equiv\,{m_h}^{2}-m_{\chi^0_3}^{2}-m_{\chi^0_{i}}^{2}$.
Utilizing the values of masses given above,
$\Delta_{h1}\equiv\,{m_h}^{2}-m_{\chi^0_3}^{2}-m_{\chi^0_{1}}^{2} \sim {\cal V}^2{m^2_\frac{3}{2}}$,
$\Delta_{h2}\equiv\,{m_h}^{2}-m_{\chi^0_3}^{2}-m_{\chi^0_{2}}^{2} \sim {\cal V}^2{m^2_\frac{3}{2}}$,
$\Delta_{h3}\equiv\,{m_h}^{2}-m_{\chi^0_3}^{2}-m_{\chi^0_{3}}^{2} \sim {\cal V}^{\frac{4}{3}}{m^2_\frac{3}{2}}$,
 and
  \begin{eqnarray}
{\hskip -0.5in} \widetilde{b}_{hh}^{(\chi^0)}& = &  \frac{1}{16\,\pi}
       (C^{\chi_{1}^0 \chi_{3}^0, h })^2
        (C^{\chi_{1}^0 \chi_{3}^0, h * })^2
         \frac{1}{\Delta_{h1}^{2}\,\Delta_{h1}^{2} }
        \times \Big[4 \,m_{\chi^0_3}^{2}\,(m_{\chi^0_3}^{2}-{m^2_h})^{2}
           +4\,m_{\chi^0_3}\,(m_{\chi^0_3}^{2}-{m^2_h})\,
           (m_{\chi^0_3}+m_{\chi^0_1})\,\Delta_{h1} \nonumber \\
       & & +3\,(m_{\chi^0_3}+m_{\chi^0_1})\,(m_{\chi^0_3}+m_{\chi^0_j})\,
             \Delta_{h1}\,\Delta_{h1}\Big]+
             (C^{\chi_{1}^0 \chi_{3}^0, h })^2
        (C^{\chi_{2}^0 \chi_{3}^0, h * })^2
         \frac{1}{\Delta_{h1}^{2}\,\Delta_{h2}^{2} }
        \times \Big[4 \,m_{\chi^0_3}^{2}\,(m_{\chi^0_3}^{2}-{m^2_h})^{2}\nonumber \\
       & &
           +4\,m_{\chi^0_3}\,(m_{\chi^0_3}^{2}-{m^2_h})\,
           (m_{\chi^0_3}+m_{\chi^0_1})\,\Delta_{h1} +3\,(m_{\chi^0_3}+m_{\chi^0_1})\,(m_{\chi^0_3}+m_{\chi^0_2})\,
             \Delta_{h1}\,\Delta_{h2}\Big]+ \nonumber\\
             &&
             (C^{\chi_{1}^0 \chi_{3}^0, h })^2
        (C^{\chi_{3}^0 \chi_{3}^0, h * })^2
         \frac{1}{\Delta_{h1}^{2}\,\Delta_{h3}^{2} }
        \times \Big[4 \,m_{\chi^0_3}^{2}\,(m_{\chi^0_3}^{2}-{m^2_h})^{2}
           +4\,m_{\chi^0_3}\,(m_{\chi^0_3}^{2}-{m^2_h})\,
           (m_{\chi^0_3}+m_{\chi^0_1})\,\Delta_{h1} \nonumber \\
       & & +3\,(m_{\chi^0_3}+m_{\chi^0_1})\,(m_{\chi^0_3}+m_{\chi^0_3})\,
             \Delta_{h1}\,\Delta_{h3}\Big]+
             (C^{\chi_{2}^0 \chi_{3}^0 h })^2
        (C^{\chi_{2}^0 \chi_{3}^0, h * })^2
         \frac{1}{\Delta_{h2}^{2}\,\Delta_{h2}^{2} }  \times \Big[4 \,m_{\chi^0_3}^{2}\,(m_{\chi^0_3}^{2}-{m^2_h})^{2}\nonumber\\
             & &
           +4\,m_{\chi^0_3}\,(m_{\chi^0_3}^{2}-{m^2_h})\,
           (m_{\chi^0_3}+m_{\chi^0_2})\,\Delta_{h2}  +3\,(m_{\chi^0_3}+m_{\chi^0_2})\,(m_{\chi^0_3}+m_{\chi^0_2})\,
             \Delta_{h2}\,\Delta_{h2}\Big] + \nonumber\\
             &&
        (C^{\chi_{2}^0 \chi_{3}^0 h })^2(C^{\chi_{3}^0 \chi_{3}^0, h * })^2
         \frac{1}{\Delta_{h2}^{2}\,\Delta_{h3}^{2} }  \times \Big[4 \,m_{\chi^0_3}^{2}\,(m_{\chi^0_3}^{2}-{m^2_h})^{2}
           +4\,m_{\chi^0_3}\,(m_{\chi^0_3}^{2}-{m^2_h})\,
           (m_{\chi^0_3}+m_{\chi^0_2})\,\Delta_{h2}
           \nonumber
\\
       & & +3\,(m_{\chi^0_3}+m_{\chi^0_3})\,(m_{\chi^0_3}+m_{\chi^0_2})\,
             \Delta_{h2}\,\Delta_{h3}\Big]
             +(C^{\chi_{3}^0 \chi_{3}^0 h })^2
        (C^{\chi_{3}^0 \chi_{3}^0, h * })^2
         \frac{1}{\Delta_{h3}^{2}\,\Delta_{h3}^{2} } \times \Big[4 \,m_{\chi^0_3}^{2}\,(m_{\chi^0_3}^{2}-{m^2_h})^{2}\nonumber\\
       & &
           +4\,m_{\chi^0_3}\,(m_{\chi^0_3}^{2}-{m^2_h})\,
           (m_{\chi^0_3}+m_{\chi^0_3})\,\Delta_{h3}   +3\,(m_{\chi^0_3}+m_{\chi^0_3})\,(m_{\chi^0_3}+m_{\chi^0_3})\,
             \Delta_{h3}\,\Delta_{h3}\Big]\nonumber
     \end{eqnarray}
     \begin{eqnarray}
     \label{bchihh}
           & & \sim \frac{1}{16\,\pi}\biggr[
       ({\tilde f}^4{\cal V}^{-\frac{4}{3}}) \frac{1}{m_{\chi^0_1}^{2}}+  ({\tilde f}^4{\cal V}^{\frac{3}{2}}\frac{v^2}{m^{2}_{pl}})
   \frac{1}{m_{\chi^0_1} m_{\chi^0_2}} +({\tilde f}^2{\cal V}^{-\frac{47}{18}}
         )
         \frac{1}{m_{\chi^0_1}m_{\chi^0_3}}+({\tilde f}^4{\cal V}^{\frac{13}{3}}
         \frac{v^4}{m^{4}_{pl}})
\frac{1}{m_{\chi^0_2}^{2}}+ \nonumber\\
       & &  ({\tilde f}^2{\cal V}^{\frac{1}{6}}
         \frac{v^2}{m^{2}_{pl}})
         \frac{1}{m_{\chi^0_2}m_{\chi^0_3}} + ({\cal V}^{-\frac{35}{9}}) \frac{1}{m_{\chi^0_1}^{2}}\biggr]\nonumber\\
       & &  \sim \frac{1}{16\,\pi} \times O(10)^{-42}{GeV}^{-2} ~{\rm for}~ {\cal V}\sim  {10}^5~{\rm and }~{\tilde f}\sim {10}^{-4}.
     \end{eqnarray}
  \\
$\bullet$
\underline{\bf {Higgs ($h,H$)--neutralino\ ($\chi_{i}^{0}$) interference term:}}
   \begin{eqnarray}
    \widetilde{a}_{hh}^{(h,H-\chi^0)}& = & 0, 
\\
    \widetilde{b}_{hh}^{(h,H-\chi^0)}& = &  \frac{1}{16\,\pi}
      \sum_{i=1}^{3} Re \left[\sum_{r=h,H} \left(\frac{C^{hhr}\:
       C_{S}^{\chi\chi r}}{4\, m_{\chi^0_3}^{2}-m_{r}^{2}
            +i\, \Gamma_{r}\, m_{r}}\right)^{*}\, C_{S}^{\chi^0_{i} \chi\, h}
       C_{S}^{\chi^0_{i} \chi\, h} \right] \nonumber \\
       & & \times \frac{[2\,m_{\chi^0_3}\,(m_{\chi^0_3}^{2}-{m^2_h})
         +3\,(m_{\chi^0_3}+m_{\chi^0_3}i)\,\Delta_{hi}]}{\Delta_{hi}^{2}}.
     \end{eqnarray}
     \begin{eqnarray}
    & & {\hskip -.5in} Re \left[\sum_{r=h,H} \left(\frac{C^{hhr}\:
       C_{S}^{\chi\chi r}}{4\, m_{\chi^0_3}^{2}-m_{r}^{2}
            +i\, \Gamma_{r}\, m_{r}}\right)^{*}\,\right]\sim \left[\left(\frac{C^{hhh}\:
       C_{S}^{\chi\chi h}}{4\, m_{\chi^0_3}^{2}-m_{h}^{2}
            +i\, \Gamma_{h}\, m_{h}}\right)^{*}\ + \left(\frac{C^{hhh}\:
       C_{S}^{\chi\chi h}}{4\, m_{\chi^0_3}^{2}-m_{H}^{2}
            +i\, \Gamma_{h}\, m_{H}}\right)^{*}\,\right] \nonumber\\
            & &  \sim \frac{10^2.{\cal V}^{-\frac{35}{36}}}{ m_{\chi^0_3}^{2}} + \frac{10^2.
       {\cal V}^{-\frac{35}{36}}}{m_{H}^{2}}\sim  \frac{
       {\cal V}^{-\frac{17}{30}}}{\, m_{\chi^0_3}^{2}}.
       \end{eqnarray}
     Expanding this, we get
      \begin{eqnarray}
       \label{bhchihh}
     \widetilde{b}_{hh}^{(h,H-\chi^0)}& = &  \frac{1}{16\,\pi}\frac{
       {\cal V}^{-\frac{17}{30}}}{\, m_{\chi^0_3}^{2}}
      \Biggl[ \frac{(C^{\chi^0_{1}\chi^{0}_{3}  h}
       )^2 (2\,m_{\chi^0_3}\,(m_{\chi^0_3}^{2}-{m^2_h})
         +3\,(m_{\chi^0_3}+m_{\chi^0_1})\,\Delta_{h1})}{\Delta_{h1}^{2}}+ \nonumber\\
         & & {\hskip -1.5in} \frac{(C^{\chi^0_{2} \chi^{0}_{3} h}
       )^2 (2\,m_{\chi^0_3}\,(m_{\chi^0_3}^{2}-{m^2_h})
         +3\,(m_{\chi^0_3}+m_{\chi^0_2})\,\Delta_{h2})}{\Delta_{h2}^{2}}+ \frac{(C^{\chi^0_{3}\chi^{0}_{3}  h}
       )^2 (2\,m_{\chi^0_3}\,(m_{\chi^0_3}^{2}-{m^2_h})
         +3\,(m_{\chi^0_3}+m_{\chi^0_3})\,\Delta_{h3})}{\Delta_{h3}^{2}}\Biggr] \nonumber\\
         & & {\hskip -1.0in} \sim  \frac{1}{16\,\pi}\frac{
       {\cal V}^{-\frac{17}{30}}}{\, m_{\chi^0_3}^{2}}
      \Biggl[ \frac{({\tilde f}{\cal V}^{-\frac{1}{3}}
       )^2 (
         3\,m_{\chi^0_1}^3\,)}{4m_{\chi^0_1}^4}+ \frac{\tilde{f}^2{\cal V}^{\frac{13}{6}}\frac{v^2}{m^{2}_{pl}}
       (
         3\,m_{\chi^0_2}^3\,)
         }{4m_{\chi^0_2}^4}+  \frac{({\cal V}^{-\frac{35}{36}}
       )^2 (5\,m_{\chi^0_3}^{3}
         \,)}{m_{\chi^0_3}^4}\Biggr] \nonumber\\
         & & {\hskip -1.0in} \sim  \frac{1}{16\,\pi}\frac{
       {\cal V}^{-\frac{17}{30}}}{\, m_{\chi^0_3}^{2}}
      \Biggl[ \frac{({\cal V}^{-\frac{35}{36}}
       )^2 (5\,m_{\chi^0_3}^{3}
         \,)}{m_{\chi^0_3}^4}\Biggr]\sim \frac{1}{16\,\pi} \times O(10)^{-26}{GeV}^{-2} ~{\rm for}~ {\cal V}\sim {10}^5.
     \end{eqnarray}
Utilizing results of equations (\ref{bhhh}), (\ref{bchihh}), (\ref{bhchihh}),
\begin{eqnarray}
\label{sumbhh}
  \widetilde{b}_{hh} &=&
 \widetilde{b}_{hh}^{(h,H)}
+\widetilde{b}_{hh}^{(\chi^{0})}
+\widetilde{b}_{hh}^{(h,H-\chi^{0})} \sim  O(10)^{-26}{GeV}^{-2}\nonumber\\
& & \widetilde{a}_{hh} =
 \widetilde{a}_{hh}^{(h,H)}
+\widetilde{a}_{hh}^{(\chi^{0})}
+\widetilde{a}_{hh}^{(h,H-\chi^{0})}=0.\nonumber\\
\end{eqnarray}

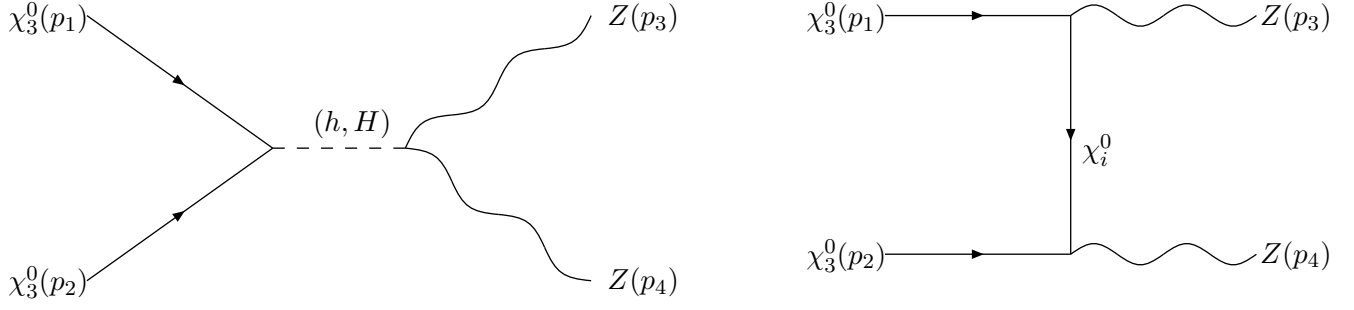
\begin{figure}
\begin{center}
\begin{picture}(150,100)(-100,100)
\ArrowLine(-100,250)(-30,200)
\Text(-115,250)[]{$\chi^0_3(p_1)$}
\ArrowLine(-100,150)(-30,200)
\Text(-115,150)[]{${\chi}^{0}_3(p_2)$}
\DashLine(-30,200)(20,200)5
\Text(0,210)[]{$(h,H)$}
\Photon(20,200)(90,250){4}{2}
\Text(110,250)[]{$Z(p_3)$}
\Photon(20,200)(90,150){4}{2}
\Text(110,150)[]{$Z(p_4)$}
\end{picture}
{\hskip 2.0in}
\begin{picture}(150,100)(-100,100)
\ArrowLine(-100,250)(-30,250)
\Text(-115,250)[]{$\chi^0_3(p_1)$}
\Photon(-30,250)(40,250){4}{2}
\Text(55,250)[]{$Z(p_3)$}
\ArrowLine(-30,250)(-30,160)
\Text(-20,200)[]{${\chi}^{0}_i$}
\ArrowLine(-100,160)(-30,160)
\Text(-115,160)[]{${\chi}^{0}_3(p_2)$}
\Photon(-30,160)(40,160){4}{2}
\Text(55,160)[]{$Z(p_4)$}
\end{picture}
\caption{Feynman diagrams for
${\chi}^{0}_3 {\chi}^{0}_3\ra ZZ$ via $s$--channel Higgs exchange and t-channel ${\chi}^{0}_i$ exchange.}
\end{center}
\end{figure}
\begin{eqnarray}
\label{eq:chichiZ}
& & C^{\chi^0_3 \chi^0_1 Z}=   (\tilde{f}{\cal V}^{\frac{5}{6}}\frac{v}{m_{pl}}) C^{\tilde{H}^0 \tilde{H}^0 Z},\nonumber\\
& & C^{\chi^0_3 \chi^0_2 Z}=  (\tilde{f}{\cal V}^{\frac{5}{6}}\frac{v}{m_{pl}}) ( C^{\lambda^0 \lambda^0 Z}+ C^{\tilde{H}^0 \tilde{H}^0 Z} ),\nonumber\\
& &  C^{\chi^0_3 \chi^0_3 Z}=  C^{\lambda^0 \lambda^0 Z}+  (\tilde{f}^2{\cal V}^{\frac{5}{3}}\frac{v^2}{m^{2}_{pl}})  C^{\tilde{H}^0 \tilde{H}^0 Z}
\end{eqnarray}

\underline{\bf{ Z Boson-Higgs-Z Boson vertex}}
The effective vertex has been calculated in \cite{dhuria+misra_EDM} by using gauge kinetic term $Re(T_B) F^2$ and the value of the same has been shown to be same as given by Standard model i.e $C^{ZZH} \sim \frac{M^{2}_{Z}}{v}$.

\underline{\bf{ Higgsino-Higgsino- Z Boson vertex}}
\begin{eqnarray}
  & &  {\cal L}= g_{{\cal Z}^I{\cal Z}^J}{\bar\chi}^{{\cal Z}^I}\slashed{Z}\ {\rm Im}\left(X^BK + i D^B\right)\chi^{{\cal Z}^I},
\end{eqnarray}
$\chi^{z_1}$ corresponds to Higgsino,
 $X=X^B\partial_B=-12i\pi\alpha^\prime\kappa_4^2\mu_7Q_B\partial_{T_B}$  corresponds to the killing isometry vector and the corresponding  $D$ term generated
 is given by:
\begin{equation}
\label{eq:D_term}
D^B=\frac{4\pi\alpha^\prime\kappa_4^2\mu_7Q_Bv^B}{\cal V}.
\end{equation}
Utilizing $g_{{\cal Z}_1{\bar {\cal Z}}_{\bar 1}}\sim  {\cal O}(1)g_{z_1{\bar z}_{\bar 1}}\sim{\cal V}^{-\frac{2}{3}}$ as given in (\ref{metric}) as well as $ v^B\sim{\cal V}^{\frac{1}{3}}$ and $Q_B\sim{\cal V}^{\frac{1}{3}}(2\pi\alpha^\prime)^2\tilde{f}$, yields value of physical Higgsino-Higgsino- Z Boson vertex as:
\begin{eqnarray}
 \label{eq:CHHZ}
& & C^{{\tilde H}^0 {\tilde H}^0 Z}\sim \frac{\left({\cal V}^{-\frac{23}{18}}\right)\tilde{f}}{\left(\sqrt{\hat{K}_{{\cal Z}_1{\bar {\cal Z}}_1}}\right)^2\sim O({10}^{-5})} \sim \left(O({10}^{5})\tilde{f}{\cal V}^{-\frac{23}{18}}\right)\sim \tilde{f}{\cal V}^{-\frac{5}{18}}.
 \end{eqnarray}

\underline{{\bf Gaugino-gaugino-Z Boson vertex}}
\begin{eqnarray}
  & &  {\cal L}= {\bar \lambda^L}\slashed{Z}\ {\rm Im}\left(X^BK + i D^B\right){\lambda^L},\nonumber\\
& & \sim {\bar \lambda^L}{\gamma}\cdot A\left\{6\kappa_4^2\mu_72\pi\alpha^\prime Q_BK+\frac{12\kappa_4^2\mu_72\pi\alpha^\prime Q_Bv^B}{\cal V}\right\}{\lambda^L}
  \end{eqnarray}
   ${\lambda^L}$ corresponds to gaugino.
Again, Utilizing values of  $ v^B\sim{\cal V}^{\frac{1}{3}}$ and $Q_B\sim{\cal V}^{\frac{1}{3}}(2\pi\alpha^\prime)^2\tilde{f}$, yields value of physical Gaugino-gaugino-Z Boson vertex as:
\begin{eqnarray}
 \label{eq:CggZ}
& & C^{{\lambda^0}{\lambda^0} Z}\sim \left({\cal V}^{-\frac{11}{18}}\right)\tilde{f} \sim \left(\tilde{f}{\cal V}^{-\frac{11}{18}}\right).
 \end{eqnarray}
 Using set of results given in equation no (\ref{eq:CHHZ}) and  (\ref{eq:CggZ}), the contribution of vertices appearing in equation (\ref{eq:NchichiZ}) are as follows:
 \begin{eqnarray}
\label{eq:NchichiZ}
& & C^{\chi^0_3 \chi^0_1 Z}=   \tilde{f^2}{\cal V}^{\frac{5}{9}}\frac{v}{m_{pl}},\nonumber\\
& & C^{\chi^0_3 \chi^0_2 Z}= \tilde{f^2}\frac{v}{m_{pl}}( {\cal V}^{\frac{5}{9}}+ {\cal V}^{\frac{2}{9}})\sim \tilde{f^2}{\cal V}^{\frac{5}{9}}\frac{v}{m_{pl}} ,\nonumber\\
& &  C^{\chi^0_3 \chi^0_3 Z}=  \tilde{f} {\cal V}^{-\frac{11}{18}}+ \tilde{f}^3 {\cal V}^{\frac{25}{18}}\frac{v^2}{m^{2}_{pl}}\sim \tilde{f} {\cal V}^{-\frac{11}{18}}.
\end{eqnarray}
Again, using (and quoting verbatim) the analytical expressions needed to calculate partial wave coefficients for $\chi^0_3\chi^0_3\rightarrow ZZ$ annihilation process, we will estimate the values of same in our set up.

$\bullet$
\underline{\bf{ Higgs--boson $(h,H)$ exchange:}}
   \begin{eqnarray}
    \widetilde{a}_{ZZ}^{(h,H)}& = & 0, 
\\
    \widetilde{b}_{ZZ}^{(h,H)}& = & \frac{3}{64\,\pi}
      \left |\sum_{r=h,H} \frac{C^{ZZr}\:
       C^{\chi^0_3 \chi^0_3 r}}{ s-m_{r}^{2}+i\, \Gamma_{r}\, m_{r}} \right |^{2}
               \frac{3\,{m_Z}^4-4\,{m_Z}^2\,{m_{\chi^0_3}}^2+4\,{m_{\chi^0_3}}^4}{{m_Z}^4}.
   \end{eqnarray}
   Expanding the summation
\begin{eqnarray}
\widetilde{b}_{ZZ}^{(h,H)}& = & \frac{3}{64\,\pi} \left |
      \frac{C^{ZZh}\: C^{\chi^0_3\chi^0_3 h}}
       {4\, m_{\chi^0_3}^{2}-m_{h}^{2}+i\, \Gamma_{h}\, m_{h}}+ \frac{C^{ZZH}\: C^{\chi^0_3\chi^0_3 H}}
       {4\, m_{\chi^0_3}^{2}-m_{H}^{2}+i\, \Gamma_{H}\, m_{H}}
       \right |^{2} \frac{3\,{m_Z}^4-4\,{m_Z}^2\,{m_{\chi^0_3}}^2+4\,{m_{\chi^0_3}}^4}{{m_Z}^4}. \nonumber\\
\end{eqnarray}
   Utilizing the value of mass $m_{h} =125 GeV, m_{H}\sim {\cal V}^{-\frac{85}{72}} M_p$, $m_{\chi^0_3}\sim {\cal V}^{-\frac{4}{3}}M_p, m_{Z}\sim 90 GeV $
and  $C^{ZZh}\sim C^{ZZH} \sim 10^2, C^{\chi^0_3 \chi^0_3 h}\sim   {\cal V}^{-\frac{35}{36}} $ from above, after simplifying, we have:
\begin{eqnarray}
\label{eq:bhZZ}
& & \widetilde{b}_{ZZ}^{(h,H)}\sim \frac{3}{64\,\pi} \left |\frac{10^2. {\cal V}^{-\frac{35}{36}}}{4.m_{\chi^0_3}^{2}}+ \frac{10^2. {\cal V}^{-\frac{35}{36}}}{m_{H}^{2}}\right |^{2} \frac{m_{\chi^0_3}^{4}}{m_{Z}^{4}}  \sim \frac{3}{64\,\pi}\left (\frac{10^2. {\cal V}^{-\frac{35}{36}}}{4.m_{\chi^0_3}^{2}}\right )^{2} \frac{m_{\chi^0_3}^{4}}{m_{Z}^{4}}  \nonumber\\
& & \sim  O(10)^{-16}{GeV}^{-2}~{\rm for}~ {\cal V}\sim {10}^5.
\end{eqnarray}
Here we assume that $\Gamma_{h,H} < m_{h,H}$  in our set up.

\vspace{0.25cm}
$\bullet$
\underline{\bf{neutralino\ ($\chi_{i}^{0}$) exchange:}}
   \begin{eqnarray}
   \label{eq:a_ZZchi0}
    \widetilde{a}_{ZZ}^{(\chi^0)} & = & \frac{1}{4\,\pi}\sum_{i,j=1}^{3}
    |C^{\chi^0_i \chi^0_3Z}|^2 |C^{\chi^0_j \chi^0_3Z}|^2
   \frac{({m_{\chi^0_3}}^2-{m_Z}^2)}{\Delta_{Zi}\Delta_{Zj}}, \nonumber\\
  &&   {\hskip -0.8in} \sim \frac{1}{4\,\pi}\biggr[|C^{\chi^0_1 \chi^0_3Z}|^2 |C^{\chi^0_1 \chi^0_3Z}|^2
   \frac{{m_{\chi^0_3}}^2}{\Delta_{Z1}\Delta_{Z1}} + |C^{\chi^0_1 \chi^0_3Z}|^2 |C^{\chi^0_2 \chi^0_3Z}|^2
   \frac{{m_{\chi^0_3}}^2}{\Delta_{Z1}\Delta_{Z2}}+  |C^{\chi^0_1 \chi^0_3Z}|^2 |C^{\chi^0_3 \chi^0_3Z}|^2
   \frac{{m_{\chi^0_3}}^2}{\Delta_{Z1}\Delta_{Z3}}+ \nonumber\\
   & &|C^{\chi^0_2 \chi^0_3Z}|^2 |C^{\chi^0_3 \chi^0_3Z}|^2
   \frac{{m_{\chi^0_3}}^2}{\Delta_{Z2}\Delta_{Z3}}+ |C^{\chi^0_3 \chi^0_3Z}|^2 |C^{\chi^0_3 \chi^0_3Z}|^2
   \frac{{m_{\chi^0_3}}^2}{\Delta_{Z3}\Delta_{Z3}}\biggr]
\end{eqnarray}
where $\Delta_{Zi}\equiv\,{m_Z}^{2}-{m^{2}_{\chi^0_3}}-{m^{2}_{\chi^0_i}}$.
For the given neutralino mass eigenstates,
$\Delta_{Z1}\equiv\,{m_Z}^{2}-{m^{2}_{\chi^0_3}}-{m^{2}_{\chi^0_1}}\sim -{\cal V}^{2}m^2_{\frac{3}{2}}$, $\Delta_{Z2}\equiv\,{m_Z}^{2}-{m^{2}_{\chi^0_3}}-{m^{2}_{\chi^0_2}}\sim - {\cal V}^{2}m^2_{\frac{3}{2}}$, $\Delta_{Z3}\equiv\,{m_Z}^{2}-{m^{2}_{\chi^0_3}}-{m^{2}_{\chi^0_3}}\sim -2{\cal V}^{\frac{4}{3}}m^2_{\frac{3}{2}}$.
Using above values and couplings given in (\ref{eq:NchichiZ}), (\ref{eq:a_ZZchi0}) reduces to
\begin{eqnarray}
\label{eq:achiZZ}
  &&  \widetilde{a}_{ZZ}^{(\chi^0)} \sim \frac{1}{4\,\pi}{m^2_{\chi^0_3}}.\frac{{\tilde f}^4{\cal V}^{-\frac{22}{9}}}{{\cal V}^{\frac{8}{3}}m^4_{\frac{3}{2}}} \sim  O(10)^{-54} {GeV}^{-2}.
\end{eqnarray}
As from \cite{Takeshi_Leszek}, the analytical expression of $\widetilde{b}_{ZZ}^{(\chi^0)}$ is defined as following:
\begin{eqnarray}
\label{eq:bZZ}
    \widetilde{b}_{ZZ}^{(\chi^0)} & = & \frac{1}{16\,\pi}
       \sum_{i,j=1}^{3} \frac{1}{{m_Z}^4\Delta_{Zi}^3 \Delta_{Zj}^3} 
\\
      & & \times  |C^{\chi^0_i \chi^0_3Z}|^2 |C^{\chi^0_j \chi^0_3Z}|^2\Big[D_{ij}^{(1)}\,\Delta_{Zi}^2
       +D_{ij}^{(2)}\,\Delta_{Zi} \Delta_{Zj}
       +D_{ij}^{(3)}\,\Delta_{Zi}^2 \Delta_{Zj}
       +D_{ij}^{(4)}\,\Delta_{Zi}^2 \Delta_{Zj}^2\Big], \nonumber
   \end{eqnarray}
where
   \begin{eqnarray}
     D_{ij}^{(1)} \! &=& \!
16\,{m_{\chi^0_3}}^2\,{m_Z}^4\,({m_Z}^2-{m_{\chi^0_3}}^2)^2,\nonumber\\
     D_{ij}^{(2)} \! &=& \!
4\,{m_{\chi^0_3}}^2\,({m_Z}^2-{m_{\chi^0_3}}^2)^2 \{3\,{m_Z}^4
                              +4\,{m_{\chi^0_3}}^2\,{m_{\chi^0_i}}\,{m_{\chi^0_j}}  \nonumber\\
& & \hspace{0.2in}
 +{m_Z}^2\,[4\,{m_{\chi^0_3}}^2+4\,{m_{\chi^0_i}}\,{m_{\chi^0_j}}-6\,{m_{\chi^0_3}}\,({m_{\chi^0_i}}+{m_{\chi^0_j}})]\}, \nonumber \\
     D_{ij}^{(3)} \! &=& \!
   -4\,{m_{\chi^0_3}}\,({m_Z}^2-{m_{\chi^0_3}}^2)
   \{4\,{m_{\chi^0_3}}^3\,{m_{\chi^0_i}}\,{m_{\chi^0_j}}+{m_Z}^4\,(7\,{m_{\chi^0_3}}+3\,{m_{\chi^0_i}}
                                                 +6\,{m_{\chi^0_j}}) \nonumber\\
& & \hspace{0.2in}
+2\,{m_Z}^2\,{m_{\chi^0_3}}\,[4\,{m_{\chi^0_3}}^2-{m_{\chi^0_i}}\,{m_{\chi^0_j}}-{m_{\chi^0_3}}\,({m_{\chi^0_i}}+5\,{m_{\chi^0_j}})]\}, \nonumber\\
D_{ij}^{(4)} \! &=& \!
    8\,{m_Z}^6+2\,{m_Z}^2\,{m_{\chi^0_3}}^2
    [4\,{m_{\chi^0_3}}^2-6\,{m_{\chi^0_i}}\,{m_{\chi^0_j}}-5\,{m_{\chi^0_3}}\,({m_{\chi^0_i}}+{m_{\chi^0_j}})] \nonumber\\
& & +4\, {m_{\chi^0_3}}^4\,[2\,{m_{\chi^0_3}}^2+3\,{m_{\chi^0_i}}\,{m_{\chi^0_j}}+
                                   2\,{m_{\chi^0_3}}\,({m_{\chi^0_i}}+{m_{\chi^0_j}})] \nonumber\\
& & +{m_Z}^4\,[3\,{m_{\chi^0_3}}^2+9\,{m_{\chi^0_i}}\,{m_{\chi^0_j}}
                                 +5\,{m_{\chi^0_3}}\,({m_{\chi^0_i}}+{m_{\chi^0_j}})].
   \end{eqnarray}
Solving above,
\begin{eqnarray}
\label{eq:Dij}
&&D_{11}^{(1)}\sim D_{12}^{(1)}\sim D_{13}^{(1)}\sim D_{23}^{(1)}\sim D_{33}^{(1)}\equiv {m_Z}^4 {\cal V}^4 m^4_{\frac{3}{2}} \nonumber\\
&& D_{11}^{(2)}\sim D_{12}^{(2)}\equiv{\cal V}^{\frac{20}{3}} m^{10}_{\frac{3}{2}},D_{13}^{(2)}\sim D_{23}^{(2)}\equiv {\cal V}^{\frac{21}{3}} m^{10}_{\frac{3}{2}}, D_{33}^{(2)}\sim {\cal V}^{\frac{20}{3}} m^{10}_{\frac{3}{2}}\nonumber\\
&&D_{11}^{(3)}\sim D_{12}^{(3)}\equiv{\cal V}^{6} m^{8}_{\frac{3}{2}},D_{13}^{(3)}\equiv D_{23}^{(3)}\sim {\cal V}^{\frac{17}{3}} m^{8}_{\frac{3}{2}},D_{33}^{(2)}\sim {\cal V}^{\frac{16}{3}} m^{8}_{\frac{3}{2}} \nonumber\\
&& D_{11}^{(4)}\sim D_{12}^{(4)}\equiv{\cal V}^{\frac{14}{3}} m^{6}_{\frac{3}{2}},D_{13}^{(4)}\equiv D_{23}^{(3)}\sim {\cal V}^{\frac{13}{3}} m^{6}_{\frac{3}{2}},D_{33}^{(4)}\sim {\cal V}^{4} m^{6}_{\frac{3}{2}}.
\end{eqnarray}
Now, incorporating results of  equations (\ref{eq:Dij}), $\Delta_{Z\ 1,2,3}$ and (\ref{eq:NchichiZ}) in (\ref{eq:bZZ}), the same reduces to:
\begin{eqnarray}
\label{eq:bchiZZ}
    \widetilde{b}_{ZZ}^{(\chi^0)}\sim  O(10)^{-14} {GeV}^{-2}.
    \end{eqnarray}

$\bullet$
\underline{\bf{Higgs $(h,H)$--neutralino\ ($\chi_{i}^{0}$) interference term:}}
   \begin{eqnarray}
  & &   \widetilde{a}_{ZZ}^{(h,H-\chi^{0})} =  0, \nonumber
\\
  & &  \widetilde{b}_{ZZ}^{(h,H-\chi^{0})} =  \frac{1}{16\,\pi}
      \sum_{i=1}^{3} Re \left[\left(\sum_{r=h,H} \frac{C^{ZZr}\:
       C_{S}^{\chi^0_3 \chi^0_3 r}}
         { 4{m_{\chi^0_3}}^2-m_{r}^{2}+i\, \Gamma_{r}\, m_{r}} \right )^{*}\right]
       \frac{1}{{m_Z}^4\,\Delta_{Zi}^2}
     C_{S}^{\chi^0_3 \chi^0_3 r} \nonumber \\
 & & \times \bigg\{
      2\,{m_{\chi^0_3}}\,({m_{\chi^0_3}}^2-{m_Z}^2)\,[-3\,{m_Z}^4
       -4\,{m_{\chi^0_3}}^3\,{m_{\chi^0_i}}+2\,{m_Z}^2\,{m_{\chi^0_3}}\,({m_{\chi^0_3}}+{m_{\chi^0_i}})]
            . \nonumber \\
 & & \hskip-0.3in  + \Delta_{Zi}
      \left[ -4\,{m_{\chi^0_3}}^4\,(2\,{m_{\chi^0_3}}+3\,{m_{\chi^0_3}})
         +2\,{m_Z}^2\,{m_{\chi^0_3}}^2\,(5\,{m_{\chi^0_3}}+6\,{m_{\chi^0_i}})    -{m_Z}^4\,(5\,{m_{\chi^0_3}}+9\,{m_{\chi^0_i}}) \right] \bigg\}.
    \end{eqnarray}
Utilizing results for $\Delta_{Z\ 1,2,3}$ and (\ref{eq:NchichiZ}), the same reduces to
\begin{eqnarray}
 & & \widetilde{b}_{ZZ}^{(h,H-\chi^{0})} =  \frac{1}{16\,\pi}\frac{{m_{\chi^0_3}}^3}{m^4_Z}.
      \sum_{i=1}^{3} Re \left[\left(\sum_{r=h,H} \frac{C^{ZZr}\:
       C^{\chi^0_3 \chi^0_3 r}}
         { 4{m_{\chi^0_3}}^2-m_{r}^{2}+i\, \Gamma_{r}\, m_{r}} \right )^{*}\right]C^{\chi^0_3 \chi^0_3 r}{M_p} \nonumber\\
         &&\hskip-0.3in \sim \frac{1}{16\,\pi}\frac{{\cal V}^2 m^3_{\frac{3}{2}}}{m^4_Z} \left (
      \frac{C^{ZZh}\: (C^{\chi^0_3\chi^0_3 h})^2}
       {4\, m_{\chi^0_3}^{2}-m_{h}^{2}+i\, \Gamma_{h}\, m_{h}}+ \frac{C^{ZZH}\: (C^{\chi^0_3\chi^0_3 H})^2}
       {4\, m_{\chi^0_3}^{2}-m_{H}^{2}+i\, \Gamma_{H}\, m_{H}}.
       \right ) {M_p}  \end{eqnarray}
       Utilizing the value of mass $m_{h} =125 GeV, m_{H}\sim {\cal V}^{-\frac{85}{72}} M_p$, $m_{\chi^0_3}\sim {\cal V}^{-\frac{4}{3}}M_p, m_{Z}\sim 90 GeV $
and  $C^{ZZh}\sim C^{ZZH} \sim 10^2,  C^{\chi^0_3 \chi^0_3 Z}\sim   {\cal V}^{-\frac{35}{36}} $ from above, after simplifying, we have:
\begin{eqnarray}
\label{eq:bhchiZZ}
& & \widetilde{b}_{ZZ}^{(h,H-\chi^{0})}\sim  \frac{1}{16\,\pi}\frac{{\cal V}^2 m^3_{\frac{3}{2}}}{m^4_Z}\left (\frac{10^2.{\cal V}^{-\frac{35}{18}}}{4.m_{\chi^0_3}^{2}}+ \frac{10^2. {\cal V}^{-\frac{35}{18}}}{m_{H}^{2}}\right)  \sim \frac{1}{16\,\pi}\frac{ m^3_{\frac{3}{2}}}{m^4_Z}\left (\frac{10^2}{4.m_{\chi^0_3}^{2}}\right )\nonumber\\
& & \sim  O(10)^{-10}{GeV}^{-2}~{\rm for}~ {\cal V}\sim {10}^5.
\end{eqnarray}
Here we assume that $\Gamma_{h,H} < m_{h,H}$  in our set up.
Utilizing results of equations (\ref{eq:bhZZ}), (\ref{eq:achiZZ}), (\ref{eq:bchiZZ}), (\ref{eq:bhchiZZ}),
\begin{eqnarray}
\label{eq:sumbZZ}
 \widetilde{b}_{ZZ} &=&
 \widetilde{b}_{ZZ}^{(h,H)}
+ \widetilde{b}_{ZZ}^{(\chi^0)}
+ \widetilde{b}_{ZZ}^{(h,H-\chi^{0})} \sim O(10)^{-10}{GeV}^{-2};
\end{eqnarray}
\begin{eqnarray}
\label{eq:sumaZZ}
& & \widetilde{a}_{ZZ} =
 \widetilde{a}_{ZZ}^{(h,H)}
+\widetilde{a}_{ZZ}^{(\chi^{0})}
+\widetilde{a}_{ZZ}^{(h,H-\chi^{0})}\sim  O(10)^{-54} {GeV}^{-2}.\nonumber\\
\end{eqnarray}

\begin{figure}
\begin{center}
\begin{picture}(150,100)(-100,100)
\ArrowLine(-100,250)(-30,200)
\Text(-115,250)[]{$\chi^0_3(p_1)$}
\ArrowLine(-100,150)(-30,200)
\Text(-115,150)[]{${\chi}^{0}_3(p_2)$}
\DashLine(-30,200)(20,200)5
\Text(0,210)[]{$(h,H,Z)$}
\ArrowLine(20,200)(90,250)
\Text(110,250)[]{$f(p_3)$}
\ArrowLine(20,200)(90,150)
\Text(110,150)[]{${\bar f}(p_4)$}
\end{picture}
{\hskip 2.0in}
\begin{picture}(150,100)(-100,100)
\ArrowLine(-100,250)(-30,250)
\Text(-115,250)[]{$\chi^0_3(p_1)$}
\ArrowLine(-30,250)(40,250)
\Text(55,250)[]{$f(p_3)$}
\DashLine(-30,250)(-30,160)5
\Text(-20,200)[]{$\tilde{f}$}
\ArrowLine(-100,160)(-30,160)
\Text(-115,160)[]{${\chi}^{0}_3(p_2)$}
\ArrowLine(-30,160)(40,160)
\Text(55,160)[]{${\bar f}(p_4)$}
\end{picture}
\caption{Feynman diagrams for
${\chi}^{0}_3 {\chi}^{0}_3\ra f {\bar f} $ via $s$--channel Higgs/Z exchange and t-channel $\tilde{f}$ exchange.}
\end{center}
\end{figure}
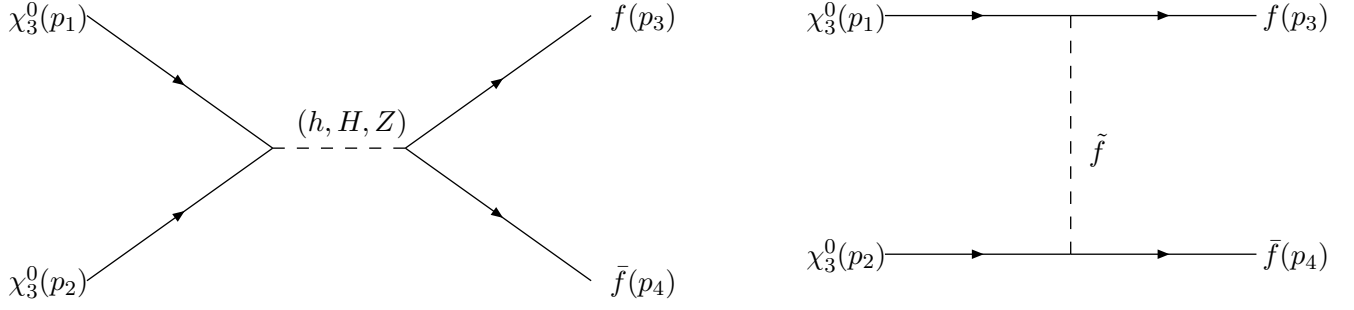

\underline{{\bf Higgs-Fermion-Fermion Interaction}}

Fig. 25  involves the $s$-channel Higgs boson ($h$, $H$)
and $Z$ boson exchange and the $t$- and $u$-channel sfermion
($\widetilde{f}_{a}$) exchange. To get the estimate of thermal cross-section for this process, one needs to evaluate the following partial wave coefficients:
\begin{eqnarray}
 \widetilde{a}_{\bar{f}f} &=&
 \widetilde{a}_{\bar{f}f}^{(h,H)}
+\widetilde{a}_{\bar{f}f}^{(Z)}
+\widetilde{a}_{\bar{f}f}^{(\widetilde{f})}
+\widetilde{a}_{\bar{f}f}^{(h,H-\widetilde{f})}
+\widetilde{a}_{\bar{f}f}^{(Z-\widetilde{f})}, \nonumber \\
 \widetilde{b}_{\bar{f}f} &=&
 \widetilde{b}_{\bar{f}f}^{(h,H)}
+\widetilde{b}_{\bar{f}f}^{(Z)}
+\widetilde{b}_{\bar{f}f}^{(\widetilde{f})}
+\widetilde{b}_{\bar{f}f}^{(h,H-\widetilde{f})}
+\widetilde{b}_{\bar{f}f}^{(Z-\widetilde{f})}.
\end{eqnarray}

We have  calculated below the interaction vertices corresponding to $s$ and $t$ channel of ${\chi}^{0}_3 {\chi}^{0}_3\ra f {\bar f} $.
\begin{equation}
\label{eq:HFF}
{\cal L}\ni ig_{{\cal A}_I{\bar {{\cal A}_I}}}{\bar\chi}^{\bar {{\cal A}_I}}_L \left[\slashed{\partial}\chi^{{\cal A}_I}_L+\Gamma^{{\cal A}_I}_{{\cal Z}_1{{\cal A}_I}}\slashed{\partial} {{\cal Z}_i}\chi^{{\cal A}_I}_L
+\frac{1}{4}\left(\partial_{{\cal Z}_i}K\slashed{\partial} {\cal Z}_i - {\rm c.c.}\right)\chi^{{\cal A}_I}_L\right]
\end{equation}
$\chi^{{\cal A}_I}_L,I=1,2 $ corresponding to first  generation of Left-handed leptons and Left-handed quarks.

\underline{Higgs-lepton-lepton interaction}

Utilizing $z_1= \delta z_1 +{\cal V}^{\frac{1}{36}}{M_p}$, for lepton mass $m_{\chi^{{\cal A}_1}} \sim {\cal O}(1) MeV $ and $m_{3/2}={\cal V}^{-2}{M_p}$, $g_{a_1{\bar {a_1}}} \ra \frac{{\cal V}^{-\frac{2}{9}}\langle z_i\rangle\delta z_i}{M_p^2}, \Gamma^{a_1}_{z_i{a_1}}\ra \frac{{\cal V}^{-\frac{2}{3}}{\langle z_i\rangle}}{M_p}, \partial_{z_i}K\ra\frac{{\cal V}^{-\frac{2}{3}}{\langle z_i\rangle}}{M_p} $, one gets:
\begin{eqnarray}
\label{eq:kinetic1}
& & {\hskip -0.2in} g_{{\cal A}_1{\bar {{\cal A}_1}}}{\bar\chi}^{\bar {{\cal A}_1}}_L \slashed{\partial}\chi^{{\cal A}_1}_L \sim {\cal O}(1) g_{a_1{\bar{a_1}}}{\bar\chi}^{\bar {{\cal A}_1}}_L \slashed{\partial}\chi^{{\cal A}_1}_L \ra \frac{{\cal V}^{-\frac{2}{9}}\langle{\cal Z}_I\rangle}{M_p}\delta{\cal Z}_I{\bar\chi}^{\bar {{\cal A}_1}}_L\slashed{p}_{\chi^{{\cal A}_1}}\chi^{{\cal A}_1}_L\sim {\cal V}^{-\frac{7}{36}}{\bar\chi}^{\bar {{\cal A}_1}}_L\delta{{\cal Z}_i}\chi^{{\cal A}_1}_L; \nonumber\\
& & {\hskip -0.2in} g_{{\cal A}_1{\bar {{\cal A}_1}}}{\bar\chi}^{\bar {{\cal A}_1}}_L\Gamma^{{\cal A}_1}_{{\cal Z}_i{{\cal A}_1}}\slashed{\partial} {{\cal Z}_i}\chi^{{\cal A}_1}_L \sim {\cal O}(1) g_{a_1{\bar {a_1}}}\Gamma^{a_i}_{z_i{a_i}}{\bar\chi}^{\bar {{\cal A}_1}}_L\slashed{\partial} {{\cal Z}_i}\chi^{{\cal A}_1}_L \ra \frac{{\cal V}^{-\frac{2}{9}}\langle{\cal Z}_I\rangle }{M_p}\delta{\cal Z}_I{\bar\chi}^{\bar {{\cal A}_1}}_L (\slashed{p}_{\chi^{{\cal A}_1}} + \slashed{p}_{\bar \chi^{{\cal A}_1}})\chi^{{\cal A}_1}_L\nonumber\\
& & \sim {\cal V}^{-\frac{7}{36}}{\bar\chi}^{\bar {{\cal A}_1}}_L\delta{{\cal Z}_i}\chi^{{\cal A}_1}_L; \nonumber\\
& & {\hskip -0.8in} g_{{\cal A}_1{\bar {{\cal A}_1}}}{\bar\chi}^{\bar {{\cal A}_1}}_L\frac{1}{4}\left(\partial_{{\cal Z}_i}K{\gamma}\cdot {\cal Z}_i - {\rm c.c.}\right)\chi^{{\cal A}_1}_L  \sim  {\cal O}(1)g_{a_1{\bar {a_1}}}\partial_{z_i}K {\bar\chi}^{\bar {{\cal A}_1}}_L\slashed{\partial} {{\cal Z}_i}\chi^{{\cal A}_i}_L \ra \frac{{\cal V}^{-\frac{2}{9}}\langle{\cal Z}_i\rangle }{M_p}\delta{\cal Z}_i{\bar\chi}^{\bar {{\cal A}_1}}_L (\slashed{p}_{\chi^{{\cal A}_1}} + \slashed{p}_{\bar \chi^{{\cal A}_1}})\chi^{{\cal A}_1}_L\nonumber\\
& & \sim {\cal V}^{-\frac{7}{36}}{\bar\chi}^{\bar {{\cal A}_1}}_L\delta{{\cal Z}_I}\chi^{{\cal A}_1}_L.
\end{eqnarray}
Incorporating results of   (\ref{eq:kinetic1}) in (\ref{eq:HFF}), the physical  Higgs-lepton-lepton  vertex will be given as
\begin{eqnarray}
\label{eq:llH}
& & C^{{{\cal A}_1}_L \bar {{\cal A}_1}_L h}= \frac{1}{{\sqrt{(\hat{K}_{{\cal Z}_i{\bar{\cal Z}}_i})^2.(\hat{K}_{{\cal A}_1{\bar{\cal A}}_1})^2}}}\left[{\cal V}^{-\frac{7}{36}}\right] \sim 10^{1} \left[ {\cal V}^{-\frac{7}{36}}\right]\sim {\cal O}(1).
\end{eqnarray}

\underline{\bf Higgs-quark-quark interaction}

Utilizing $z_1 = \delta z_1 +{\cal V}^{\frac{1}{36}}{M_p}$, for quark mass $m_{\chi^{{\cal A}_2}} \sim
{\cal O}(5) MeV $ and $m_{3/2}={\cal V}^{-2}{M_p}$, $g_{a_2{\bar {a_2}}} \ra \frac{{\cal V}^{-\frac{11}{9}}{z_i}\langle z_i\rangle}{M_p^2}, \Gamma^{a_2}_{z_i{a_2}}\ra \frac{{\cal V}^{-\frac{2}{3}}{\langle z_i\rangle}}{M_p}, \partial_{z_i}K\ra\frac{{\cal V}^{-\frac{2}{3}}{\langle z_i\rangle}}{M_p} $, one gets:
\begin{eqnarray}
\label{eq:kinetic2}
& & {\hskip -0.2in} g_{{\cal A}_2{\bar {{\cal A}_2}}}{\bar\chi}^{\bar {{\cal A}_2}}_L \slashed{\partial}\chi^{{\cal A}_2}_L \sim {\cal O}(1) g_{a_2{\bar{a_2}}}{\bar\chi}^{\bar {{\cal A}_2}}_L \slashed{\partial}\chi^{{\cal A}_2}_L \ra \frac{{\cal V}^{-\frac{11}{9}}\langle{\cal Z}_i\rangle }{M_p}\delta{\cal Z}_i{\bar\chi}^{\bar {{\cal A}_2}}_L\slashed{p}_{\chi^{{\cal A}_2}}\chi^{{\cal A}_2}_L\sim {\cal V}^{-\frac{43}{36}}{\bar\chi}^{\bar {{\cal A}_2}}_L\delta{{\cal Z}_i}\chi^{{\cal A}_2}_L; \nonumber\\
& & {\hskip -0.2in} g_{{\cal A}_2{\bar {{\cal A}_2}}}{\bar\chi}^{\bar {{\cal A}_2}}_L\Gamma^{{\cal A}_2}_{{\cal Z}_i{{\cal A}_2}}\slashed{\partial} {{\cal Z}_i}\chi^{{\cal A}_2}_L \sim {\cal O}(1) g_{a_2{\bar {a_2}}}\Gamma^{a_2}_{z_i{a_2}}{\bar\chi}^{\bar {{\cal A}_2}}_L\slashed{\partial} {{\cal Z}_i}\chi^{{\cal A}_2}_L \ra \frac{{\cal V}^{-\frac{11}{9}}\langle{\cal Z}_I\rangle }{M_p}\delta{\cal Z}_I{\bar\chi}^{\bar {{\cal A}_2}}_L (\slashed{p}_{\chi^{{\cal A}_2}} + \slashed{p}_{\bar \chi^{{\cal A}_2}})\chi^{{\cal A}_2}_L\nonumber\\
& & \sim {\cal V}^{-\frac{43}{36}}{\bar\chi}^{\bar {{\cal A}_2}}_L\delta{{\cal Z}_i}\chi^{{\cal A}_2}_L; \nonumber\\
& & {\hskip -0.8in} g_{{\cal A}_2{\bar {{\cal A}_2}}}{\bar\chi}^{\bar {{\cal A}_2}}_L.\frac{1}{4}\left(\partial_{{\cal Z}_i}K{\gamma}\cdot {\cal Z}_i - {\rm c.c.}\right)\chi^{{\cal A}_2}_L  \sim  {\cal O}(1)g_{a_2{\bar {a_2}}}\partial_{z_i}K {\bar\chi}^{\bar {{\cal A}_2}}_L\slashed{\partial} {{\cal Z}_i}\chi^{{\cal A}_2}_L \ra \frac{{\cal V}^{-\frac{11}{9}}\langle{\cal Z}_i\rangle }{M_p}\delta{\cal Z}_i{\bar\chi}^{\bar {{\cal A}_2}}_L (\slashed{p}_{\chi^{{\cal A}_2}} + \slashed{p}_{\bar \chi^{{\cal A}_2}})\chi^{{\cal A}_2}_L\nonumber\\
& & \sim  {\cal V}^{-\frac{43}{36}} {\bar\chi}^{\bar {{\cal A}_2}}_L\delta{{\cal Z}_i}\chi^{{\cal A}_2}_L.
\end{eqnarray}
Incorporating results of  (\ref{eq:kinetic2}) in (\ref{eq:HFF}), the physical  Higgs-quark-quark  vertex will be given as
\begin{eqnarray}
\label{eq:qqH}
& & C^{{{\cal A}_2}_L \bar {{\cal A}_2}_L h}= \frac{1}{{\sqrt{(\hat{K}_{{\cal Z}_i{\bar{\cal Z}}_i})^2.(\hat{K}_{{\cal A}_2{\bar{\cal A}}_2})^2}}}\left[{\cal V}^{-\frac{43}{36}}\right] \nonumber\\
& & \sim 10^{-7} \left[ {\cal V}^{-\frac{43}{36}} \right]\sim  {\cal V}^{-\frac{13}{5}} .
\end{eqnarray}
We would estimate:
\begin{equation}
C^{f \bar f  h}= {\rm Max}\left[C^{{{\cal A}_1}_L \bar {{\cal A}_1}_L h}, C^{{{\cal A}_2}_L \bar {{\cal A}_2}_L h}\right] \sim {\cal O}(1).
\end{equation}

\underline{{\bf fermion-fermion- Z Boson vertex}}
\begin{eqnarray}
  & &  {\cal L}\ni g_{{\cal A}_1{\cal A}_1}{\bar\chi}^{{\cal A}_1}{\gamma}\cdot A\ {\rm Im}\left(X^BK + i D^B\right)\chi^{{\cal A}_1}+ g_{{\cal A}^2{\cal A}_2}{\bar\chi}^{{\cal A}_2}{\gamma}\cdot A\ {\rm Im}\left(X^BK + i D^B\right)\chi^{{\cal A}_2},
\end{eqnarray}
$\chi^{{\cal A}_{1,2}}$  correspond to first  generation of leptons and quarks,
 $X=X^B\partial_B=-12i\pi\alpha^\prime\kappa_4^2\mu_7Q_B\partial_{T_B}$  corresponds to the killing isometry vector and  $D$ term generated
 is given by:
\begin{equation}
\label{eq:D_term}
D^B=\frac{4\pi\alpha^\prime\kappa_4^2\mu_7Q_Bv^B}{\cal V}.
\end{equation}
Utilizing $g_{{\cal A}_1{\bar {\cal A}}_{\bar 1}}\sim  {\cal O}(1)g_{a_1{\bar a}_{1}}\sim{\cal V}^{\frac{4}{9}}$, $g_{{\cal A}_2{\bar {\cal A}}_{ 2}}\sim  {\cal O}(1)g_{a_2{\bar a}_{2}}\sim{\cal V}^{-\frac{5}{9}}$  as given in \ref{metric} as well as $ v^B\sim{\cal V}^{\frac{1}{3}}$ and $Q_B\sim{\cal V}^{\frac{1}{3}}(2\pi\alpha^\prime)^2\tilde{f}$, the contribution of physical fermion-fermion- Z Boson vertex  $C^{f \bar f Z}\sim {\cal O}(1)$( see detailed explanation as given in \cite{dhuria+misra_EDM}).

Quoting the analytical expressions as given in \cite{Takeshi_Leszek}:

$\bullet$
\underline{\bf {Higgs-boson $(h,H)$ exchange:}}
   \begin{eqnarray}
   \widetilde{a}_{\bar{f}f}^{(h,H)}& = & 0, 
\\
   \widetilde{b}_{\bar{f}f}^{(h,H)}& = & \frac{3}{4\, \pi}
     \left | \sum_{r=h,H} \frac{C^{ffr}\:
      C^{\chi^0_3 \chi^0_3 r}}{4\, {m_{\chi^0_3}}^{2}-m_{r}^{2}
        +i\, \Gamma_{r}\, m_{r}} \right |^{2} ({m_{\chi^0_3}}^{2}-m_{f}^{2}).
    \end{eqnarray}
Expanding the summation
\begin{eqnarray}
\widetilde{b}_{ff}^{(h,H)}& = & \frac{3}{64\,\pi} \left |
      \frac{C^{ffh}\: C^{\chi^0_3\chi^0_3 h}}
       {4\, m_{\chi^0_3}^{2}-m_{h}^{2}+i\, \Gamma_{h}\, m_{h}}+ \frac{C^{hhH}\: C^{\chi^0_3\chi^0_3 H}}
       {4\, m_{\chi^0_3}^{2}-m_{H}^{2}+i\, \Gamma_{H}\, m_{H}}
                 \right |^{2}({m_{\chi^0_3}}^{2}-m_{f}^{2}).
\end{eqnarray}
Again, utilizing the value of masses as given in above cases and  $C^{ffh}\sim{\cal O}(1), C^{\chi^0_3 \chi^0_3 h}\sim C^{\chi^0_3 \chi^0_3 H}\sim  {\cal V}^{-\frac{35}{36}} $ from above, after simplifying, we have:
\begin{eqnarray}
\label{eq:bhff}
& & \widetilde{b}_{ff}^{(h,H)}\sim \frac{3}{64\,\pi} \left |\frac{{\cal V}^{-\frac{35}{36}}}{4.{\cal V}^{-\frac{8}{3}}m^2_{pl}}+ \frac{{\cal V}^{-\frac{35}{36}}}{{\cal V}^{-\frac{85}{36}}m^2_{pl}}\right |^{2} ({m_{\chi^0_3}}^{2}-m_{f}^{2})\sim \frac{3}{64\,\pi}\left (\frac{{\cal V}^{-\frac{35}{36}}}{{\cal V}^{-\frac{8}{3}}m^{2}_{pl}}\right )^{2} {\cal V}^{-\frac{8}{3}}M_p^2 \nonumber\\
& & \sim  O(10)^{-46}GeV^{-2}~{\rm for}~ {\cal V}\sim {10}^5.
\end{eqnarray}

$\bullet$
\underline{\bf {$Z$--boson exchange:}}
  \begin{eqnarray}
      \widetilde{a}_{\bar{f}f}^{(Z)} &=& \frac{1}{2\, \pi}
       \left |  \frac{C^{ffZ}\: C_{A}^{\chi\chi Z}}
         {4 {m_{\chi^0_3}}^{2}-{m_Z}^{2}+i\, \Gamma_{Z}\, {m_Z}} \right |^{2}
       \frac{m_{f}^{2}\,({m_Z}^{2}-4\, {m_{\chi^0_3}}^{2})^{2}}{{m_Z}^{4}}.
       \end{eqnarray}
Utilizing the value of mass as given above,
and  $C^{ffZ}\sim {\cal O}(1), C^{\chi^0_3 \chi^0_3 Z}\sim  {\cal V}^{-\frac{35}{36}} $ from above, after simplifying, we have
\begin{eqnarray}
\label{aZff}
& & \widetilde{a}_{{\bar f}f}^{(h,H)}\sim \frac{3}{64\,\pi} \left |\frac{ {\cal V}^{-\frac{35}{36}}}{4.{\cal V}^{-\frac{8}{3}}m^2_{pl}}+ \frac{ {\cal V}^{-\frac{35}{36}}}{{\cal V}^{-\frac{85}{36}}m^2_{pl}}\right |^{2} \frac{m_{f}^{2}\,({m_Z}^{2}-4\, {m_{\chi^0_3}}^{2})^{2}}{{m_Z}^{4}}\sim \frac{3}{64\,\pi}\left (\frac{{\cal V}^{-\frac{35}{36}}}{{\cal V}^{-\frac{8}{3}}m^2_{pl}}\right )^{2} \frac{m_{f}^{2}\,(4\, {m_{\chi^0_3}})^{4}}{{m_Z}^{4}} \nonumber\\
& & \sim  O(10)^{-25}{GeV}^{-2}~{\rm for}~ {\cal V}\sim {10}^5;
\end{eqnarray}

\begin{eqnarray}
\label{eq:bZff}
     \widetilde{b}_{\bar{f}f}^{(Z)} &=& \frac{1}{2\, \pi}
      \left |  \frac{C^{ffZ}\: C^{\chi^0_3 \chi^0_3 Z}}
       {4\, {m_{\chi^0_3}}^{2}-{m_Z}^{2}+i\, \Gamma_{Z}\, {m_Z}} \right |^{2}
        \frac{1}{{m_Z}^{2}\,((4\, {m_{\chi^0_3}}^{2}-{m_Z}^{2})^{2}
          +(\Gamma_{Z}\, {m_Z})^{2})} \nonumber \\
      & &  \times \bigg[2 \,|C^{ffZ}|^{2}\, \Big\{{m_Z}^2\,
             ({m_{\chi^0_3}}^{2}-m_{f}^{2})\,({m_Z}^{2}-4 {m_{\chi^0_3}}^{2})^{2}  \nonumber \\
      & & \hspace{1.1in} +\Gamma_{Z}^{2} [{m_{\chi^0_3}}^{2}\, {m_Z}^{4} + m_{f}^{2}\,
           (24\, {m_{\chi^0_3}}^{4}-6\, {m_Z}^{2}\, {m_{\chi^0_3}}^{2}-{m_Z}^{4})] \Big\}
             \nonumber \\
         & & \hspace{0.2in} +{m_Z}^{2}\, |C^{ffZ}|^{2}\,
        \Big\{(2\, {m_{\chi^0_3}}^{2}+m_{f}^{2})\,[(4\, {m_{\chi^0_3}}^{2}-{m_Z}^{2})^{2}
             + {m_Z}^{2}\,\Gamma_{Z}^{2}]\Big\} \bigg] \nonumber \\
        & &  {\hskip -0.5in} \sim   \frac{1}{2\, \pi}
      \left |  \frac{C^{ffZ}\: C^{\chi^0_3 \chi^0_3 Z}}
       {4\, {m_{\chi^0_3}}^{2}-{m_Z}^{2}+i\, \Gamma_{Z}\, {m_Z}} \right |^{2}
        \frac{1}{{m_Z}^{2}\,(16\, {m_{\chi^0_3}}^{4})
       }  \times \bigg[2.16 \,  {m_Z}^2\,{m_{\chi^0_3}}^{6}
              + {m_Z}^{2}\,
        (16\, {m_{\chi^0_3}}^{6})\, \bigg]\nonumber\\
        & & {\hskip -0.5in} \sim \frac{1}{2\, \pi}  {m_{\chi^0_3}}^{2}.
      \left |  \frac{C^{ffZ}\: C^{\chi^0_3 \chi^0_3 Z}}
       {4\, {m_{\chi^0_3}}^{2}-{m_Z}^{2}+i\, \Gamma_{Z}\, {m_Z}} \right |^{2}  \sim \frac{1}{2\, \pi}.{\cal V}^{-\frac{8}{3}}m^2_{pl}\left (\frac{{\cal V}^{-\frac{35}{36}}}{4.{\cal V}^{-\frac{8}{3}}M_p^2}\right )^{2}  \nonumber\\
& & \sim  {\cal O}(10)^{-34}{GeV}^{-2}~{\rm for}~ {\cal V}\sim {10}^5.
\end{eqnarray}

$\bullet$
\underline{\bf {sfermion\ ($\widetilde{f}_{a}$) exchange:}}
   \begin{eqnarray}
   \label{eq:atildefff}
    \widetilde{a}_{\bar{f}f}^{(\widetilde{f})} &=& \frac{1}{32\, \pi}
      \sum_{a,b}\frac{(m_{f}\,C_{+}^{a}
           +{m_{\chi^0_3}}\,D_{+}^{a})\,(m_{f}\,C_{+}^{b}
           +{m_{\chi^0_3}}\,D_{+}^{b})}
         {\Delta_{{\widetilde f}_{a}}\, \Delta_{{\widetilde f}_{b}}},
             \end{eqnarray}
a is the index for sfermion mass eigenstates so that $a= 1,..6$ for  squark and charged sleptons and $a=1,..3$ corresponds for sneutrino, and:
\begin{eqnarray}
C_{\pm}^{a} &=& |\Lambda_{fL}^{a}|^{2}\pm|\Lambda_{fR}^{a}|^{2}, \\
\label{ff-coupl2:eq}
D_{\pm}^{a} &=& \Lambda_{fL}^{a}(\Lambda_{fR}^{a})^{*}
                \pm(\Lambda_{fL}^{a})^{*}\Lambda_{fR}^{a};
\end{eqnarray}
$\Lambda_{fL}^{a}$ corresponds to the Neutralino-fermion-sfermion interactions mediated by L- handed
 squarks/sleptons and $\Lambda_{fR}^{a}$ corresponds to the Neutralino-fermion-fermion interactions
 mediated by R-handed squarks/sleptons. Using results from {\bf section 3}:
 $C^{\chi^0_3 {l_L}\tilde {l_L}}=\tilde{f} {\cal V}^{-\frac{1}{2}}$, $C^{{\chi^{0}_3} {u_L}\tilde {u_L}}\sim \tilde{f}{\cal V}^{-\frac{4}{5}}$.
 With exactly similar procedure, we find: $C^{\chi^0_3 {l_L}\tilde {l_R}}=\tilde{f} {\cal V}^{-\frac{12}{15}}, C^{{\chi^{0}_3} {u_L}\tilde {u_R}}\sim \tilde{f}{\cal V}^{-\frac{25}{36}} $.
Utilizing above

\begin{eqnarray}
\label{CaDa}
\sum_{a} C_{\pm}^{a}& = & Max \left({\left|C^{\chi^0_3 {l_L}\tilde {l_L}}\right|^2  \pm  \left|C^{\chi^0_3 {l_L}\tilde {l_R}} \right|^2}, {\left| C^{{\chi^{0}_3} {u_L}\tilde {u_L}}\right|^2  \pm  \left| C^{{\chi^{0}_3} {u_L}\tilde {u_R}}\right|^2}\right) \sim   \tilde{f^2} {\cal V}^{-1} \nonumber\\
& & {\hskip -0.8in}\sum_{a}D_{\pm}^{a} = Max\left( C^{\chi^0_3 {l_L}\tilde {l_L}}(C^{\chi^0_3 {l_L}\tilde {l_R}})^{*},C^{{\chi^{0}_3} {u_L}\tilde {u_L}}(C^{{\chi^{0}_3} {u_L}\tilde {u_R}})^{*}\right) \pm  c.c.. \sim    \tilde{f^2}{\cal V}^{-\frac{13}{10}}
\end{eqnarray}
and
$$\Delta_{{\widetilde f}_{a}}\equiv
m_{f}^{2}-{m_{\chi^0_3}}^{2}-m_{\widetilde{f}_{a}}^{2} \sim -{m_{\chi^0_3}}^{2}.$$
Considering only  first two generations quarks/sleptons and assuming the universality in  scalar masses for first two generations quarks/sleptons, equation (\ref{eq:atildefff}) reduces to the following simplified form:
\begin{eqnarray}
  \widetilde{a}_{\bar{f}f}^{(\widetilde{f})} &=& \frac{1}{ \pi}
      (\frac{m_{f}\, \tilde{f^2} {\cal V}^{-1}
           +{m_{\chi^0_3}}\,\tilde{f^2}{\cal V}^{-\frac{13}{10}})\,(m_{f}\,\tilde{f^2} {\cal V}^{-1}
           +{m_{\chi^0_3}}\,\tilde{f^2}{\cal V}^{-\frac{13}{10}})}
         {m_{\chi^0_3}^{4}} \nonumber\\
        & &  \sim  {10}^{-52}{GeV}^{-2}, {\rm for}~ {m_{\chi^0_3}}\sim {\cal V}^{-\frac{4}{3}}M_p ~{\rm and}~ {\cal V}\sim 10^5.
             \end{eqnarray}
Now, utilizing the numerical estimates of coupling summed over first two generation of squarks/sleptons given above, we will evaluate the value of $\widetilde{b}_{\bar{f}f}^{(\widetilde{f})}$. Using (and quoting verbatim) the form of expression from \cite{Takeshi_Leszek}:
\begin{eqnarray}
     \widetilde{b}_{\bar{f}f}^{(\widetilde{f})} &=& \frac{1}{64\, \pi}
      \sum_{a,b} \frac{1}{\Delta_{{\widetilde f}_{a}}^{3}\,
         \Delta_{{\widetilde f}_{b}}^{3}}
     \bigg\{C_{+}^{a}\,C_{+}^{b}\Big[8\, m_{f}^{2}\, {m_{\chi^0_3}}^{2}\,
       ({m_{\chi^0_3}}^{2}-m_{f}^{2})\, \Delta_{{\widetilde f}_{a}}^{2} \nonumber \\
      & & \hspace{0.4in}- 4\, {m_{\chi^0_3}}^{2}\, (m_{f}^{4}+{m_{\chi^0_3}}^{2}\, m_{f}^{2}
       -2\, {m_{\chi^0_3}}^{4})\, \Delta_{{\widetilde f}_{a}}\,
         \Delta_{{\widetilde f}_{b}} \nonumber \\
     & & \hspace{0.4in}+4\,  {m_{\chi^0_3}}^{2}\, ( m_{f}^{2}+2\, {m_{\chi^0_3}}^{2})
        \Delta_{{\widetilde f}_{a}}^{2}\, \Delta_{{\widetilde f}_{b}}
         +4\, ( {m_{\chi^0_3}}^{2}-m_{f}^{2})\,\Delta_{{\widetilde f}_{a}}^{2}\,
          \Delta_{{\widetilde f}_{b}}^{2} \Big] \nonumber \\
     & & +D_{+}^{a}\, D_{+}^{b} \Big[ 8\, {m_{\chi^0_3}}^{4}\,
          ( {m_{\chi^0_3}}^{2}-m_{f}^{2})\,  \Delta_{{\widetilde f}_{a}}^{2}
        +4\, {m_{\chi^0_3}}^{2}\, ({m_{\chi^0_3}}^{4}+{m_{\chi^0_3}}^{2}\, m_{f}^{2}
        -2\, m_{f}^{4})\Delta_{{\widetilde f}_{a}}\,
             \Delta_{{\widetilde f}_{b}} \nonumber \\
  & &  \hspace{0.4in} +4\, {m_{\chi^0_3}}^{2} (5\, {m_{\chi^0_3}}^{2}-2\, m_{f}^{2})\,
      \Delta_{{\widetilde f}_{a}}^{2}\, \Delta_{{\widetilde f}_{b}}
        -3\, (m_{f}^{2}-3\,  {m_{\chi^0_3}}^{2})\, \Delta_{{\widetilde f}_{a}}^{2}\,
          \Delta_{{\widetilde f}_{b}}^{2} \Big] \nonumber \\
  & & +C_{-}^{a}\,C_{-}^{b} \Big[8\, {m_{\chi^0_3}}^{2}\, ( {m_{\chi^0_3}}^{2}
     -m_{f}^{2})^{2}\,\Delta_{{\widetilde f}_{a}}\,\Delta_{{\widetilde f}_{b}}
     +8\, {m_{\chi^0_3}}^{2}\, ({m_{\chi^0_3}}^2-m_{f}^{2})\Delta_{{\widetilde f}_{a}}^{2}\,
      \Delta_{{\widetilde f}_{b}} \nonumber \\
    & & \hspace{0.4in}+2\, (m_{f}^{2}+2\, {m_{\chi^0_3}}^{2})\,
       \Delta_{{\widetilde f}_{a}}^{2}\,\Delta_{{\widetilde f}_{b}}^{2} \Big]
       \nonumber \\
     & & + C_{+}^{a}\,D_{+}^{b}\,2\,m_{f}\, {m_{\chi^0_3}}\,
      \Big[ 8 \,  {m_{\chi^0_3}}^{2} ({m_{\chi^0_3}}^{2}-m_{f}^{2})\,
      \Delta_{{\widetilde f}_{a}}^{2} +12 \,{m_{\chi^0_3}}^{2} ({m_{\chi^0_3}}^{2}-m_{f}^{2})\,
       \Delta_{{\widetilde f}_{a}}\,\Delta_{{\widetilde f}_{b}}
         \nonumber \\
   & & \hspace{0.4in} +3 \,
        \Delta_{{\widetilde f}_{a}}^{2}\,\Delta_{{\widetilde f}_{b}}^{2}
           -2\,(m_{f}^{2}-4\, {m_{\chi^0_3}}^{2})\,
        \Delta_{{\widetilde f}_{a}}\,\Delta_{{\widetilde f}_{b}}^{2}
          -6\,(m_{f}^{2}-2\, {m_{\chi^0_3}}^{2})\,\Delta_{{\widetilde f}_{a}}^{2}\,
             \Delta_{{\widetilde f}_{b}} \Big] \bigg\}, \nonumber
\end{eqnarray}
for ${m_{\chi^0_3}}\sim {\cal V}^{-\frac{4}{3}}M_p, \Delta_{{\widetilde f}_{a}}
 \sim -{m_{\chi^0_3}}^{2}$, after solving, one gets:
\begin{eqnarray}
\label{eq:btildefff}
 \widetilde{b}_{\bar{f}f}^{(\widetilde{f})} &=& \frac{1}{\pi}\times\frac{{\tilde f}^4 {\cal V}^{-2}}{m^2_{\chi^0_3}} \sim O(10)^{-48} {GeV}^{-2}.
\end{eqnarray}

$\bullet$
\underline{\bf{Higgs $(h,H)$--sfermion\ ($\widetilde{f}_{a}$) interference term:}}
   \begin{eqnarray}
     \widetilde{a}_{\bar{f}f}^{(h,H-\widetilde{f})}& = & 0, 
\\
     \widetilde{b}_{\bar{f}f}^{(h,H-\widetilde{f})}& = &-\frac{1}{8\,\pi}\,
       \sum_{a} Re\,\left[ \sum_{r=h,H} \frac{C_{S}^{ffr}\:
          C_{S}^{\chi\chi r}}{4\, {m_{\chi^0_3}}^{2}-m_{r}^{2}
          +i\, \Gamma_{r}\, m_{r}} \right]
          \frac{({m_{\chi^0_3}}^2-m_{f}^{2})}{\Delta_{{\widetilde f}_{a}}^2}
            \nonumber \\
      & & \times \left[C_{+}^{a}\,2\,m_{f}\,{m_{\chi^0_3}}
            + D_{+}^{a}\,(2\,{m_{\chi^0_3}}^{2}+3\,\Delta_{{\widetilde f}_{a}})
                  \right];
                  \end{eqnarray}
  expanding summation
  \begin{eqnarray}
  \label{eq:bhfff}
& & \sim \frac{1}{8\,\pi} \left [
  \frac{C^{ffh}\: C^{\chi^0_3\chi^0_3 h}}
{4\, m_{\chi^0_3}^{2}-m_{h}^{2}+i\, \Gamma_{h}\, m_{h}}+ \frac{C^{hhH}\: C^{\chi^0_3\chi^0_3 H}}
{4\, m_{\chi^0_3}^{2}-m_{H}^{2}+i\, \Gamma_{H}\, m_{H}}
 \right ]\frac{({m_{\chi^0_3}}^2-m_{f}^{2})}{\Delta_{{\widetilde f}_{a}}^2}
 \nonumber \\
& & \times \left[C_{+}^{a}\,2\,m_{f}\,{m_{\chi^0_3}}
+ D_{+}^{a}\,(2\,{m_{\chi^0_3}}^{2}+3\,\Delta_{{\widetilde f}_{a}}) \right] \nonumber \\
  & & \sim  \frac{1}{\pi} \left [\frac{{\cal V}^{-\frac{35}{36}}}{4.{\cal V}^{-\frac{8}{3}}m^2_{pl}}+ \frac{{\cal V}^{-\frac{35}{36}}}{{\cal V}^{-\frac{85}{36}}m^2_{pl}}\right] {\cal V}^{-\frac{13}{10}}{\tilde f}^2 \sim \frac{1}{\pi}{\tilde f}^2{\cal V}^{-\frac{13}{10}} \left (\frac{{\cal V}^{\frac{5}{3}}}{m^{2}_{pl}}\right )  \nonumber\\
& & \sim  O(10)^{-47}{GeV}^{-2}~{\rm for}~ {\cal V}\sim {10}^5.
\end{eqnarray}

$\bullet$
\underline{\bf {$Z$--sfermion\ ($\widetilde{f}_{a}$) interference term:}}
    \begin{eqnarray}
     & & \hskip-0.3in \widetilde{a}_{\bar{f}f}^{(Z-\widetilde{f})}=-\frac{1}{4\, \pi}
        \sum_{a} Re \left [  \frac{C^{ffZ}\:
           C^{\chi^0_3 \chi^0_3 Z}}{4\, {m_{\chi^0_3}}^{2}-{m_Z}^{2}
             +i\, \Gamma_{Z}\, {m_Z}} \right] \frac{m_{f}\,
                 ({m_Z}^{2}-4\,{m_{\chi^0_3}}^{2})}{{m_Z}^{2}} \frac{(m_{f}\,C_{+}^{a}
           +{m_{\chi^0_3}}\,D_{+}^{a})}{\Delta_{{\widetilde f}_{a}}};
\end{eqnarray}
utilizing the value of mass $m_{Z} =90 GeV$, $m_{\chi^0_3}\sim {\cal V}^{\frac{2}{3}}m_{\frac{3}{2}}\sim {\cal V}^{-\frac{4}{3}}M_p$;  $C^{ffZ}\sim {\cal O}(1), C^{\chi^0_3 \chi^0_3 Z}\sim {\tilde f} {\cal V}^{-\frac{11}{18}} $ and equation no (\ref{CaDa}) , after simplifying, we have
\begin{eqnarray}
\label{eq:afZff}
\label{afZff}
 \widetilde{a}_{\bar{f}f}^{(Z-\widetilde{f})}&=&-\frac{1}{4\, \pi}
  \left [  \frac{ 
           {\tilde f} {\cal V}^{-\frac{11}{18}} }{4\, {m_{\chi^0_3}}^{2}} \right] \frac{m_{f}\,
                 ({m_Z}^{2}-4\,{m_{\chi^0_3}}^{2})}{{m_Z}^{2}}  \frac{(m_{f}\,C_{+}^{a}
           +{m_{\chi^0_3}}\,D_{+}^{a})}{\Delta_{{\widetilde f}_{a}}}, \nonumber\\
           & & \sim  O(10)^{-38} GeV^{-2}
 \end{eqnarray}
     \begin{eqnarray}
     & & \hskip -0.6in \widetilde{b}_{\bar{f}f}^{(Z-\widetilde{f})} =-\frac{1}{8\, \pi}
        \sum_{a} Re \left [ \left( \frac{ C^{\chi^0_3 \chi^0_3 Z}}
          {(4\, {m_{\chi^0_3}}^{2}-{m_Z}^{2}+i\, \Gamma_{Z}\, {m_Z})^{2}} \right) \,
           \frac{1}{{m_Z}^{2}\, \Delta_{{\widetilde f}_{a}}^{3}}\right.
             \nonumber \\
      & & \hskip -0.6in \times \bigg [C^{ffZ}\,C_{-}^{a}\,
           \Big\{2\,{m_Z}^{2}\,P_Z\,\Delta_{{\widetilde f}_{a}}
            [2\,{m_{\chi^0_3}}^{2}\,({m_{\chi^0_3}}^{2}+\Delta_{{\widetilde f}_{a}})
             +m_{f}^{2}\,(-2\,{m_{\chi^0_3}}^{2}+\Delta_{{\widetilde f}_{a}})]\,\Big\}
          \nonumber \\
      & & \hskip -0.6in+\,C^{ffZ}\Big\{C_{+}^{a}
         \Big[2\,m_{f}^{2}\,{m_{\chi^0_3}}^{2}\,({m_{\chi^0_3}}^2-m_{f}^{2})\,
            ({m_Z}^{2}-4\,{m_{\chi^0_3}}^{2})\,P_Z \, {m_{\chi^0_3}}^{2} [m_{f}^{2} {m_Z}^{2}
         +\,2\, {m_{\chi^0_3}}^{2}\,({m_Z}^{2}-6\,m_{f}^{2})]\,P_Z\,
         \Delta_{{\widetilde f}_{a}} \nonumber \\
      & & \hskip -0.6in  + \,2\,{m_Z}\,\{-{m_Z}\,({m_{\chi^0_3}}^2-m_{f}^{2})\,
         ({m_Z}^{2}-4\,{m_{\chi^0_3}}^{2})  +\,i\,\Gamma_{Z}\,[{m_Z}^{2}\,{m_{\chi^0_3}}^{2}
         -m_{f}^{2}({m_Z}^{2}+3\,{m_{\chi^0_3}}^{2})]\}\,
            \Delta_{{\widetilde f}_{a}}^{2}\Big] \nonumber \\
      & &\hskip -0.6in + m_{f}\, {m_{\chi^0_3}}\,D_{+}^{a}\Big[4\,{m_{\chi^0_3}}^{2}\,
         ({m_{\chi^0_3}}^2-m_{f}^{2})\,({m_Z}^{2}-4\,{m_{\chi^0_3}}^{2})\,P_Z+2\,[6\,{m_Z}^{2}\,{m_{\chi^0_3}}^{2}
        -16\ {m_{\chi^0_3}}^{4}-m_{f}^{2}\,(3\,{m_Z}^{2}
        -4\,{m_{\chi^0_3}}^{2})]\,P_Z\,\Delta_{{\widetilde f}_{a}} \nonumber \\
      & &\hskip -0.6in  -3\,[({m_Z}^{2}-4\,{m_{\chi^0_3}}^{2})^{2}
        -i\, {m_Z}\,\Gamma_{Z}\,({m_Z}^{2}-8\,{m_{\chi^0_3}}^{2})]\,
         \Delta_{{\widetilde f}_{a}}^{2}\Big]\bigg\}\Bigg] \Bigg]; \nonumber \\
\end{eqnarray}
where $P_Z\equiv 4\,{m_{\chi^0_3}}^{2}-{m_Z}^{2}+i\, \Gamma_{Z}\, {m_Z}$.
Again using the numerical values of masses and relevant couplings and assuming $\Gamma_{Z}{m_Z}<< {m_{\chi^0_3}}^{2}$, the above expression reduces to
\begin{eqnarray}
\label{eq:bfZff}
\widetilde{b}_{\bar{f}f}^{(Z-\widetilde{f})} &=& {\cal O}(10)^{-46} {GeV}^{-2}.
\end{eqnarray}

Utilizing the results from equation no (\ref{eq:bhff}), (\ref{eq:bZff}), (\ref{eq:btildefff}), (\ref{eq:bhfff}), (\ref{eq:bfZff}) and  (\ref{aZff}), (\ref{eq:afZff}):
\begin{eqnarray}
\label{eq:sumbZZ}
 \widetilde{b}_{\bar{f}f} &=&
 \widetilde{b}_{\bar{f}f}^{(h,H)}
+ \widetilde{b}_{\bar{f}f}^{(Z)} +
\widetilde{b}_{\bar{f}f}^{(\widetilde{f})}+
+ \widetilde{b}_{\bar{f}f}^{(h,H-\widetilde{f})}+
\widetilde{b}_{\bar{f}f}^{(Z-\widetilde{f})}
\sim  O(10)^{-34}{GeV}^{-2};
\end{eqnarray}
\begin{eqnarray}
\label{eq:sumaZZ}
 \widetilde{a}_{\bar{f}f} &=&
 \widetilde{a}_{\bar{f}f}^{(h,H)}+
\widetilde{a}_{\bar{f}f}^{(Z)} +
\widetilde{a}_{\bar{f}f}^{(\widetilde{f})}+
+\widetilde{a}_{\bar{f}f}^{(h,H-\widetilde{f})}+
\widetilde{a}_{\bar{f}f}^{(Z-\widetilde{f})}\sim   {\cal O}(10)^{-25} {GeV}^{-2}.\nonumber\\
\end{eqnarray}
Relative velocity $v_{f_1 f_2}$ is defined as
\begin{eqnarray}
\label{eq:vf1f2}
v_{f_1 f_2}\equiv
\left[1-\frac{(m_{f_1}+m_{f_2})^2}{4m^2_{\chi^0_3}}\right]^{1/2}
\left[1-\frac{(m_{f_1}-m_{f_2})^2}{4m^2_{\chi^0_3}}\right]^{1/2}.
\end{eqnarray}
For $f_1, f_2 = hh, v_{h h} \equiv 1$, $ f_1, f_2 = ZZ, v_{ZZ} \equiv 1$, $ f_1, f_2 = f{\bar f}, v_{f{\bar f}} \equiv 1$.
Having estimated the partial wave coefficients for each possible annihilation processes, summing up their contribution as according to (\ref{eq:abf1f2}):
\begin{eqnarray}
a  &=&
\widetilde{a}_{hh}+ \widetilde{a}_{ZZ}+ \widetilde{a}_{f {\bar f}} \equiv  O(10)^{-25} {GeV}^{-2} \nonumber \\
b  &=& \widetilde{b}_{hh}+ \widetilde{b}_{ZZ}+ \widetilde{b}_{f {\bar f}} \equiv  O(10)^{-10} {GeV}^{-2}
\end{eqnarray}
\begin{equation}
\jxf\equiv \int_0^{x_f}dx \langle\sigma v_{\rm M\o l}\rangle(x)= \int_0^{x_f}dx  (a + b{x_f})= a{x_f} + b\frac{{x^2_f}}{2}
\label{jxfdef:eq}
\end{equation}
where
$x=T/m_\chi$. The value of $x_f$ can be calculated by solving iteratively the equation
\begin{eqnarray}
x_f^{-1} & = & \ln \left( \frac{m_\chi}{2 \pi^3} \sqrt{\frac{45}{2g_* G_N}}
\langle\sigma v_{\rm M\o l}\rangle({x_f})\, x_f^{1/2} \right),
\label{freeze-out-temperature:eq}
\end{eqnarray}
where $g_*$ represents the effective
number of degrees of freedom at freeze-out ($\sqrt{g_*}\simeq 9$).
Solving above equation, $x_f\equiv T_f/m_{\chi^0_3}$ comes out to be around $1/33$

The analytical expression of relic abundance  is given as \cite{Jame_D_wells}
\begin{eqnarray}
\label{eq:relic-density}
\Omega_\chi h^2 & = & \frac{1}{{\mu}^2 \sqrt{g_*}J({x_F})}
\end{eqnarray}
where ${\mu}= 1.2 \times 10^5 GeV$
For $J(x_f) \sim a(x_f) + b\frac{{x_f}^2}{2}\sim  10^{-10}\frac{{x_f}^2}{2}{GeV}^{-2}$ and $x_f= \frac{1}{33}$, $ \sqrt{g_*}=9$,
\begin{eqnarray}
\Omega_{\chi^0_3}h^2\sim\frac{2\cdot(33)^2}{1.44 \times 10^{10}\cdot 9\cdot 10^{-10}}\equiv 168.
\end{eqnarray}
For $m_{\frac{3}{2}}\sim {\cal V}^{-2} m_{pl}\sim 10^8 GeV$ and $m_{\chi^0_3}\sim {\cal V}^{\frac{2}{3}}m_{\frac{3}{2}}\sim 10^{11} GeV$ as given in table \ref{table:mass scales}, relic abundance of gravitino is given as :
\begin{eqnarray}
\label{eq:omega1}
\Omega_{\tilde G}h^2& = & \Omega_{\chi^0_3}h^2\times \frac{m_{\frac{3}{2}}}{m_{\chi^0_3}}=\Omega_{\chi^0_3}\times {\cal V}^{-\frac{2}{3}}  \sim 0.16,
\end{eqnarray}
clearly a very desirable value!

\subsection{Slepton relic density calculations}

For the case of slepton NLSPs, the dominant annihilation channel possible in our set up are: ${\slp{a}\slp{b}^*\rightarrow ZZ}$, ${\slp{a}\slp{b}^*\rightarrow Zh}$,${\slp{a}\slp{b}^*\rightarrow hh}$, ${\slp{a}\slp{b}^*\rightarrow \gm\gm}$, ${\slp{a}\slp{b}^*\rightarrow \gm h}$, ${\slp{a}\slp{b}^*\rightarrow ll}$.



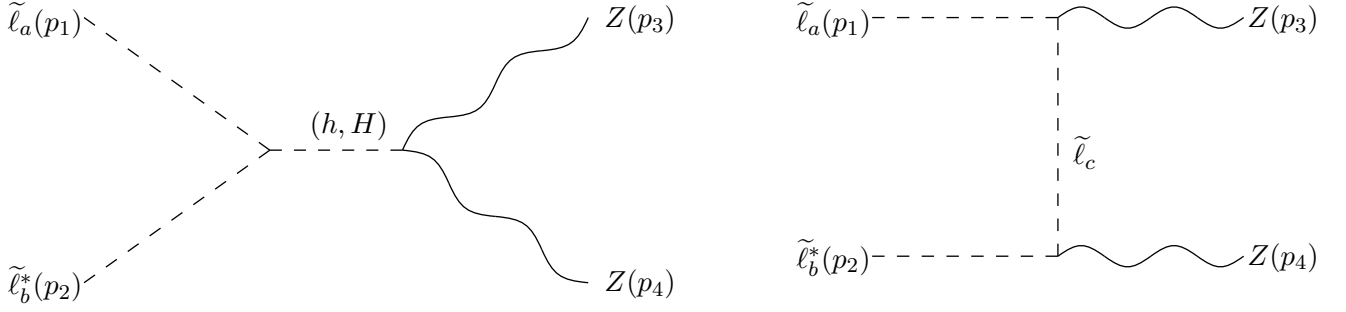
\begin{figure}
\begin{center}
\begin{picture}(150,100)(-100,100)
\DashLine(-100,250)(-30,200)5
\Text(-115,250)[]{$\slp{a}(p_1)$}
\DashLine(-100,150)(-30,200)5
\Text(-115,150)[]{$\slp{b}^*(p_2)$}
\DashLine(-30,200)(20,200)5
\Text(0,210)[]{$(h,H)$}
\Photon(20,200)(90,250){4}{2}
\Text(110,250)[]{$Z(p_3)$}
\Photon(20,200)(90,150){4}{2}
\Text(110,150)[]{$Z(p_4)$}
\end{picture}
\hskip 2.0in
\begin{picture}(150,100)(-100,100)
\DashLine(-100,250)(-30,250)5
\Text(-115,250)[]{$\slp{a}(p_1)$}
\Photon(-30,250)(40,250){4}{2}
\Text(55,250)[]{$Z(p_3)$}
\DashLine(-30,250)(-30,160)5
\Text(-20,200)[]{$\slp{c}$}
\DashLine(-100,160)(-30,160)5
\Text(-115,160)[]{${\slp{b}^*}(p_2)$}
\Photon(-30,160)(40,160){4}{2}
\Text(55,160)[]{$Z(p_4)$}
\end{picture}
\caption{Feynman diagrams for
$\slp{a} \slp{b}^{*}\ra ZZ$ via $s$--channel Higgs exchange and t-channel $\slp{c}$ exchange.}
\end{center}
\end{figure}
The  analytical expressions for  $\widetilde{w}(s)$  are given  in \cite{takeshi_coannihilation}.
Once again, the approach is to first calculate required vertices in the context of ${\cal N}=1$ gauged supergravity and then utilize the same to calculate partial wave coefficients.

${\bf (a) {\slp{a}\slp{b}^*\rightarrow ZZ}}$

\underline{\bf{Slepton-slepton- Higgs vertex}}

Expanding effective supergravity potential  $V= \exp^K G^{T_S T_S}|D_{T_S}W|^2$ in the fluctuations around  ${\cal Z}_i \rightarrow {\cal Z}_i +{\cal V}^{\frac{1}{36}}M_p$ , ${\cal A}_1 \rightarrow {\cal A}_1 +{\cal V}^{-\frac{2}{9}}M_p$,  contribution of term quadratic in ${\cal A}_1$ as well as ${\cal Z}_i$ is of the order $ {\cal V}^{\frac{-89}{36}}<{\cal Z}_i> $, which after giving VEV to  one of the ${\cal Z}_i$, will be given as:
\begin{equation}
\label{eq:Clalbh}
 C^{\slp{a} \slp{b} h}\sim \frac{1}{{\sqrt{(\hat{K}_{{\cal Z}_i{\bar{\cal Z}}_i})^2 (\hat{K}_{{\cal A}_1{\bar{\cal A}}_1})^2}}}\left[{\cal V}^{-\frac{89}{36}}<{{\cal Z}^i}>{{\cal \bar {Z}}^i}{ {\cal A}^1}{ {\cal A}^{*1}}\right]\sim O({\cal V}^{-\frac{34}{15}}).
\end{equation}
Similarly
\begin{equation}
\label{eq:Clalbh1}
 C^{\slp{a} \slp{b} hh }\sim \frac{1}{{\sqrt{(\hat{K}_{{\cal Z}_i{\bar{\cal Z}}_i})^2 (\hat{K}_{{\cal A}_1{\bar{\cal A}}_1})^2}}}\left[{\cal V}^{-\frac{89}{36}}{{\cal Z}^i}{{\cal \bar{Z}}^j}{ {\cal A}^1}{ {\cal A}^{*1}}\right]\sim O({\cal V}^{-\frac{41}{18}}).
\end{equation}

\underline{\bf{ Slepton-slepton-Z Boson- Z Boson vertex}}

In the context of supergravity action, contribution of required vertex will be given by:\\
${\bar{\partial}}_{{\bar {\cal A}}_{1}}\partial_{{\cal A}_1}G_{T_B{\bar T}_B} X^{T_B} X^{\bar {T_B}} {A}^{\mu} A_{\nu}$. As given in equation (\ref{eq:da1da1barGBBbar})
\begin{equation}
{\bar{\partial}}_{{\bar {\cal A}}_{1}}\partial_{{\cal A}_1}G_{T_B{\bar T}_B}\sim {\cal O}(1) {\bar{\partial}}_{{\bar a}_{\bar 1}}\partial_{a_1}G_{T_B{\bar T}_B}\sim {\cal V}^{-\frac{8}{9}}{\cal A}^{*}_{1}{\cal A}_{1},
\end{equation}
$X=X^B\partial_B=-12i\pi\alpha^\prime\kappa_4^2\mu_7Q_B\partial_{T_B} \sim {\cal V}^{-\frac{2}{3}}$.

Incorporating values from above, the physical Slepton-slepton-Z Boson- Z Boson vertex is proportional to
\begin{equation}
\label{eq:CllZZ}
C^{\slp{a} \slp{b}^{*}ZZ}\sim \frac{{\cal V}^{-\frac{8}{9}}{\tilde f}^2 {\cal V}^{-\frac{4}{3}}}{\sqrt{(K_{A_1 A_1})^2}}\sim \frac{{\tilde f}^2 {\cal V}^{-\frac{20}{9}}}{O(10)^{4}}\sim {\tilde f}^2 {\cal V}^{-3}.
\end{equation}

\underline{\bf{ Slepton- slepton- Z Boson vertex}}

The gauge kinetic term for slepton-slepton- Z Boson vertex, relevant to the second Feynman graph in  Fig.26 will be given by $ \frac{e^K}{\kappa_4^2}G^{T_B\bar{T}_B}\tilde{\bigtriangledown}_\mu T_B\tilde{\bigtriangledown}^\mu {\bar T}_{\bar B}$. This implies that the following term generates the required slepton-slepton-gauge boson vertex:
\begin{eqnarray}
\label{eq:sq sq gl}
& & \hskip-0.63in {\frac{6i\kappa_4^2\mu_72\pi\alpha^\prime Q_BG^{T_B{\bar T}_B}}{\kappa_4^2{\cal V}^2}\kappa_4^2}A^\mu\partial_\mu\left(\kappa_4^2\mu_7(2\pi\alpha^\prime)^2C_{1{\bar 1}}{\cal A}_1{\bar {\cal A}}_{\bar 1}\right)\xrightarrow[{\small \kappa_4^2\mu_7(2\pi\alpha^\prime)^2C_{1{\bar 1}}\sim{\cal V}^{\frac{10}{9}}}]{\small G^{T_B{\bar T}_B}\sim{\cal V}^{\frac{7}{3}},}\frac{{\cal V}^{\frac{7}{9}}}{\left(\sqrt{\hat{K}_{{\cal A}_1{\bar {\cal A}}_1}}\right)^2}\nonumber\\
& & {\hskip -0.5in} \sim \tilde{f} {\cal V}, {\rm for~{\cal V}\sim {10}^5}
\end{eqnarray}
Quoting directly the analytical expressions given in \cite{takeshi_coannihilation},
we  obtain a numerical estimate of {$\widetilde w$}:
\begin{eqnarray}
\label{eq:wZZ}
 {\widetilde{w}}_{\slp{a}\slp{b}^*\rightarrow ZZ}=
 {\widetilde{w}}_{ZZ}^{(h,H,P)} + {\widetilde{w}}_{ZZ}^{(\slp{})} + {\widetilde{w}}_{ZZ}^{(h,H,P-\slp{})}.
\end{eqnarray}

$\bullet$
 \underline{Higgs ($h,H$) exchange ($+$ Point interaction):}
\begin{eqnarray}
{\widetilde{w}}_{ZZ}^{(h,H,P)} & = &
\left|
\sum_{r=h,H} \frac{C^{ZZr}C^{\slp{b}^*\slp{a}r}}{\prop{h}}
-C^{\slp{b}^*\slp{a}ZZ}
\right|^2
\frac{s^2-4 \mz^2 s+12\mz^4}{8\mz^4}\nonumber\\
 & \sim &
\left|
 \frac{(10^2 V^{-\frac{34}{15}})m_{pl}}{\prop{r}}+ \frac{(10^2 V^{-\frac{34}{15}})m_{pl}}{\prop{H}}
-{\tilde f}^2 {\cal V}^{-3}
\right|^2
\frac{s^2-4 \mz^2 s+12\mz^4}{8\mz^4};
\end{eqnarray}

$\bullet$
 \underline{slepton\ ($\slp{c}$) exchange:}
\begin{eqnarray}
\label{eq:w_ZZ_sl}
& & {\widetilde{w}}_{ZZ}^{(\slp{})}  =
\frac{1}{\mz^4}\sum_{c,d=1}^{2}
C^{\slp{b}^*\slp{c}Z} C^{\slp{c}^*\slp{a}Z}
C^{\slp{b}^*\slp{d}Z *} C^{\slp{d}^*\slp{a}Z *} \nonumber \\
 & & \times \left[
       \T_4 - 2(\slpsl{2}+2\mz^2) \T_3 \right.  +[ \slpsl{4} +4\slsl{2} + 2\mz^2(\slpsl{2}+3\mz^2) ]\T_2 \nonumber \\
 & & -2[ (\slpsl{2})(\slsl{2}-\mz^4)
         + \mz^2(\slpsl{4} -4\slsl{2} +2\mz^4) ]\T_1 \nonumber \\
 & & +(\msli^2-\mz^2)^2(\mslj^2-\mz^2)^2\T_0  -\Y_4
      +[ s(\slpsl{2}-2\mz^2)
               -2(\msli^2-\mz^2)(\mslj^2-\mz^2)  ]\Y_2 \nonumber\\
               & & -[ s^2(\msli^2-\mz^2)(\mslj^2-\mz^2)
  +s\{ -\slsl{2}(\slpsl{2})+3\mz^2(\slpsl{4})-3\mz^4(\slpsl{2})+2\mz^6 \}
                   \nonumber \\
 & & \left.
 + (\msli^2-\mz^2)^2(\mslj^2-\mz^2)^2
]\Y_0
\right].
\end{eqnarray}
Considering only first-generation squarks,
\begin{equation}
\sum_{c,d=1}^{2}
C^{\slp{b}^*\slp{c}Z} C^{\slp{c}^*\slp{a}Z}
C^{\slp{b}^*\slp{d}Z *} C^{\slp{d}^*\slp{a}Z *}\sim \tilde{f^4}{\cal V}^4
\end{equation}
(\ref{eq:w_ZZ_sl}) yields:
\begin{eqnarray}
{\widetilde{w}}_{ZZ}^{(\slp{})} & = &
\frac{{f^4}{\cal V}^4}{\mz^4} \bigg[
       \T_4 - 4 \msli^2\T_3 +  6 \msli^4 \T_2  -4 \msli^6 \T_1 \ + \msli^8 \T_0  -\Y_4
      +[2 s\msli^2 -2 \msli^4 ]\Y_2 -  \nonumber\\
      && [ s^2 \msli^4  - 2s\msli^6 + \msli^8 ]\Y_0 \bigg];
\end{eqnarray}
$\T_4, \T_3,\Y_0...$ etc are auxiliary functions defined in the appendix of \cite{takeshi_coannihilation} . Given that $ \msli^2 \sim {\cal V}^{-3}M^{2}_p$ from  equation (\ref{eq:m_A_1}) and $ m_Z\sim 90 GeV $, on solving and simplifying the same, we get
\begin{eqnarray}
\T_0 &=& \frac{1}{-{\cal V}^{\frac{42}{5}}-4 {\cal V}^5+\left({\cal V}^{\frac{21}{5}}+s\right) {\cal V}^{\frac{21}{5}}+{\cal V}^{\frac{8}{5}}}
\nonumber
\end{eqnarray}
\begin{eqnarray}
&& \T_1 \equiv\frac{\sqrt{s-4 {\cal V}^{\frac{4}{5}}} \sqrt{s-4 {\cal V}^{\frac{21}{5}}} {\cal V}^{\frac{13}{5}}+\left(-4 {\cal V}^{\frac{17}{5}}+s {\cal V}^{\frac{13}{5}}+1\right) \log \left(\frac{s-2 {\cal V}^{\frac{4}{5}}-\sqrt{s-4 {\cal V}^{\frac{4}{5}}} \sqrt{s-4 {\cal V}^{\frac{21}{5}}}}{s-2
   {\cal V}^{\frac{4}{5}}+\sqrt{s-4 {\cal V}^{\frac{4}{5}}} \sqrt{s-4 {\cal V}^{\frac{21}{5}}}}\right)}{\sqrt{s-4 {\cal V}^{\frac{4}{5}}} \left(-4 {\cal V}^{\frac{17}{5}}+s {\cal V}^{\frac{13}{5}}+1\right) \sqrt{s-4 {\cal V}^{\frac{21}{5}}}};\nonumber
   \end{eqnarray}
   \begin{eqnarray}
 && {\hskip -0.5in}\T_2 \equiv -\frac{4 \left(\sqrt{s-4 {\cal V}^{\frac{4}{5}}} \sqrt{s-4 {\cal V}^{\frac{21}{5}}} \left({\cal V}^{\frac{34}{5}}+s {\cal V}^{\frac{13}{5}}\right) {\cal V}^{\frac{8}{5}}+\left(-8 {\cal V}^{\frac{46}{5}}+2 s {\cal V}^{\frac{42}{5}}\right) \log \left(\frac{s-2 {\cal V}^{\frac{4}{5}}-\sqrt{s-4 {\cal V}^{\frac{4}{5}}} \sqrt{s-4 {\cal V}^{\frac{21}{5}}}}{s-2 {\cal V}^{\frac{4}{5}}+\sqrt{s-4 {\cal V}^{\frac{4}{5}}} \sqrt{s-4 {\cal V}^{\frac{21}{5}}}}\right)\right)}{\sqrt{s-4 {\cal V}^{\frac{4}{5}}} \sqrt{s-4
   {\cal V}^{\frac{21}{5}}} \left(16 {\cal V}^5-4 s {\cal V}^{\frac{21}{5}}-4 {\cal V}^{\frac{8}{5}}\right)}; \nonumber
   \end{eqnarray}
   \begin{eqnarray}
  &&{\hskip -1in}\T_3 \equiv  \frac{2 \left(6 \left(4 {\cal V}^{\frac{17}{5}}-s {\cal V}^{\frac{13}{5}}\right) \log \left(\frac{s-2 {\cal V}^{\frac{4}{5}}-\sqrt{s-4 {\cal V}^{\frac{4}{5}}} \sqrt{s-4 {\cal V}^{\frac{21}{5}}}}{s-2 {\cal V}^{\frac{4}{5}}+\sqrt{s-4 {\cal V}^{\frac{4}{5}}} \sqrt{s-4
   {\cal V}^{\frac{21}{5}}}}\right) {\cal V}^{10}+\sqrt{s-4 {\cal V}^{\frac{4}{5}}} \sqrt{s-4 {\cal V}^{\frac{21}{5}}} \left(-2 {\cal V}^{11}-6 s {\cal V}^{\frac{34}{5}}+s^2 {\cal V}^{\frac{13}{5}}\right)
   {\cal V}^{\frac{8}{5}}\right)}{\sqrt{s-4 {\cal V}^{\frac{4}{5}}} \sqrt{s-4 {\cal V}^{\frac{21}{5}}} \left(16 {\cal V}^5-4 s {\cal V}^{\frac{21}{5}}-4 {\cal V}^{\frac{8}{5}}\right)};\nonumber
   \end{eqnarray}
   \begin{eqnarray}
  &&{\hskip -0.3in}\T_4 \equiv \frac{4 \log \left(\frac{s-2 {\cal V}^{\frac{4}{5}}-\sqrt{s-4 {\cal V}^{\frac{4}{5}}} \sqrt{s-4 {\cal V}^{\frac{21}{5}}}}{s-2 {\cal V}^{\frac{4}{5}}+\sqrt{s-4 {\cal V}^{\frac{4}{5}}} \sqrt{s-4 {\cal V}^{\frac{21}{5}}}}\right) {\cal V}^{\frac{63}{5}}}{\sqrt{s-4 {\cal V}^{\frac{4}{5}}}
   \sqrt{s-4 {\cal V}^{\frac{21}{5}}}}+\frac{1}{3} \left(\frac{3 {\cal V}^{\frac{76}{5}}}{-4 {\cal V}^{\frac{17}{5}}+s {\cal V}^{13/5}+1}+18 {\cal V}^{42/5}-7s {\cal V}^{17/5} {\cal V}^{4/5}+s^2\right);
   \nonumber
   \end{eqnarray}
   \begin{eqnarray}
   &&\Y_0 \equiv \frac{2 \log \left(\frac{s-2 {\cal V}^{\frac{4}{5}}-\sqrt{s-4 {\cal V}^{\frac{4}{5}}} \sqrt{s-4 {\cal V}^{\frac{21}{5}}}}{s-2 {\cal V}^{\frac{4}{5}}+\sqrt{s-4 {\cal V}^{\frac{4}{5}}} \sqrt{s-4 {\cal V}^{\frac{21}{5}}}}\right)}{\sqrt{s-4 {\cal V}^{\frac{4}{5}}}
   \left(s-2 {\cal V}^{\frac{4}{5}}\right) \sqrt{s-4 {\cal V}^{\frac{21}{5}}}};
\nonumber
   \end{eqnarray}
   \begin{eqnarray}
  &&{\hskip -0.3in}\Y_2 \equiv \frac{2 \left(s-{\cal V}^{\frac{4}{5}} \left({\cal V}^{\frac{17}{5}}+2\right)\right) \log \left(\frac{s-2 {\cal V}^{\frac{4}{5}}-\sqrt{s-4 {\cal V}^{\frac{4}{5}}} \sqrt{s-4 {\cal V}^{\frac{21}{5}}}}{s-2 {\cal V}^{\frac{4}{5}}+\sqrt{s-4 {\cal V}^{4/5}} \sqrt{s-4
   {\cal V}^{\frac{21}{5}}}}\right) {\cal V}^{\frac{21}{5}}}{\sqrt{s-4 {\cal V}^{\frac{4}{5}}} \left(s-2 {\cal V}^{\frac{4}{5}}\right) \sqrt{s-4 {\cal V}^{\frac{21}{5}}}}+1;
    \nonumber
   \end{eqnarray}
   \begin{eqnarray}
  && \Y_4 \equiv \frac{1}{6} \left(-s^2+2 {\cal V}^{\frac{4}{5}} \left(5 {\cal V}^{\frac{17}{5}}+2\right) s-2 \left(6 {\cal V}^{\frac{42}{5}}+8 {\cal V}^5+3 {\cal V}^{\frac{8}{5}}\right)\right).   \nonumber
   \end{eqnarray}

$\bullet$
 \underline{Higgs ($h,H$) ($+$ Point) -- slepton\ ($\slp{c}$) interference:}
\begin{eqnarray}
{\widetilde{w}}_{ZZ}^{(h,H,P-\slp{})} & = &
\frac{1}{2\mz^4}\sum_{c=1}^2 \Re
\left[ \left( \sum_{r=h,H}
\frac{C^{ZZr}C^{\slp{b}^*\slp{a}r}}{\prop{r}}-C^{\slp{b}^*\slp{a}ZZ}
\right)^* C^{\slp{b}^*\slp{c}Z} C^{\slp{c}^*\slp{a}Z} \right]
\nonumber \\
 & & \times \left[
s^2 + s (\slpsl{2}-2\mslk^2-4\mz^2)
+ 2\mz^2(\slpsl{2}-2\mslk^2+2\mz^2) \right. \nonumber \\
 & &
-2 [ s(\msli^2-\mslk^2-\mz^2)(\mslj^2-\mslk^2-\mz^2) \nonumber \\
 & &
+2\mz^2 \{ \slpsl{4}-\slsl{2}
+\mslk^2(\mslk^2-\msli^2-\mslj^2-2\mz^2) \nonumber \\
 & &
\left. -\mz^2 (\slpsl{2}-\mz^2) \} ]\F \right].
\end{eqnarray}
Using the universality in squark masses as obtained in Appendix B, the simplification leads to:
\begin{eqnarray}
&& {\widetilde{w}}_{ZZ}^{(h,H,P-\slp{})} \equiv
\frac{1}{2\mz^4} \Re
\left[ \left( \sum_{r=h,H}
\frac{(10^2 {\cal V}^{-\frac{34}{15}}m_{pl})}{\prop{r}}-{\tilde f}^2 {\cal V}^{-3}
\right)^* \tilde{f^2}{\cal V}^2 \right]
\nonumber \\
 & & \times \left[
(s^2 -4s \mz^2 + 4 \mz^4) -2 (s\mz^4 -8 \mz^4\msli^2)\F  \right]
\end{eqnarray}
where
\begin{eqnarray}
&& \F\equiv \frac{\log \left(\frac{s-2 {\cal V}^{\frac{4}{5}}-\sqrt{s-4 {\cal V}^{\frac{4}{5}}} \sqrt{s-4 {\cal V}^{\frac{21}{5}}}}{s-2 {\cal V}^{\frac{4}{5}}+\sqrt{s-4 {\cal V}^{\frac{4}{5}}} \sqrt{s-4 {\cal V}^{\frac{21}{5}}}}\right)}{\sqrt{s-4 {\cal V}^{\frac{4}{5}}} \sqrt{s-4
   {\cal V}^{\frac{21}{5}}}} .\nonumber
\end{eqnarray}
Summing up the contribution of ${\widetilde{w}}$ for all s, t and u-channels as according to equation (\ref{eq:wZZ}):
\begin{eqnarray}
\widetilde{w}_{ZZ}|_{s=(4 \msli^2)}\sim {\cal V}^{\frac{34}{5}}, \widetilde{w}^{\prime}_{ZZ}|_{s=(4 \msli^2)}\sim {\cal V}^{\frac{13}{5}}.
\end{eqnarray}
The ``reduced'' coefficients $\widetilde{a}_{ZZ}$ and
$\widetilde{b}_{ZZ}$ will be given by \footnote{Here, we are considering:  ${\tilde f}\sim 10^{-5}$}
\begin{eqnarray}
\label{eq:aZZ}
&& \widetilde{a}_{ZZ}  \sim  \frac{1}{32 \pi \msli^2} {\cal V}^{\frac{34}{5}}\sim 10^{11} {GeV}^{-2};
\widetilde{b}_{ZZ} \sim  \frac{3}{64 \pi \msli^2}{\cal V}^{\frac{13}{5}}\sim  10^{-10} {GeV}^{-2}.
\end{eqnarray}

${\bf (b) {\slp{a}\slp{b}^*\rightarrow Zh}}$

The process ${\slp{a}\slp{b}^*\rightarrow Zh}$ involves
the $s$--channel $Z$--boson exchange,and the $t$-- and $u$--channel slepton ($\slp{a}$, $a=1,2$) exchange:
\begin{eqnarray}
\label{eq:wzh}
 {\widetilde{w}}_{\slp{a}\slp{b}^*\rightarrow Zh}=
  {\widetilde{w}}_{Zh}^{(Z)} + {\widetilde{w}}_{Zh}^{(\slp{})}
+  {\widetilde{w}}_{Zh}^{(Z-\slp{})}.
\end{eqnarray}

\begin{figure}
\begin{center}
\begin{picture}(150,100)(-100,100)
\DashLine(-100,250)(-30,200)5
\Text(-115,250)[]{$\slp{a}(p_1)$}
\DashLine(-100,150)(-30,200)5
\Text(-115,150)[]{$\slp{b}^*(p_2)$}
\DashLine(-30,200)(20,200)5
\Text(0,210)[]{$(h,H)$}
\Photon(20,200)(90,250){4}{2}
\Text(110,250)[]{$Z(p_3)$}
\DashLine(20,200)(90,150)5
\Text(110,150)[]{$h(p_4)$}
\end{picture}
\hskip 2.0in
\begin{picture}(150,100)(-100,100)
\DashLine(-100,250)(-30,250)5
\Text(-115,250)[]{$\slp{a}(p_1)$}
\Photon(-30,250)(40,250){4}{2}
\Text(55,250)[]{$Z(p_3)$}
\DashLine(-30,250)(-30,160)5
\Text(-20,200)[]{$\slp{c}$}
\DashLine(-100,160)(-30,160)5
\Text(-115,160)[]{${\slp{b}^*}(p_2)$}
\DashLine(-30,160)(40,160)5
\Text(55,160)[]{$h(p_4)$}
\end{picture}
\caption{Feynman diagrams for
$\slp{a} \slp{b}^{*}\ra Zh$ via $s$--channel Higgs exchange and t-channel $\slp{c}$ exchange.}
\end{center}
\end{figure}
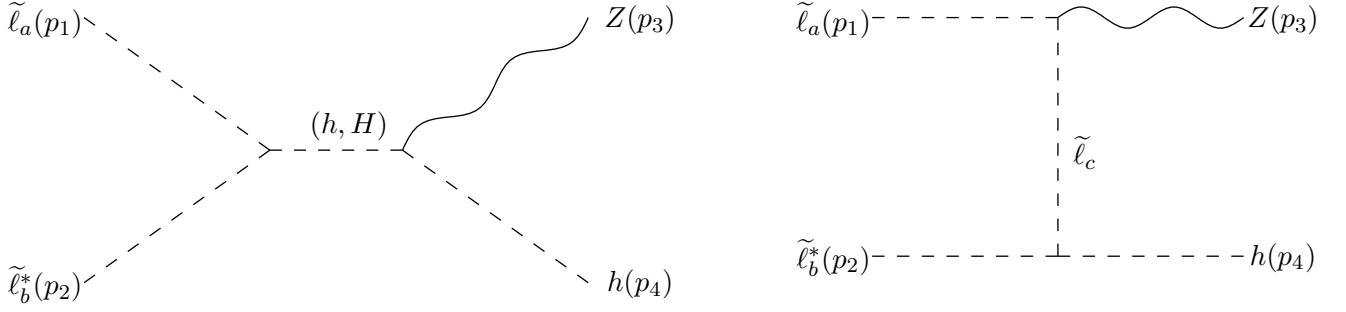

$\bullet$
 \underline{$Z$ exchange:}
\begin{eqnarray}
& &\hskip-0.8in {\widetilde{w}}_{Zh}^{(Z)}  =
\frac{1}{12\mz^6}
\left| \frac{C^{ZZh}C^{\slp{b}^*\slp{a}Z}}{\prop{Z}} \right|^2\nonumber\\
 & & \times\Bigg[ s^2 \Big\{ 3(\slmsl{2})^2+\mz^4 \Big\} -2s \Big\{ 3(\slmsl{2})^2(\mh^2+2\mz^2)
             + \mz^4(\slpsl{2}+\mh^2-5\mz^2) \Big\} \nonumber \\
 & & \hskip -0.8in +\mz^4 \Big\{ (\mh^2-\mz^2)^2 +4(\slpsl{2})(\mh^2-5\mz^2) \Big\}
 +(\slmsl{2})^2(3\mh^4+6\mz^2\mh^2+19\mz^4) \nonumber \\
 & & \hskip -0.8in - \frac{2}{s}\mz^2 \Big\{ (\slmsl{2})^2(3\mh^4-2\mz^2\mh^2+\mz^4)
      +\mz^2(\mh^2-\mz^2)^2(\slpsl{2}) \Big\}    + \frac{4}{s^2}\mz^4(\mh^2-\mz^2)^2(\slmsl{2})^2 \Bigg].
      \nonumber\\
      & &
\end{eqnarray}

For $ C^{ZZh} \sim \frac{m^{2}_{Z}}{v}, C^{\slp{b}^*\slp{a}Z}\sim {\tilde f}{\cal V}$ and assuming universality in slepton masses; after simplification,above expression reduces to:
\begin{eqnarray}
{\widetilde{w}}_{Zh}^{(Z)} & = &
\frac{1}{12\mz^6}
\left| \frac{10^{2}.{\tilde f}{\cal V}}{\prop{Z}} \right|^2
 \times \Bigg[ s^2 \mz^4 -4s \msli^2\mz^4 +8 \mz^4 \msli^2 {m^2_h}- \frac{2}{s}\mz^4 \msli^2 {m^4_h}\Bigg].
\end{eqnarray}

$\bullet$
 \underline{slepton\ ($\slp{c}$) exchange:}
\begin{eqnarray}
\label{eq:wZhl}
& & \hskip -1in{\widetilde{w}}_{Zh}^{(\slp{})}  =
\frac{1}{\mz^2}\sum_{c,d=1}^{2}
C^{\slp{c}^*\slp{a}Z} C^{\slp{b}^*\slp{c}h}
C^{\slp{d}^*\slp{a}Z *} C^{\slp{b}^*\slp{d}h *}  \left[ \Tt_2 - 2(\msli^2+\mz^2) \Tt_1
                  +(\msli^2-\mz^2)^2 \Tt_0 \right] \nonumber \\
& & \hskip -1in +
\frac{1}{\mz^2}\sum_{c,d=1}^{2}
C^{\slp{c}^*\slp{a}h} C^{\slp{b}^*\slp{c}Z}
C^{\slp{d}^*\slp{a}h *} C^{\slp{b}^*\slp{d}Z *}  \left[ \Tu_2 - 2(\mslj^2+\mz^2) \Tu_1
                  +(\mslj^2-\mz^2)^2 \Tu_0 \right] \nonumber \\
 & & \hskip -1in +
\frac{1}{\mz^2} \Re \sum_{c,d=1}^{2}
C^{\slp{c}^*\slp{a}Z} C^{\slp{b}^*\slp{c}h}
C^{\slp{d}^*\slp{a}h *} C^{\slp{b}^*\slp{d}Z *} \nonumber \\
 & &\hskip -1in \times \left[ -2\Y_2
+\frac{1}{s}(s-\mh^2+\mz^2)(\slmsl{2}) \Y_1 \right. + \Big\{ s(\slpsl{2}-2\mz^2)-(\slpsl{4}) -2\mz^2\mh^2 \nonumber \\
 & & \hskip -1in   +(3\mz^2-\mh^2)(\slpsl{2})
        +\frac{1}{s}(\mh^2-\mz^2)(\slmsl{2})^2 \left. - \frac{1}{2s^2}(\mh^2-\mz^2)^2(\slmsl{2})^2 \Big\} \Y_0 \right].
\end{eqnarray}

Assuming universality in slepton masses of first two generations
\begin{eqnarray}
&&\hskip -1in \sum_{c,d=1}^{2} C^{\slp{c}^*\slp{a}h} C^{\slp{b}^*\slp{c}Z}
C^{\slp{d}^*\slp{a}h *} C^{\slp{b}^*\slp{d}Z *}\sim \sum_{c,d=1}^{2}
C^{\slp{c}^*\slp{a}h} C^{\slp{b}^*\slp{c}Z}
C^{\slp{d}^*\slp{a}h *} C^{\slp{b}^*\slp{d}Z *}\sim
 \sum_{c,d=1}^{2}
C^{\slp{c}^*\slp{a}Z} C^{\slp{b}^*\slp{c}h}
C^{\slp{d}^*\slp{a}h *} C^{\slp{b}^*\slp{d}Z *} \equiv
{\tilde f}^2 {\cal V}^{-\frac{38}{15}} {M_p^2}.\nonumber
\end{eqnarray}
Therefore, (\ref{eq:wZhl}) reduces to:
\begin{eqnarray}
\label{eq:wZhfin}
&& {\widetilde{w}}_{Zh}^{(\slp{})} \sim
\frac{{\tilde f}^2 {\cal V}^{-\frac{38}{15}} {m^2_{pl}}}{\mz^2}\times \left[(\Tt_2 +\Tu_2) - \msli^2(\Tt_1 +\Tu_1)+ \msli^4 (\Tt_0 +\Tu_0) -2\Y_2 + (2s msli^2-2 \msli^4)\Y_0 \right]. \nonumber\\
\end{eqnarray}

$\bullet$
 \underline{$Z$ -- slepton\ ($\slp{c}$) interference:}
\begin{eqnarray}
\label{eq:w_Zhslp}
& & {\widetilde{w}}_{Zh}^{(Z-\slp{})}  =
\frac{1}{\mz^4}\Re \sum_{c=1}^{2}
\left( \frac{C^{ZZh}C^{\slp{b}^*\slp{a}Z}}{\prop{Z}} \right)^* \Bigg[
C^{\slp{c}^*\slp{a}Z} C^{\slp{b}^*\slp{c}h}
\Big\{ -(s - \mh^2)(\slmsl{2}) \nonumber \\
 & &  +2\mz^2(\msli^2-\mslk^2+\mz^2)
     -\frac{1}{s}\mz^2(\mh^2-\mz^2)(\slmsl{2})  + \Big[ s(\slmsl{2}+\mz^2)(\msli^2-\mslk^2-\mz^2)   \nonumber\\
     & &  + \mz^2 \{ -2\mslk^4+\mslk^2(3\msli^2+\mslj^2+3\mz^2)-3\msli^4
                                                   \nonumber \\
 & &  +3\slsl{2}-2\mslj^4+\mz^2(4\msli^2+\mslj^2)-\mz^4 \}     -\mh^2(\slmsl{2}+\mz^2)(\msli^2-\mslk^2+\mz^2) \Big] \Ft \Big\} \nonumber\\
 & & + C^{\slp{c}^*\slp{a}h} C^{\slp{b}^*\slp{c}Z}
\Big\{ (s - \mh^2)(\slmsl{2})  +2\mz^2(\mslj^2-\mslk^2+\mz^2)
     +\frac{1}{s}\mz^2(\mh^2-\mz^2)(\slmsl{2})\nonumber\\
      & & + \Big[ -s(\slmsl{2}-\mz^2)(\mslj^2-\mslk^2-\mz^2) +\mz^2 \{ -2\mslk^4+\mslk^2(3\mslj^2+\msli^2+3\mz^2)-3\mslj^4
 +3\slsl{2}\nonumber\\
 & & -2\msli^4+\mz^2(4\mslj^2+\msli^2)-\mz^4 \} +\mh^2(\slmsl{2}-\mz^2)(\mslj^2-\mslk^2+\mz^2) \Big] \Fu \Big\}
 \Bigg].
\end{eqnarray}
For $ C^{ZZh}\sim10^2, C^{\slp{c}^*\slp{a}Z}\sim {\tilde f}{\cal V}, C^{\slp{b}^*\slp{c}h}\sim {\cal V}^{-\frac{34}{15}}$, after simplification, (\ref{eq:w_Zhslp}) reduces to
\begin{eqnarray}
&& {\widetilde{w}}_{Zh}^{(Z-\slp{})} \equiv
\frac{1}{\mz^4}\Re \sum_{c=1}^{2}
\left( \frac{(10^2. {\tilde f}{\cal V})M_p}{\prop{Z}} \right)^*
 C^{\slp{c}^*\slp{a}Z} C^{\slp{b}^*\slp{c}h}\Bigg[
4\mz^4 + (-2s \mz^4 + \msli^2 \mz^4)(\Ft + \Fu) \Bigg] \nonumber
\end{eqnarray}
where
\begin{eqnarray}
&& \Ft \sim \Fu \equiv \frac{3 \log \left(\frac{3 s-8 V^{\frac{4}{5}}-\sqrt{\frac{3 s+2 \left(-4+\sqrt{15}\right) V^{\frac{4}{5}}}{s}} \sqrt{3 s-2 \left(4+\sqrt{15}\right) V^{\frac{4}{5}}} \sqrt{s-4 V^{\frac{21}{5}}}}{3
   s-8 V^{\frac{4}{5}}+\sqrt{\frac{3 s+2 \left(-4+\sqrt{15}\right) V^{\frac{4}{5}}}{s}} \sqrt{3 s-2 \left(4+\sqrt{15}\right) V^{\frac{4}{5}}} \sqrt{s-4 V^{\frac{21}{5}}}}\right)}{\sqrt{\frac{3 s+2
   \left(-4+\sqrt{15}\right) V^{\frac{4}{5}}}{s}} \sqrt{3 s-2 \left(4+\sqrt{15}\right) V^{\frac{4}{5}}} \sqrt{s-4 V^{\frac{21}{5}}}}. \nonumber
\end{eqnarray}
Summing up the contribution of ${\widetilde{w}}$ for all s, t and u-channels as according to equation (\ref{eq:w_Zhslp})
\begin{eqnarray}
\widetilde{w}_{Zh}|_{s=(4 \msli^2)}\sim {\cal V}^{\frac{34}{15}}, \widetilde{w}^{\prime}_{Zh}|_{s=(4 \msli^2)}\sim {\cal V}^{\frac{29}{15}}.
\end{eqnarray}
The ``reduced'' coefficients $\widetilde{a}_{Zh}$ and
$\widetilde{b}_{Zh}$ will be given by
\begin{eqnarray}
\label{eq:aZh}
&& \widetilde{a}_{Zh}  \sim  \frac{1}{32 \pi \msli^2} {\cal V}^{\frac{34}{15}}\sim 10^{-11} {GeV}^{-2};
\widetilde{b}_{Zh} \sim  \frac{3}{64 \pi \msli^2}{\cal V}^{\frac{29}{15}}\sim  10^{-21} {GeV}^{-2}.
\end{eqnarray}

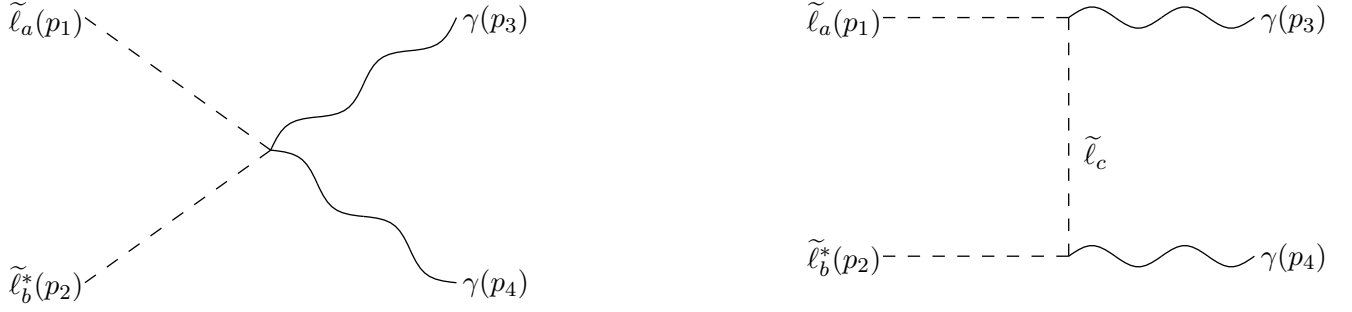
\begin{figure}
\begin{center}
\begin{picture}(150,100)(-100,100)
\DashLine(-100,250)(-30,200)5
\Text(-115,250)[]{$\slp{a}(p_1)$}
\DashLine(-100,150)(-30,200)5
\Text(-115,150)[]{$\slp{b}^*(p_2)$}
\Photon(-30,200)(40,250){4}{2}
\Text(55,250)[]{$\gm(p_3)$}
\Photon(-30,200)(40,150){4}{2}
\Text(55,150)[]{$\gm(p_4)$}
\end{picture}
{\hskip 2.0in}
\begin{picture}(150,100)(-100,100)
\DashLine(-100,250)(-30,250)5
\Text(-115,250)[]{$\slp{a}(p_1)$}
\Photon(-30,250)(40,250){4}{2}
\Text(55,250)[]{$\gm(p_3)$}
\DashLine(-30,250)(-30,160)5
\Text(-20,200)[]{$\slp{c}$}
\DashLine(-100,160)(-30,160)5
\Text(-115,160)[]{$\slp{b}^*(p_2)$}
\Photon(-30,160)(40,160){4}{2}
\Text(55,160)[]{$\gm(p_4)$}
\end{picture}
\caption{Feynman diagrams for
$\slp{a} \slp{b}^*\ra \gm\gm $ via point interaction and t-channel $\slp{c}^*$ exchange.}
\end{center}
\end{figure}

${\bf (c) {\slp{a}\slp{b}^*\rightarrow \gm\gm}}$

The four--point contact interaction needs to be included along with
the $t$-- and $u$--channel slepton ($\slp{a}$, $a=1,2$) exchange:
\begin{eqnarray}
\label{eq:wgmgm}
 {\widetilde{w}}_{\slp{a}\slp{b}^*\rightarrow \gm\gm} = {\widetilde{w}}_{\gm\gm}^{(P)}
             + {\widetilde{w}}_{\gm\gm}^{(\slp{})} + {\widetilde{w}}_{\gm\gm}^{(P-\slp{})}.\nonumber
\end{eqnarray}

$\bullet$
 \underline{Contact interaction:}
\begin{eqnarray}
{\widetilde{w}}_{\gm\gm}^{(P)} \sim   (C^{\slp{a} \slp{b}^* \gm\gm})^2 \sim e^4 ({\tilde f}^2 {\cal V}^{-3})^2 \delta_{ab}
\end{eqnarray}
%
$\bullet$
 \underline{slepton\ ($\slp{a}$) exchange:}
\begin{eqnarray}
{\widetilde{w}}_{\gm\gm}^{(\slp{})} & = &
 (C^{\slp{a} \slp{b}^* \gm})^4 \delta_{ab} \left[
 4(\T_2 + 2 \msli^2 \T_1 + \msli^4 \T_0) - (s-4\msli^2)^2 \Y_0
\right]\nonumber\\
&& \sim {\tilde f}^4{\cal V}^4\left[
 4(\T_2 + 2 \msli^2 \T_1 + \msli^4 \T_0) - (s-4\msli^2)^2 \Y_0
\right]
\end{eqnarray}
where
\begin{eqnarray}
&& \T_0\equiv  \frac{1}{V^{21/5} \left(V^{21/5}+s\right)-V^{42/5}}; \T_1\equiv  \frac{\log \left(\frac{\sqrt{s}-\sqrt{s-4 V^{21/5}}}{\sqrt{s}+\sqrt{s-4 V^{21/5}}}\right)}{\sqrt{s} \sqrt{s-4 V^{21/5}}}+\frac{1}{s}\nonumber
\end{eqnarray}
\begin{eqnarray}
&& \T_2\equiv  \frac{2 \log \left(\frac{\sqrt{s}-\sqrt{s-4 V^{21/5}}}{\sqrt{s}+\sqrt{s-4 V^{21/5}}}\right) V^{21/5}}{\sqrt{s} \sqrt{s-4 V^{21/5}}}+\frac{V^{21/5}}{s}+1; \Y_0\equiv \frac{2 \log \left(\frac{\sqrt{s}-\sqrt{s-4 V^{21/5}}}{\sqrt{s}+\sqrt{s-4 V^{21/5}}}\right)}{s^{3/2} \sqrt{s-4 V^{21/5}}}\nonumber
\end{eqnarray}
%
$\bullet$
 \underline{Contact -- slepton\ ($\slp{c}$) interference:}
\begin{eqnarray}
\hskip-1in{\widetilde{w}}_{\gm\gm}^{(P-\slp{})} & = &
2 C^{\slp{a} \slp{b}^* \gm\gm} (C^{\slp{a} \slp{b}^* \gm})^2 e^4 \delta_{ab}
\left[ - 4 + (s-8\msli^2)\F \right] \sim ({\tilde f}^4 {\cal V}^{-1}) e^4 \delta_{ab}
\left[ - 4 + (s-8\msli^2)\F \right]
\end{eqnarray}
and
\begin{eqnarray}
&& \F \equiv \frac{\log \left(\frac{\frac{1}{2} \sqrt{s} \sqrt{s-4 V^{21/5}}-\frac{s}{2}}{-\frac{s}{2}-\frac{1}{2} \sqrt{s-4 V^{21/5}} \sqrt{s}}\right)}{\sqrt{s} \sqrt{s-4
   V^{21/5}}}
\end{eqnarray}
Summing up the contribution of ${\widetilde{w}}$ for all channels: (\ref{eq:wgmgm})
\begin{eqnarray}
\widetilde{w}_{\gm\gm}|_{s=(4 \msli^2)}\sim {\cal V}^{\frac{21}{5}}, \widetilde{w}^{\prime}_{\gm\gm}|_{s=(4 \msli^2)}\sim {\cal O}(1),
\end{eqnarray}
the ``reduced'' coefficients $\widetilde{a}_{\gm\gm}$ and
$\widetilde{b}_{\gm\gm}$ will be given by
\begin{eqnarray}
\label{eq:agmgm}
&& \widetilde{a}_{\gm\gm}  \sim  \frac{1}{32 \pi \msli^2} {\cal V}^{\frac{21}{5}}\sim 10^{-2} {GeV}^{-2};
\widetilde{b}_{\gm\gm} \sim  \frac{3}{64 \pi \msli^2}{\cal I}(1)\sim  10^{-23} {GeV}^{-2}.
\end{eqnarray}

\begin{figure}
\begin{center}
\begin{picture}(-200,100)(100,100)
\DashLine(-100,250)(-30,250)5
\Text(-115,250)[]{$\slp{a}(p_1)$}
\DashLine(-30,250)(40,250)5
\Text(55,250)[]{$h(p_3)$}
\DashLine(-30,250)(-30,160)5
\Text(-20,200)[]{$\slp{c}$}
\DashLine(-100,160)(-30,160)5
\Text(-115,160)[]{$\slp{b}^*(p_2)$}
\Photon(-30,160)(40,160){4}{2}
\Text(55,160)[]{$\gm(p_4)$}
\end{picture}
\end{center}
\vskip -4.5in
\begin{center}
\caption{Feynman diagrams for
$\slp{a} \slp{b}^*\ra h\gm $ via  t-channel $\slp{c}^*$ exchange.}
\end{center}
\end{figure}

${\bf (c) {\slp{a}\slp{b}^*\rightarrow \gm h}}$

$\bullet$
 \underline{slepton\ ($\slp{c}$) exchange:}

 With

\begin{eqnarray}
 {\widetilde{w}}_{\slp{a}\slp{b}^*\rightarrow \gm h}= {\widetilde{w}}_{\gm h}^{(\slp{})}:
\end{eqnarray}
\begin{eqnarray}
{\widetilde{w}}_{\gm h}^{(\slp{})} & = &
-2 \left| C^{\slp{b}^*\slp{a}\gm} \right|^2\left| C^{\slp{b}^*\slp{a}h} \right|^2
\left[ (\Tt_1+\msli^2\Tt_0)+(\Tu_1+\mslj^2\Tu_0) \right. \nonumber \\
 & & \left. + (s+\mh^2-2\msli^2-2\mslj^2)\Y_0 \right]\nonumber\\
 && \sim ({\tilde f} {\cal V}^{-19/15})^2 \left[ (\Tt_1+\msli^2\Tt_0)+(\Tu_1+\mslj^2\Tu_0) + (s+\mh^2-4\msli^2)\Y_0 \right] \nonumber
\end{eqnarray}
where
\begin{eqnarray}
&& \Tt_0\equiv{\cal V} \frac{s}{s {\cal V}^{\frac{42}{5}}+s \left({\cal V}^{21/5}-{\cal V}^{\frac{4}{5}}+s\right) {\cal V}^{\frac{21}{5}}+\left(-2 s {\cal V}^{\frac{21}{5}}+{\cal V}^{\frac{8}{5}}-s {\cal V}^{\frac{4}{5}}\right) {\cal V}^{21/5}}, \nonumber
\end{eqnarray}
\begin{eqnarray}
&& \Tu_0 \equiv{\cal V} \frac{s}{s {\cal V}^{\frac{42}{5}}+s \left({\cal V}^{\frac{21}{5}}-{\cal V}^{\frac{4}{5}}+s\right) {\cal V}^{\frac{21}{5}}+\left(-2 s {\cal V}^{\frac{21}{5}}+{\cal V}^{\frac{8}{5}}-s {\cal V}^{\frac{4}{5}}\right) {\cal V}^{\frac{21}{5}}}, \nonumber
\end{eqnarray}
\begin{eqnarray}
&& \Tt_1\equiv{\cal V} \frac{\log \left(\frac{s-{\cal V}^{\frac{4}{5}}-\sqrt{s-{\cal V}^{\frac{4}{5}}} \sqrt{1-\frac{{\cal V}^{\frac{4}{5}}}{s}} \sqrt{s-4 {\cal V}^{\frac{21}{5}}}}{s-{\cal V}^{\frac{4}{5}}+\sqrt{s-{\cal V}^{\frac{4}{5}}} \sqrt{1-\frac{{\cal V}^{\frac{4}{5}}}{s}} \sqrt{s-4
   {\cal V}^{\frac{21}{5}}}}\right) \left(s-{\cal V}^{\frac{4}{5}}\right)^2+s \sqrt{1-\frac{{\cal V}^{\frac{4}{5}}}{s}} \sqrt{s-4 {\cal V}^{\frac{21}{5}}} \sqrt{s-{\cal V}^{\frac{4}{5}}}}{\left(s-{\cal V}^{\frac{4}{5}}\right)^{5/2}
   \sqrt{1-\frac{{\cal V}^{\frac{4}{5}}}{s}} \sqrt{s-4 {\cal V}^{\frac{21}{5}}}}, \nonumber
\end{eqnarray}
\begin{eqnarray}
&& \Tu_1 \equiv{\cal V} \frac{\log \left(\frac{s-{\cal V}^{\frac{4}{5}}-\sqrt{s-{\cal V}^{\frac{4}{5}}} \sqrt{1-\frac{{\cal V}^{\frac{4}{5}}}{s}} \sqrt{s-4 {\cal V}^{\frac{21}{5}}}}{s-{\cal V}^{\frac{4}{5}}+\sqrt{s-{\cal V}^{\frac{4}{5}}} \sqrt{1-\frac{{\cal V}^{\frac{4}{5}}}{s}}
   \sqrt{s-4 {\cal V}^{\frac{21}{5}}}}\right) \left(s-{\cal V}^{\frac{4}{5}}\right)^2+s \sqrt{1-\frac{{\cal V}^{\frac{4}{5}}}{s}} \sqrt{s-4 {\cal V}^{\frac{21}{5}}} \sqrt{s-{\cal V}^{\frac{4}{5}}}}{\left(s-{\cal V}^{\frac{4}{5}}\right)^{5/2}
   \sqrt{1-\frac{{\cal V}^{\frac{4}{5}}}{s}} \sqrt{s-4 {\cal V}^{\frac{21}{5}}}},\nonumber
\end{eqnarray}
\begin{eqnarray}
&& \Y_0 \equiv{\cal V} \frac{2 \log \left(\frac{s-{\cal V}^{\frac{4}{5}}-\sqrt{s-{\cal V}^{\frac{4}{5}}} \sqrt{1-\frac{{\cal V}^{\frac{4}{5}}}{s}} \sqrt{s-4 {\cal V}^{\frac{21}{5}}}}{s-{\cal V}^{\frac{4}{5}}+\sqrt{s-{\cal V}^{\frac{4}{5}}} \sqrt{1-\frac{{\cal V}^{\frac{4}{5}}}{s}} \sqrt{s-4
   {\cal V}^{\frac{21}{5}}}}\right)}{\left(s-{\cal V}^{\frac{4}{5}}\right)^{3/2} \sqrt{1-\frac{{\cal V}^{\frac{4}{5}}}{s}} \sqrt{s-4 {\cal V}^{\frac{21}{5}}}}.\nonumber
\end{eqnarray}
For $s= 4 \msli^2$
\begin{eqnarray}
\widetilde{w}_{\gm h}|_{s=(4 \msli^2)}\sim {\cal V}^{-\frac{61}{5}}, \widetilde{w}^{\prime}_{ZZ}|_{s=(4 \msli^2)}\sim {\cal V}^{-13}.
\end{eqnarray}
The ``reduced'' coefficients $\widetilde{a}_{\gm h}$ and
$\widetilde{b}_{\gm h}$ will be given by
\begin{eqnarray}
\label{eq:agmh}
&& \widetilde{a}_{\gm h}  \sim  \frac{1}{32 \pi\msli^2} {\cal V}^{-\frac{61}{5}}\sim 10^{-83} {GeV}^{-2};
\widetilde{b}_{\gm h} \sim  \frac{3}{64 \pi \msli^2}{\cal V}^{-13}\sim  10^{-87} {GeV}^{-2}.
\end{eqnarray}

\begin{figure}
\begin{center}
\begin{picture}(150,100)(-100,100)
\DashLine(-100,250)(-30,200)5
\Text(-115,250)[]{$\slp{a}(p_1)$}
\DashLine(-100,150)(-30,200)5
\Text(-115,150)[]{$\slp{b}^*(p_2)$}
\DashLine(-30,200)(20,200)5
\Text(0,210)[]{$(h,H)$}
\DashLine(20,200)(90,250)5
\Text(110,250)[]{$h(p_3)$}
\DashLine(20,200)(90,150)5
\Text(110,150)[]{$h(p_4)$}
\end{picture}
{\hskip 2.0in}
\begin{picture}(150,100)(-100,100)
\DashLine(-100,250)(-30,250)5
\Text(-115,250)[]{$\slp{a}(p_1)$}
\DashLine(-30,250)(40,250)5
\Text(55,250)[]{$h(p_3)$}
\DashLine(-30,250)(-30,160)5
\Text(-20,200)[]{$\slp{c}$}
\DashLine(-100,160)(-30,160)5
\Text(-115,160)[]{$\slp{b}^*(p_2)$}
\DashLine(-30,160)(40,160)5
\Text(55,160)[]{$h(p_4)$}
\end{picture}
\caption{Feynman diagrams for
$\slp{a} \slp{b}^*\ra hh$ via $s$--channel Higgs exchange and t-channel $\slp{c}^*$ exchange.}
\end{center}
\end{figure}
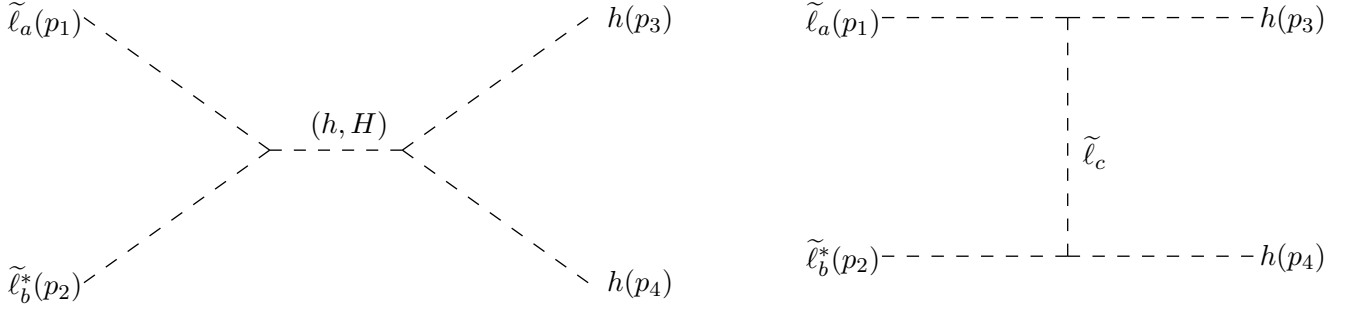

${\bf (d) {\slp{a}\slp{b}^*\rightarrow hh}}$

 Now,
\begin{eqnarray}
\label{eq:whh}
 {\widetilde{w}}_{\slp{a}\slp{b}^*\rightarrow hh}=
 {\widetilde{w}}_{hh}^{(h,H,P)} + {\widetilde{w}}_{hh}^{(\slp{})} + {\widetilde{w}}_{hh}^{(h,H,P-\slp{})}
\end{eqnarray}

where

$\bullet$
 \underline{Higgs ($h,H$) exchange ($+$ Point interaction):}

\begin{eqnarray}
{\widetilde{w}}_{hh}^{(h,H,P)} & = &
\left|
\sum_{r=h,H} \frac{C^{hHr}C^{\slp{b}^*\slp{a}r}}{\prop{r}}
-C^{\slp{b}^*\slp{a}hh}
\right|^2.
\end{eqnarray}

For $m_{h} =125 GeV, m_{H}\sim {\cal V}^{\frac{59}{72}} m_{\frac{3}{2}}$, $m_{\chi^0_3}\sim {\cal V}^{\frac{2}{3}}m_{\frac{2}{3}}\sim {\cal V}^{-\frac{4}{3}}M_p$,
and  $C^{hhh}\sim \frac{m^{2}_{h}}{v}, C^{\slp{a} \slp{b} h}\sim  C^{\slp{a} \slp{b} H}\sim
{\cal V}^{-\frac{34}{15}}, C^{\slp{a} \slp{b}hh}\sim
{\cal V}^{-\frac{41}{18}}$ from above, after simplifying, we have
\begin{eqnarray}
{\widetilde{w}}_{hh}^{(h,H,P)} & = &
\left|
 \frac{(10^2.{\cal V}^{-\frac{34}{15}}){m_{pl}}}{\prop{h}} + \frac{(10^2.{\cal V}^{-\frac{34}{15}})m_{pl}}{\prop{H}}
-{\cal V}^{-\frac{41}{18}}
\right|^2.
\end{eqnarray}

$\bullet$
 \underline{slepton\ ($\slp{c}$) exchange:}
\begin{eqnarray}
{\widetilde{w}}_{hh}^{(\slp{})} & = &
\sum_{c,d=1}^2
C^{\slp{c}^*\slp{a}h} C^{\slp{b}^*\slp{c}H}
C^{\slp{d}^*\slp{a}h *} C^{\slp{b}^*\slp{d}H *} \Tt_0
     + \sum_{c,d=1}^2 C^{\slp{c}^*\slp{a}H} C^{\slp{b}^*\slp{c}h}
C^{\slp{d}^*\slp{a}H *} C^{\slp{b}^*\slp{d}h *} \Tu_0 \nonumber \\
 & & - 2 \Re \sum_{c,d=1}^2
C^{\slp{c}^*\slp{a}h} C^{\slp{b}^*\slp{c}H}
C^{\slp{d}^*\slp{a}H *} C^{\slp{b}^*\slp{d}h *} \Y_0.
\end{eqnarray}
Strictly speaking we are considering first  generation of squarks which get identified with same modulus ${\cal A}_1$.
Therefore,
$$\sum_{c,d=1}^2
C^{\slp{c}^*\slp{a}h} C^{\slp{b}^*\slp{c}H}
C^{\slp{d}^*\slp{a}h *} C^{\slp{b}^*\slp{d}H *} \sim ({\cal V}^{-\frac{34}{15}})^4 \sim {\cal V}^{-9}{M_p^4}.$$
Hence
\begin{eqnarray}
{\widetilde{w}}_{hh}^{(\slp{})} & = & {\cal V}^{-9}(\Tt_0 +  \Tu_0 -\Y_0)M_p^4.\nonumber\\
\end{eqnarray}
The contribution of $\Tt_0, \Tu_0, \Y_0$ for  $\slp{a} \slp{b}^*\ra  hh$ is of the same order as $\slp{a} \slp{b}^* \ra ZZ $.

$\bullet$
 \underline{Higgs ($h,H$) ($+$ Point) -- slepton ($\slp{c}$) interference:}
\begin{eqnarray}
& & \hskip -1in {\widetilde{w}}_{hh}^{(h,H,P-\slp{})}  =
2 \Re \sum_{c=1}^2
   \left(
\sum_{r=h,H} \frac{C^{hHr}C^{\slp{b}^*\slp{a}r}}{\prop{r}}
-C^{\slp{b}^*\slp{a}hh}
\right)^*  \Big[
C^{\slp{c}^*\slp{a}h} C^{\slp{b}^*\slp{c}H} \Ft
+ C^{\slp{c}^*\slp{a}H} C^{\slp{b}^*\slp{c}h} \Fu \Big],
\end{eqnarray}
which expanding and incorporating in the values for the couplings,
\begin{eqnarray}
& & {\widetilde{w}}_{hh}^{(h,H,P-\slp{})}  =
2{\cal V}^{-\frac{68}{15}}
   \left(
 \frac{(10^2.{\cal V}^{-\frac{34}{15}})m_{pl}}{\prop{r}}+ \frac{(10^2.{\cal V}^{-\frac{34}{15}})m_{pl}}{\prop{r}}
-{\cal V}^{-\frac{41}{18}}
\right)^*  \Big[\Ft +  \Fu \Big]\nonumber\\
& & {\rm where} ~ \Ft=\Fu\equiv \frac{\log \left(\frac{-4 {\cal V}^2-2 {\cal V}^{2/5}+s-\sqrt{s-4 {\cal V}^{2/5}} \sqrt{s-12 {\cal V}^2}}{-4 {\cal V}^2-2
   {\cal V}^{2/5}+s+\sqrt{s-4 {\cal V}^{2/5}} \sqrt{s-12 {\cal V}^2}}\right)}{\sqrt{s-4 {\cal V}^{2/5}} \sqrt{s-12 {\cal V}^2}}.
   \end{eqnarray}
Summing up the contribution of ${\widetilde{w}}$ for all channels:
\begin{eqnarray}
\widetilde{w}_{hh}|_{s=(4 \msli^2)}\sim {\cal V}^{-3}, \widetilde{w}^{\prime}_{hh}|_{s=(4 \msli^2)}\sim {\cal V}^{-\frac{36}{5}},
\end{eqnarray}
and the ``reduced'' coefficients $\widetilde{a}_{hh}$ and
$\widetilde{b}_{hh}$ will be given by
\begin{eqnarray}
\label{eq:ahh}
&& \widetilde{a}_{hh}  \sim  \frac{1}{32 \pi \msli^2} {\cal V}^{-3}\sim 10^{-37} {GeV}^{-2};
\widetilde{b}_{hh} \sim  \frac{3}{64 \pi \msli^2}{\cal V}^{-\frac{36}{5}}\sim  10^{-48} {GeV}^{-2}
\end{eqnarray}

{\boldmath (e) ${\slp{a}\slp{b}\rightarrow \ell\ell}$}

This process proceeds only  $t$-- and $u$--channel neutralino ($\neu{i}$, $i=1,2,3$) exchange ($
 {\widetilde{w}}_{\slp{a}\slp{b}\rightarrow \ell\ell}={\widetilde{w}}_{\ell\ell}^{(\neu{})}
$).
\begin{figure}[t!]
\begin{center}
\begin{picture}(-200,100)(100,100)
\DashLine(-100,250)(-30,250)5
\Text(-115,250)[]{$\slp{a}(p_1)$}
\Line(-30,250)(40,250)
\Text(55,250)[]{$l(p_3)$}
\Line(-30,250)(-30,160)
\Text(-20,200)[]{$\neu{i}$}
\DashLine(-100,160)(-30,160)5
\Text(-115,160)[]{$\slp{b}^*(p_2)$}
\Line(-30,160)(40,160)
\Text(55,160)[]{$l(p_4)$}
\end{picture}
\end{center}
\vskip -0.5in
\caption{Feynman diagram for
$\slp{a} \slp{b}^*\ra hh$ via  t-channel $\slp{c}^*$ exchange.}
\end{figure}
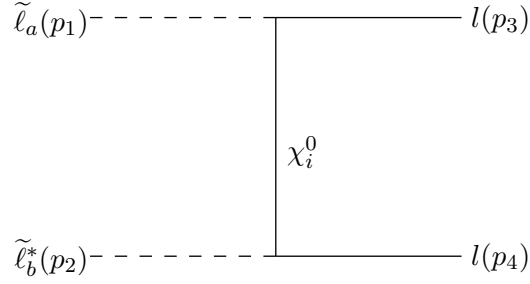

$\bullet$
\underline{neutralino\ ($\neu{i}$) exchange:}
{\small
\begin{eqnarray}
& & \hskip -1in {\widetilde{w}}_{\slp{a}\slp{b}\rightarrow \ell\ell}^{(\neu{})}  =
\sum_{i,j=1}^3 \Bigg[
\Dijqp{LLLL} \mnmn (s-2\ml^2)\T_0  -2\ml^2 \Big[ \Dijqp{LLRR}\mnmn\T_0+\Dijqp{LRRL}\T_1 \Big] + \Dijqp{LRLR}\Big[ -\T_2
  -(s-\msli^2-\mslj^2)\T_1\nonumber\\
  & &\hskip -1in -(\msli^2-\ml^2)(\mslj^2-\ml^2)\T_0 \Big]   -\ml\mnq \Big[ \Dijqp{LLLR} \{ \T_1-(\msli^2-\ml^2)\T_0 \}
     + \Dijqp{LLRL} \{ \T_1-(\mslj^2-\ml^2)\T_0 \} \Big] \nonumber \\
 & & \hskip-1in -\ml\mnp \Big[ \Dijqp{LRLL} \{ \T_1-(\msli^2-\ml^2)\T_0 \}
     + \Dijqp{RLLL} \{ \T_1-(\mslj^2-\ml^2)\T_0 \} \Big] \Bigg]\nonumber\\
& & \hskip -1in  + \frac{1}{2}\sum_{i,j=1}^3
\Bigg[
-2 \Dijqp{LLLL} \mnmn (s-2\ml^2)\Y_0  +4\Dijqp{LLRR}\mnmn\ml^2\Y_0
     -2\Dijqp{LRLR}\ml^2(\slpsl{2}-2\ml^2)\Y_0 \nonumber\\
      & & + 2 \Dijqp{LRRL}\Big[ -\Y_2 -(\slsl{2}-\ml^4)\Y_0 \Big] -\ml\mnq \Big[ \Dijqp{LLLR} \{ \Y_1+(s+\slmsl{2}-4\ml^2)\Y_0 \}
      \nonumber \\
 & & \hskip -1in
     + \Dijqp{LLRL} \{ \Y_1+(s-\msli^2+\mslj^2-4\ml^2) \Y_0 \}
     \Big]
     +\ml\mnp \Big[ \Dijqp{LRLL} \{ \Y_1-(s+\slmsl{2}-4\ml^2)\Y_0 \}\nonumber\\
& &   \hskip -1in     + \Dijqp{RLLL} \{ \Y_1-(s-\msli^2+\mslj^2-4\ml^2) \Y_0 \}
\Big] \Bigg],
\end{eqnarray}}
where
\begin{eqnarray}
\Dijqp{LLLL} & = &
           C^{\neu{i}\slp{a}^*\ell *}_L C^{\neu{i}\slp{b}^*\ell *}_L
           C^{\neu{j}\slp{b}^*\ell  }_L C^{\neu{j}\slp{a}^*\ell  }_L
                    + (L \rightarrow R), \nonumber \\
\Dijqp{LLRR} & = &
           C^{\neu{i}\slp{a}^*\ell *}_L C^{\neu{i}\slp{b}^*\ell *}_L
           C^{\neu{j}\slp{b}^*\ell  }_R C^{\neu{j}\slp{a}^*\ell  }_R
                    + (L \leftrightarrow R), \nonumber \\
\Dijqp{LRRL} & = &
           C^{\neu{i}\slp{a}^*\ell *}_L C^{\neu{i}\slp{b}^*\ell *}_R
           C^{\neu{j}\slp{b}^*\ell  }_R C^{\neu{j}\slp{a}^*\ell  }_L
                    + (L \leftrightarrow R), \nonumber \\
\Dijqp{LRLR} & = &
           C^{\neu{i}\slp{a}^*\ell *}_L C^{\neu{i}\slp{b}^*\ell *}_R
           C^{\neu{j}\slp{b}^*\ell  }_L C^{\neu{j}\slp{a}^*\ell  }_R
                    + (L \leftrightarrow R), \nonumber \\
\Dijqp{LLLR} & = &
           C^{\neu{i}\slp{a}^*\ell *}_L C^{\neu{i}\slp{b}^*\ell *}_L
           C^{\neu{j}\slp{b}^*\ell  }_L C^{\neu{j}\slp{a}^*\ell  }_R
                    + (L \leftrightarrow R), \nonumber \\
\Dijqp{LLRL} & = &
           C^{\neu{i}\slp{a}^*\ell *}_L C^{\neu{i}\slp{b}^*\ell *}_L
           C^{\neu{j}\slp{b}^*\ell  }_R C^{\neu{j}\slp{a}^*\ell  }_L
                    + (L \leftrightarrow R), \nonumber \\
\Dijqp{LRLL} & = &
           C^{\neu{i}\slp{a}^*\ell *}_L C^{\neu{i}\slp{b}^*\ell *}_R
           C^{\neu{j}\slp{b}^*\ell  }_L C^{\neu{j}\slp{a}^*\ell  }_L
                    + (L \leftrightarrow R), \nonumber \\
\Dijqp{RLLL} & = &
           C^{\neu{i}\slp{a}^*\ell *}_R C^{\neu{i}\slp{b}^*\ell *}_L
           C^{\neu{j}\slp{b}^*\ell  }_L C^{\neu{j}\slp{a}^*\ell  }_L
                    + (L \leftrightarrow R).
\end{eqnarray}
Given that $m_{\chi^0_3}\sim {\cal V}^{-\frac{4}{3}}M_p, m_{\chi^0_1}=  m_{\chi^0_2}\sim {\cal V}^{-1}M_p$, $ C^{\neu{i}\slp{a}^*\ell *}_R \sim {\tilde f} {\cal V}^{-\frac{1}{2}}, C^{\neu{i}\slp{a}^*\ell *}_R \sim {\tilde f} {\cal V}^{-\frac{12}{15}}$, by expanding summation, the above expression reduces to
\begin{eqnarray}
&& {\widetilde{w}}_{\slp{a}\slp{b}\rightarrow \ell\ell}^{(\neu{})} \sim
\frac{\sqrt{s} \sqrt{s-4 {\cal V}^{\frac{21}{5}}} \left(8s {\cal V}^2- 160 {\cal V}^{\frac{7}{5}}\right)- 2 \left(-160
{\cal V}^{\frac{7}{5}} {\cal V}^{\frac{21}{5}}+9 s^2+s 8 {\cal V}^{\frac{31}{5}}\right) \log
   \left(\frac{\sqrt{s}-\sqrt{s-4 {\cal V}^{\frac{21}{5}}}}{\sqrt{s}+\sqrt{s-4 {\cal V}^{\frac{21}{5}}}}\right)}{2 s^{3/2} {\cal V}^{\frac{29}{5}} \sqrt{s-4 {\cal V}^{\frac{21}{5}}}}. \nonumber
\end{eqnarray}

For $s=4\msli^2$
\begin{eqnarray}
\widetilde{w}_{\ell\ell}|_{s=(4\msli^2)}\sim {\cal V}^{-\frac{19}{5}}, \widetilde{w}^{\prime}_{\ell\ell}|_{s=(4 \msli^2)}\sim {\cal V}^{-8}.
\end{eqnarray}
The ``reduced'' coefficients $\widetilde{a}_{\ell\ell}$ and
$\widetilde{b}_{\ell\ell}$ will be given by
\begin{eqnarray}
\label{eq:aellell}
&& \widetilde{a}_{\ell\ell}  \sim  \frac{1}{32 \pi \msli^2} {\cal V}^{-\frac{19}{5}}\sim 10^{-41} {GeV}^{-2};
\widetilde{b}_{\ell\ell} \sim  \frac{3}{64 \pi \msli^2}{\cal V}^{-8}\sim  10^{-62} {GeV}^{-2}.
   \end{eqnarray}

 Summing up the contribution of partial wave coefficients for each possible annihilation processes:
\begin{eqnarray}
a_{{\slp{a}\slp{b}^*\rightarrow f_1f_2}}  &=&
\widetilde{a}_{ZZ}+ \widetilde{a}_{hZ}+ \widetilde{a}_{h\gm}+ \widetilde{a}_{\gm\gm}+ \widetilde{a}_{hh}+ \widetilde{a}_{ll} \equiv  O(10)^{11} {GeV}^{-2} \nonumber \\
b_{{\slp{a}\slp{b}^*\rightarrow f_1f_2}}  &=& \widetilde{b}_{ZZ}+ \widetilde{b}_{hZ}+ \widetilde{b}_{h\gm}+ \widetilde{b}_{\gm\gm}+ \widetilde{b}_{hh}+ \widetilde{b}_{ll} \equiv  O(10)^{-10} {GeV}^{-2},
\end{eqnarray}
and
\begin{equation}
\jxf\equiv \int_0^{x_f}dx \langle\sigma v_{\rm M\o l}\rangle(x)= \int_0^{x_f}dx  (a + b{x_f})= a{x_f} + b\frac{{x^2_f}}{2},\nonumber
\end{equation}
where $x=T/\msli$.

Using the analytical expression of relic abundance as given in (\ref{eq:relic-density})
\begin{eqnarray}
\label{eq:relic-density1}
\Omega_{\slp{a}} & = & \frac{1}{{\mu}^2 \sqrt{g_*}J({x_F})}
\end{eqnarray}
where ${\mu}= 1.2 \times 10^5 GeV$. For $J(x_f) \sim a(x_f) + b\frac{{x_f}^2}{2}\sim  10^{11} {x_f}{GeV}^{-2}$ and $x_f= \frac{1}{33}$, $ \sqrt{g_*}=9$
\begin{eqnarray}
\Omega_{\slp{a}}\sim\frac{33}{1.44 \times 10^{10}\cdot 9\cdot 10^{11}}\equiv 10^{-20},
\end{eqnarray}
for $m_{\frac{3}{2}}\sim 10^8 GeV$ and $\msli\sim {\cal V}^{\frac{1}{2}}m_{\frac{3}{2}}$, relic abundance of gravitino will be given as :
\begin{eqnarray}
\label{eq:omega2}
\Omega_{\tilde G}& = & \Omega_{\slp{a}}\times \frac{m_{\frac{3}{2}}}{\msli}=10^{-20} \times {\cal V}^{-\frac{1}{2}} \sim 10^{-22}~{\rm  for}~ {\cal V}\sim 10^5.
\end{eqnarray}
From (\ref{eq:omega1}) and (\ref{eq:omega2}), it appears that relic abundance of gravitino turns out to be extremely suppressed in case of slepton (NLSP) (co-)annihilations as compared to relic abundance of gravitino in case of neutralino (NLSP) aannihilations, for almost similar value of thermal cross-section. 

\section{Results and Discussion}

In most of the supersymmetric models, gravitino appears as most natural candidate of dark matter, the stability of which is governed by R-parity conservation that was initially proposed to explain the non-observation of proton decays in collider physics. However in split SUSY models, the constraints appearing from R-parity conservation are relaxed because of the fact that high scalar masses help to avoid the fast proton decay. Therefore it is interesting to see if gravitino appears as a stable dark matter candidate by taking into consideration trilinear R-parity violating couplings.

Large volume Swiss-Cheese type IIB compactifications \cite{LVS} have been of wide interest, both from cosmological as well as phenomenological points of view. In this paper, we have investigated, in detail, the possibility of gravitino as a viable dark matter candidate in the framework  of ``L(arge) V(olume) $\mu$- split SUSY" scenario. The presented scenario includes four Wilson line moduli on the world volume of space-time filling $D7$-branes wrapped around the ``big divisor" and two position moduli of a mobile space-time filling $D3$-brane restricted to (nearly) a special Lagrangian sub-manifold.   We have shown that the ${\cal N}=1$ gauged supergravity  fermionic mass term corresponding to fermionic super-partners of two (${\cal A}_1$ and ${\cal A}_3$) of the aforementioned four Wilson line moduli  matches very well with the order of Dirac mass of the electron, and the ${\cal N}=1$ gauged supergravity  fermionic mass term corresponding to fermionic super-partners of the remaining two (${\cal A}_2$ and ${\cal A}_4$) of the aforementioned four Wilson line moduli,  matches very well with the order of Dirac masses of the first generation SM-like quarks, such that the fermionic superpartners of ${\cal A}_1$ and ${\cal A}_3$ get identified, respectively with $e_L$ and $e_R$, and the fermionic superpartners of ${\cal A}_2$ and ${\cal A}_4$ get identified, respectively with the first generation quarks: $u/d_L$ and $u/d_R$. In doing so we also show that the RG-flow equations for the ``effective" Yukawa couplings relevant to the fermionic mass calculations, change by ${\cal O}(1)$ in flowing from the string to the EW scale; it has been shown in \cite{ferm_masses_MS} that with some fine tuning, it is possible to RG flow the value of $\langle z_i\rangle$ (identified with Higgs VEV) $\sim V^{\frac{1}{36}}M_p$ at string scale to $\langle z_i\rangle \sim 246MeV$ at the EW scale. In fact, using the RG-flow arguments of \cite{ferm_masses_MS},  one can also show that the  Weinberg-type dimension-five Majorana-mass generating operator: ${\cal O}(\langle z_i\rangle^2)$ coefficient in $
\frac{e^{\frac{K}{2}}
\frac{\partial^2W}{\partial{\cal A}_1^2}}{\sqrt{K_{{\cal Z}_i\bar{\cal Z}_i}^2K_{{\cal A}_1\bar{\cal A}_1}^2}}\left(\bar{\chi}_L^{{\cal A}_1}{\cal Z}_i\right)^2$ or in fact $\frac{e^{\frac{K}{2}}{\cal D}_{\bar{{\cal A}}_1}D_{{\cal A}_1}\bar{W}}{\sqrt{K_{{\cal Z}_i\bar{\cal Z}_i}^2K_{{\cal A}_1\bar{\cal A}_1}^2}}\left(\bar{\chi}_L^{{\cal A}_1}{\cal Z}_i\right)^2$ produces the correct first-generation neutrino mass scale of slightly less than $1eV$ for $\langle z_i\rangle\sim{\cal O}(1){\cal V}^{\frac{1}{36}}$.

 We also show that it is possible to generate the massive gauge boson mass scales corresponding to the W/Z-bosons, under a similar RG flow.  Building up on this set up, we have also evaluated masses of all supersymmetric as well as soft SUSY breaking parameters obtainable from bulk $F$-terms in the context of gravity mediation. Using  RG-flow  solutions of scalar masses, similar to the ones for a single Wilson line modulus  set up of \cite{Sparticles_Misra_Shukla}, one can again show that values of same do not change from string scale down to EW scale. On evaluation of the soft supersymmetry breaking parameters, we find a universality in the squarks/sleptons masses and assuming
non-universality in the $D3$-brane position moduli ( to be identified with the neutral components of the Higgs doublets) masses, on diagonalizing the Higgs mass matrix we obtain, at the EW scale, one light Higgs of the order $125 GeV$ and one heavy Higgs  while the Higgsino mass parameter comes out to be heavy. The lightest neutralino is predominantly gaugino(Wino/Bino)-type  formed by linear a combination of gaugino and  Higgsino. The mass scales of various SM and superpartners are given in table \ref{table:mass scales}. After calculating the masses of various SM and their superpartners, it appears that gravitino is the Lightest Supersymmetric particle (LSP) for Calabi-Yau volume ${\cal V} \sim 10^5$ which motivates the query: can we have gravitino DM in gravity mediation scenarios ?

In big bang cosmology, gravitino population depends on two kinds of mechanisms: thermal as well as non-thermal production. We assume that reheating temperature is low enough to produce the effective  relic abundance of gravitino in agreement with experimental observations, therefore almost all of the gravitinos are produced by electromagnetic as well as hadronic decays of unstable NLSP. The scale of masses of various superpartners suggest that because of an ${\cal O}(1)$ difference between the masses of the sleptons/squarks and the lightest neutralino and given that in our calculations we have not bothered about such ${\cal O}(1)$ factors, both can exist as valid  NLSP candidate if life-times of the same are  such that they do not spoil the bounds given by Big Bang Nucleosynthesis(BBN). Based on this fact, by considering the contributions of all required couplings in ${\cal N}=1$ gauged supergravity action and assuming that similar to the aforementioned effective Yukawa couplings, there is only an ${\cal O}(1)$ multiplicative change in the three-point interaction vertices in RG-flowing  from the string to the EW scales, in section {\bf 3}, we have estimated the mean lifetime by calculating two-body and three-body decay widths of Gravitino decaying into SM particles, which in case of two-body decays, comes out to be $10^{17}$ seconds and in case of three-body decays, comes out to be $10^{21}$ seconds, i.e., greater than the age of the universe and hence satisfies the requirement of  an appropriate DM candidate. In section {\bf 4.1}  we have discussed in detail the leading two body decays of gaugino/neutralino $\tilde{B}\rightarrow\psi_\mu Z, \mu \gamma $ and three body  $\lambda^0_{1,2}\rightarrow\psi_\mu u {\bar u}, \psi_\mu W^+W^-$ decays and  determine that life-time of these decays comes out to be too short to affect the prediction of BBN. In  {\bf 4.1.2}, we have (re)calculated the tree-level as well as one-loop gluino decays into neutralino as well as Gravitino/Goldstino in four-Wilson-line-moduli set up, large life time(s) of which satisfy one of the important phenomenological features of ``$\mu$- split SUSY" but can not be considered as an appropriate NLSP's because late decays of the same into Goldstino(longitudinal component of Gravitino) can surely elude the constraints coming from BBN. Furthermore,  since we are considering  R-parity violating couplings into account, in addition to NLSP decaying into LSP, there might be chances that neutralino(NLSP) directly decays into SM particles via R-parity violating couplings and hence affect the  relic abundance of gravitino produced by neutralino. To ensure that this does not happen, in {\bf 4.2}, we have studied in detail the three-body decay width of three-body decay of neutralino into SM particles ($\chi^0_3\rightarrow u_L \bar d_R l_L$), which comes out to be very much less than the two- and three-body decay widths of ordinary neutralino decays into LSP. Therefore, we show that R-parity violating couplings do not affect relic abundance of gravitino.  On a similar ground, we have calculated three body decay width of sleptons $ \tilde{l}\rightarrow \tilde{G}l, \tilde{l}\rightarrow l^\prime \tilde{G} V$, which similar to neutralino decays are short enough to affect bounds on BBN.  The numerical estimates of various N(LSP)'s are provided in table \ref{table:decay_lifetime}. In short, the explicit calculation of life times of both, LSP and NLSP, confirms that gravitino can be considered as an appropriate dark matter candidate. However this is not the end of the story. The viable dark matter candidate should also have the right order of relic abundance in the range provided by WMAP data and other direct as well as indirect experiments. With the assumption that the next-lightest supersymmetric particle (NLSP) freezes out with its thermal relic density before decaying to the gravitino and then eventually decays to gravitino, relic density of gravitino will be completely given in terms of relic density of NLSP. Therefore, in section {\bf 5}, utilizing the partial  wave expansion approach, we explicitly calculate thermal annihilation cross-section of  sleptons and neutralino which ultimately give gravitino relic abundance from sleptons to be extremely suppressed while the same from neutralino comes out to be $0.16$, almost in agreement with the value of $\Omega_{C}h^2$ suggested by the WMAP 7-year CMB anisotropy observation \cite{C.L.Bennett}.

We repeatedly mention that all the above results are simply dependent on dilute  $\tilde f$-fluxes and Calabi-Yau volume ${\cal V}$ which we have fixed to be around $10^{-4}$ and $10^{5}$ respectively in the paper. The lower bound on $\tilde f\sim 10^{-4}$ appears by imposing that the flux-dependent $D$-term potential is sub-dominant as compared to the $F$-term potential in the dilute flux approximation (see \cite{Dhuria+Misra_mu_Split SUSY}). Though one could have chosen a viable range of $\tilde f$s satisfying the lower bound of $\tilde f\sim 10^{-4}$, the justification behind choosing a particular value of $\tilde f \sim 10^{-4}$  is because of getting the right amount of relic abundance of Gravitino. The justification behind constraining  a value of Calabi-Yau ${\cal V}$ to be of the order $10^5$ is based on the right identification of Wilson line moduli and position moduli with Standard Model particle spectrum. The masses of first-generation leptons and quarks are generated from Yukawa couplings only for a particular value of Calabi-Yau ${\cal V}\sim 10^5$ in the set-up.

An interesting outcome of our current project is that for trilinear R-parity violating couplings $\lambda'_{ijk}$ (to be interpreted as an effective Yukawa couplings in ${\cal N}=1$ gauged supergravity evaluated in {\bf 4}) $\sim {\cal V}^{-\frac{5}{3}}$ and $\lambda''_{ijk}$ (to be interpreted as an effective Yukawa couplings in ${\cal N}=1$ gauged supergravity evaluated in {\bf 4}) $\sim {\cal V}^{-\frac{43}{30}}$, according to the analytic expression given in \cite{susy_primer_S.P.Martin}, the rough estimate of proton decay width comes out to be
\begin{equation}
\Gamma_{p\rightarrow e^{+}{\pi}^0}\sim {m^{5}_{proton}}\frac{\left|\lambda'_{ijk}\lambda''_{ijk}\right|^2}{m^4_{{\tilde q}_I}}\sim \frac{{\cal V}^{-6.1}}{m^4_{{\tilde q}_I}}\sim 10^{-73} GeV,
\end{equation} thus giving life time of about $10^{42}$ years, which explicitly governs the stability of proton in the presence of small R-parity violating couplings discussed in $\mu$-split SUSY scenario.

To summarize, we conclude that the gravitino qualifies as a potential dark matter candidate in  Large volume``$\mu$ split SUSY" scenario.

\section*{Acknowledgements}

MD is supported by a CSIR Senior Research Fellowship. One of us (MD) would like to thank DESY, Hamburg, for financial support and hospitality where part of this work was done, as well as the opportunity to present the results at the International School on Strings and Fundamental Physics, DESY, Hamburg, July 1 - 13, 2012. AM would like to thank Purdue University, McGill University and the Abdus Salam ICTP (under the regular associateship program) for their kind hospitality and support where part of this work was done, and would also thank P.Ouyang for informative discussions. We would also acknowledge participation of J.Beuria in earlier stages of the project.

\appendix
\setcounter{equation}{0} \seceqaa
\section{Geometric K\"{a}hler Potential}

The crux of the Donaldson's algorithm is that the sequence
$$\frac{1}{k\pi}\partial_i{\bar\partial}_{\bar j}\left(ln\sum_{\alpha,\beta}h^{\alpha{\bar\beta}}s_\alpha{\bar s}_{\bar\beta}\right)$$
on $P(\{z_i\})$, in the $k\rightarrow\infty$-limit - which in practice implies $k\sim10$ - converges to a unique Calabi-Yau metric for the given K\"{a}hler class and complex structure; $h_{\alpha{\bar\beta}}$ is a balanced metric on the line bundle ${\cal O}_{P(\{z_i\})}(k)$ (with sections $s_\alpha$) for any $k\geq1$, i.e., $$T(h)_{\alpha{\bar\beta}}\equiv \frac{N_k}{\sum_{j=1}w_j}\sum_{i}\frac{s_\alpha(p_i)\overline{s_\beta(p_i)}w_i}
{h^{\gamma{\bar\delta}}
s_\gamma(p_i)\overline{s_\delta(p_i)}}=h_{\alpha{\bar\beta}},$$ where the weight at point $p_i$, $w_i\sim\frac{i^*(J_{GLSM}^3)}{\Omega\wedge{\bar\Omega}}$ with the embedding map $i:P(\{z_i\})\hookrightarrow{\bf WCP}^4$ and the number of sections is denoted by $N_k$.  The defining hypersurface of the Swiss-Cheese Calabi-Yau in the $x_2=1$-coordinate patch in ${\bf WCP}^4[1,1,1,6,9]$ is given by:
\begin{equation}
\label{eq:hyp_def}
1+z_1^{18}+z_2^{18}+z_3^3+z_4^2-\psi z_1z_2z_3z_4-3\phi z_1^6z_2^6=0.
\end{equation}
In the large volume limit, the above can be satisfied if, e.g., $1+z_1^{18}+z_2^{18}\sim-z_3^3$,
$z_4^2\sim\psi z_1z_2z_3z_4+3\phi z_1^6z_2^6$. For $z_{1,2}\sim{\cal V}^{\frac{1}{36}}$, one sees the same are satisfied for $z_{3,4}\sim {\cal V}^{\frac{1}{6}}$ provided $\psi {\cal V}^{\frac{1}{18}}\sim 3\phi$. Therefore:
\begin{eqnarray}
\label{eq:sec_metr_h}
& & h_{1{\bar z}_i}\sim\frac{{\cal V}^{\frac{1}{36}}}{h^{z_4^2{\bar z}_4^2}{\cal V}^{\frac{2}{3}}},\  h_{1{\bar z}_4}\sim\frac{{\cal V}^{\frac{1}{6}}}{h^{z_4^2{\bar z}_4^2}{\cal V}^{\frac{2}{3}}}\nonumber\\
& & h_{1{\bar z}_i^2}\sim\frac{{\cal V}^{\frac{1}{18}}}{h^{z_4^2{\bar z}_4^2}{\cal V}^{\frac{2}{3}}},\  h_{1{\bar z}_4^2}\sim\frac{{\cal V}^{\frac{1}{3}}}{h^{z_4^2{\bar z}_4^2}{\cal V}^{\frac{2}{3}}}\nonumber\\
& & h_{1{\bar z}_i{\bar z}_4}\sim\frac{{\cal V}^{\frac{1}{6}+\frac{1}{36}}}{h^{z_4^2{\bar z}_4^2}{\cal V}^{\frac{2}{3}}},\  h_{z_i{\bar z}_j}\sim\frac{{\cal V}^{\frac{1}{18}}}{h^{z_4^2{\bar z}_4^2}{\cal V}^{\frac{2}{3}}}\nonumber\\
& & h_{z_i{\bar z}_4}\sim\frac{{\cal V}^{\frac{1}{36}+\frac{1}{6}}}{h^{z_4^2{\bar z}_4^2}{\cal V}^{\frac{2}{3}}},\  h_{z_i{\bar z}_j{\bar z}_k}\sim\frac{{\cal V}^{\frac{1}{12}}}{h^{z_4^2{\bar z}_4^2}{\cal V}^{\frac{2}{3}}}\nonumber\\
& & h_{z_i{\bar z}_4^2}\sim\frac{{\cal V}^{\frac{1}{36}+\frac{1}{3}}}{h^{z_4^2{\bar z}_4^2}{\cal V}^{\frac{2}{3}}},\  h_{z_i{\bar z}_j{\bar z}_4}\sim\frac{{\cal V}^{\frac{1}{18}+\frac{1}{6}}}{h^{z_4^2{\bar z}_4^2}{\cal V}^{\frac{2}{3}}}\nonumber\\
& & h_{z_4{\bar z}_4}\sim\frac{{\cal V}^{\frac{1}{3}}}{h^{z_4^2{\bar z}_4^2}{\cal V}^{\frac{2}{3}}},\  h_{z_4{\bar z}_i^2}\sim\frac{{\cal V}^{\frac{1}{18}+\frac{1}{6}}}{h^{z_4^2{\bar z}_4^2}{\cal V}^{\frac{2}{3}}}\nonumber\\
& & h_{z_4{\bar z}_4^2}\sim\frac{\sqrt{\cal V}}{h^{z_4^2{\bar z}_4^2}{\cal V}^{\frac{2}{3}}},\ h_{z_4{\bar z}_i{\bar z}_4}\sim\frac{{\cal V}^{\frac{1}{36}+\frac{1}{3}}}{h^{z_4^2{\bar z}_4^2}{\cal V}^{\frac{2}{3}}}\nonumber\\
& & h_{z_iz_j{\bar z}_k{\bar z}_l}\sim\frac{{\cal V}^{\frac{1}{9}}}{h^{z_4^2{\bar z}_4^2}{\cal V}^{\frac{2}{3}}},\  h_{z_iz_j{\bar z}_4^2}\sim\frac{{\cal V}^{\frac{1}{18}+\frac{1}{3}}}{h^{z_4^2{\bar z}_4^2}{\cal V}^{\frac{2}{3}}}\nonumber\\
& & h_{z_iz_j{\bar z}_k{\bar z}_4}\sim\frac{{\cal V}^{\frac{1}{12}+\frac{1}{6}}}{h^{z_4^2{\bar z}_4^2}{\cal V}^{\frac{2}{3}}},\  h_{z_4^2{\bar z}_4^2}\sim\frac{1}{h^{z_4^2{\bar z}_4^2}}\nonumber\\
& & h_{z_4^2{\bar z}_i{\bar z}_4}\sim\frac{{\cal V}^{\frac{1}{36}+\frac{1}{2}}}{h^{z_4^2{\bar z}_4^2}{\cal V}^{\frac{2}{3}}},\  h_{z_iz_4{\bar z}_i{\bar z}_4}\sim\frac{{\cal V}^{\frac{1}{18}+\frac{1}{3}}}{h^{z_4^2{\bar z}_4^2}{\cal V}^{\frac{2}{3}}},
\end{eqnarray}
which on being inverted gives:
\begin{eqnarray}
\label{eq:metsecinv}
& h^{\alpha{\bar\beta}}\sim & \left(\begin{array}{cccccc}
h^{11} & h^{1{\bar z}_i} & h^{1{\bar z}_4} & h^{1{\bar z}_i^2} & h^{1{\bar z}_4^2} & h^{1{\bar z}_i{\bar z}_4}\\
h^{z_i1} & h^{z_i{\bar z}_j} & h^{z_i{\bar z}_4} & h^{z_i{\bar z}_j{\bar z}_k} & h^{z_i{\bar z}_4^2} & h^{z_i{\bar z}_j{\bar z}_4}\\
h^{z_41} & h^{z_4{\bar z}_i} & h^{z_4{\bar z}_4} & h^{z_4{\bar z}_i{\bar z}_j} & h^{z_4{\bar z}_4^2} & h^{z_4{\bar z}_i{\bar z}_4}\\
h^{z_iz_j1} & h^{z_iz_j{\bar z}_k} & h^{z_iz_j{\bar z}_4} & h^{z_iz_j{\bar z}_k{\bar z}_l} & h^{z_iz_j{\bar z}_4^2} & h^{z_iz_j{\bar z}_k{\bar z}_4}\\
 h^{z_4^21}& h^{z_4^2{\bar z}_i} & h^{z_4^2{\bar z}_4} & h^{z_4^2{\bar z}_i{\bar z}_j} & h^{z_4^2{\bar z}_4^2} & h^{z_4^2{\bar z}_i{\bar z}_4}\\
h^{z_iz_41} & h^{z_iz_4{\bar z}_k} & h^{z_iz_4{\bar z}_4} & h^{z_iz_4{\bar z}_j{\bar z}_k} & h^{z_iz_4{\bar z}_4^2} & h^{z_iz_4{\bar z}_j{\bar z}_4}
\end{array}\right)\nonumber\\
&  & \left(
\begin{array}{llllll}
 h^{z_4^2{\bar z}_4^2} V^{2/3} & h^{z_4^2{\bar z}_4^2} V^{23/36} & h^{z_4^2{\bar z}_4^2} \sqrt{V} & h^{z_4^2{\bar z}_4^2} V^{11/18} & h^{z_4^2{\bar z}_4^2} \sqrt[3]{V} & h^{z_4^2{\bar z}_4^2} V^{17/36} \\
 h^{z_4^2{\bar z}_4^2} V^{23/36} & h^{z_4^2{\bar z}_4^2} V^{11/18} & h^{z_4^2{\bar z}_4^2} V^{17/36} & h^{z_4^2{\bar z}_4^2} V^{7/12} & h^{z_4^2{\bar z}_4^2} V^{11/36} & h^{z_4^2{\bar z}_4^2} V^{4/9} \\
 h^{z_4^2{\bar z}_4^2} \sqrt{V} & h^{z_4^2{\bar z}_4^2} V^{17/36} & h^{z_4^2{\bar z}_4^2} \sqrt[3]{V} & h^{z_4^2{\bar z}_4^2} V^{4/9} & h^{z_4^2{\bar z}_4^2} \sqrt[6]{V} & h^{z_4^2{\bar z}_4^2} V^{11/36} \\
 h^{z_4^2{\bar z}_4^2} V^{11/18} & h^{z_4^2{\bar z}_4^2} V^{7/12} & h^{z_4^2{\bar z}_4^2} V^{4/9} & h^{z_4^2{\bar z}_4^2} V^{5/9} & h^{z_4^2{\bar z}_4^2} V^{5/18} & h^{z_4^2{\bar z}_4^2} V^{5/12} \\
 h^{z_4^2{\bar z}_4^2} \sqrt[3]{V} & h^{z_4^2{\bar z}_4^2} V^{11/36} & h^{z_4^2{\bar z}_4^2} \sqrt[6]{V} & h^{z_4^2{\bar z}_4^2} V^{5/18} & h^{z_4^2{\bar z}_4^2} & h^{z_4^2{\bar z}_4^2} V^{5/36} \\
 h^{z_4^2{\bar z}_4^2} V^{17/36} & h^{z_4^2{\bar z}_4^2} V^{4/9} & h^{z_4^2{\bar z}_4^2} V^{11/36} & h^{z_4^2{\bar z}_4^2} V^{5/12} & h^{z_4^2{\bar z}_4^2} V^{5/36} & h^{z_4^2{\bar z}_4^2} V^{5/18}
\end{array}
\right).\nonumber\\
& &
\end{eqnarray}
Using (\ref{eq:metsecinv}), one hence obtains the  K\"{a}hler potential ansatz of (\ref{eq:K}).

\setcounter{equation}{0} \seceqbb

\section{Evaluation of intersection numbers $C^{I{\bar J}}$}
The contribution of various intersection numbers $C^{I{\bar J}}$ corresponding to four wilson line moduli formed by including four D7- branes overlapping the big "Divisor" can be calculated by evaluating  $C^{I{\bar J}}_\alpha=\int_{\Sigma^B}i^*\omega_\alpha\wedge A^I\wedge A^{\bar J}$ where values of $A^{I=1,2,3,4}$'s are given in equation no (\ref{eq:A_123}). Utilizing their expressions, resultant contribution of various $C_{I\bar J}$'s are given as below:
\begin{equation}
\label{eq:C11bar_i}
C_{1{\bar 1}}\sim\frac{1}{{\cal V}}\int z_1^{18}z_2^{19}{\bar z}_1^{18}{\bar z}_2^{19}
 \delta\left(|z_1|-{\cal V}^{\frac{1}{36}}\right)\delta\left(|z_2|-{\cal V}^{\frac{1}{36}}\right)
 dz_1\wedge d{\bar z}_1\wedge dz_2\wedge d{\bar z}_2,
 \end{equation}
 which using $dz_1\wedge d{\bar z}_1\wedge dz_2\wedge d{\bar z}_2\sim |z_1|d|z_1|\wedge d(arg z_1)\wedge  |z_2|d|z_2|\wedge d(arg z_2)$ yields:
 \begin{eqnarray*}
 \label{eq:CIJbar_I}
& & C_{1{\bar 1}}\sim\frac{1}{\cal V}\int |z_1|^{36}|z_1|\delta\left(|z_1|-{\cal V}^{\frac{1}{36}}\right)d|z_1|\int |z_2|^{38}|z_2|
\delta\left(|z_2|-{\cal V}^{\frac{1}{36}}\right)d|z_2|\nonumber\\
& & \sim {\cal V}^{\frac{37}{36}+\frac{39}{36}-1}={\cal V}^{\frac{10}{9}};\nonumber\\
& & C_{2{\bar 2}}\sim\frac{1}{{\cal V}}\int z_1^{18}z_2{\bar z}_1^{18}{\bar z}_2
 \delta\left(|z_1|-{\cal V}^{\frac{1}{36}}\right)\delta\left(|z_2|-{\cal V}^{\frac{1}{36}}\right)
 dz_1\wedge d{\bar z}_1\wedge dz_2\wedge d{\bar z}_2 \nonumber\\
& & \sim\frac{1}{\cal V}\int |z_1|^{36}|z_1|\delta\left(|z_1|-{\cal V}^{\frac{1}{36}}\right)d|z_1|\int |z_2|^{2}|z_2|
\delta\left(|z_2|-{\cal V}^{\frac{1}{36}}\right)d|z_2|\nonumber\\
& & \sim {\cal V}^{\frac{37}{36}+\frac{3}{36}-1}={\cal V}^{\frac{1}{9}};
\end{eqnarray*}

\begin{eqnarray}
& & C_{1{\bar 2}}\sim \frac{1}{{\cal V}}\int z_1^{18}z_2^{19}{\bar z}_1^{18}{\bar z}_2
 \delta\left(|z_1|-{\cal V}^{\frac{1}{36}}\right)\delta\left(|z_2|-{\cal V}^{\frac{1}{36}}\right)
 dz_1\wedge d{\bar z}_1\wedge dz_2\wedge d{\bar z}_2 \nonumber\\
& & \sim\frac{1}{\cal V}\int |z_1|^{36}|z_1|\delta\left(|z_1|-{\cal V}^{\frac{1}{36}}\right)d|z_1|\int |z_2|^{20}|z_2|
\delta\left(|z_2|-{\cal V}^{\frac{1}{36}}\right)d|z_2|\nonumber\\
& & \sim {\cal V}^{\frac{37}{36}+\frac{21}{36}-1}={\cal V}^{\frac{11}{18}};\nonumber\\
& & C_{1{\bar 3}}\sim \frac{1}{{\cal V}}\int z_1^{18}z_2^{19}{\bar z}_1^{18}{\bar z}_2^{37}
 \delta\left(|z_1|-{\cal V}^{\frac{1}{36}}\right)\delta\left(|z_2|-{\cal V}^{\frac{1}{36}}\right)
 dz_1\wedge d{\bar z}_1\wedge dz_2\wedge d{\bar z}_2 \nonumber\\
& & \sim\frac{1}{\cal V}\int |z_1|^{36}|z_1|\delta\left(|z_1|-{\cal V}^{\frac{1}{36}}\right)d|z_1|\int |z_2|^{56}|z_2|
\delta\left(|z_2|-{\cal V}^{\frac{1}{36}}\right)d|z_2|\nonumber\\
& & \sim {\cal V}^{\frac{37}{36}+\frac{57}{36}-1}={\cal V}^{\frac{29}{18}};\nonumber\\
& & C_{3{\bar 3}}\sim \frac{1}{{\cal V}}\int z_1^{18}z_2^{37}{\bar z}_1^{18}{\bar z}_2^{37}
 \delta\left(|z_1|-{\cal V}^{\frac{1}{36}}\right)\delta\left(|z_2|-{\cal V}^{\frac{1}{36}}\right)
 dz_1\wedge d{\bar z}_1\wedge dz_2\wedge d{\bar z}_2 \nonumber\\
& & \sim\frac{1}{\cal V}\int |z_1|^{36}|z_1|\delta\left(|z_1|-{\cal V}^{\frac{1}{36}}\right)d|z_1|\int |z_2|^{74}|z_2|
\delta\left(|z_2|-{\cal V}^{\frac{1}{36}}\right)d|z_2|\nonumber\\
& & \sim {\cal V}^{\frac{37}{36}+\frac{75}{36}-1}={\cal V}^{\frac{19}{9}};\nonumber\\
& & C_{2{\bar 3}}\sim  \frac{1}{{\cal V}}\int z_1^{18}z_2{\bar z}_1^{18}{\bar z}_2^{37}
 \delta\left(|z_1|-{\cal V}^{\frac{1}{36}}\right)\delta\left(|z_2|-{\cal V}^{\frac{1}{36}}\right)
 dz_1\wedge d{\bar z}_1\wedge dz_2\wedge d{\bar z}_2 \nonumber\\
& & \sim\frac{1}{\cal V}\int |z_1|^{36}|z_1|\delta\left(|z_1|-{\cal V}^{\frac{1}{36}}\right)d|z_1|\int |z_2|^{38}|z_2|
\delta\left(|z_2|-{\cal V}^{\frac{1}{36}}\right)d|z_2|\nonumber\\
& & \sim {\cal V}^{\frac{37}{36}+\frac{39}{36}-1}={\cal V}^{\frac{10}{9}};\nonumber\\
& & C_{4{\bar 4}}\sim  \frac{1}{{\cal V}}\int z_1^{36}z_2^{37}{\bar z}_1^{36}{\bar z}_2^{37}
 \delta\left(|z_1|-{\cal V}^{\frac{1}{36}}\right)\delta\left(|z_2|-{\cal V}^{\frac{1}{36}}\right)
 dz_1\wedge d{\bar z}_1\wedge dz_2\wedge d{\bar z}_2 \nonumber\\
& & \sim\frac{1}{\cal V}\int |z_1|^{72}|z_1|\delta\left(|z_1|-{\cal V}^{\frac{1}{36}}\right)d|z_1|\int |z_2|^{74}|z_2|
\delta\left(|z_2|-{\cal V}^{\frac{1}{36}}\right)d|z_2|\nonumber\\
& & \sim {\cal V}^{\frac{73}{36}+\frac{75}{36}-1}={\cal V}^{\frac{28}{9}};\nonumber\\
& & C_{1{\bar 4}}\sim  \frac{1}{{\cal V}}\int z_1^{18}z_2^{19}{\bar z}_1^{36}{\bar z}_2^{37}
 \delta\left(|z_1|-{\cal V}^{\frac{1}{36}}\right)\delta\left(|z_2|-{\cal V}^{\frac{1}{36}}\right)
 dz_1\wedge d{\bar z}_1\wedge dz_2\wedge d{\bar z}_2 \nonumber\\
& & \sim\frac{1}{\cal V}\int |z_1|^{54}|z_1|\delta\left(|z_1|-{\cal V}^{\frac{1}{36}}\right)d|z_1|\int |z_2|^{56}|z_2|
\delta\left(|z_2|-{\cal V}^{\frac{1}{36}}\right)d|z_2|\nonumber\\
& & \sim {\cal V}^{\frac{55}{36}+\frac{57}{36}-1}={\cal V}^{\frac{19}{9}};
\nonumber\\
& &  C_{2{\bar 4}}\sim  \frac{1}{{\cal V}}\int z_1^{18}z_2{\bar z}_1^{36}{\bar z}_2^{37}
 \delta\left(|z_1|-{\cal V}^{\frac{1}{36}}\right)\delta\left(|z_2|-{\cal V}^{\frac{1}{36}}\right)
 dz_1\wedge d{\bar z}_1\wedge dz_2\wedge d{\bar z}_2 \nonumber\\
& & \sim\frac{1}{\cal V}\int |z_1|^{54}|z_1|\delta\left(|z_1|-{\cal V}^{\frac{1}{36}}\right)d|z_1|\int |z_2|^{38}|z_2|
\delta\left(|z_2|-{\cal V}^{\frac{1}{36}}\right)d|z_2|\nonumber\\
& & \sim {\cal V}^{\frac{55}{36}+\frac{39}{36}-1}={\cal V}^{\frac{29}{18}};\nonumber\\
& & C_{3{\bar 4}}\sim  \frac{1}{{\cal V}}\int z_1^{18}z_2^{37}{\bar z}_1^{36}{\bar z}_2^{37}
 \delta\left(|z_1|-{\cal V}^{\frac{1}{36}}\right)\delta\left(|z_2|-{\cal V}^{\frac{1}{36}}\right)
 dz_1\wedge d{\bar z}_1\wedge dz_2\wedge d{\bar z}_2 \nonumber\\
& & \sim\frac{1}{\cal V}\int |z_1|^{54}|z_1|\delta\left(|z_1|-{\cal V}^{\frac{1}{36}}\right)d|z_1|\int |z_2|^{74}|z_2|
\delta\left(|z_2|-{\cal V}^{\frac{1}{36}}\right)d|z_2|\nonumber\\
& & \sim {\cal V}^{\frac{55}{36}+\frac{75}{36}-1}={\cal V}^{\frac{47}{18}}.\nonumber\\
\end{eqnarray}
\section{Moduli space metric}
On expanding moduli space metric in the fluctuations linear in $z_{i=1,2}$ and $a_{i=1,2,3,4}$ about their stabilized values given above in the paper, the expressions for moduli space metric, its inverse and its derivatives w.r.t each moduli are quoted below.

\footnotesize
\begin{eqnarray}
\label{metric}
 & &
 {\hskip -0.2in} g_{A{\bar B}}\sim \left(
\begin{array}{lllll}
 \frac{1}{{\cal V}^{2/3}} & \frac{1}{{\cal V}^{5/12}} & \frac{1}{{\cal V}^{11/12}} & \sqrt[12]{{\cal V}} & {\cal V}^{7/12} \\
 \frac{1}{{\cal V}^{5/12}} & {\cal V}^{4/9} & \frac{1}{\sqrt[18]{{\cal V}}} & {\cal V}^{17/18} & {\cal V}^{13/9} \\
 \frac{1}{{\cal V}^{11/12}} & \frac{1}{\sqrt[18]{{\cal V}}} & \frac{1}{{\cal V}^{5/9}} & {\cal V}^{4/9} & {\cal V}^{17/18} \\
 \sqrt[12]{{\cal V}} & {\cal V}^{17/18} & {\cal V}^{4/9} & {\cal V}^{13/9} & {\cal V}^{35/18} \\
 {\cal V}^{7/12} & {\cal V}^{13/9} & {\cal V}^{17/18} & {\cal V}^{35/18} & {\cal V}^{22/9} \nonumber\\
\end{array}
\right) {\rm + }
   \left(
\begin{array}{lllll}
 \frac{1}{{\cal V}} & \frac{1}{{\cal V}^{4/9}} & \frac{1}{{\cal V}^{17/18}} & \sqrt[18]{{\cal V}} & {\cal V}^{5/9} \\
 \frac{1}{{\cal V}^{4/9}} & \frac{1}{\sqrt{{\cal V}}} & -\frac{1}{{\cal V}^{25/36}} & {\cal V}^{11/36} & {\cal V}^{29/36} \\
 \frac{1}{{\cal V}^{17/18}} & \frac{1}{{\cal V}^{25/36}} & \frac{1}{{\cal V}^{43/36}} & \frac{1}{{\cal V}^{7/36}} & {\cal V}^{11/36} \\
 \sqrt[18]{{\cal V}} & {\cal V}^{11/36} & \frac{1}{{\cal V}^{7/36}} & \sqrt{{\cal V}} & {\cal V}^{47/36} \\
 {\cal V}^{5/9} & {\cal V}^{29/36} & {\cal V}^{11/36} & {\cal V}^{47/36} & {\cal V}^{65/36} \nonumber\\
\end{array}
\right) {\delta z_1~ +}
\end{eqnarray}

\begin{eqnarray}\left(
\begin{array}{lllll}
 \frac{1}{{\cal V}^{4/9}} & \frac{1}{{\cal V}^{7/36}} & \frac{1}{{\cal V}^{25/36}} & {\cal V}^{11/36} & {\cal V}^{29/36} \\
 \frac{1}{{\cal V}^{7/36}} & {\cal V}^{2/3} & \sqrt[6]{{\cal V}} & {\cal V}^{7/6} & {\cal V}^{5/3} \\
 \frac{1}{{\cal V}^{25/36}} & \sqrt[6]{{\cal V}} & \frac{1}{\sqrt[3]{{\cal V}}} & {\cal V}^{2/3} & {\cal V}^{7/6} \\
 {\cal V}^{11/36} & {\cal V}^{7/6} & {\cal V}^{2/3} & {\cal V}^{5/3} & {\cal V}^{13/6} \\
 {\cal V}^{29/36} & {\cal V}^{5/3} & {\cal V}^{7/6} & {\cal V}^{13/6} & {\cal V}^{8/3} \nonumber\\
\end{array}
\right)
 {\delta a_1 +}
 \left(
\begin{array}{lllll}
 \frac{1}{{\cal V}^{17/18}} & \frac{1}{{\cal V}^{25/36}} & \frac{1}{{\cal V}^{43/36}} & \frac{1}{{\cal V}^{7/36}} & {\cal V}^{11/36} \\
 \frac{1}{{\cal V}^{25/36}} & \sqrt[6]{{\cal V}} & \frac{1}{\sqrt[3]{{\cal V}}} & {\cal V}^{2/3} & {\cal V}^{7/6} \\
 \frac{1}{{\cal V}^{43/36}} & \frac{1}{\sqrt[3]{{\cal V}}} & \frac{1}{{\cal V}^{5/6}} & \sqrt[6]{{\cal V}} & {\cal V}^{2/3} \\
 \frac{1}{{\cal V}^{7/36}} & {\cal V}^{2/3} & \sqrt[6]{{\cal V}} & {\cal V}^{7/6} & {\cal V}^{5/3} \\
 {\cal V}^{11/36} & {\cal V}^{7/6} & {\cal V}^{2/3} & {\cal V}^{5/3} & {\cal V}^{13/6} \nonumber\\
\end{array}
\right)
{\delta a_2}
\end{eqnarray}
\begin{eqnarray}\left(
\begin{array}{lllll}
 \sqrt[18]{{\cal V}} & {\cal V}^{11/36} & \frac{1}{{\cal V}^{7/36}} & {\cal V}^{29/36} & {\cal V}^{47/36} \\
 {\cal V}^{11/36} & {\cal V}^{7/6} & {\cal V}^{2/3} & {\cal V}^{5/3} & {\cal V}^{13/6} \\
 \frac{1}{{\cal V}^{7/36}} & {\cal V}^{2/3} & \sqrt[6]{{\cal V}} & {\cal V}^{7/6} & {\cal V}^{5/3} \\
 {\cal V}^{29/36} & {\cal V}^{5/3} & {\cal V}^{7/6} & {\cal V}^{13/6} & {\cal V}^{8/3} \\
 {\cal V}^{47/36} & {\cal V}^{13/6} & {\cal V}^{5/3} & {\cal V}^{8/3} & {\cal V}^{19/6}
\end{array}
\right)
 {\delta a_3 +}
 \left(
\begin{array}{lllll}
 {\cal V}^{5/9} & {\cal V}^{29/36} & {\cal V}^{11/36} & {\cal V}^{47/36} & {\cal V}^{65/36} \\
 {\cal V}^{29/36} & {\cal V}^{5/3} & {\cal V}^{7/6} & {\cal V}^{13/6} & {\cal V}^{8/3} \\
 {\cal V}^{11/36} & {\cal V}^{7/6} & {\cal V}^{2/3} & {\cal V}^{5/3} & {\cal V}^{13/6} \\
 {\cal V}^{47/36} & {\cal V}^{13/6} & {\cal V}^{5/3} & {\cal V}^{8/3} & {\cal V}^{19/6} \\
 {\cal V}^{65/36} & {\cal V}^{8/3} & {\cal V}^{13/6} & {\cal V}^{19/6} & {\cal V}^{11/3}
\end{array}
\right)
{\delta a_4}
\end{eqnarray}
\begin{eqnarray}
\label{metricinv}
 g^{A{\bar B}}\sim
 \left(
\begin{array}{lllll}
 {\cal V}^{2/3} & \frac{1}{{\cal V}^{7/36}} & {\cal V}^{11/36} & \frac{1}{{\cal V}^{25/36}} & \frac{1}{{\cal V}^{43/36}} \\
 \frac{1}{{\cal V}^{7/36}} & \frac{1}{{\cal V}^{4/9}} & \sqrt[18]{{\cal V}} & \frac{1}{{\cal V}^{17/18}} & \frac{1}{{\cal V}^{13/9}} \\
 {\cal V}^{11/36} & \sqrt[18]{{\cal V}} & {\cal V}^{5/9} & \frac{1}{{\cal V}^{4/9}} & \frac{1}{{\cal V}^{17/18}} \\
 \frac{1}{{\cal V}^{25/36}} & \frac{1}{{\cal V}^{17/18}} & \frac{1}{{\cal V}^{4/9}} & \frac{1}{{\cal V}^{13/9}} & \frac{1}{{\cal V}^{35/18}} \\
 \frac{1}{{\cal V}^{43/36}} & \frac{1}{{\cal V}^{13/9}} & \frac{1}{{\cal V}^{17/18}} & \frac{1}{{\cal V}^{35/18}} & \frac{1}{{\cal V}^{22/9}} \nonumber\\
\end{array}
\right) {\rm +}
  \left(
\begin{array}{lllll}
 \sqrt[3]{{\cal V}} & \frac{1}{{\cal V}^{2/9}} & {\cal V}^{5/18} & \frac{1}{{\cal V}^{13/18}} & \frac{1}{{\cal V}^{11/9}} \\
 \frac{1}{{\cal V}^{2/9}} & \frac{1}{{\cal V}^{13/12}} & \frac{1}{{\cal V}^{7/12}} & \frac{1}{{\cal V}^{19/12}} & \frac{1}{{\cal V}^{25/12}} \\
 {\cal V}^{5/18} & \frac{1}{{\cal V}^{7/12}} & \frac{1}{\sqrt[12]{{\cal V}}} & \frac{1}{{\cal V}^{13/12}} & \frac{1}{{\cal V}^{19/12}} \\
 \frac{1}{{\cal V}^{13/18}} & \frac{1}{{\cal V}^{19/12}} & \frac{1}{{\cal V}^{13/12}} & \frac{1}{{\cal V}^{25/12}} & \frac{1}{{\cal V}^{31/12}} \\
 \frac{1}{{\cal V}^{11/9}} & \frac{1}{{\cal V}^{25/12}} & \frac{1}{{\cal V}^{19/12}} & \frac{1}{{\cal V}^{31/12}} & \frac{1}{{\cal V}^{37/12}} \nonumber\\
\end{array}
\right){\delta z_1~+ }
\end{eqnarray}
\begin{eqnarray}
  \left(
\begin{array}{lllll}
 {\cal V}^{8/9} & \sqrt[36]{{\cal V}} & {\cal V}^{19/36} & \frac{1}{{\cal V}^{17/36}} & 0 \\
 \sqrt[36]{{\cal V}} & \frac{1}{{\cal V}^{2/9}} & {\cal V}^{5/18} & \frac{1}{{\cal V}^{13/18}} & \frac{1}{{\cal V}^{11/9}} \\
 {\cal V}^{19/36} & {\cal V}^{5/18} & {\cal V}^{7/9} & \frac{1}{{\cal V}^{2/9}} & 0 \\
 \frac{1}{{\cal V}^{17/36}} & \frac{1}{{\cal V}^{13/18}} & \frac{1}{{\cal V}^{2/9}} & \frac{1}{{\cal V}^{11/9}} & \frac{1}{{\cal V}^{31/18}} \\
 0 & \frac{1}{{\cal V}^{11/9}} & 0 & \frac{1}{{\cal V}^{31/18}} & 0 \nonumber\\
\end{array}
\right) {\delta a_1~+}
 \left(
\begin{array}{lllll}
 {\cal V}^{7/18} & \frac{1}{{\cal V}^{17/36}} & \sqrt[36]{{\cal V}} & \frac{1}{{\cal V}^{35/36}} & \frac{1}{{\cal V}^{53/36}} \\
 \frac{1}{{\cal V}^{17/36}} & \frac{1}{{\cal V}^{13/18}} & \frac{1}{{\cal V}^{2/9}} & \frac{1}{{\cal V}^{11/9}} & \frac{1}{{\cal V}^{31/18}} \\
 \sqrt[36]{{\cal V}} & \frac{1}{{\cal V}^{2/9}} & {\cal V}^{5/18} & \frac{1}{{\cal V}^{13/18}} & \frac{1}{{\cal V}^{11/9}} \\
 \frac{1}{{\cal V}^{35/36}} & \frac{1}{{\cal V}^{11/9}} & \frac{1}{{\cal V}^{13/18}} & \frac{1}{{\cal V}^{31/18}} & \frac{1}{{\cal V}^{20/9}} \\
 \frac{1}{{\cal V}^{53/36}} & \frac{1}{{\cal V}^{31/18}} & \frac{1}{{\cal V}^{11/9}} & \frac{1}{{\cal V}^{20/9}} & \frac{1}{{\cal V}^{49/18}} \nonumber\\
\end{array}
\right){a_2}
\end{eqnarray}
\begin{eqnarray}
  \left(
\begin{array}{lllll}
 {\cal V}^{25/18} & {\cal V}^{19/36} & {\cal V}^{37/36} & 0 & 0 \\
 {\cal V}^{19/36} & {\cal V}^{5/18} & {\cal V}^{7/9} & 0 & 0 \\
 {\cal V}^{37/36} & {\cal V}^{7/9} & {\cal V}^{23/18} & {\cal V}^{5/18} & \frac{1}{{\cal V}^{2/9}} \\
 0 & 0 & {\cal V}^{5/18} & {\cal V}^{13/18} & 0 \\
 0 & 0 & \frac{1}{{\cal V}^{2/9}} & 0 & \frac{1}{{\cal V}^{31/18}}
\end{array}
\right) {\delta a_3~+}
\left(
\begin{array}{lllll}
 {\cal V}^{17/9} & {\cal V}^{37/36} & {\cal V}^{55/36} & {\cal V}^{19/36} & \sqrt[36]{{\cal V}} \\
 {\cal V}^{37/36} & {\cal V}^{7/9} & {\cal V}^{23/18} & {\cal V}^{5/18} & \frac{1}{{\cal V}^{2/9}} \\
 {\cal V}^{55/36} & {\cal V}^{23/18} & {\cal V}^{16/9} & {\cal V}^{7/9} & {\cal V}^{5/18} \\
 {\cal V}^{19/36} & {\cal V}^{5/18} & {\cal V}^{7/9} & \frac{1}{{\cal V}^{2/9}} & \frac{1}{{\cal V}^{13/18}} \\
 \sqrt[36]{{\cal V}} & \frac{1}{{\cal V}^{2/9}} & {\cal V}^{5/18} & \frac{1}{{\cal V}^{13/18}} & \frac{1}{{\cal V}^{11/9}}
\end{array}
\right){a_4}
\end{eqnarray}
  \begin{eqnarray}
  \label{metricdervz1}
 \partial_{z_1}g_{A{\bar B}}\sim \left(
\begin{array}{lllll}
 \frac{1}{{\cal V}} & \frac{1}{{\cal V}^{4/9}} & \frac{1}{{\cal V}^{17/18}} & \sqrt[18]{{\cal V}} & {\cal V}^{5/9} \\
 \frac{1}{{\cal V}^{4/9}} & \frac{1}{\sqrt{{\cal V}}} & \frac{1}{{\cal V}^{25/36}} & 60 & {\cal V}^{29/36} \\
 \frac{1}{{\cal V}^{17/18}} & \frac{1}{{\cal V}^{25/36}} & \frac{1}{{\cal V}^{43/36}} & \frac{1}{{\cal V}^{7/36}} & {\cal V}^{11/36} \\
 \sqrt[18]{{\cal V}} & 60 & \frac{1}{{\cal V}^{7/36}} & \sqrt{{\cal V}} & {\cal V}^{47/36} \\
 {\cal V}^{5/9} & {\cal V}^{29/36} & {\cal V}^{11/36} & {\cal V}^{47/36} & {\cal V}^{65/36} \nonumber\\
\end{array}
\right){\rm +}\left(
\begin{array}{lllll}
 \frac{1}{{\cal V}^{37/36}} & \frac{1}{{\cal V}^{7/9}} & \frac{1}{{\cal V}^{23/18}} & \frac{1}{{\cal V}^{5/18}} & {\cal V}^{2/9} \\
 \frac{1}{{\cal V}^{7/9}} & \frac{1}{{\cal V}^{19/36}} & \frac{1}{{\cal V}^{13/18}} & \frac{1}{\sqrt[36]{{\cal V}}} & {\cal V}^{7/9} \\
 \frac{1}{{\cal V}^{23/18}} & \frac{1}{{\cal V}^{13/18}} & \frac{1}{{\cal V}^{11/9}} & \frac{1}{{\cal V}^{2/9}} & {\cal V}^{5/18} \\
 \frac{1}{{\cal V}^{5/18}} & \frac{1}{\sqrt[36]{{\cal V}}} & \frac{1}{{\cal V}^{2/9}} & {\cal V}^{17/36} & {\cal V}^{23/18} \\
 {\cal V}^{2/9} & {\cal V}^{7/9} & {\cal V}^{5/18} & {\cal V}^{23/18} & {\cal V}^{16/9}
\end{array}
\right)
 {\delta z_1}
\end{eqnarray}
\begin{eqnarray}
\left(
\begin{array}{lllll}
 \frac{1}{{\cal V}^{7/9}} & \frac{1}{{\cal V}^{2/9}} & \frac{1}{{\cal V}^{13/18}} & {\cal V}^{5/18} & {\cal V}^{7/9} \\
 \frac{1}{{\cal V}^{2/9}} & \sqrt[36]{{\cal V}} & \frac{1}{{\cal V}^{17/36}} & {\cal V}^{19/36} & {\cal V}^{37/36} \\
 \frac{1}{{\cal V}^{13/18}} & \frac{1}{{\cal V}^{17/36}} & \frac{1}{{\cal V}^{35/36}} & \sqrt[36]{{\cal V}} & {\cal V}^{19/36} \\
 {\cal V}^{5/18} & {\cal V}^{19/36} & \sqrt[36]{{\cal V}} & {\cal V}^{37/36} & {\cal V}^{55/36} \\
 {\cal V}^{7/9} & {\cal V}^{37/36} & {\cal V}^{19/36} & {\cal V}^{55/36} & {\cal V}^{73/36} \nonumber\\
\end{array}
\right) {\delta a_1~+}
\left(
\begin{array}{lllll}
 \frac{1}{{\cal V}^{23/18}} & \frac{1}{{\cal V}^{13/18}} & \frac{1}{{\cal V}^{11/9}} & \frac{1}{{\cal V}^{2/9}} & {\cal V}^{5/18} \\
 \frac{1}{{\cal V}^{13/18}} & \frac{1}{{\cal V}^{17/36}} & \frac{1}{{\cal V}^{35/36}} & \sqrt[36]{{\cal V}} & {\cal V}^{19/36} \\
 \frac{1}{{\cal V}^{11/9}} & \frac{1}{{\cal V}^{35/36}} & \frac{1}{{\cal V}^{53/36}} & \frac{1}{{\cal V}^{17/36}} & \sqrt[36]{{\cal V}}
   \\
 \frac{1}{{\cal V}^{2/9}} & \sqrt[36]{{\cal V}} & \frac{1}{{\cal V}^{17/36}} & {\cal V}^{19/36} & {\cal V}^{37/36} \\
 {\cal V}^{5/18} & {\cal V}^{19/36} & \sqrt[36]{{\cal V}} & {\cal V}^{37/36} & {\cal V}^{55/36}\nonumber\\
\end{array}
\right){\delta a_2}
\end{eqnarray}
\begin{eqnarray}
 \left(
\begin{array}{lllll}
 \frac{1}{{\cal V}^{5/18}} & {\cal V}^{5/18} & \frac{1}{{\cal V}^{2/9}} & {\cal V}^{7/9} & {\cal V}^{23/18} \\
 {\cal V}^{5/18} & {\cal V}^{19/36} & \sqrt[36]{{\cal V}} & {\cal V}^{37/36} & {\cal V}^{55/36} \\
 \frac{1}{{\cal V}^{2/9}} & \sqrt[36]{{\cal V}} & \frac{1}{{\cal V}^{17/36}} & {\cal V}^{19/36} & {\cal V}^{37/36} \\
 {\cal V}^{7/9} & {\cal V}^{37/36} & {\cal V}^{19/36} & {\cal V}^{55/36} & {\cal V}^{73/36} \\
 {\cal V}^{23/18} & {\cal V}^{55/36} & {\cal V}^{37/36} & {\cal V}^{73/36} & {\cal V}^{91/36}
\end{array}
\right){\delta a_3~+}
\left(
\begin{array}{lllll}
 {\cal V}^{2/9} & {\cal V}^{7/9} & {\cal V}^{5/18} & {\cal V}^{23/18} & {\cal V}^{16/9} \\
 {\cal V}^{7/9} & {\cal V}^{37/36} & {\cal V}^{19/36} & {\cal V}^{55/36} & {\cal V}^{73/36} \\
 {\cal V}^{5/18} & {\cal V}^{19/36} & \sqrt[36]{{\cal V}} & {\cal V}^{37/36} & {\cal V}^{55/36} \\
 {\cal V}^{23/18} & {\cal V}^{55/36} & {\cal V}^{37/36} & {\cal V}^{73/36} & {\cal V}^{91/36} \\
 {\cal V}^{16/9} & {\cal V}^{73/36} & {\cal V}^{55/36} & {\cal V}^{91/36} & {\cal V}^{109/36}
\end{array}
\right){\delta a_4}
\end{eqnarray}
\begin{eqnarray}
\label{metricderva1}
   \partial_{a_1}g_{A{\bar B}}\sim \left(
\begin{array}{lllll}
 \frac{1}{{\cal V}^{4/9}} & \frac{1}{\sqrt{{\cal V}}} & \frac{1}{{\cal V}^{25/36}} & 81 & {\cal V}^{29/36} \\
 \frac{1}{\sqrt{{\cal V}}} & {\cal V}^{2/3} & \sqrt[6]{{\cal V}} & {\cal V}^{7/6} & {\cal V}^{5/3} \\
 \frac{1}{{\cal V}^{25/36}} & \sqrt[6]{{\cal V}} & \frac{1}{\sqrt[3]{{\cal V}}} & {\cal V}^{2/3} & {\cal V}^{7/6} \\
 81 & {\cal V}^{7/6} & {\cal V}^{2/3} & {\cal V}^{5/3} & {\cal V}^{13/6} \\
 {\cal V}^{29/36} & {\cal V}^{5/3} & {\cal V}^{7/6} & {\cal V}^{13/6} & {\cal V}^{8/3}\nonumber\\
\end{array}
\right){\rm +}
 \left(
\begin{array}{lllll}
 \frac{1}{{\cal V}^{7/9}} & \frac{1}{{\cal V}^{19/36}} & \frac{1}{{\cal V}^{13/18}} & \frac{1}{\sqrt[36]{{\cal V}}} & {\cal V}^{7/9} \\
 \frac{1}{{\cal V}^{19/36}} & \sqrt[36]{{\cal V}} & \frac{1}{{\cal V}^{17/36}} & {\cal V}^{19/36} & {\cal V}^{37/36} \\
 \frac{1}{{\cal V}^{13/18}} & \frac{1}{{\cal V}^{17/36}} & \frac{1}{{\cal V}^{35/36}} & \sqrt[36]{{\cal V}} & {\cal V}^{19/36} \\
 \frac{1}{\sqrt[36]{{\cal V}}} & {\cal V}^{19/36} & \sqrt[36]{{\cal V}} & {\cal V}^{37/36} & {\cal V}^{55/36} \\
 {\cal V}^{7/9} & {\cal V}^{37/36} & {\cal V}^{19/36} & {\cal V}^{55/36} & {\cal V}^{73/36}\nonumber\\
\end{array}
\right){\delta z_1 +}
\end{eqnarray}
\begin{eqnarray}
\left(
\begin{array}{lllll}
 \frac{1}{{\cal V}^{2/9}} & \sqrt[36]{{\cal V}} & \frac{1}{{\cal V}^{17/36}} & {\cal V}^{19/36} & {\cal V}^{37/36} \\
 \sqrt[36]{{\cal V}} & {\cal V}^{8/9} & {\cal V}^{7/18} & {\cal V}^{25/18} & {\cal V}^{25/18} \\
 \frac{1}{{\cal V}^{17/36}} & {\cal V}^{7/18} & \frac{1}{\sqrt[9]{{\cal V}}} & {\cal V}^{8/9} & {\cal V}^{25/18} \\
 {\cal V}^{19/36} & {\cal V}^{25/18} & {\cal V}^{8/9} & {\cal V}^{17/9} & {\cal V}^{43/18} \\
 {\cal V}^{37/36} & {\cal V}^{25/18} & {\cal V}^{25/18} & {\cal V}^{43/18} & {\cal V}^{26/9}\nonumber\\
\end{array}
\right){\delta a_1 +}
 \left(
\begin{array}{lllll}
 \frac{1}{{\cal V}^{13/18}} & \frac{1}{{\cal V}^{17/36}} & \frac{1}{{\cal V}^{17/36}} & \sqrt[36]{{\cal V}} & {\cal V}^{19/36} \\
 \frac{1}{{\cal V}^{17/36}} & {\cal V}^{7/18} & \frac{1}{\sqrt[9]{{\cal V}}} & {\cal V}^{8/9} & {\cal V}^{25/18} \\
 \frac{1}{{\cal V}^{17/36}} & \frac{1}{\sqrt[9]{{\cal V}}} & \frac{1}{{\cal V}^{11/18}} & {\cal V}^{7/18} & {\cal V}^{8/9} \\
 \sqrt[36]{{\cal V}} & {\cal V}^{8/9} & {\cal V}^{7/18} & {\cal V}^{25/18} & {\cal V}^{17/9} \\
 {\cal V}^{19/36} & {\cal V}^{25/18} & {\cal V}^{8/9} & {\cal V}^{17/9} & {\cal V}^{43/18}\nonumber\\
\end{array}
\right){\delta a_2 }
\end{eqnarray}
\begin{eqnarray}
\left(
\begin{array}{lllll}
 {\cal V}^{5/18} & {\cal V}^{19/36} & \sqrt[36]{{\cal V}} & {\cal V}^{37/36} & {\cal V}^{55/36} \\
 {\cal V}^{19/36} & {\cal V}^{25/18} & {\cal V}^{8/9} & {\cal V}^{17/9} & {\cal V}^{43/18} \\
 \sqrt[36]{{\cal V}} & {\cal V}^{8/9} & {\cal V}^{7/18} & {\cal V}^{25/18} & {\cal V}^{17/9} \\
 {\cal V}^{37/36} & {\cal V}^{17/9} & {\cal V}^{25/18} & {\cal V}^{43/18} & {\cal V}^{26/9} \\
 {\cal V}^{55/36} & {\cal V}^{43/18} & {\cal V}^{17/9} & {\cal V}^{26/9} & {\cal V}^{61/18}
\end{array}
\right){\delta a_3 +}
 \left(
\begin{array}{lllll}
 {\cal V}^{7/9} & {\cal V}^{37/36} & {\cal V}^{19/36} & {\cal V}^{55/36} & {\cal V}^{73/36} \\
 {\cal V}^{37/36} & {\cal V}^{17/9} & {\cal V}^{25/18} & {\cal V}^{43/18} & {\cal V}^{26/9} \\
 {\cal V}^{19/36} & {\cal V}^{25/18} & {\cal V}^{8/9} & {\cal V}^{17/9} & {\cal V}^{43/18} \\
 {\cal V}^{55/36} & {\cal V}^{43/18} & {\cal V}^{17/9} & {\cal V}^{26/9} & {\cal V}^{61/18} \\
 {\cal V}^{73/36} & {\cal V}^{26/9} & {\cal V}^{43/18} & \text{Null}^{61/18} & {\cal V}^{35/9}
\end{array}
\right){\delta a_4 }
\end{eqnarray}
\begin{eqnarray}
\label{metricderva2}
 \partial_{a_2}g_{A{\bar B}}\sim \left(
\begin{array}{lllll}
 \frac{1}{{\cal V}^{17/18}} & \frac{1}{{\cal V}^{25/36}} & \frac{1}{{\cal V}^{43/36}} & \frac{1}{{\cal V}^{7/36}} & {\cal V}^{11/36} \\
 \frac{1}{{\cal V}^{25/36}} & \sqrt[6]{{\cal V}} & \frac{1}{\sqrt[3]{{\cal V}}} & {\cal V}^{2/3} & {\cal V}^{7/6} \\
 \frac{1}{{\cal V}^{43/36}} & \frac{1}{\sqrt[3]{{\cal V}}} & \frac{1}{{\cal V}^{5/6}} & \sqrt[6]{{\cal V}} & {\cal V}^{2/3} \\
 \frac{1}{{\cal V}^{7/36}} & {\cal V}^{2/3} & \sqrt[6]{{\cal V}} & {\cal V}^{7/6} & {\cal V}^{5/3} \\
 {\cal V}^{11/36} & {\cal V}^{7/6} & {\cal V}^{2/3} & {\cal V}^{5/3} & {\cal V}^{13/6}\nonumber\\
\end{array}
\right){\rm +}
 \left(
\begin{array}{lllll}
 \frac{1}{{\cal V}^{23/18}} & \frac{1}{{\cal V}^{13/18}} & \frac{1}{{\cal V}^{11/9}} & \frac{1}{{\cal V}^{2/9}} & {\cal V}^{5/18} \\
 \frac{1}{{\cal V}^{13/18}} & \frac{1}{{\cal V}^{17/36}} & \frac{1}{{\cal V}^{35/36}} & \sqrt[36]{{\cal V}} & {\cal V}^{19/36} \\
 \frac{1}{{\cal V}^{11/9}} & \frac{1}{{\cal V}^{35/36}} & \frac{1}{{\cal V}^{53/36}} & \frac{1}{{\cal V}^{17/36}} & \sqrt[36]{{\cal V}} \\
 \frac{1}{{\cal V}^{2/9}} & \sqrt[36]{{\cal V}} & \frac{1}{{\cal V}^{17/36}} & {\cal V}^{19/36} & {\cal V}^{37/36} \\
 {\cal V}^{5/18} & {\cal V}^{19/36} & \sqrt[36]{{\cal V}} & {\cal V}^{37/36} & {\cal V}^{55/36}\nonumber\\
\end{array}
\right){\delta z_1~+}
\end{eqnarray}
\begin{eqnarray}\left(
\begin{array}{lllll}
 \frac{1}{{\cal V}^{13/18}} & \frac{1}{{\cal V}^{17/36}} & \frac{1}{{\cal V}^{35/36}} & \sqrt[36]{{\cal V}} & \sqrt[36]{{\cal V}} \\
 \frac{1}{{\cal V}^{17/36}} & {\cal V}^{7/18} & \frac{1}{\sqrt[9]{{\cal V}}} & {\cal V}^{8/9} & {\cal V}^{25/18} \\
 \frac{1}{{\cal V}^{35/36}} & \frac{1}{\sqrt[9]{{\cal V}}} & \frac{1}{{\cal V}^{11/18}} & {\cal V}^{7/18} & {\cal V}^{8/9} \\
 \sqrt[36]{{\cal V}} & {\cal V}^{8/9} & {\cal V}^{7/18} & {\cal V}^{25/18} & {\cal V}^{17/9} \\
 \sqrt[36]{{\cal V}} & {\cal V}^{25/18} & {\cal V}^{8/9} & {\cal V}^{17/9} & {\cal V}^{43/18}\nonumber\\
\end{array}
\right){\delta a_1~+}
\left(
\begin{array}{lllll}
 \frac{1}{{\cal V}^{11/9}} & \frac{1}{{\cal V}^{35/36}} & \frac{1}{{\cal V}^{53/36}} & \frac{1}{{\cal V}^{17/36}} & \sqrt[36]{{\cal V}}
   \\
 \frac{1}{{\cal V}^{35/36}} & \frac{1}{\sqrt[9]{{\cal V}}} & \frac{1}{{\cal V}^{11/18}} & {\cal V}^{7/18} & {\cal V}^{8/9} \\
 \frac{1}{{\cal V}^{53/36}} & \frac{1}{{\cal V}^{11/18}} & \frac{1}{{\cal V}^{10/9}} & \frac{1}{\sqrt[9]{{\cal V}}} & {\cal V}^{7/18}
   \\
 \frac{1}{{\cal V}^{17/36}} & {\cal V}^{7/18} & \frac{1}{\sqrt[9]{{\cal V}}} & {\cal V}^{8/9} & {\cal V}^{25/18} \\
 \sqrt[36]{{\cal V}} & {\cal V}^{8/9} & {\cal V}^{7/18} & {\cal V}^{25/18} & {\cal V}^{17/9}\nonumber\\
\end{array}
\right){\delta a_2~}
\end{eqnarray}
\begin{eqnarray}\left(
\begin{array}{lllll}
 \frac{1}{{\cal V}^{2/9}} & \sqrt[36]{{\cal V}} & \frac{1}{{\cal V}^{17/36}} & {\cal V}^{19/36} & {\cal V}^{37/36} \\
 \sqrt[36]{{\cal V}} & {\cal V}^{8/9} & {\cal V}^{7/18} & {\cal V}^{25/18} & {\cal V}^{17/9} \\
 \frac{1}{{\cal V}^{17/36}} & {\cal V}^{7/18} & \frac{1}{\sqrt[9]{{\cal V}}} & {\cal V}^{8/9} & {\cal V}^{25/18} \\
 {\cal V}^{19/36} & {\cal V}^{25/18} & {\cal V}^{8/9} & {\cal V}^{17/9} & {\cal V}^{43/18} \\
 {\cal V}^{37/36} & {\cal V}^{17/9} & {\cal V}^{25/18} & {\cal V}^{43/18} & {\cal V}^{26/9}
\end{array}
\right){\delta a_3~+}
\left(
\begin{array}{lllll}
 {\cal V}^{5/18} & {\cal V}^{19/36} & \sqrt[36]{{\cal V}} & {\cal V}^{37/36} & {\cal V}^{55/36} \\
 {\cal V}^{19/36} & {\cal V}^{25/18} & {\cal V}^{8/9} & {\cal V}^{17/9} & {\cal V}^{43/18} \\
 \sqrt[36]{{\cal V}} & {\cal V}^{8/9} & {\cal V}^{7/18} & {\cal V}^{25/18} & {\cal V}^{17/9} \\
 {\cal V}^{37/36} & {\cal V}^{17/9} & {\cal V}^{25/18} & {\cal V}^{43/18} & {\cal V}^{26/9} \\
 {\cal V}^{55/36} & {\cal V}^{43/18} & {\cal V}^{17/9} & {\cal V}^{26/9} & {\cal V}^{61/18}
\end{array}
\right){\delta a_4~}
\end{eqnarray}
\begin{eqnarray}
\label{metricderva3}
{\hskip -0.3in}  \partial_{a_3}g_{A{\bar B}}(z_1)\sim \left(
\begin{array}{lllll}
 \sqrt[18]{{\cal V}} & 81 & \frac{1}{{\cal V}^{7/36}} & \sqrt{{\cal V}} & {\cal V}^{47/36} \\
 81 & {\cal V}^{7/6} & {\cal V}^{2/3} & {\cal V}^{5/3} & {\cal V}^{13/6} \\
 \frac{1}{{\cal V}^{7/36}} & {\cal V}^{2/3} & \sqrt[6]{{\cal V}} & {\cal V}^{7/6} & {\cal V}^{5/3} \\
 \sqrt{{\cal V}} & {\cal V}^{5/3} & {\cal V}^{7/6} & {\cal V}^{13/6} & {\cal V}^{8/3} \\
 {\cal V}^{47/36} & {\cal V}^{13/6} & {\cal V}^{5/3} & {\cal V}^{8/3} & {\cal V}^{19/6}\nonumber\\
\end{array}
\right){\rm +}
\left(
\begin{array}{lllll}
 \frac{1}{{\cal V}^{5/18}} & \frac{1}{\sqrt[36]{{\cal V}}} & \frac{1}{{\cal V}^{2/9}} & {\cal V}^{17/36} & {\cal V}^{23/18} \\
 \frac{1}{\sqrt[36]{{\cal V}}} & {\cal V}^{19/36} & \sqrt[36]{{\cal V}} & {\cal V}^{37/36} & {\cal V}^{55/36} \\
 \frac{1}{{\cal V}^{2/9}} & \sqrt[36]{{\cal V}} & \frac{1}{{\cal V}^{17/36}} & {\cal V}^{19/36} & {\cal V}^{37/36} \\
 {\cal V}^{17/36} & {\cal V}^{37/36} & {\cal V}^{19/36} & {\cal V}^{55/36} & {\cal V}^{73/36} \\
 {\cal V}^{23/18} & {\cal V}^{55/36} & {\cal V}^{37/36} & {\cal V}^{73/36} & {\cal V}^{91/36}\nonumber\\
\end{array}
\right) {\delta z_1 +}
\end{eqnarray}
\begin{eqnarray}
\left(
\begin{array}{lllll}
 {\cal V}^{5/18} & {\cal V}^{19/36} & \sqrt[36]{{\cal V}} & {\cal V}^{37/36} & {\cal V}^{55/36} \\
 {\cal V}^{19/36} & {\cal V}^{25/18} & {\cal V}^{8/9} & {\cal V}^{17/9} & {\cal V}^{43/18} \\
 \sqrt[36]{{\cal V}} & {\cal V}^{8/9} & {\cal V}^{7/18} & {\cal V}^{25/18} & {\cal V}^{17/9} \\
 {\cal V}^{37/36} & {\cal V}^{17/9} & {\cal V}^{25/18} & {\cal V}^{43/18} & {\cal V}^{26/9} \\
 {\cal V}^{55/36} & {\cal V}^{43/18} & {\cal V}^{17/9} & {\cal V}^{26/9} & {\cal V}^{61/18}\nonumber\\
\end{array}
\right){\delta a_1~+}
\left(
\begin{array}{lllll}
 {\cal V}^{2/9} & \sqrt[36]{{\cal V}} & \frac{1}{{\cal V}^{17/36}} & {\cal V}^{19/36} & {\cal V}^{37/36} \\
 \sqrt[36]{{\cal V}} & {\cal V}^{8/9} & {\cal V}^{7/18} & {\cal V}^{25/18} & {\cal V}^{17/9} \\
 \frac{1}{{\cal V}^{17/36}} & {\cal V}^{7/18} & \frac{1}{\sqrt[9]{{\cal V}}} & {\cal V}^{8/9} & {\cal V}^{25/18} \\
 {\cal V}^{19/36} & {\cal V}^{25/18} & {\cal V}^{8/9} & {\cal V}^{17/9} & {\cal V}^{43/18} \\
 {\cal V}^{37/36} & {\cal V}^{17/9} & {\cal V}^{25/18} & {\cal V}^{43/18} & {\cal V}^{26/9}\nonumber\\
\end{array}
\right) {\delta a_2}
\end{eqnarray}
\begin{eqnarray}
\left(
\begin{array}{lllll}
 {\cal V}^{7/9} & {\cal V}^{37/36} & {\cal V}^{19/36} & {\cal V}^{55/36} & {\cal V}^{73/36} \\
 {\cal V}^{37/36} & {\cal V}^{17/9} & {\cal V}^{25/18} & {\cal V}^{43/18} & {\cal V}^{26/9} \\
 {\cal V}^{19/36} & {\cal V}^{25/18} & {\cal V}^{8/9} & {\cal V}^{17/9} & {\cal V}^{43/18} \\
 {\cal V}^{55/36} & {\cal V}^{43/18} & {\cal V}^{17/9} & {\cal V}^{26/9} & {\cal V}^{61/18} \\
 {\cal V}^{73/36} & {\cal V}^{26/9} & {\cal V}^{43/18} & {\cal V}^{61/18} & {\cal V}^{35/9}
\end{array}
\right){\delta a_3~+}
 \left(
\begin{array}{lllll}
 {\cal V}^{23/18} & {\cal V}^{55/36} & {\cal V}^{37/36} & {\cal V}^{73/36} & {\cal V}^{91/36} \\
 {\cal V}^{55/36} & {\cal V}^{43/18} & {\cal V}^{17/9} & {\cal V}^{26/9} & {\cal V}^{61/18} \\
 {\cal V}^{37/36} & {\cal V}^{17/9} & {\cal V}^{25/18} & {\cal V}^{43/18} & {\cal V}^{26/9} \\
 {\cal V}^{73/36} & {\cal V}^{26/9} & {\cal V}^{43/18} & {\cal V}^{61/18} & {\cal V}^{35/9} \\
 {\cal V}^{91/36} & {\cal V}^{61/18} & {\cal V}^{26/9} & {\cal V}^{35/9} & {\cal V}^{79/18}
\end{array}
\right){\delta a_4}
\end{eqnarray}
\begin{eqnarray}
\label{metricderva4}
  \partial_{a_4}g_{A{\bar B}}\sim \left(
\begin{array}{lllll}
 {\cal V}^{5/9} & {\cal V}^{29/36} & {\cal V}^{11/36} & {\cal V}^{47/36} & {\cal V}^{65/36} \\
 {\cal V}^{29/36} & {\cal V}^{5/3} & {\cal V}^{7/6} & {\cal V}^{13/6} & {\cal V}^{8/3} \\
 {\cal V}^{11/36} & {\cal V}^{7/6} & {\cal V}^{2/3} & {\cal V}^{5/3} & {\cal V}^{13/6} \\
 {\cal V}^{47/36} & {\cal V}^{13/6} & {\cal V}^{5/3} & {\cal V}^{8/3} & {\cal V}^{19/6} \\
 {\cal V}^{65/36} & {\cal V}^{8/3} & {\cal V}^{13/6} & {\cal V}^{19/6} & {\cal V}^{11/3}\nonumber\\
\end{array}
\right){\rm +}
\left(
\begin{array}{lllll}
 {\cal V}^{2/9} & {\cal V}^{7/9} & {\cal V}^{5/18} & {\cal V}^{23/18} & {\cal V}^{16/9} \\
 {\cal V}^{7/9} & {\cal V}^{37/36} & {\cal V}^{19/36} & {\cal V}^{55/36} & {\cal V}^{73/36} \\
 {\cal V}^{5/18} & {\cal V}^{19/36} & \sqrt[36]{{\cal V}} & {\cal V}^{37/36} & {\cal V}^{55/36} \\
 {\cal V}^{23/18} & {\cal V}^{55/36} & {\cal V}^{37/36} & {\cal V}^{73/36} & {\cal V}^{91/36} \\
 {\cal V}^{16/9} & {\cal V}^{73/36} & {\cal V}^{55/36} & {\cal V}^{91/36} & {\cal V}^{109/36}\nonumber\\
\end{array}
\right){\delta z_1 }
\end{eqnarray}
\begin{eqnarray}
\left(
\begin{array}{lllll}
 {\cal V}^{7/9} & {\cal V}^{37/36} & {\cal V}^{19/36} & {\cal V}^{55/36} & {\cal V}^{73/36} \\
 {\cal V}^{37/36} & {\cal V}^{17/9} & {\cal V}^{25/18} & {\cal V}^{43/18} & {\cal V}^{26/9} \\
 {\cal V}^{19/36} & {\cal V}^{25/18} & {\cal V}^{8/9} & {\cal V}^{17/9} & {\cal V}^{43/18} \\
 {\cal V}^{55/36} & {\cal V}^{43/18} & {\cal V}^{17/9} & {\cal V}^{26/9} & {\cal V}^{61/18} \\
 {\cal V}^{73/36} & {\cal V}^{26/9} & {\cal V}^{43/18} & {\cal V}^{61/18} & {\cal V}^{35/9}\nonumber\\
\end{array}
\right){\delta a_1~+}
\left(
\begin{array}{lllll}
 {\cal V}^{5/18} & {\cal V}^{19/36} & \sqrt[36]{{\cal V}} & {\cal V}^{37/36} & {\cal V}^{55/36} \\
 {\cal V}^{19/36} & {\cal V}^{25/18} & {\cal V}^{8/9} & {\cal V}^{17/9} & {\cal V}^{43/18} \\
 \sqrt[36]{{\cal V}} & {\cal V}^{8/9} & {\cal V}^{7/18} & {\cal V}^{25/18} & {\cal V}^{17/9} \\
 {\cal V}^{37/36} & {\cal V}^{17/9} & {\cal V}^{25/18} & {\cal V}^{43/18} & {\cal V}^{26/9} \\
 {\cal V}^{55/36} & {\cal V}^{43/18} & {\cal V}^{17/9} & {\cal V}^{26/9} & {\cal V}^{61/18}\nonumber\\
\end{array}
\right){\delta a_2}
\end{eqnarray}
\begin{eqnarray}
\left(
\begin{array}{lllll}
 {\cal V}^{23/18} & {\cal V}^{55/36} & {\cal V}^{37/36} & {\cal V}^{73/36} & {\cal V}^{91/36} \\
 {\cal V}^{55/36} & {\cal V}^{43/18} & {\cal V}^{17/9} & {\cal V}^{26/9} & {\cal V}^{61/18} \\
 {\cal V}^{37/36} & {\cal V}^{17/9} & {\cal V}^{25/18} & {\cal V}^{43/18} & {\cal V}^{26/9} \\
 {\cal V}^{73/36} & {\cal V}^{26/9} & {\cal V}^{43/18} & {\cal V}^{61/18} & {\cal V}^{35/9} \\
 {\cal V}^{91/36} & {\cal V}^{61/18} & {\cal V}^{26/9} & {\cal V}^{35/9} & {\cal V}^{79/18}
\end{array}
\right){\delta a_3~+}
\left(
\begin{array}{lllll}
 {\cal V}^{16/9} & {\cal V}^{73/36} & {\cal V}^{55/36} & {\cal V}^{91/36} & {\cal V}^{109/36} \\
 {\cal V}^{73/36} & {\cal V}^{26/9} & {\cal V}^{43/18} & {\cal V}^{61/18} & {\cal V}^{35/9} \\
 {\cal V}^{55/36} & {\cal V}^{43/18} & {\cal V}^{17/9} & {\cal V}^{26/9} & {\cal V}^{61/18} \\
 {\cal V}^{91/36} & {\cal V}^{61/18} & {\cal V}^{26/9} & {\cal V}^{35/9} & {\cal V}^{79/18} \\
 {\cal V}^{109/36} & {\cal V}^{35/9} & {\cal V}^{61/18} & {\cal V}^{79/18} & {\cal V}^{44/9}
\end{array}
\right){\delta a_4}
\end{eqnarray}
\normalsize

\section{Evaluation of matrix elements corresponding to $\tilde{l}\rightarrow l^\prime \tilde{G} V$ decay}
Expressions for $\calO_{i,i}$, $\calO_{i,j}^{\text re}$,
$\calO_{i,8}^{\text im}$ ($i<8$) and $\calO_{8,j}^{\text im}$ ($j>8$), and the  corresponding contributions to
$\int\int ds_{13}ds_{23}\frac{|{\cal M}|^2}{m^3_{\tilde{l}}}$ ( using $m_{\tilde G}\sim m_{\frac{3}{2}}\sim {\cal V}^{-2}m_{pl} $ and $m_{\tilde l}\sim {\cal V}^{\frac{1}{2}}m_{\frac{3}{2}}$ from equations (\ref{eq:gaugino_mass}) and (\ref{eq:m_A_1}))  are given as under:
\begin{eqnarray}
\label{eq:decay width_I}
\calO_{1,1} &=& -\frac{4}{3\mg^2m_V^2}
\left[\mg^2\msl^2 - \left(\pgpsl\right)^2\right]
\left[\mg^2 - \pgpsl + \pgpZ\right] \nonumber\\
&& \!\!\!\!\!\!\!\! \times \left[\mg^2m_V^2 + \msl^2m_V^2 - 2m_V^2\pgpsl
- \left(\pgpZ\right)^2 + 2\pgpZ\pslpZ
- \left(\pslpZ\right)^2\right],
\nonumber\\
& & \int_{m^2_{3/2}}^{2m^2_{\tilde{l}}}ds_{13}\int_{m_V^2}^{m^2_{\tilde{l}}}ds_{23}\frac{|{\cal M}_1|^2{\cal O}_{1,1}}{m^3_{\tilde{l}}}\sim\left(\tilde{f}{\cal V}^{-\frac{11}{18}}\right)^2{\cal V}^{-\frac{1}{2}}M_p\sim10^{-16.5}M_p;\nonumber\\
\calO_{2,2}&=& -\frac{4}{3\mg^2m_V^2}
\left[\mg^2 - \pgpsl + \pgpZ\right]
\left[\msl^2m_V^2 - \left(\pslpZ\right)^2\right] \nonumber \\
&& \!\!\!\!\!\!\!\! \times \left[-\left(\pgpsl\right)^2
+ 2\pgpsl\pgpZ - \left(\pgpZ\right)^2 +  \mg^2\left(\msl^2 + m_V^2
- 2\pslpZ \right) \right],\nonumber\\
& & \int_{m^2_{3/2}}^{2m^2_{\tilde{l}}}ds_{13}\int_{m_V^2}^{m^2_{\tilde{l}}}ds_{23}\frac{|{\cal M}_2|^2{\cal O}_{2,2}}{m^3_{\tilde{l}}}\sim\left(\frac{\tilde{f}{\cal V}^{-2}}{10^4}\right)^2{\cal V}^{-\frac{1}{2}}M_p\sim10^{-21}M_p;\nonumber\\
\calO_{3,3} &=& {{0}};\nonumber\\
\calO_{4,4} &=&
\frac{4}{3\mg^2m_V^2}
\left[\mg^2m_V^2 - \left(\pgpZ\right)^2\right] \nonumber \\
&& \times \left[\mg^2m_V^2 - m_V^2\pgpsl + 2\left(\pgpZ\right)^2
+ \pgpZ \left(3m_V^2 - 2\pslpZ\right)\right],\nonumber\\
& & \int_{m^2_{3/2}}^{2m^2_{\tilde{l}}}ds_{13}\int_{m_V^2}^{m^2_{\tilde{l}}}ds_{23}\frac{|{\cal M}_4|^2{\cal O}_{4,4}}{m^3_{\tilde{l}}}\sim\left(10\tilde{f}{\cal V}^{-\frac{3}{2}}\right)^2{\cal V}^{\frac{5}{9}}M_p\sim10^{-18}M_p;
\end{eqnarray}

\begin{eqnarray}
\calO_{5,5} &=& -\frac{4}{3\mg^2}
\left[\mg^2\msl^2 - \left(\pgpsl\right)^2\right] \nonumber \\
&& \times
\left[\mg^2m_V^2 - m_V^2\pgpsl - 4\left(\pgpZ\right)^2 + \pgpZ
      \left(-3m_Z^2 + 4\pslpZ\right)\right],\nonumber\\
      & & \int_{m^2_{3/2}}^{2m^2_{\tilde{l}}}ds_{13}\int_{m_V^2}^{m^2_{\tilde{l}}}ds_{23}\frac{|{\cal M}_5|^2{\cal O}_{5,5}}{m^3_{\tilde{l}}}\sim\left(10\tilde{f}{\cal V}^{-\frac{11}{18}}\right)^2{\cal V}^{-\frac{1}{2}}M_p\sim10^{-16.5}M_p;\nonumber\\
      \calO_{6,6} &=& -\frac{4}{3\mg^2}
\left[\mg^2m_Z^2 - \left(\pgpZ\right)^2\right] \nonumber \\
&& \times
\left[\mg^2m_V^2 - m_V^2\pgpsl -  4\left(\pgpZ\right)^2 + \pgpZ
\left(-3m_V^2 + 4\pslpZ\right)\right],\nonumber\\
& & \int_{m^2_{3/2}}^{2m^2_{\tilde{l}}}ds_{13}\int_{m_V^2}^{m^2_{\tilde{l}}}ds_{23}\frac{|{\cal M}_6|^2{\cal O}_{6,6}}{m^3_{\tilde{l}}}\sim\left(10\tilde{f}{\cal V}^{-\frac{3}{2}}\right)^2{\cal V}^{-\frac{25}{9}}M_p\sim10^{-35}M_p;\nonumber\\
\calO_{7,7} &=& \frac{4}{3\mg^2m_V^2}
\left[ \mg^2m_V^2 - \left(\pgpZ\right)^2\right] \nonumber \\
&& \left\{ \mg^4m_V^2 + 2\mg^2\left(\pgpZ\right)^2
+ 4\pgpZ^3 - \pgpsl \left[ \mg^2m_V^2
+ 4\left(\pgpZ\right)^2\right] \right. \nonumber\\
&& \left. - \mg^2\pgpZ \left(m_V^2 - 2\pslpZ\right)\right\},\nonumber\\
& & \int_{m^2_{3/2}}^{2m^2_{\tilde{l}}}ds_{13}\int_{m_V^2}^{m^2_{\tilde{l}}}ds_{23}\frac{|{\cal M}_7|^2{\cal O}_{7,7}}{m^3_{\tilde{l}}}\sim\left(10\tilde{f}{\cal V}^{-\frac{3}{2}}\right)^2{\cal V}^{\frac{8}{9}}M_p\sim10^{-16.6}M_p;\nonumber\\
\calO_{8,8} &=&-\frac{4}{3\mg^2m_V^2}
\left[\mg^2 - \pgpsl + \pgpZ\right]
\left[2\mg^2m_V^2 + \left(\pgpZ\right)^2\right],\nonumber\\
& & \int_{m^2_{3/2}}^{2m^2_{\tilde{l}}}ds_{13}\int_{m_V^2}^{m^2_{\tilde{l}}}ds_{23}\frac{|{\cal M}_8|^2{\cal O}_{8,8}}{m^3_{\tilde{l}}}\sim10^{-27.5}M_p;\nonumber\\
\calO_{9,9} &=& \frac{4}{3\mg^2m_V^2}
\left[ 2\mg^2m_V^2 + \left(\pgpZ\right)^2\right] \nonumber \\
&& \times
\left[\mg^2m_V^2 - m_V^2\pgpsl -  2\left(\pgpZ\right)^2 - \pgpZ
      \left(m_V^2 - 2\pslpZ\right)\right],\nonumber\\
      & & \int_{m^2_{3/2}}^{2m^2_{\tilde{l}}}ds_{13}\int_{m_V^2}^{m^2_{\tilde{l}}}ds_{23}\frac{|{\cal M}_9|^2{\cal O}_{9,9}}{m^3_{\tilde{l}}}\sim\left(10\tilde{f}{\cal V}^{-\frac{3}{2}}\right)^2{\cal V}^{\frac{5}{9}}M_p\sim10^{-18}M_p;
\nonumber\\
\calO_{10,10}&=&\frac{4}{3\mg^2m_V^2}
\left[2\mg^2m_V^2 + \left(\pgpZ\right)^2\right] \nonumber \\
&& \left\{ \mg^4m_V^2 - 2\mg^2\left(\pgpZ\right)^2
- 4\left(\pgpZ\right)^3 \right. \nonumber\\
&& \left. + \pgpsl\left[-\mg^2m_V^2
+ 4\left(\pgpZ\right)^2\right]
+ \mg^2\pgpZ\left(3m_V^2 -  2\pslpZ\right)\right\},\nonumber\\
& & \int_{m^2_{3/2}}^{2m^2_{\tilde{l}}}ds_{13}\int_{m_V^2}^{m^2_{\tilde{l}}}ds_{23}\frac{|{\cal M}_{10}|^2{\cal O}_{10,10}}{m^3_{\tilde{l}}}\sim\left(10\tilde{f}{\cal V}^{-\frac{3}{2}}\right)^2{\cal V}^{\frac{8}{9}}M_p\sim10^{-16.6}M_p;\end{eqnarray}

\begin{eqnarray}
\calO_{1,2}^{\text re}&=&\frac{8}{3\mg^2m_V^2}
\left[\mg^2 - \pgpsl + \pgpZ \right];\nonumber\\
&& \times \left[ - \left(\pgpsl\right)^2 + \pgpsl \pgpZ
+ \mg^2\left(\msl^2 - \pslpZ\right)\right] \nonumber \\
&&\times \left[ -\msl^2m_V^2 + m_Z^2\pgpsl -
\pgpZ \pslpZ + \left(\pslpZ\right)^2\right],\nonumber\\
& & \int_{m^2_{3/2}+m^2_V}^{2m^2_{\tilde{l}}+m_V^2}ds_{13}\int_{m_V^2}^{m^2_{\tilde{l}}}ds_{23}\frac{\Re\left({\cal M}_1{\cal M}_2\right){\cal O}^{\rm re}_{1,2}}{m^3_{\tilde{l}}}\sim10^{-26}M_p;\nonumber\\
\calO_{1,3}^{\text re} &=& \calO_{1,4}^{\text re} = {{0}}
\\
\calO_{1,5}^{\text re} &=& \frac{8}{3\mg^2}
\left[\mg^2\msl^2 - \left(\pgpsl\right)^2\right]
\left[ -\left(\pgpZ\right)^2 +
\mg^2\left(m_V^2 - \pslpZ\right) \right. \nonumber\\
&&
\left. +  \pgpZ\left(-\msl^2 + \pslpZ\right)
+ \pgpsl \left(-m_V^2 + \pgpZ + \pslpZ\right)\right],\nonumber\\
& & \int_{m^2_{3/2}}^{2m^2_{\tilde{l}}}ds_{13}\int_{m_V^2}^{m^2_{\tilde{l}}}ds_{23}\frac{\Re\left({\cal M}_1{\cal M}_5\right){\cal O}^{\rm re}_{1,5}}{m^3_{\tilde{l}}}\sim\tilde{f}^2{\cal V}^{-\frac{11}{9}-\frac{1}{2}}\sim10^{-16.5}M_p;\nonumber\\
{\cal{O}}_{1,6}^{\text re} &=& {\cal{O}}_{1,7}^{\text
re}={\cal{O}}_{1,8}^{\text re}={\cal{O}}_{1,9}^{\text
re}={\cal{O}}_{1,10}^{\text re}=0\nonumber\\
{\cal{O}}_{2,3}^{\text re}&=&{\cal{O}}_{2,4}^{\text re}={{0}}
 \\
{\cal{O}}_{2,5}^{\text re}&=&\frac{4}{3\mg^2}
\left\{
\left(\pgpsl\right)^2\left[\mg^2m_V^2
+ 2\left(\msl^2- m_V^2\right)\pgpZ\right] \right.
\nonumber \\ &&
+ 2\left(\pgpsl\right)^3\left(m_V^2 - \pslpZ\right)
- 2\pgpsl \left[ \left(\pgpZ\right)^2
\left(\msl^2 - \pslpZ\right) \right. \nonumber \\
&&\left. + \mg^2\left(\msl^2 -\pslpZ\right)
\left(m_V^2 - \pslpZ\right) +  \mg^2\pgpZ\pslpZ\right]
\nonumber \\ &&
+\mg^2\left[\msl^2\left(\pgpZ\right)^2
- 2\pgpZ \left(\msl^2 - \pslpZ\right)^2 \right. \nonumber \\
&& \left. \left. + \mg^2\left(-\msl^2m_V^2
+ \left(\pslpZ\right)^2\right)\right]\right\},\nonumber\\
& & \int_{m^2_{3/2}}^{2m^2_{\tilde{l}}}ds_{13}\int_{m_V^2}^{m^2_{\tilde{l}}}ds_{23}\frac{\Re\left({\cal M}_2{\cal M}_5\right){\cal O}^{\rm re}_{2,5}}{m^3_{\tilde{l}}}\sim10^{-37}M_p;
\nonumber\\
{\cal{O}}_{2,6}^{\text re}&=&{\cal{O}}_{2,7}^{\text
re}={\cal{O}}_{2,8}^{\text re}={\cal{O}}_{2,9}^{\text
re}={\cal{O}}_{2,10}^{\text re}={{0}}
 \\
{\cal{O}}_{3,4}^{\text re}&=&{\cal{O}}_{3,5}^{\text
re}={\cal{O}}_{3,6}^{\text re}={\cal{O}}_{3,7}^{\text
re}={\cal{O}}_{3,8}^{\text re}={\cal{O}}_{3,9}^{\text
re}={\cal{O}}_{3,10}^{\text re}={{0}}
 \\
{\cal{O}}_{4,5}^{\text re}&=&{{0}};\nonumber\\
{\cal{O}}_{4,6}^{\text re}&=& \frac{8}{\mg}
\left[\mg^2m_V^2 - \left(\pgpZ\right)^2\right] \left[m_V^2 +
     \pgpZ - \pslpZ\right],\nonumber\\
     & & \int_{m^2_{3/2}}^{2m^2_{\tilde{l}}}ds_{13}\int_{m_V^2}^{m^2_{\tilde{l}}}ds_{23}\frac{\Re\left({\cal M}_4{\cal M}_6\right){\cal O}^{\rm re}_{4,6}}{m^3_{\tilde{l}}}\sim\tilde{f}^2{\cal V}^{-\frac{65}{18}}M_p\sim10^{-42}M_p;\nonumber\\
     \calO_{4,7}^{\text re}&=& \frac{8}{\mg}
\left[\mg^2 - \pgpsl + \pgpZ\right]
\left[\mg^2m_V^2 - \left(\pgpZ\right)^2\right],\nonumber\\
& & \int_{m^2_{3/2}}^{2m^2_{\tilde{l}}}ds_{13}\int_{m_V^2}^{m^2_{\tilde{l}}}ds_{23}\frac{\Re\left({\cal M}_4{\cal M}_7\right){\cal O}^{\rm re}_{4,7}}{m^3_{\tilde{l}}}\sim\left(\tilde{f}{\cal V}^{-\frac{3}{2}}\right){\cal V}^{-\frac{28}{9}}M_p\sim10^{-36}M_p;
\end{eqnarray}

 \begin{eqnarray}
\calO_{4,8}^{\text re}&=& \frac{8}{3\mg m_V^2}
\left[-m_V^2\left(\mg^2 - 2\pgpsl\right)\pgpZ + \pgpZ^ 3 \right. \nonumber\\
&&
\left. + \mg^2m_V^2\left(m_V^2 - \pslpZ\right) -
       \left(\pgpZ\right)^2\left(m_V^2 + \pslpZ\right)\right],\nonumber\\
       & & \int_{m^2_{3/2}}^{2m^2_{\tilde{l}}}ds_{13}\int_{m_V^2}^{m^2_{\tilde{l}}}ds_{23}\frac{\Re\left({\cal M}_4{\cal M}_8\right){\cal O}^{\rm re}_{4,8}}{m^3_{\tilde{l}}}\sim10^{-15}M_p;
    \nonumber\\
\calO_{4,9}^{\text re}&=& \frac{8}{3\mg^2 m_V^2}
 \left[\mg^4m_V^4 + \mg^2m_V^4\pgpZ +
      \mg^2m_V^2\left(\pgpZ\right)^2 -  2\left(\pgpZ\right)^4 \right.
\nonumber \\ &&
\left. -  m_V^2\pgpsl\left(\mg^2m_V^2 +  \left(\pgpZ\right)^2\right)
- \left(\pgpZ\right)^3 \left(m_V^2 - 2\pslpZ\right)\right],\nonumber\\
& & \int_{m^2_{3/2}}^{2m^2_{\tilde{l}}}ds_{13}\int_{m_V^2}^{m^2_{\tilde{l}}}ds_{23}\frac{\Re\left({\cal M}_4{\cal M}_9\right){\cal O}^{\rm re}_{4,9}}{m^3_{\tilde{l}}}\sim\left(10\tilde{f}{\cal V}^{-\frac{3}{2}}\right)^2{\cal V}^{\frac{2}{9}}M_p\sim10^{-20}M_p;\nonumber\\
\calO_{4,10}^{\text re}&=&
\frac{8}{3\mg}
\left\{\mg^4m_V^2 - \mg^2\left(\pgpZ\right)^2 -
      3\pgpZ^3 +  \pgpsl\left[-\mg^2m_V^2 +
      3\left(\pgpZ\right)^2\right] \right. \nonumber\\
&& \left. + \mg^2\pgpZ\left(3m_V^2 -  2\pslpZ\right)\right\},\nonumber\\
& & \int_{m^2_{3/2}}^{2m^2_{\tilde{l}}}ds_{13}\int_{m_V^2}^{m^2_{\tilde{l}}}ds_{23}\frac{\Re\left({\cal M}_4{\cal M}_{10}\right){\cal O}^{\rm re}_{4,10}}{m^3_{\tilde{l}}}\sim\left(10\tilde{f}{\cal V}^{-\frac{3}{2}}\right)^2{\cal V}^{-\frac{28}{9}}M_p\sim10^{-24}M_p;\nonumber\\
{\cal{O}}_{5,6}^{\text re}&=&{\cal{O}}_{5,7}^{\text
re}={\cal{O}}_{5,8}^{\text re}={\cal{O}}_{5,9}^{\text
re}={\cal{O}}_{5,10}^{\text re}={{0}}\nonumber\\
\calO_{6,7}^{\text re}&=& \frac{8}{3\mg^2}
\left[\mg^2m_V^2 - \left(\pgpZ\right)^2\right] \nonumber \\
&& \times
\left[\left(3\mg^2 - 2\pgpsl\right)\pgpZ
+ 2\left(\pgpZ\right)^2 +
\mg^2\left(m_V^2 - \pslpZ\right)\right],\nonumber\\
& & \int_{m^2_{3/2}}^{2m^2_{\tilde{l}}}ds_{13}\int_{m_V^2}^{m^2_{\tilde{l}}}ds_{23}\frac{\Re\left({\cal M}_6{\cal M}_7\right){\cal O}^{\rm re}_{6,7}}{m^3_{\tilde{l}}}\sim\left(10\tilde{f}{\cal V}^{-\frac{3}{2}}\right)^2{\cal V}^{-\frac{22}{9}}M_p\sim10^{-33}M_p;\nonumber\\
\calO_{6,8}^{\text re}&=&-\frac{8}{3\mg^2}
 \left[\mg^4m_V^2 - \mg^2\left(\pgpZ\right)^2
+ \left(\pgpZ\right)^3 \right.\nonumber\\
&& \left. - \pgpsl \left(\mg^2m_V^2 + \left(\pgpZ\right)^2\right)
- \mg^2\pgpZ \left(m_V^2 - 2\pslpZ\right)\right],\nonumber\\
& & \int_{m^2_{3/2}}^{2m^2_{\tilde{l}}}ds_{13}\int_{m_V^2}^{m^2_{\tilde{l}}}ds_{23}\frac{\Re\left({\cal M}_6{\cal M}_8\right){\cal O}^{\rm re}_{6,8}}{m^3_{\tilde{l}}}\sim10^{-27}M_p;\nonumber\\
\calO_{6,9}^{\text re}&=&\frac{8}{3\mg}
\left[m_V^2\left(3\mg^2 - 2\pgpsl\right) \pgpZ - 3\left(\pgpZ\right)^3
\right. \nonumber\\
&&
\left. - \left(\pgpZ\right)^2 \left(m_V^2 - 3\pslpZ\right)
+ \mg^2m_V^2\left(m_V^2 - \pslpZ\right)\right],\nonumber\\
& & \int_{m^2_{3/2}}^{2m^2_{\tilde{l}}}ds_{13}\int_{m_V^2}^{m^2_{\tilde{l}}}ds_{23}\frac{\Re\left({\cal M}_6{\cal M}_9\right){\cal O}^{\rm re}_{6,9}}{m^3_{\tilde{l}}}\sim\left(10\tilde{f}{\cal V}^{-\frac{3}{2}}\right)^2{\cal V}^{-\frac{31}{9}}M_p\sim10^{-38}M_p;
\end{eqnarray}

\begin{eqnarray}
\calO_{6,10}^{\text re}&=&\frac{8}{3\mg^2}
\left[\mg^4m_V^2\pgpZ -  \left(\mg^2 - 2\pgpsl\right) \left(\pgpZ\right)^3 -
      2\left(\pgpZ\right)^4
\right. \nonumber \\
&& \left. + \mg^4m_V^2\left(m_V^2 - \pslpZ\right)
+ \mg^2\left(\pgpZ\right)^2 \left(m_V^2 - \pslpZ\right)\right],\nonumber\\
& & \int_{m^2_{3/2}}^{2m^2_{\tilde{l}}}ds_{13}\int_{m_V^2}^{m^2_{\tilde{l}}}ds_{23}\frac{\Re\left({\cal M}_6{\cal M}_{10}\right){\cal O}^{\rm re}_{6,10}}{m^3_{\tilde{l}}}\sim\left(10\tilde{f}{\cal V}^{-\frac{3}{2}}\right)^2{\cal V}^{-\frac{22}{9}}M_p\sim10^{-33}M_p;
\nonumber\\
\calO_{7,8}^{\text re}&=&\frac{8}{3\mg^2m_V^2}
 \left[\mg^4m_V^2\pgpZ -  \left(\mg^2 - 2\pgpsl\right)\left(\pgpZ\right)^3 -
      2\left(\pgpZ\right)^4
\right. \nonumber \\ &&
\left. + \mg^4m_V^2\left(m_V^2 - \pslpZ\right)
+ \mg^2\left(\pgpZ\right)^2 \left(m_V^2 - \pslpZ\right)\right]\nonumber\\
& & \int_{m^2_{3/2}}^{2m^2_{\tilde{l}}}ds_{13}\int_{m_V^2}^{m^2_{\tilde{l}}}ds_{23}\frac{\Re\left({\cal M}_7{\cal M}_8\right){\cal O}^{\rm re}_{7,8}}{m^3_{\tilde{l}}}\sim10^{-17.5}M_p;
\nonumber\\
\calO_{7,9}^{\text re}&=&\frac{8}{3\mg}
\left[\mg^4m_V^2 - \mg^2\left(\pgpZ\right)^2 +  \left(\pgpZ\right)^3
\right. \nonumber\\
&&
\left. - \pgpsl \left(\mg^2m_V^2 + \left(\pgpZ\right)^2\right)
     -  \mg^2\pgpZ \left(m_V^2 - 2\pslpZ\right)\right],\nonumber\\
     & & \int_{m^2_{3/2}}^{2m^2_{\tilde{l}}}ds_{13}\int_{m_V^2}^{m^2_{\tilde{l}}}ds_{23}\frac{\Re\left({\cal M}_7{\cal M}_9\right){\cal O}^{\rm re}_{7,9}}{m^3_{\tilde{l}}}\sim\left(10\tilde{f}{\cal V}^{-\frac{3}{2}}\right)^2{\cal V}^{-\frac{28}{9}}M_p\sim10^{-24}M_p;\nonumber\\
\calO_{7,10}^{\text re}&=&
\frac{8}{3\mg^2m_Z^2}
\left\{\mg^6m_V^4 + \mg^4m_V^4\pgpZ +  \mg^4m_V^2\left(\pgpZ\right)^2
\right. \nonumber \\ &&
-  2\mg^2\left(\pgpZ\right)^4      -  4\left(\pgpZ\right)^5
\nonumber \\
&& \left. -  \pgpsl\left[\mg^4m_V^4 +  \mg^2m_V^2\left(\pgpZ\right)^2 -
  4\left(\pgpZ\right)^4\right] \right.
\nonumber\\
&&
\left. + \mg^2\left(\pgpZ\right)^3 \left(3m_V^2 - 2\pslpZ\right)\right\},\nonumber\\
& & \int_{m^2_{3/2}}^{2m^2_{\tilde{l}}}ds_{13}\int_{m_V^2}^{m^2_{\tilde{l}}}ds_{23}\frac{\Re\left({\cal M}_7{\cal M}_{10}\right){\cal O}^{\rm re}_{7,10}}{m^3_{\tilde{l}}}\sim\left(10\tilde{f}{\cal V}^{-\frac{3}{2}}\right)^2{\cal V}^{\frac{8}{9}}M_p\sim10^{-16.6}M_p;\nonumber\\
\calO_{8,9}^{\text re}&=&\frac{8}{3\mg m_V^2}
\left[ 2\mg^2m_V^2 + \left(\pgpZ\right)^2\right]
\left[m_V^2 + \pgpZ -\pslpZ\right],\nonumber\\
& & \int_{m^2_{3/2}}^{2m^2_{\tilde{l}}}ds_{13}\int_{m_V^2}^{m^2_{\tilde{l}}}ds_{23}\frac{\Re\left({\cal M}_8{\cal M}_9\right){\cal O}^{\rm re}_{8,9}}{m^3_{\tilde{l}}}\sim10^{-25.5}M_p;\nonumber\\
\calO_{8,10}^{\text re}&=&\frac{8}{3\mg^2 m_V^2}
\left[2\mg^2m_V^2 + \left(\pgpZ\right)^2\right] \nonumber \\
&& \times
\left[-\left(\mg^2 - 2\pgpsl\right)\pgpZ
- 2\left(\pgpZ\right)^2 +\mg^2\left(m_V^2 - \pslpZ\right)\right],\nonumber\\
& & \int_{m^2_{3/2}}^{2m^2_{\tilde{l}}}ds_{13}\int_{m_V^2}^{m^2_{\tilde{l}}}ds_{23}\frac{\Re\left({\cal M}_8{\cal M}_{10}\right){\cal O}^{\rm re}_{8,10}}{m^3_{\tilde{l}}}\sim10^{-18.5}M_p;\nonumber\\
\calO_{9,10}^{\text re}&=&\frac{8}{3\mg}
\left[\mg^2 - \pgpsl+ \pgpZ\right]
\left[2\mg^2m_V^2 + \left(\pgpZ\right)^2\right],\nonumber\\
& & \int_{m^2_{3/2}}^{2m^2_{\tilde{l}}}ds_{13}\int_{m_V^2}^{m^2_{\tilde{l}}}ds_{23}\frac{\Re\left({\cal M}_9{\cal M}_{10}\right){\cal O}^{\rm re}_{9,10}}{m^3_{\tilde{l}}}\sim\left(10\tilde{f}{\cal V}^{-\frac{3}{2}}\right)^2{\cal V}^{-\frac{28}{9}}M_p\sim10^{-24}M_p;\nonumber\\
\end{eqnarray}

\begin{eqnarray}
\calO_{1,8}^{\text im}&=&
-\frac{8}{3\mg^2m_V^2}(\mg^2-\pgpsl+\pgpZ)
\left\{\left[\msl^2m_V^2-(\pslpZ)^2+(\pgpZ)(\pslpZ)\right]\mg^2
\right. \nonumber \\
&& \left. -m_V^2(\pgpsl)^2
+(\pgpsl)(\pgpZ)(\pslpZ-\pgpZ)\right\},\nonumber\\
& & \int_{m^2_{3/2}}^{2m^2_{\tilde{l}}}ds_{13}\int_{m_V^2}^{m^2_{\tilde{l}}}ds_{23}\frac{\Imag\left({\cal M}_1{\cal M}_8^*\right){\cal O}^{\rm im}_{1,8}}{m^3_{\tilde{l}}}\sim10^{-44}M_p;
\nonumber\\
\calO_{2,8}^{\text im}&=&
-\frac{8}{3\mg^2m_V^2}(\mg^2-\pgpsl+\pgpZ)
\left[(\msl^2m_V^2-(\pslpZ)^2)\mg^2-m_V^2(\pgpsl)^2\right.
\nonumber \\
&& \left.-(\pgpZ)^2\pslpZ
+(\pgpsl)(\pgpZ)(m_V^2+\pslpZ)\right],\nonumber\\
& & \int_{m^2_{3/2}}^{2m^2_{\tilde{l}}}ds_{13}\int_{m_V^2}^{m^2_{\tilde{l}}}ds_{23}\frac{\Imag\left({\cal M}_2{\cal M}_8^*\right){\cal O}^{\rm im}_{1,8}}{m^3_{\tilde{l}}}\sim10^{-33}M_p;\nonumber\\
\calO_{3,8}^{\text im}&=&\calO_{4,8}^{\text im}=0
\nonumber\\
\calO_{5,8}^{\text im}&=&
\frac{8}{3\mg^2}\left\{
(\pgpsl-\mg^2)\pslpZ \mg^2 -\pgpsl(\pgpZ)^2\right.
\nonumber \\
&& \left.
+\pgpZ\left[\pgpsl \mg^2
+(\pslpZ-2 \msl^2)\mg^2+(\pgpsl)^2\right]\right\},\nonumber\\
& & \int_{m^2_{3/2}}^{2m^2_{\tilde{l}}}ds_{13}\int_{m_V^2}^{m^2_{\tilde{l}}}ds_{23}\frac{\Imag\left({\cal M}_5{\cal M}_8^*\right){\cal O}^{\rm im}_{5,8}}{m^3_{\tilde{l}}}\sim10^{-38}M_p;\nonumber\\
\calO_{6,8}^{\text im}&=&\calO_{7,8}^{\text im}
=\calO_{8,9}^{\text im}=\calO_{8,10}^{\text im}=0
\ .
\end{eqnarray}
\section{$K_{T_B{\bar T}_B}$, the Massive Gauge Boson mass and Soft SUSY Parameters Miscellania}

The relevant term in the ${\cal N}=1$ SUGRA action of \cite{Wess_Bagger}, assuming that for multiple $D7$-brane stacks, the non-abelian killing isometry of the moduli space gets identified with the SM gauge group, the mass squared of the $W/Z$-bosons are given by: $g_{T_B{\bar T}_B}\left(X^{T_B}\right)^2$. Now,
\begin{eqnarray}
\label{eq:dBdBbarK}
& & G_{T_B{\bar T}_B}\sim\frac{1}{\sqrt{T_B+{\bar T}_B - C_{I{\bar J}}a_I{\bar a}_{\bar J} - \mu_3l^2{\cal V}^{\frac{1}{18}}}\Sigma} - \frac{3\left(T_B+{\bar T}_B - C_{I{\bar J}}a_I{\bar a}_{\bar J} - \mu_3l^2{\cal V}^{\frac{1}{18}}\right)}{\Sigma^2},
\end{eqnarray}
where $\Sigma\equiv \left(T_B+{\bar T}_B - C_{I{\bar J}}a_I{\bar a}_{\bar J} - \mu_3l^2{\cal V}^{\frac{1}{18}}\right)^{\frac{3}{2}}$ $-\left(T_B+{\bar T}_B  - \mu_3l^2{\cal V}^{\frac{1}{18}}\right)^{\frac{3}{2}}
+\sum_{\beta}n^0_\beta(...)$.
So, to get a mass of $90 GeV$ or so at the EW scale, one needs to show that (\ref{eq:dBdBbarK}) can RG-flow down to the required very small value of the square of the same; lets call this small value as $G^{EW}_{T_B{\bar T}_B}\sim 10^{-20}$. This would imply the following relation between ${\cal T}_B\equiv T_B+{\bar T}_B - C_{I{\bar J}}a_I{\bar a}_{\bar J} - \mu_3l^2{\cal V}^{\frac{1}{18}}$ and $\Sigma$:
\begin{eqnarray}
\label{eq:small gauge mass_I}
& & \frac{1}{\sqrt{{\cal T}_B}} - \frac{3{\cal T}_B}{\Sigma^2}=G^{EW}_{T_B{\bar T}_B}\nonumber\\
& & {\rm or}: \Sigma^2 - \frac{\Sigma}{\delta\sqrt{{\cal T}_B}} + \frac{3{\cal T}_B}{\Sigma}=0,
\end{eqnarray}
whose valid solution is given by:
\begin{equation}
\label{eq:small_gauge_mass_II}
\hskip -0.3in \Sigma=\frac{1}{2}\left(\frac{1}{G^{EW}_{T_B{\bar T}_B}\sqrt{{\cal T}_B}}-\sqrt{1-\frac{12{\cal T}_B}{G^{EW}_{T_B{\bar T}_B}}}\right)\approx  3 {\cal T}_B^{\frac{3}{2}} + 9 {\cal T}_B^{\frac{7}{2}}G^{EW}_{T_B{\bar T}_B} + {\cal O}(G^{EW}_{T_B{\bar T}_B}\ ^2).
\end{equation}
The above is equivalent to:
\begin{equation}
\label{eq:small_gauge_mass_III}
{\cal T}_B(EW)\sim0.6{\cal V}^{\frac{2}{3}},
\end{equation}
which can be satisfied, e.g., if
\begin{equation}
\label{eq:small_gauge_mass_IV}
\langle a_I\rangle(EW)\sim\frac{1}{{\cal O}(1)}\langle a_I\rangle(M_S).
\end{equation}
Now,
\begin{equation}
\label{eq:daIGBBbar}
\partial_{a^I}G_{T_B{\bar T}_B}\sim -\frac{C_{I{\bar J}}{\bar a}^{\bar J}}{2{\cal T}_B^{\frac{3}{2}}\Sigma} - \frac{9C_{I{\bar J}}{\bar a}^{\bar J}}{2\Sigma^2} + 9\frac{{\cal T}_B^{\frac{3}{2}}C_{I{\bar J}}{\bar a}^{\bar J}}{\Sigma^3}.
\end{equation}
Similarly
\begin{equation}
\label{eq:dziGBBbar}
\partial_{z^i}G_{T_B{\bar T}_B}\sim -\frac{\mu_3l^2{\bar z}^{\bar j}}{2{\cal T}_B^{\frac{3}{2}}\Sigma} - \frac{9\mu_3l^2{\bar z}^{\bar j}}{2\Sigma^2} + 9\frac{{\cal T}_B^{\frac{3}{2}}\mu_3l^2{\bar z}^{\bar j}}{\Sigma^3}.
\end{equation}
Therefore,
\begin{eqnarray}
\label{eq:dadabarGBBbar}
& & \hskip-0.4in {\bar{\partial}}_{{\bar a}_{\bar K}}\partial_{a_I}G_{T_B{\bar T}_B}\sim\frac{C_{I{\bar K}}}{2{\cal T}_B^{\frac{3}{2}}\Sigma} + \frac{3}{4}\frac{C_{I{\bar J}}{\bar a}^{\bar J}C_{L{\bar K}}a^L}{{\cal T}_B^{\frac{5}{2}}}
 + \frac{9}{4}\frac{C_{I{\bar J}}{\bar a}^{\bar J}C_{L{\bar K}}a^L}{\Sigma^2{\cal T}_B} - \frac{9C_{I{\bar K}}}{2\Sigma^2} + \frac{27}{2}\frac{C_{I{\bar J}}{\bar a}^{\bar J}C_{L{\bar K}}a^L\sqrt{{\cal T}_B}}{\Sigma^3} + \frac{9C_{I{\bar K}}{\cal T}_B^{\frac{3}{2}}}{\Sigma^3} \nonumber\\
& &  - \frac{81}{2}\frac{{\cal T}_B^2C_{I{\bar J}}{\bar a}^{\bar J}C_{L{\bar K}}a^L}{\Sigma^4} + \frac{27}{2}\frac{\sqrt{{\cal T}_B}C_{I{\bar J}}{\bar a}^{\bar J}C_{L{\bar K}}a^L}{\Sigma^3}.
\end{eqnarray}
Now, for $I=K=a_1$,  $C_{L{\bar a}_1}\langle a^L\rangle(M_S)\sim{\cal V}^{\frac{8}{9}}$; which because of  (\ref{eq:small_gauge_mass_IV}), implies that at the $EW$ scale,
\begin{equation}
\label{eq:da1da1barGBBbar}
{\bar{\partial}}_{{\bar a}_{\bar K}}\partial_{a_I}G_{T_B{\bar T}_B}\Biggr|_{\Sigma(EW)\sim3{\cal T}_B^{\frac{3}{2}}(EW),{\cal T}_B(EW)\sim0.6{\cal V}^{\frac{2}{3}}}\sim{\cal V}^{-\frac{8}{9}}.
\end{equation}
Similarly
\begin{eqnarray}
\label{eq:dzdzbarGBBbar}
& & \hskip-0.4in {\bar{\partial}}_{{\bar z}_{\bar k}}\partial_{z_i}G_{T_B{\bar T}_B}\sim\frac{\mu_3l^2}{2{\cal T}_B^{\frac{3}{2}}\Sigma} + \frac{3}{4}\frac{(\mu_3l^2)^2{\bar z}^{\bar j}z^l}{{\cal T}_B^{\frac{5}{2}}}
 + \frac{9}{4}\frac{(\mu_3l^2)^2{\bar z}^{\bar j}z^l}{\Sigma^2\sqrt{{\cal T}_B}} - \frac{\mu_3l^2}{\Sigma^2} + \frac{27}{2}\frac{(\mu_3l^2)^2{\bar z}^{\bar j}z^L}{\Sigma^2{\cal T}_B} - \frac{{\cal T}_B^{\frac{3}{2}}\mu_3l^2}{\Sigma^3} \nonumber\\
& & +9\frac{{\cal T}_B^{\frac{3}{2}}\mu_3l^2}{\Sigma^3} - \frac{81}{2}\frac{{\cal T}_B^2(\mu_3l^2)^2{\bar z}^{\bar j}z^l}{\Sigma^4} + \frac{27}{2}\frac{\sqrt{{\cal T}_B}(\mu_3l^2)^2{\bar z}^{\bar j}z^l}{\Sigma^3}\nonumber\\
\end{eqnarray}
Now, for $i=k=z_1$,  at the $EW$ scale will be
\begin{equation}
\label{eq:dz1dz1barGBBbar}
{\bar{\partial}}_{{\bar z}_{\bar k}}\partial_{z_i}G_{T_B{\bar T}_B}\Biggr|_{{\cal T}_B(EW)\sim0.6{\cal V}^{\frac{2}{3}}}\sim \frac{\mu_3l^2}{2{\cal T}_B^{\frac{3}{2}}\Sigma}\Biggr|_{{\cal T}_B(EW)\sim0.6{\cal V}^{\frac{2}{3}}}\sim{\cal V}^{-2}.
\end{equation}

We will briefly describe evaluation of various soft supersymmetry breaking parameters in our current setup involving four Wilson line moduli. To begin with, in order to evaluate the gaugino masses, one needs to evaluate the bulk $F$-terms which in turn entails evaluating the bulk metric. Writing the K\"{a}hler sector of the K\"{a}hler potential as: $K\sim-2 ln\left[\left(\sigma_B + \bar{\sigma}_B - \gamma K_{\rm geom}\right)^{\frac{3}{2}} - \left(\sigma_S + \bar{\sigma}_S - \gamma K_{rm geom}\right)^{\frac{3}{2}}\right.$ \\$\left. + \sum_{\beta\in H_2^-(CY_3)}n^0_\beta\sum_{(n,m)} cos\left( i n k\cdot(G - \bar{G})g_s + m k\cdot (G + \bar{G}) \right)\right]$, and working near  \\ $sin \left( i n k\cdot(G - \bar{G})g_s + m k\cdot (G + \bar{G}) \right)=0$ - corresponding to a local minimum - generates the following components of the bulk metric: $G_{m\bar{n}}\sim
\left(\begin{array}{cccc} {\cal V}^{-\frac{37}{36}} & {\cal V}^{-\frac{59}{36}} & 0 & 0\\
{\cal V}^{-\frac{59}{36}} & {\cal V}^{-\frac{4}{3}} & 0 & 0 \\
0 & 0 & {\cal O}(1) & {\cal O}(1) \\
0 & 0 & {\cal O}(1) & {\cal O}(1)
\end{array}\right),$ which therefore produces the following inverse: $G^{m\bar{n}}\sim\left(\begin{array}{cccc} {\cal V}^{\frac{37}{36}} & {\cal V}^{\frac{13}{18}} & 0 & 0 \\
{\cal V}^{\frac{13}{18}} & {\cal V}^{\frac{4}{3}} & 0 & 0 \\
0 & 0 & {\cal O}(1) & {\cal O}(1) \\
0 & 0 & {\cal O}(1) & {\cal O}(1)
\end{array}\right)$. Given that bulk $F$-terms are defined as: $F^m=e^{\frac{K}{2}}G^{m\bar{n}}D_{\bar{n}}\bar{W}$, one obtains:
\begin{equation}
\label{eq:bulk_F}
F^{\sigma_S}\sim{\cal V}^{-\frac{n^s}{2} + \frac{1}{36}}M_p^2,\ F^{\sigma_B}\sim{\cal V}^{-\frac{n^s}{2} - \frac{5}{18}}M_p^2,\ F^{G^a}\sim{\cal V}^{-\frac{n^s}{2}-1}M_p^2,
\end{equation}
implying that the gaugino mass is:
\begin{equation}
\label{eq:gaugino_mass}
m_{\tilde{g}}=\frac{F^m\partial_m T_B}{Re T_B}\stackrel{<}{\sim}{\cal V}^{\frac{2}{3}}m_{3/2},
\end{equation}
where the gravitino mass, $m_{3/2}\sim{\cal V}^{-\frac{n^s}{2}-1}M_p$.

To calculate the mixed double derivatives of the K\"{a}hler potential (\ref{eq:K}) with respect to (\ref{eq:eigenvectors}), we use (\ref{eq:a_I+z_i}) and:
\begin{eqnarray}
\label{eq:ders_K}
& & \partial_{z_i}K\sim\frac{\sqrt{{\cal T}_B}\mu_3l^2z_i}{\Sigma} - \frac{\sqrt{{\cal T}_S}\mu_3l^2z_i}{\Sigma},\nonumber\\
& & \partial_{z_i}\bar{\partial}_{\bar{z}_j}K\sim\frac{\mu_3^2l^4 z_iz_j}{\sqrt{{\cal T}_B}\Sigma} -
\frac{\sqrt{{\cal T}_B}\sqrt{{\cal T}_S}\mu_3^2z_iz_j}{\Sigma^2} + \frac{\sqrt{{\cal T}_B}\mu_3\delta_{ij}}{\Sigma},\nonumber\\
& & \partial_{z_i}\bar{\partial}_{\bar{a}_I}K\sim\frac{\mu_3l^2C^{J\bar{I}}a_J}{\Sigma\sqrt{{\cal T}_B}}
-\frac{\sqrt{{\cal T}_B}\sqrt{{\cal T}_S}\mu_3l^2z_iC^{J\bar{I}}a_J}{\Sigma^2} + \frac{\sqrt{{\cal T}_S}\sqrt{{\cal T}_B}C^{J\bar{I}}a_J}{\Sigma^2};\nonumber\\
& & \partial_{a_I}K\sim\frac{\sqrt{{\cal T}_B}C^{I\bar{J}}a_{\bar{J}}}{\Sigma},\nonumber\\
& & \partial_{a_I}\bar{\partial}_{\bar{a}_J}K\sim\frac{C^{I\bar{K}}\bar{a}_{\bar{K}}C^{L\bar{J}}a_L}{\Sigma\sqrt{{\cal T}_B}} - \frac{{\cal T}_BC^{I\bar{K}}\bar{a}_{\bar{K}}C^{L\bar{J}}a_L}{\Sigma^2} - \frac{\sqrt{{\cal T}_B}C^{I\bar{J}}}{\Sigma},
\end{eqnarray}
we see that:
\begin{eqnarray}
\label{eq:KZiZibar}
& & \hskip-0.8in \partial_{{\cal Z}_i}\bar{\partial}_{\bar{\cal Z}_j}K=\frac{\partial z_k}{\partial{\cal Z}_i}\frac{\bar{\partial}\bar{z}_l}{\bar{\partial}\bar{\cal Z}_j}\partial_{z_k}\bar{\partial}_{\bar{z}_l}K
+ \frac{\partial a_I}{\partial{\cal Z}_i}\frac{\bar{\partial}\bar{a}_J}{\bar{\partial}\bar{\cal Z}_j}
\partial_{a_I}\bar{\partial}_{\bar{a}_J}K + \frac{\partial a_I}{\partial{\cal Z}_i}\frac{\bar{\partial}\bar{z}_l}{\bar{\partial}\bar{\cal Z}_j}\partial_{a_I}\bar{\partial}_{\bar{z}_l}K
+ \frac{\partial z_l}{\partial{\cal Z}_i}\frac{\bar{\partial}\bar{a}_J}{\bar{\partial}\bar{\cal Z}_j}
\partial_{z_l}\bar{\partial}_{\bar{a}_J}K\nonumber\\
& & \sim\frac{\sqrt{{\cal T}_B}\mu_3l^2}{\Sigma}\sim{\cal V}^{-\frac{2}{3}},
\end{eqnarray}
implying:
\begin{eqnarray}
\label{eq:ds_KZiZibar}
& & \partial_{\sigma_B}\left(\partial_{{\cal Z}_i}\bar{\partial}_{\bar{\cal Z}_j}K\right)\sim\frac{{\cal T}_B\mu_3l^2}{\Sigma^2}\sim{\cal V}^{-\frac{4}{3}},\nonumber\\
& & \partial_{\sigma_B}\bar{\partial}_{\sigma_B}\left(\partial_{{\cal Z}_i}\bar{\partial}_{\bar{\cal Z}_j}K\right)
\sim{\cal V}^{-2}\mu_3l^2,\nonumber\\
& & \partial_{\sigma_B}\bar{\partial}_{\bar{\sigma}_S}\left(\partial_{{\cal Z}_i}\bar{\partial}_{\bar{\cal Z}_j}K\right)\sim{\cal V}^{-\frac{43}{36}}\mu_3l^2,\nonumber\\
& & \partial_{\sigma_S}\bar{\partial}_{\bar{\sigma}_S}\left(\partial_{{\cal Z}_i}\bar{\partial}_{\bar{\cal Z}_j}K\right)\sim{\cal V}^{-\frac{11}{12}}\mu_3l^2,\nonumber\\
& & \partial_{G^a}\bar{\partial}_{\sigma_{B,S}}\left(\partial_{{\cal Z}_i}\bar{\partial}_{\bar{\cal Z}_j}K\right)\Biggr|_{sin \left( i n k\cdot(G - \bar{G})g_s + m k\cdot (G + \bar{G}) \right)=0}=0,\nonumber\\
& & \partial_{G^a}\bar{\partial}_{\bar{G}^b}\left(\partial_{{\cal Z}_i}\bar{\partial}_{\bar{\cal Z}_j}K\right)\Biggr|_{sin \left( i n k\cdot(G - \bar{G})g_s + m k\cdot (G + \bar{G}) \right)=0}\sim{\cal V}^{-\frac{2}{3}}.
\end{eqnarray}
From (\ref{eq:bulk_F}) and (\ref{eq:ds_KZiZibar}), one obtains:
\begin{eqnarray}
\label{eq:Fs.ddlnKZ1Z1bar}
& & |F^{\sigma_S}|^2\partial_{\sigma_S}\bar{\partial}_{\bar{\sigma}_S} ln\left(\partial_{{\cal Z}_i}\bar{\partial}_{\bar{\cal Z}_j}K\right)\sim{\cal V}^{-\frac{85}{36}},\nonumber\\
& & |F^{\sigma_B}|^2\partial_{\sigma_B}\bar{\partial}_{\bar{\sigma}_B} ln\left(\partial_{{\cal Z}_i}\bar{\partial}_{\bar{\cal Z}_j}K\right)\sim{\cal V}^{-\frac{35}{9}},\nonumber\\
& & F^{\sigma_S}\bar{F}^{\bar{\sigma}_B} \partial_{\sigma_S}\bar{\partial}_{\bar{\sigma}_B}ln\left(\partial_{{\cal Z}_i}\bar{\partial}_{\bar{\cal Z}_j}K\right)\sim{\cal V}^{-\frac{59}{18}},\nonumber\\
& & F^{G^a}\bar{F}^{\bar{G}^b}\partial_{G^a}\bar{\partial}_{\bar{G}^b}ln\left(\partial_{{\cal Z}_i}\bar{\partial}_{\bar{\cal Z}_j}K\right)\Biggr|_{sin \left( i n k\cdot(G - \bar{G})g_s + m k\cdot (G + \bar{G}) \right)=0}\sim{\cal V}^{-4},
\end{eqnarray}
implying that
\begin{equation}
\label{eq:mass_Zi}
m_{{\cal Z}_i}\sim {\cal V}^{\frac{59}{72}}m_{3/2}.
\end{equation}
Similarly, utilizing:
\begin{eqnarray}
\label{eq:KA1A1bar}
& & \hskip-0.6in\partial_{{\cal A}_1}\bar{\partial}_{\bar{\cal A}_1}K=\frac{\partial a_I}{\partial{\cal A}_1}\frac{\bar{\partial}\bar{a}_J}{\bar{\partial}\bar{\cal A}_1}\partial_{a_I}\bar{\partial}_{\bar{a}_J}K
+ \frac{\partial a_I}{\partial{\cal A}_1}\frac{\bar{\partial}\bar{z}_j}{\bar{\partial}\bar{\cal A}_1}\partial_{a_I}\bar{\partial}_{\bar{z}_j}K + \frac{\partial z_i}{\partial{\cal A}_1}\frac{\bar{\partial}\bar{a}_J}{\bar{\partial}\bar{\cal A}_1}\partial_{z_i}\bar{\partial}_{\bar{a}_J}K + \frac{\partial z_i}{\partial{\cal A}_1}\frac{\bar{\partial}\bar{z}_j}{\bar{\partial}\bar{\cal A}_1}\partial_{z_i}\bar{\partial}_{\bar{z}_j}K\nonumber\\
& & \sim\frac{{\cal V}^{\frac{10}{9}}\sqrt{{\cal T}_B}}{\Sigma},
\end{eqnarray}
along with:
\begin{eqnarray}
\label{eq:Fs.ddlnKA1A1bar}
& & |F^{\sigma_S}|^2\partial_{\sigma_S}\bar{\partial}_{\bar{\sigma}_S} ln\left(\partial_{{\cal A}_1}\bar{\partial}_{\bar{\cal A}_1}K\right)\sim{\cal V}^{-\frac{107}{36}},\nonumber\\
& & |F^{\sigma_B}|^2\partial_{\sigma_B}\bar{\partial}_{\bar{\sigma}_B} ln\left(\partial_{{\cal A}_1}\bar{\partial}_{\bar{\cal A}_2}K\right)\sim{\cal V}^{-\frac{35}{9}},\nonumber\\
& & F^{\sigma_S}\bar{F}^{\bar{\sigma}_B} \partial_{\sigma_S}\bar{\partial}_{\bar{\sigma}_B}ln\left(\partial_{{\cal A}_1}\bar{\partial}_{\bar{\cal A}_2}K\right)\sim{\cal V}^{-\frac{35}{9}},\nonumber\\
& & F^{G^a}\bar{F}^{\bar{G}^b}\partial_{G^a}\bar{\partial}_{\bar{G}^b}ln\left(\partial_{{\cal Z}_i}\bar{\partial}_{\bar{\cal Z}_j}K\right)\Biggr|_{sin \left( i n k\cdot(G - \bar{G})g_s + m k\cdot (G + \bar{G}) \right)=0}\sim{\cal V}^{-4},
\end{eqnarray}
yields:
\begin{equation}
\label{eq:m_A_1}
m_{{\cal A}_1}\sim\sqrt{{\cal V}}m_{3/2}.
\end{equation}
Due to the logarithmic derivatives in the definition of the open moduli masses, in fact, one can show that there is a universality in the open moduli masses and that $m_{{\cal A}_I}\sim\sqrt{{\cal V}}m_{3/2}$.

We now show that the universality in the trilinear $A$-couplings that was seen in the case of the $D3$-brane position moduli and a single Wilson line modulus in \cite{D3_D7_Misra_Shukla}, is preserved even for the current four-Wilson-line moduli setup.

Using,
\begin{equation}
\label{ew:A}
A_{{\cal I}{\cal J}{\cal K}}=F^m\left(\partial_mK + \partial_m ln Y_{{\cal I}{\cal J}{\cal K}}
+ \partial_m ln\left(K_{{\cal I}\bar{\cal I}}K_{{\cal J}\bar{\cal J}}K_{{\cal K}\bar{\cal K}}\right)\right),
\end{equation}
and noting:
\begin{eqnarray}
\label{eq:A_constituents}
& & F^m\partial_mK \sim {\cal V}^{-\frac{35}{18}},\nonumber\\
& & F^m\partial_m ln Y_{{\cal Z}_i{\cal Z}_j{\cal Z}_k}\Biggr|_{sin \left( i n k\cdot(G - \bar{G})g_s + m k\cdot (G + \bar{G}) \right)=0}\sim{\cal V}^{-\frac{35}{36}},\nonumber\\
& & F^m \partial_m ln\left(K_{{\cal Z}_i\bar{\cal Z}_i}K_{{\cal Z}_j\bar{\cal Z}_j}K_{{\cal Z}_k\bar{\cal Z}_k}\right)\Biggr|_{sin \left( i n k\cdot(G - \bar{G})g_s + m k\cdot (G + \bar{G}) \right)=0}\sim {\cal V}^{-\frac{4}{3}},\nonumber\\
& & F^m\partial_m ln K_{{\cal A}_1\bar{\cal A}_1} = F^m\partial_m ln K_{{\cal A}_2\bar{\cal A}_2} = F^m\partial_m ln K_{{\cal A}_3\bar{\cal A}_3} = {\cal V}^{-2},
\end{eqnarray}
one sees that one obtains a universal trilinear $A$-coupling:
\begin{equation}
\label{eq:A-univ}
A_{{\cal I}{\cal J}{\cal K}}\sim{\cal V}^{\frac{37}{36}}m_{3/2}\sim\hat{\mu}_{{\cal Z}_1{\cal Z}_2}.
\end{equation}

One can show that
\begin{equation}
\label{eq:muhat_Z1Z2}
\hat{\mu}_{{\cal Z}_1{\cal Z}_2}=\frac{e^{\frac{K}{2}}\mu_{{\cal Z}_1{\cal Z}_2}}{\sqrt{K_{{\cal Z}_1\bar{\cal Z}_1}K_{{\cal Z}_2\bar{\cal Z}_2}}}\sim{\cal V}^{\frac{19}{18}}m_{3/2}.
\end{equation}
 Further,
\begin{eqnarray}
\label{eq:muhatB_Z1Z2}
& & \left(\hat{\mu}B\right)_{{\cal Z}_1{\cal Z}_2}=\frac{e^{-i arg(W)+\frac{K}{2}}}{\sqrt{K_{{\cal Z}_1\bar{\cal Z}_1}K_{{\cal Z}_2\bar{\cal Z}_2}}}F^m\left(\partial_mK \mu_{{\cal Z}_1{\cal Z}_2} + \partial_m\mu_{{\cal Z}_1{\cal Z}_2} - \mu_{{\cal Z}_1{\cal Z}_2}\partial_m ln\left(K_{{\cal Z}_1\bar{\cal Z}_1}K_{{\cal Z}_2\bar{\cal Z}_2}\right)\right)\nonumber\\
& & \sim\hat{\mu}_{{\cal Z}_1{\cal Z}_2}\left(F^m\partial_mK+F^{\sigma_S} - F^m\partial_m ln\left(K_{{\cal Z}_1\bar{\cal Z}_1}K_{{\cal Z}_2\bar{\cal Z}_2}\right)\right)\nonumber\\
& & \sim{\cal V}^{\frac{19}{18}+\frac{37}{36}}m^2_{3/2}\sim\hat{\mu}^2_{{\cal Z}_1{\cal Z}_2},
\end{eqnarray}
an observation which was very useful in obtaining a light Higgs of mass $125GeV$ for a two Wilson line moduli set up of \cite{Dhuria+Misra_mu_Split SUSY} - a conclusion which is still true using the same as shown below.

The Higgs masses after soft supersymmetry breaking is given by $(m_{Z_i}^{2}+\hat{\mu}_{Z_i}^{2})^{1/2}$ (where $m_z$'s correspond to  mobile $D3$- Brane position moduli masses (to be identified with soft Higgs scalar mass parameter)) and the Higgsino mass is given by  $\hat{\mu}_{Z_i}$. Had the supersymmetry been unbroken, Higgs(sino) masses would have had been degenerate with coefficient $\hat{\mu}_{Z_i}$. Nevertheless we have defined SUSY breaking but we are still justified to use RG flow equation' solutions because $\hat{\mu}_{Z_i}>>m_{Z_i}$.
 However, due to lack of universality in moduli masses but universality in trilinear $A_{ijk}$ couplings, we need to use solution of RG flow equation for moduli masses as given in \cite{Nath+Arnowitt}.
\begin{equation}
\label{eq:heavy_H_I}
m_{{\cal Z}_{1}}^{2}(t)=m_{o}^{2}(1+\delta_1)+m_{1/2}^{2} g(t)+\frac{3}{5}S_0p,
\end{equation}
where
\begin{eqnarray}
\label{eq:S_0_def}
& & S_0=Tr(Ym^2)=m_{{\cal Z}_2}^2-m_{{\cal Z}_1}^2+\sum_{i=1}^{n_g}(m_{\tilde q_L}^2-2
m_{\tilde u_R}^2 +m_{\tilde d_R}^2 - m_{\tilde l_L}^2 + m_{\tilde e_R}^2)
\end{eqnarray}
in which all the masses are at the string scale and
 $n_g$ is the number of generations. $p$ is defined by
$p=\frac{5}{66}[1-(\frac{\tilde\alpha_1(t)}
{\tilde\alpha_1(M_s)})]$
where  $\tilde\alpha_1\equiv g_1^2/(4\pi)^2$ and  $g_1$ is the
$U(1)_Y$ gauge coupling constant. Further,
\begin{equation}
\label{eq:light_H_I}
 m_{{\cal Z}_{2}}^{2}(t) = m_0^2\Delta_{{\cal Z}_2} + m_{1/2}^{2}e(t) + A_{o}m_{1/2}f(t) + m_{o}^{2}h(t) -k(t)A_{o}^{2} - \frac{3}{5}S_0p
\end{equation}
where $\Delta_{{\cal Z}_2}$ is given by
\begin{eqnarray}
\label{eq:Delta_def}
\Delta_{{\cal Z}_2}=\frac{(D_0-1)}{2}(\delta_2+\delta_3+\delta_4)+ \delta_2;  D_0=1-6 {\cal Y}_t \frac{F(t)}{E(t)}
\end{eqnarray}
Here ${\cal Y}_t\equiv\hat{Y}_t^2(M_s)/(4\pi)^2$ where $\hat{Y}_t(M_s)$ is the physical top Yukawa coupling at the string scale which following \cite{Ibanez_et_al} will be set to 0.08, and
\begin{equation}
\label{eq:E_def}
E(t)=(1+\beta_3t)^{\frac{16}{3b_3}}(1+\beta_2t)^{\frac{3}{b_2}}
(1+\beta_1t)^{\frac{13}{9b_1}}
\end{equation}
In equation (\ref{eq:E_def})  $\beta_i\equiv\alpha_i(M_s)b_i/4\pi$
($\alpha_1=(5/3)\alpha_Y$), $b_i$ are
the one loop
beta function coefficients defined  by $(b_1,b_2,b_3)=
(33/5,1,-3)$,
and $F(t)=\int_0^t E(t)dt$.

In the dilute flux approximation,
$g_1^2(M_S)=g_2^2(M_S)=g_3^2(M_S)$. To ensure $E(t)\in\bf{R}$, the $SU(3)$-valued $1+\beta_3t>0$, which for $t=57$ implies that $g_3^2(M_s)<\frac{(4\pi)^2}{3\times57}\sim{\cal O}(1)$. Hence,  $g_3^2(M_s)=0.4$ is what we will be using (See \cite{Dhuria+Misra_mu_Split SUSY}).From (\ref{eq:heavy_H_I}), the previous results in this appendix, one sees that:
\begin{equation}
\label{eq:heavy_H_II}
\hskip-0.1in m_{{\cal Z}_1}^2(M_{EW})\sim m_{{\cal Z}_1}^2(M_s)+{(0.39)}{\cal V}^{\frac{4}{3}}m_{3/2}^2+\frac{1}{22}\times\frac{19\pi}{100}\times S_0,
\end{equation}
and
\begin{equation}
\label{eq:light_H_II}
m_{{\cal Z}_2}^2(M_{EW})\sim  {(0.32)}{\cal V}^{\frac{4}{3}}m_{3/2}^2+{(-0.03)}n^s\hat{\mu}_{{\cal Z}_1{\cal Z}_2}{\cal V}^{\frac{2}{3}}m_{3/2}+{(0.96)} m_0^2-{(0.01)}(n^s)^2\hat{\mu}_{{\cal Z}_1{\cal Z}_2}-\frac{19\pi}{2200}\times S_0,
\end{equation}
where we used $A_{{\cal Z}_i{\cal Z}_i{\cal Z}_i}\sim n^s\hat{\mu}_{{\cal Z}_1{\cal Z}_2}$ (\ref{eq:A-univ}).
The solution for RG flow equation for $\hat{\mu}^2$ to one loop order is given by \cite{Nath+Arnowitt}:
\begin{eqnarray}
\label{eq:muhat_I}
& & \hat{\mu}^2_{{\cal Z}_i{\cal Z}_i}=-\biggl[m_0^2 C_1+A_0^2 C_2 +m_{\frac{1}{2}}^2C_3+m_{\frac{1}{2}}
A_0C_4-\frac{1}{2}M_Z^2 +\frac{19\pi}{2200}\left(\frac{tan^2\beta+1}{tan^2\beta-1}\right)S_0\biggr],
\end{eqnarray}
wherein
\begin{eqnarray}
\label{eq:defs_muhat_1LRG}
& & C_1=\frac{1}{tan^2\beta-1}(1-\frac{3 D_0-1}{2}tan^2\beta) +
\frac{1}{tan^2\beta-1}\left(\delta_1-\delta_2tan^2\beta-\frac{ D_0-1}{2}(\delta_2 +
\delta_3+\delta_4)tan^2\beta\right);\nonumber\\
 & & C_2=-\frac{tan^2\beta}{tan^2\beta-1}k(t); C_3=-\frac{1}{tan^2\beta-1}\left(g(t)- tan^2\beta e(t)\right);~C_4=-\frac{tan^2\beta}{tan^2\beta-1}f(t),
\end{eqnarray}
and where the functions $e(t),f(t),g(t),k(t)$ are as defined
in the appendix of \cite{Dhuria+Misra_mu_Split SUSY}.  In the large $tan\beta$ (but less than 50)-limit and assuming $\delta_1=\delta_2=0$, one sees that:
\begin{equation}
\label{eq:muhat_II}
\hat{\mu}_{{\cal Z}_1{\cal Z}_2}^2\sim-\biggl[-m_0^2-{(0.01)}(n^s)^2\hat{\mu}_{{\cal Z}_1{\cal Z}_2}^2+{(0.32)}{\cal V}^{\frac{4}{3}}m_{3/2}^2-1/2 M_{EW}^2+{(0.03)}{\cal V}^{\frac{2}{3}}n^s\hat{\mu}_{{\cal Z}_1{\cal Z}_2}m_{3/2}+\frac{19\pi}{2200}S_0\biggr].
\end{equation}
From (\ref{eq:light_H_II}) and (\ref{eq:muhat_II}) one therefore sees that the mass-squared of one of the two Higgs doublets, $m_{H_2}^2$, at the $EW$ scale is given by:
\begin{equation}
\label{eq:muhat_III}
\hskip -0.3in m^2_{H_2}=m^2_{{\cal Z}_2}+\hat{\mu}^2_{{\cal Z}_i{\cal Z}_i}= 2m^2_{0}-{(0.06)}{\cal V}^{\frac{2}{3}}n^s\hat{\mu}_{{\cal Z}_1{\cal Z}_2}m_{3/2}+\frac{1}{2}M_{EW}^2-\frac{19\pi}{1100}S_0.
\end{equation}
From \cite{D3_D7_Misra_Shukla}, we notice:
\begin{equation}
\label{eq:mu_m_Z_sq}
{\cal V}^{\frac{2}{3}}\hat{\mu}_{{\cal Z}_1{\cal Z}_2}m_{3/2}\sim m_0^2,
\end{equation}
using which in (\ref{eq:muhat_III}), one sees that for an ${\cal O}(1)\ n^s$,
\begin{equation}
\label{eq:muhat_IV}
m^2_{H_2}(M_{EW})\sim  1.94 m_0^2 + \frac{1}{2}M_{EW}^2-\frac{19\pi}{1100}S_0
\end{equation}
We have assumed at $ m^2_{{\cal Z}_2}(M_s)\sim m^2_0$ (implying $\delta_2=0$ but $ \delta_{1,3,4}\neq0$).   Further,
\begin{eqnarray}
\label{eq:heavy_Higgs}
& & {\hskip -0.5in}m^2_{H_1}(M_{EW})=\left(m^2_{{\cal Z}_1}+\hat{\mu}_{{\cal Z}_1{\cal Z}_2}^2\right)(M_{EW})
 \sim m^2_{{\cal Z}_1}(M_s)(1+ \delta_1)+\frac{1}{2}M_{EW}^2 + m_0^2 -{(0.03)}{\cal V}^{\frac{2}{3}}n^s\hat{\mu}_{{\cal Z}_1{\cal Z}_2}m_{3/2}\nonumber\\
  & & + {(0.01)}(n^s)^2\hat{\mu}_{{\cal Z}_1{\cal Z}_2}^2 \sim (1.97+ \delta_1) m^2_0 +\frac{1}{2}M_{EW}^2 + {(0.01)}(n^s)^2\hat{\mu}_{{\cal Z}_1{\cal Z}_2}^2
\end{eqnarray}
 By assuming $\left(\hat{\mu}B\right)_{{\cal Z}_1{\cal Z}_2}\sim\hat{\mu}^2_{{\cal Z}_1{\cal Z}_2}$ - see (\ref{eq:muhatB_Z1Z2}) -  to be valid at the string and EW scales, the Higgs mass matrix at the $EW$-scale can thus be expressed as:
\begin{eqnarray}
\label{eq:Higss_mass_matrix}
& & \left(\begin{array}{cc}
m^2_{H_1} & \hat{\mu}B\\
\hat{\mu}B & m^2_{H_2}\end{array}\right) \sim\left(\begin{array}{cc}
m^2_{H_1} & \xi\hat{\mu}^2\\
\xi\hat{\mu}^2 & m^2_{H_2}
\end{array}\right),
\end{eqnarray}
$\xi$ being an appropriately chosen ${\cal O}(1)$ constant - see (\ref{eq:evs_3}). The eigenvalues are given by:
\begin{eqnarray}
\label{eq:eigenvalues}
& & \frac{1}{2}\biggl(m^2_{H_1}+m^2_{H_2}\pm\sqrt{\left(m^2_{H_1}-m^2_{H_1}\right)^2+4\xi^2\hat{\mu}^4}\biggr).
\end{eqnarray}
As (for ${\cal O}(1)\ n^s$)
\begin{eqnarray}
\label{eq:evs_1}
& & m^2_{H_1}+m^2_{H_2}\sim (3.91 +\delta_1) m_0^2-0.06S_0+... \nonumber\\
& & m^2_{H_1}-m^2_{H_2}\sim (0.03+ \delta_1) m_0^2+0.06S_0+...,\nonumber\\
& & \hat{\mu}_{{\cal Z}_1{\cal Z}_2}^2\sim 0.97 m_0^2 -0.03S_0+...
\end{eqnarray}
one sees that the eigenvalues are:
\begin{eqnarray}
\label{eq:evs_2}
& ( 3.91 + \delta_1)  m_0^2 -0.06S_0+.... &\pm\sqrt{\left((0.03+ \delta_1) m_0^2 +0.06S_0+...\right)^2+\xi^2\left(1.94 m_0^2- 0.06S_0\right)^2}.\nonumber\\
& &
\end{eqnarray}
Hence, assuming non universality w.r.t. to both $D3$-brane position moduli masses ($m_{Z_{1,2}}$) and squark/slepton masses, if $S_0$ and $\xi$ are fine tuned as follows:
\begin{equation}
\label{eq:evs_3}
(0.03+ \delta_1) m_0^2\sim -0.06S_0\ {\rm and}\ \xi\sim 2 + \frac{1}{8}\frac{m^2_{EW}}{m_0^2},
\end{equation}
one sees that one obtains one light Higgs doublet (corresponding to the negative sign of the square root)
and one heavy Higgs doublet (corresponding to the positive sign of the square root). Note, however, the squared Higgsino mass parameter $\hat{\mu}_{{\cal Z}_1{\cal Z}_2}$ then turns out to be heavy with a value, at the EW scale of around $0.01{\cal V}m_{3/2}$  i.e to the order of squark/slepton mass squared scale which is possible in case of $\mu$ split SUSY scenario discussed above.  This shows the possibility of realizing  $\mu$ split SUSY scenario in the context of LVS phenomenology named as large volume ``$\mu$-split SUSY" scenario.


\newpage
\begin{thebibliography}{0}
  \bibitem{review-iii} {\it Local D3/D7 $\mu$-Split SUSY, 125 GeV Higgs and Large Volume Ricci-Flat Swiss-Cheese Metrics: A Brief Review}, (Invited Review)  Mod.\ Phys.\ Lett.\ A {\bf 27}, 1230013 (2012)  [arXiv:hep-th/1203.2948 ].
\bibitem{C.L.Bennett} N. Jarosik et al {\it Seven-Year Wilkinson Microwave Anisotropy Probe (WMAP) Observations:Sky Maps, Systematic Errors, and Basic Results},
    Astrophys.J.Suppl. 192 (2011) 14  [arXiv:astro-ph/1001.4744]
\bibitem{GMoreau} G.~Moreau and M.~Chemtob, {\it R-parity violation and the cosmological gravitino
problem}, Phys.Rev. D65 (2002) 024033
[arXiv:hep-ph/0107286]
\bibitem{MBolz}M.~Bolz, A.~Brandenburg, W.~Buchm\"{u}ller, {\it Thermal Production of Gravitinos},
Nucl.Phys.B606:518-544,2001; Erratum-ibid.B790:336-337,2008
[arXiv:hep-ph/0107286]
\bibitem{V.S.Rychkov} Vyacheslav S. Rychkov, Alessandro Strumia, {\it Thermal production of gravitinos}
Phys.Rev.D75:075011,2007
[arXiv:hep-ph/0701104]
\bibitem{R.Rangarajan} Raghavan Rangarajan and Narendra Sahu, {\it Gravitino production in an in?ationary Universe and implications for leptogenesis},  Mod.Phys.Lett.A23:427-436,2008
    [arXiv:hep-ph/0606228]
    \bibitem{Fweng} Fei Wang, Wenyu Wang, Jin Min Yang, {\it Gravitino dark matter from gluino late decay in split supersymmetry}, 	Phys.Rev. D72 (2005) 077701
       [arXiv:hep-ph/0507172]
\bibitem{Jonathan.L.Feng}Jonathan L.~Feng, Shufang Su, Fumihiro Takayama, {\it Supergravity with a Gravitino LSP}, Phys.Rev.D70:075019,2004
    [arXiv:hep-ph/0404231]
\bibitem{D3_D7_Misra_Shukla}A.~Misra, P.~Shukla, {\it Swiss Cheese D3-D7 Soft SUSY Breaking}, Nuclear Physics B 827 (2010) 112 [arXiv:hep-th/0906.4517].
\bibitem{ferm_masses_MS}A.~Misra and P.~Shukla,{\it On 'Light' Fermions and Proton Stability in 'Big Divisor' D3/D7 Swiss Cheese Phenomenology}, Eur. Phys. J. C (2011) 71:1662  [arXiv:hep-th/1007.1157 ].
\bibitem{Dhuria+Misra_mu_Split SUSY} M.~Dhuria and A.~Misra,
{\it Towards Large Volume Big Divisor D3-D7 'mu-Split Supersymmetry' and
  Ricci-Flat Swiss-Cheese Metrics, and Dimension-Six Neutrino Mass Operators},
  Nucl.\ Phys.\  B {\bf 855}, 439 (2012)
  [arXiv:hep-th/1106.5359 ].
   \bibitem{Wess_Bagger}J.~Wess and J.~Bagger,
  {\it Supersymmetry and supergravity},
{\it  Princeton, USA: Univ. Pr. (1992) 259 p}.
\bibitem{Takeshi_Leszek} T.~Nihei, L.~Roszkowski, R.Ruiz de Austri,
  {\it Exact Cross Sections for the Neutralino WIMP Pair-Annihilation},
  JHEP 0203 (2002) 031
  [arXiv:hep-ph/0202009]
\bibitem{takeshi_coannihilation} T.~Nihei, L.~Roszkowski,  R.~Ruiz de Austri
  {\it Exact Cross Sections for the Neutralino-Slepton Coannihilation},
  	JHEP0207:024,2002
  [arXiv:hep-ph/0206266]
  \bibitem{DDF} F.~Denef, M.~R.~Douglas and B.~Florea, {\it Building a better racetrack}, JHEP {\bf 0406}, 034 (2004),  [arXiv:hep-th/0404257].
 \bibitem{Jockers_thesis}H.~Jockers,
 {\it The effective action of D-branes in Calabi-Yau orientifold
  compactifications},
  Fortsch.\ Phys.\  {\bf 53}, 1087 (2005)
  [arXiv:hep-th/0507042].
\bibitem{BBHL} K.~Becker, M.~Becker, M.~Haack and J.~Louis,
{\it Supersymmetry breaking and alpha'-corrections to flux induced potentials}, JHEP {\bf 0206}, (2002) 060
[arXiv:hep-th/0204254].
\bibitem{Grimm}T.~W.~Grimm, {\it Non-Perturbative Corrections and Modularity in N=1 Type IIB Compactifications},
  JHEP {\bf 0710} (2007) 004, [arXiv:hep-th/0705.3253 ]
  \bibitem{Klemm_GV} M.~-x.~Huang, A.~Klemm and S.~Quackenbush,
{\it Topological string theory on compact Calabi-Yau: Modularity and boundary conditions},  Lect.\ Notes Phys.\  {\bf 757}, 45 (2009)  [hep-th/0612125]
  \bibitem{Donaldson_i} V.~Braun, T.~Brelidze, M.~R.~Douglas and B.~A.~Ovrut, {\it Calabi-Yau Metrics for Quotients and Complete Intersections}, JHEP {\bf 0805}, 080 (2008), [arXiv:hep-th/0712.3563 ].
  \bibitem{SYZ}  A.~Strominger, S.~-T.~Yau and E.~Zaslow,
{\it Mirror symmetry is T duality},  Nucl.\ Phys.\ B {\bf 479}, 243 (1996)  [hep-th/9606040].
\bibitem{BBS} K.~Becker, M.~Becker and A.~Strominger,
{\it Five-branes, membranes and nonperturbative string theory},  Nucl.\ Phys.\ B {\bf 456}, 130 (1995)  [hep-th/9507158].
\bibitem{appl_confs_uplift_dasguptaetal} F.~Chen, K.~Dasgupta, P.~Franche, S.~Katz and R.~Tatar,
{\it Supersymmetric Configurations, Geometric Transitions and New Non-Kahler Manifolds},  Nucl.\ Phys.\ B {\bf 852}, 553 (2011)  [arXiv: hep-th/1007.5316].
\bibitem{Kachru+McGreevy_slag}S.~Kachru and J.~McGreevy,
{\it Supersymmetric three cycles and supersymmetry breaking},
  Phys.\ Rev.\ D {\bf 61}, 026001 (2000)
  [hep-th/9908135].
 \bibitem{Grefe_i}
  M.~Grefe, {\it Neutrino signals from gravitino dark matter with broken R-parity},
  DESY-THESIS-2008-043.
\bibitem{Gmoreau} G.~Moreau and M.~Chemtob, {\it R-parity violation and the cosmological gravitino problem,}
  Phys.\ Rev.\ D {\bf 65}, 024033 (2002)
  [hep-ph/0107286].

  \bibitem{kohri_BBN} M.~Kawasaki, K.~Kohri, T.~Moroi
  {\it Big-Bang Nucleosynthesis and Hadronic Decay of Long-Lived Massive Particles}
  	Phys.Rev. D71 (2005) 083502
  [astro-ph/0408426]
\bibitem{Hasenkamp}J.~Hasenkamp, {\it General neutralino NLSP with gravitino dark matter vs. big bang nucleosynthesis},
  DESY-THESIS-2009-016.
  \bibitem{dhuria+misra_EDM} M. Dhuria, A.~Misra, {\it  A Healthy Electron/Neutron EDM in D3/D7 $\mu$-Split SUSY},  [arXiv:hep-ph/1308.3233].
\bibitem{Manuel Toharia}M.~Toharia, James D.~Wells, {\it Gluino decays with heavier scalar superpartners},
  JHEP0602:015,2006, [arXiv:hep-ph/0503175].
\bibitem{Guidice_et_al}P.~Gambino, G.~F.~Giudice, P.~Slavich,{\it Gluino Decays in Split Supersymmetry}, Nucl.\ Phys.\  {\bf B726}, 35-52 (2005).
\bibitem{2comp}H.~K.~Dreiner, H.~E.~Haber and S.~P.~Martin, {\it Two-component spinor techniques and Feynman rules for quantum field theory and supersymmetry}, Phys.\ Rept.\  {\bf 494}, 1 (2010),
  [arXiv:hep-ph/0812.1594 ].
\bibitem{Pass+Velt} G.~Passarino and M.~J.~G.~Veltman,
{\it One Loop Corrections for e+ e- Annihilation Into mu+ mu- in the Weinberg Model},  Nucl.\ Phys.\  B {\bf 160}, 151 (1979).
 \bibitem{Griffiths_particle}D.~Griffiths, {\it Introduction to Elementary Particles}, John Wiley $\&$ Sons Inc (March 15, 1987).
  \bibitem{gravitinomodexp}  S.~M.~Carroll, D.~Z.~Freedman, M.~E.~Ortiz and D.~N.~Page,
{\it Physical states in canonically quantized supergravity},
  Nucl.\ Phys.\  B {\bf 423}, 661 (1994)
  [arXiv:hep-th/9401155].
  \bibitem{Feng_et_al_sleptondecay} J.~L.~Feng, S.~-f.~Su and F.~Takayama,
  {\it SuperWIMP gravitino dark matter from slepton and sneutrino decays},
  Phys.\ Rev.\ D {\bf 70}, 063514 (2004)
  [arXiv:hep-ph/0404198].

  \bibitem{KHLOPOV_LINDE} M.~Yu.~Khlopov, A.~D.~Linde, {\it Is it easy to save gravitino?} Phys. Lett. (1984), V. 138B, PP. 265-268.
  \bibitem{Keith.a.olive} M.~Srednicki, R.~Watkins, Keith A.~Olive
{\it Calculations of Relic Densities in the Early Universe}, Nucl.Phys. B310 (1988) 693.
\bibitem {Jame_D_wells} James D. Wells,
  {\it In *Kane, G.L. (ed.): Perspectives on supersymmetry II* 269-287}
  [arXiv:hep-ph/9708285]

\bibitem{Sparticles_Misra_Shukla}A.~Misra and P.~Shukla,
{\it Soft SUSY breaking parameters and RG running of squark and slepton masses in large volume Swiss Cheese compactifications},  Phys.\ Lett.\ B {\bf 685}, 347 (2010)  [arXiv:hep-th/0909.0087 ].

\bibitem{susy_primer_S.P.Martin} S.~P.~Martin, {\it A Supersymmetry primer},
  In *Kane, G.L. (ed.): Perspectives on supersymmetry II* 1-153
  [hep-ph/9709356].

  \bibitem{LVS}V.~Balasubramanian, P.~Berglund, J.~P.~Conlon and F.~Quevedo,
  {\it Systematics of moduli stabilisation in Calabi-Yau flux compactifications},
  JHEP {\bf 0503}, 007 (2005)  [hep-th/0502058]; A.~Misra and P.~Shukla,
  {\it Moduli stabilization, large-volume dS minimum without D3-bar branes, (non-)supersymmetric black hole attractors and two-parameter Swiss cheese Calabi-Yau's},  Nucl.\ Phys.\ B {\bf 799}, 165 (2008)  [arXiv:0707.0105 [hep-th]].




\bibitem{Nath+Arnowitt}P.~Nath, R.~Arnowitt, {\it Non-universal Soft SUSY Breaking and Dark Matter}, [arXiv:hep-ph/9801259 ]
\bibitem{Ibanez_et_al}L.~E.~Ib\`{a\~{n}}ez, C.~L\'{o}pez and C.~Mu\~{n}oz, {\it The low-energy supersymmetric spectrum according to N = 1 supergravity guts},
  Nuclear Physics B Volume 256, 1985





       \end{thebibliography}
   \end{document}